\documentclass[a4paper,finnish,7pt]{article}

\usepackage[utf8]{inputenc}
\usepackage[T1]{fontenc}
\usepackage{graphicx}
\usepackage{braket}
\usepackage{algorithm}
\usepackage{float}
\usepackage{enumerate}
\usepackage{titlesec}
\titlespacing*{\section}
{2pt}{0.5cm}{0.5cm}

\usepackage{amsmath}
\usepackage{amsfonts,amssymb,amsbsy,amsmath}  
\usepackage{fancyhdr, lastpage}
\usepackage[a]{esvect}
\usepackage[colorlinks=true, urlcolor=black]{hyperref}
\usepackage{float}

\newcommand{\R}{\mathbb{R}}

\usepackage{cancel}

\begin{document}

\title{\bf Top Mass Shift Caused by the Recalibration of Flavor-Dependent Jet Energy Corrections in the D\O\ Lepton+Jets Top Mass Measurement}
\author{Hannu Siikonen}
\maketitle

This paper introduces and applies a method for propagating changes in jet energy corrections into a D\O\ top quark mass measurement.
It will serve as a source for the PRL presenting it's final results, and will be later published in a journal, or as a part of the author's PhD thesis.

Motivation for the present measurement arises from the recalibration of the D\O\ flavor-dependent jet energy corrections, presented in Ref. \cite{ref:tonimaster}.
The recalibration was driven by the observation that the corrections for D\O\ Run~IIb differ notably from those of Run~IIa.
The significant conclusion of the numerical analysis of Ref. \cite{ref:tonimaster} was that the Run~IIb corrections should have resembled more those of Run~IIa.

The recalibrated jet energy corrections have an influence on the reconstructed top quarks through the jets that are produced in top quark decays.
Hence, a change in the flavor-dependent jet corrections implies the need to revise the top mass measurement, performed by D\O\ in the lepton+jets channel \cite{ref:d02015}.
The D\O\ top mass average is almost completely determined by this measurement \cite{ref:d0comb}, and hence it has a direct impact on the top mass world average.

A complete revision of the D\O\ top mass measurement is unlikely to occur.
This paper demonstrates the resulting top mass shift using lighter methodology.

\newpage

\section{The D\O\ lepton+jets Top Mass Measurement}

The lepton+jets (l+jets) collision events have a distinct topology: a t-tbar pair decays almost exclusively producing two b quarks and two W bosons.
One of the W bosons decays into two light quarks (flavors udsc) and the other one into a charged lepton and a neutrino.
Each of the two light quarks and the two b quarks are expected to produce a jet.
Separate flavor dependent jet energy corrections ($F_{\text{corr}}$) are applied for light quark jets and b jets.

On the Matrix Element (ME) level the l+jets event includes four resonances: two W bosons and two top quarks.
The D\O\ l+jets top mass measurement relies on the ME method, where the top quark mass is evaluated on the ME level.
In this way, the method makes use of the information from the mass resonances.
In this section, the ME method is analyzed in detail, based on the D\O\ paper of Ref.~\cite{ref:d02015}.

\subsection{The Matrix Element Method}

The ME method makes a connection from the measured detector data down to the ME-level, from where the top quark mass is extracted.
It assumes that the measured momentum of the charged lepton is close to that at the ME-level, while the neutrino momentum is determined indirectly.
The transverse component of the neutrino momentum is determined from the transverse momentum of the t-tbar system.
In contrast, the $z$-component is found by employing the leptonic W resonance, and the W boson mass.

As the lepton momenta are thus determined, significant freedom is left in the non-trivial connection between quarks and jets.
The energies of reconstructed jets ($E_j$) are mapped to the corresponding ME-level quarks ($E_q$) using so-called Transfer Functions (TF's).
These operate as probability distributions projected from $E_j$ to $E_q$.
We have chosen to use the subscripts $q$ and $j$ to increase clarity, in contrast to the corresponding D\O\ notation $y$ and $x$.

As the masses of quarks are known, the quark four-momentum is determined solely by the energy and the direction of the quark.
The D\O\ measurement assumes that the quark directions are equal to those of the corresponding jets.
Hence an energy-based TF is sufficient for determining the quark properties.

The TF's are calibrated using Monte Carlo (MC) simulations.
We denote the functional form of a TF by $\text{TF}_{\text{era}}(E_j,E_q)$, where 
\begin{equation}
\int_\R~\text{TF}_{\text{era}}(E_j,E_q)~dE_j~=~1
\end{equation}
and $\text{era} \in \{\text{IIa,IIb1,IIb2,IIb34}\}$ refers to the D\O\ Run II eras.
It is important to note that the D\O\ jet energy $E_j$ excludes neutrinos and muons.
Hence, the TF's for different types of quarks are quite different, as for instance b-jets are enriched in neutrinos and muons.

The TF's act as weights while connecting the measured final state to the feasible ME-level states.
This limits the allowed phase space of the ME-level states, and permits an integral over them.
Such an integral gives a probabilistic view of the ME-state for each event candidate.
The related event probability ($P_{evt}$) is dependent on the underlying ME-level top mass ($m_t$).

To improve the measurement accuracy, D\O\ introduces a slack-parameter $K_{JES}$, which is used for an in-situ jet energy calibration:
\begin{equation} \label{eq:kjesdivision}
 E_j' = \frac{E_j}{K_{JES}}.
\end{equation}
The calibration is applied by replacing the measured jet energy $E_j$ with the interpreted jet energy $E_j'$.
This is performed exclusively in the top mass measurement, and does not affect the calibration of the TF's.
As Eq.~\eqref{eq:kjesdivision} shifts the difference between the jet and quark energies, the event probability $P_{evt}$ is dependent on $K_{JES}$ in addition to $m_t$.

The measured collision dataset corresponds to a set of $P_{evt}$ values.
A combined likelihood function is constructed from these in the $m_t - K_{\text{JES}}$ plane:
\begin{equation} \label{eq:mainlikeli}
 \mathcal{L} (m_t,K_{\text{JES}}) \propto \prod_{i \in \text{Data}} P_{evt}^{ i} (m_t,K_{\text{JES}}).
\end{equation}
D\O\ uses this likelihood as an unnormalized probability distribution.
It attains approximately a 2D-Gaussian form as a function of $m_t$ and $K_{\text{JES}}$.

In the likelihood $\mathcal{L} (m_t,K_{\text{JES}})$ there are two parameters, but the underlying ME has three degrees of freedom: two top resonances and the hadronic W boson resonance.
The leptonic W resonance is completely consumed by neutrino reconstruction.
The top resonances are dependent on $m_t$, but also on $K_{\text{JES}}$.
On the other hand, the hadronic W resonance depends only on $K_{\text{JES}}$.

The global maximum likelihood value of $\mathcal{L} (m_t,K_{\text{JES}})$ must be such that the hadronic W boson data agrees best with the W mass world average.
On the quark level, this argument implies that the average hadronic diquark mass equals to W boson mass:
\begin{equation} \label{eq:meanmw}
 \left<m_{q_1q_2}\right> =  m_W.
\end{equation}
As the hadronic W resonance depends only on $K_{\text{JES}}$, this means that the maximum likelihood value of $K_{\text{JES}}$ and the hadronic W resonance are directly connected.
Hence the optimal value of $K_{\text{JES}}$ can be fixed without referring to the top mass resonances.

At a given value of $K_{\text{JES}}$, the relationship between the likelihood function and the two top mass resonances is purely determined by $m_t$.
If separate values $m_t^{had}$ and $m_t^{lep}$ were used for the hadronic and leptonic resonances in the ME,
it is likely that the global maximum likelihood value would be reached at separate $m_t$ values: $m_t^{had} \neq m_t^{lep}$.
Therefore, the full $m_t$ dependence of the likelihood function is determined by the combination of the two resonances.

Assuming that the 2D-Gaussian hypothesis is accurate, the probability distribution can be projected onto the $m_t$ ($K_{\text{JES}}$) axis to obtain a 1D distribution.
It retains the correct maximum likelihood position of the full 2D likelihood.
For $m_t$, the projected likelihood is
\begin{equation} \label{eq:mtlikeli}
 \mathcal{L} (m_t) = \int_\mathbb{R} dK_{\text{JES}} \mathcal{L} (m_t,K_{\text{JES}}).
\end{equation}
Treating the likelihood $\mathcal{L} (m_t)$ as an unscaled probability, we find the basic formula for mass extraction at the D\O:
\begin{equation} \label{eq:mtfit}
 m_t^{fit} = \left< m_t \right>_0 = \frac{\int_\mathbb{R} dm_t m_t \mathcal{L} (m_t)}{\int_\mathbb{R} dm_t \mathcal{L} (m_t)}.
\end{equation}
The Eqs. (\ref{eq:mtlikeli},\ref{eq:mtfit}) are valid also for $K_{\text{JES}}$ by interchanging it with $m_t$.
Thus, the ME method combines the theoretical ME with the information extracted from MC simulations.
The latter is completely compressed into the TF's.

To perform a numerical integration of $\mathcal{L}(m_t,K_{\text{JES}})$, a grid of $(m_t,K_{\text{JES}})$ values is used.
If the grid is sufficiently wide and tight, the whole $\mathbb{R}^2$ is effectively covered by the integral.
Each event-wise probability $P_{evt}^{ i}$ needs to be evaluated at each grid point.

\subsection{Calibration of the Matrix Element Method}

Like most top mass measurements, the D\O\ measurement utilizes MC simulations and the underlying generator level top mass $m_t^{gen}$.
Eq.~\eqref{eq:mtfit} is based on the assumption that the theoretical ME integral with a top quark pole mass agrees directly with the MC-based TF's.
D\O\ studies this assumption by assuming that $m_t^{gen}$ is equal to the ME-level pole mass.
Thus, the $m_t^{fit}$ value extracted from simulation results is expected to be equal to $m_t^{gen}$.
To find the true connection between $m_t^{fit}$ and $m_t^{gen}$, D\O\ employs MC pseudo-experiments, in which multiple values of $m_t^{gen}$ are used.
The connection is parametrized as a linear mapping:
\begin{equation} \label{eq:gunction}
 g : m_t^{gen} \rightarrow m_t^{fit}.
\end{equation}
D\O\ denotes the mapping parameters as
\begin{equation} \label{eq:mtfitappro}
 m_t^{fit} = g(m_t^{gen}) = \mathcal{S} \left( m_t^{gen} - 172.5 \right) + \mathcal{O} + 172.5,
\end{equation}
where $\mathcal{S}$ stands for scale and $\mathcal{O}$ for offset.
Typically $\mathcal{S}$ differs slightly from unity and $\mathcal{O}$ is a small non-zero number.
This shows the necessity of the calibration.

The D\O\ $m_t$ extraction procedure goes as follows: the likelihood function is calculated from the measured data, and from this $m_t^{fit}$ is computed.
Using the inverse mapping $g^{-1}$ of Eq.~\eqref{eq:gunction}, $m_t^{fit}$ can be mapped to a value analogous to $m_t^{gen}$.
For clarity, this result for data is denoted by $m_t^{\text{calib}}$ instead of $m_t^{gen}$:
\begin{equation} \label{eq:mtgenfitfit}
 m_t^{\text{calib}} = g^{-1}(m_t^{fit}).
\end{equation}
Thus, the D\O\ analysis chain assumes that $m_t^{gen}$ and $m_t^{\text{calib}}$ correspond to the top pole mass.
In reality, a subtle difference appears to exist between the pole mass and $m_t^{gen}$~\cite{ref:moritz}.

For the calibration of $K_{\text{JES}}$, similar steps are taken.
The statement of Eq.~\eqref{eq:kjesdivision} can be multiplied by a \emph{generator level} energy scale, $K_{\text{JES}}^{gen}$.
Ideally, $K_{\text{JES}}^{fit}$ should be equal to $K_{\text{JES}}^{gen}$ to keep the hadronic W resonance at a constant position.
By applying multiple $K_{\text{JES}}^{gen}$ gen values in the MC pseudo-experiments, a linear mapping similar to that in Eq.~\eqref{eq:gunction} is found:
\begin{equation} \label{eq:function}
 f : K_{\text{JES}}^{gen} \rightarrow K_{\text{JES}}^{fit}.
\end{equation}
For this function, the following parametrization is used:
\begin{equation} \label{eq:calib}
 K_{\text{JES}}^{\text{fit}} = f(K_{\text{JES}}^{gen}) = 1 + \mathcal{S}_K (K_{\text{JES}}^{gen}-1) + \mathcal{O}_K.
\end{equation}
The calibration procedure of $K_{\text{JES}}$ has no practical impact on the measured $m_t$ value, as it is simply a change of variables on the $K_{\text{JES}}$ axis.
Nevertheless, D\O\ does not give a complete picture of the effects that are compensated by this mapping.
One could argue that the main importance is in reducing any disagreement between the MC simulations and the ME integral.
In the D\O\ measurements, $f(1)$ can differ notably from the value 1, which implies that the mapping $f$ may be used to compensate errors in the shifts imposed by the TF's.

It is not necessary to calculate $m_t^{\text{calib}}$ and $K_{\text{JES}}^{\text{calib}}$ manually using the inverse functions of Eqs.~(\ref{eq:gunction},\ref{eq:function}).
It can be shown (see Appendix~\ref{app:likeli}) that by replacing $m_t$ by $g(m_t)$ and $K_{\text{JES}}$ by $f(K_{\text{JES}})$ in the likelihood functions,
the fits yield automatically correct results.

However, the D\O\ papers show some ambiguity about this subject, as is further discussed in Appendix~\ref{app:likeli}.
D\O\ indicates that instead of the functions $f$ and $g$, their inverse functions are inserted into the likelihood.
The author of this paper has been in correspondence with D\O\ authors about the subject, but the investigation has been inconclusive.
Correcting such a misplacement would lower the D\O\ top mass value by 0.1~GeV and make some $\chi^2$ values more sensible.
As this is a matter of fine-tuning, we will not concentrate on it outside Appendix~\ref{app:likeli}.

\subsection{Steps Between the Matrix Element and Detector Data}

To understand the D\O\ measurement, we need to understand the relationship between jets and quarks.
The full difference between $E_j$ and $E_q$ in simulation stands as
\begin{equation} \label{eq:exmc}
 E_j = F_{\text{Corr}} \otimes \text{JEC} \otimes \text{Detector Simulation} \otimes \text{Hadronization} \otimes \text{Showering} \otimes E_q,
\end{equation}
where we have taken into account Jet Energy Corrections (JEC) and all simulation steps.
The symbol $\otimes$ underlines that some of the steps are more complicated than simple multiplications, but can be approximated as such.
For data, the corresponding equation reads as
\begin{equation} \label{eq:exdt}
 E_j = \text{JEC} \otimes \text{Detector} \otimes \text{Physics} \otimes E_q.
\end{equation}
In general, the step $\text{Physics} \otimes E_q$ is only partially understood.
To gain further understanding of it, the MC-calibrated TF's are used.
In terms of a TF, the transition between jet and quark energies reads as
\begin{equation} \label{eq:extf}
 E_q = \text{TF} \otimes E_j,
\end{equation}
Eq.~\eqref{eq:extf} can be inverted to produce
\begin{equation} \label{eq:cinversion}
 E_j = \text{TF}^{-1} \otimes E_q.
\end{equation}
The TF's are calibrated purely using MC simulations.
Hence, comparing Eqs.~\eqref{eq:exmc} and \eqref{eq:cinversion}, we have
\begin{equation} \label{eq:tfform}
\text{TF}^{-1} = F_{\text{Corr}} \otimes \text{JEC} \otimes \text{Detector Simulation} \otimes \text{Hadronization} \otimes \text{Showering}.
\end{equation}
Eq.~\eqref{eq:tfform} is valid for both data and MC, as it is the written-out form of the TF's.
The steps within it are interpreted as those of the TF calibration MC sample.
This underlines the fact that for data, we do not certainly know what $E_q$ is.
The analysis is based on the assumption that after all the calibrations, the quark level in data corresponds to that in MC.

In the top mass measurement, jet energies $E_j$ are replaced with $E_j'$ from Eq.~\eqref{eq:kjesdivision} using the calibration from Eq.~\eqref{eq:function},
so that Eq.~\eqref{eq:cinversion} reaches the form
\begin{equation} \label{eq:cinversion2}
 E_j = f\left(K_{\text{JES}}\right) \otimes \text{TF}^{-1} \otimes E_q.
\end{equation}
By utilizing Eq.~\eqref{eq:tfform}, Eq.~\eqref{eq:cinversion2} can be expressed as
\begin{equation} \label{eq:tfintro}
E_j = f\left(K_{\text{JES}}\right) \otimes F_{\text{Corr}} \otimes \text{JEC} \otimes \text{Detector Simulation} \otimes E_j^{gen},
\end{equation}
where we have made the substitution
\begin{equation} \label{eq:ejgen}
E_j^{gen} = \text{Hadronization} \otimes \text{Showering} \otimes E_q.
\end{equation}
$E_j^{gen}$ is the so-called \emph{generator level} jet energy.
By using the same simulation settings as D\O, $E_j^{gen}$ values analogous to those of D\O\ can be produced.

\newpage

\section{Phenomenological Propagation of Changes in $F_{\text{Corr}}$ to Top Mass}

In this section, the impact of a $F_{\text{Corr}}$ recalibration on the l+jets top mass measurement is shown on a phenomenological level.
The tag D\O\ is used for the original $F_{\text{Corr}}$ values, and the recalibrated ones are expressed without a tag.

\subsection{$F_{\text{Corr}}$ and the Interpreted Quark Energy}

As was shown in Eq.~\eqref{eq:tfform}, the values of $F_{\text{Corr}}$ are propagated to the corresponding TF's.
According to Eq.~\eqref{eq:cinversion2}, we can write the interpreted $E_q$ for a given fixed measurement $E_j$:
\begin{equation} \label{eq:eypropto}
 E_q \equiv E_q\left(f\left(K_{\text{JES}}\right)\right) = \frac{\text{TF} \otimes E_j}{f\left(K_{\text{JES}}\right)}.
\end{equation}
That is, by fixing $E_j$, $E_q$ becomes a function of $f\left(K_{\text{JES}}\right)$.
Considering Eq.~\eqref{eq:tfform} with the original and the new $F_{\text{Corr}}$ values,
and equating the terms independent of $F_{\text{Corr}}$, we find how $F_{\text{Corr}}$ transforms the TF's
\begin{equation} \label{eq:tfproblem}
F_{\text{Corr}} \otimes \text{TF} = F_{\text{Corr}}^{\text{D\O}} \otimes \text{TF}^{\text{D\O}},
\end{equation}
from which it follows that
\begin{equation} \label{eq:invtfd0}
 \text{TF}^{\text{D\O}} = \left( \frac{F_{\text{Corr}}}{F_{\text{Corr}}^{\text{D\O}}} \right) \otimes \text{TF}.
\end{equation}
Combining Eqs.~(\ref{eq:eypropto},\ref{eq:invtfd0}) gives a relationship between the quark energy interpretations:
\begin{equation} \label{eq:eycorr}
 E_q^{\text{D\O}}
 = \frac{\text{TF}^{\text{D\O}} \otimes E_j}{f^{\text{D\O}}\left(K_{\text{JES}}^{\text{D\O}}\right)}
 = \frac{f\left(K_{\text{JES}}\right)}{f^{\text{D\O}}\left(K_{\text{JES}}^{\text{D\O}}\right)} \otimes \left( \frac{F_{\text{Corr}}}{F_{\text{Corr}}^{\text{D\O}}} \right) \otimes \frac{\text{TF} \otimes E_j}{f\left(K_{\text{JES}}\right)}
 = \frac{f\left(K_{\text{JES}}\right)}{f^{\text{D\O}}\left(K_{\text{JES}}^{\text{D\O}}\right)} \otimes \left( \frac{F_{\text{Corr}}}{F_{\text{Corr}}^{\text{D\O}}} \right) \otimes E_q.
\end{equation}
This indicates that for a fixed jet energy measurement $E_j$, a change in the $F_{\text{Corr}}$ values leads to a change in the interpreted quark energy.
To fully understand this result, it is necessary to understand how the change in $F_{\text{Corr}}$ is reflected in $f\left(K_{\text{JES}}\right)$.

\subsection{$F_{\text{Corr}}$ and $K_{\text{JES}}$ Scaling}

In Eq.~\eqref{eq:meanmw} it was argued that the maximum likelihood solution extracted by D\O\ makes the hadronic diquark system agree in average with the W boson mass.
This mechanism sets the value of $K_{\text{JES}}$, and it can be used to find the transformation law for $f\left(K_{\text{JES}}\right)$.
The invariant mass of the two zero-mass light quarks can be expressed as
\begin{equation} \label{eq:invmass}
 m_{q_1q_2} = \sqrt{2E_{q_1}E_{q_2}(1-\cos\phi_{12})},
\end{equation}
where $E_{q_1}$ and $E_{q_2}$ are the quark energies and $\phi_{12}$ is the opening angle between the quark directions.
Employing Eq.~\eqref{eq:eycorr}, and considering Eq.~\eqref{eq:invmass} for the interpreted quark energies with the D\O\ and the recalibrated $F_{\text{Corr}}$ values, it is found
\begin{equation} \label{eq:massrels}
 m_{q_1q_2}^{\text{D\O}}
 = \frac{f\left(K_{\text{JES}}\right)}{f^{\text{D\O}}\left(K_{\text{JES}}^{\text{D\O}}\right)} \otimes \left( \frac{F_{\text{Corr}}^{lq}}{F_{\text{Corr}}^{lq,\text{D\O}}} \right) \otimes m_{q_1q_2}.
\end{equation}
In this expression, we have fused the average $F_{\text{Corr}}^{lq}$ value of the two light quarks into a single factor.
The following abbreviations for the light quark and b quark $F_{\text{Corr}}$ values were adopted:
\begin{eqnarray}
 F_{\text{Corr}}^{lq} & \equiv & F_{\text{Corr,u/d/s/c quark}} \\
 F_{\text{Corr}}^{b} & \equiv & F_{\text{Corr,b quark}}
\end{eqnarray}
Averaging over Eq.~\eqref{eq:massrels} and applying Eq.~\eqref{eq:meanmw} for both the D\O\ and the recalibrated results, it is found
\begin{equation} \label{eq:wresult}
 m_W = \frac{f\left(K_{\text{JES}}\right)}{f^{\text{D\O}}\left(K_{\text{JES}}^{\text{D\O}}\right)} \otimes \left( \frac{F_{\text{Corr}}^{lq}}{F_{\text{Corr}}^{lq,\text{D\O}}} \right) \otimes m_W.
\end{equation}
Here, the $F_{\text{Corr}}^{lq}$ factors are interpreted as average values.
Canceling out the $m_W$ terms, Eq.~\eqref{eq:wresult} implies
\begin{equation} \label{eq:kjescorr}
 \frac{f\left(K_{\text{JES}}\right)}{f^{\text{D\O}}\left(K_{\text{JES}}^{\text{D\O}}\right)} = 
 \frac{F_{\text{Corr}}^{lq,\text{D\O}}}{F^{lq}_{\text{Corr}}}.
\end{equation}
Hence, the light quark $F_{\text{Corr}}$ values have a scaling effect on the $K_{\text{JES}}$ terms.

\subsection{The Full Impact of $F_{\text{Corr}}$ on Quarks}

Utilizing Eqs.~\eqref{eq:kjescorr} and \eqref{eq:eycorr}, the full impact of a change in $F_{\text{Corr}}$ on the interpreted quark energies can be written:
\begin{equation} \label{eq:thetruth}
 E_q^{\text{D\O}}
 = \left( \frac{F_{\text{Corr}}^{lq,\text{D\O}}}{F^{lq}_{\text{Corr}}} \right) \otimes \left( \frac{F_{\text{Corr}}}{F_{\text{Corr}}^{\text{D\O}}} \right) \otimes E_q.
\end{equation}
The result of Eq.~\eqref{eq:thetruth} can be separately applied for b quarks and light quarks:
\begin{eqnarray}
 E_{lq}^{\text{D\O}} & = & \label{eq:thetruthl} \label{eq:liteq}
 \cancel{ \left( \frac{F_{\text{Corr}}^{lq,\text{D\O}}}{F_{\text{Corr}}^{lq}} \right)} \otimes \cancel{\left( \frac{F_{\text{Corr}}^{lq}}{F_{\text{Corr}}^{lq,\text{D\O}}} \right)} \otimes E_{lq} = E_{lq} \\
 E_{b}^{\text{D\O}} & = & \label{eq:thetruthb} \label{eq:bq}
 \left( \frac{F_{\text{Corr}}^{lq,\text{D\O}}}{F_{\text{Corr}}^{lq}} \right) \otimes \left( \frac{F_{\text{Corr}}^{b}}{F_{\text{Corr}}^{b,\text{D\O}}} \right) \otimes E_{b}.
\end{eqnarray}
Thus, the interpreted light quark energies of a fixed measurement are the same on the average level, even if $F_{\text{Corr}}$ calibration changes.
This is natural, as the invariant mass of the two light quarks is forced to be equal with the W boson mass.
All the changes in the $F_{\text{Corr}}$ values are hence conveyed to the interpretation of b quarks.
In our studies the trend is that $F_{\text{Corr}}^{lq} < F_{\text{Corr}}^{lq,\text{D\O}}$ and $F_{\text{Corr}}^{b} > F_{\text{Corr}}^{b,\text{D\O}}$.
Therefore, both of the coefficients in Eq.~\eqref{eq:thetruthb} make the D\O\ reconstruction of the b quark energy of a corresponding jet energy $E_j$ higher it should be.

The reconstructed b quark energies have a great impact on the top quark measurement for both the hadronic and the leptonic top mass resonances.
On the average, approximately half of the energy of a decaying top quark goes to a b quark.
For a reconstructed top quark, a systematically amplified b quark energy shows as a systematically amplified top quark mass.
Thus, we have a reason to believe that the top mass value produced with the original D\O\ $F_{\text{Corr}}$ calibration was systematically shifted upwards.
For the accurate numerical evaluation of this shift, an involved analysis is necessary.

\newpage

\section{Analysis Planning}

In the previous section it was phenomenologically confirmed that the a recalibration of the $F_{\text{Corr}}$ values leads to a shift in the measured top mass value.
In this section, a method is designed for the numerical evaluation of this shift.

First, we aim to replicate what the D\O\ data looked like before the original D\O\ $F_{\text{Corr}}$ values were applied.
Then, the recalibrated $F_{\text{Corr}}$ values are applied.
As a result, the D\O\ top mass estimate is shifted downwards, as the interpreted b quark energies become lower according to Eq.~\eqref{eq:bq}.

\subsection{A Fully $F_{\text{Corr}}$ Dependent Pseudo-Experiment}

It is not feasible to replicate all the steps of the D\O\ measurement, as given in Eq.~\eqref{eq:tfintro}.
That is, the D\O\ data cannot be replicated on the level of the jet energies $E_j$.
Nevertheless, we will attempt to capture the essential $F_{\text{Corr}}$ features using a more compact analysis chain.
To reach this goal, it is essential to reconstruct jet energies on a level where the same $F_{\text{Corr}}$ behavior is observable, as for the full $E_j$ energies.
The terms independent of $F_{\text{Corr}}$ can then be left out.

By Eq.~\eqref{eq:kjescorr} we know that variations of $F_{\text{Corr}}^{lq}$ are propagated into the $K_{\text{JES}}$ term.
To account for this, the $f\left(K_{\text{JES}}\right)$ term can be factorized into two parts:
\begin{equation} \label{eq:kjesparts}
 f\left(K_{\text{JES}}\right) = f\left(K_{\text{JES}}\right)_0 \otimes K_{\text{JES}}^{\text{Res}}.
\end{equation}
All the $F_{\text{Corr}}$ dependence is pressed into the residual $K_{\text{JES}}^{\text{Res}}$ term, while $f\left(K_{\text{JES}}\right)_0$ holds all the other dependencies.
Combining these developments, Eq.~\eqref{eq:tfintro} transforms to
\begin{equation} \label{eq:tfintro3}
E_j = K_{\text{JES}}^{\text{Res}} \otimes f\left(K_{\text{JES}}\right)_0 \otimes F_{\text{Corr}} \otimes \text{JEC} \otimes \text{Detector Simulation} \otimes E_j^{gen}.
\end{equation}
Here, the term $f\left(K_{\text{JES}}\right)_0 \otimes\text{JEC} \otimes \text{Detector Simulation}$ is in average canceled out.
In other words, the jet energy corrections cancel out the average detector response.
For a single event, this collective term has a smearing effect.
As we do not have access to the D\O\ JEC nor Detector simulation,
and the relationship between $f\left(K_{\text{JES}}\right)_0$ and $K_{\text{JES}}^{\text{Res}}$ is not known,
we are encouraged to make the inversion
\begin{equation} \label{eq:fullcircle}
E_j'' = \left(f\left(K_{\text{JES}}\right)_0 \otimes \text{JEC} \otimes \text{Detector Simulation}\right)^{-1} \otimes E_j = K_{\text{JES}}^{\text{Res}} \otimes F_{\text{Corr}} \otimes E_j^{gen}. 
\end{equation}
The result of Eq.~\eqref{eq:fullcircle} is used in the main analyses.
It is the minimal jet energy definition that takes into account all $F_{\text{Corr}}$-related effects.
Using the $E_j''$-level jet energies instead of the $E_j$ ones is sufficient, as we are interested only in the $F_{\text{Corr}}$-dependent $m_t$ shift.
To capture the full $F_{\text{Corr}}$-dependence in $E_j''$, it was crucial not to cancel out the $K_{\text{JES}}$ dependence completely.
Through $K_{\text{JES}}^{\text{Res}}$, any changes in the $F_{\text{Corr}}^{lq}$ values can be compensated in the desired way.

\subsection{Resonance Position Estimators}

The D\O\ likelihood function of Eq.~\eqref{eq:mainlikeli} focuses on mass resonances.
To emulate its behavior, we will need to evaluate the positions of the invariant mass resonances on the level of $E_j''$ energies.
For this purpose, \emph{resonance position estimators} are designed.
For the three mass resonances of interest, these are denoted by $\hat{m}_W$, $\hat{m}_t^{had}$ and $\hat{m}_t^{lep}$.

The most complicated resonance position estimator is an intricate fit on the resonance mass distribution (\textbf{fit}).
Generally, the fits are made using complicated generalizations of Gaussian functions - e.g. Voigtian functions - and the resonance position is evaluated from the fit parameters.
Moreover, estimators that pick the maximum (\textbf{max}) and median (\textbf{med}) value from the mass histogram are used.
In addition, two different integral-based estimators are used.
One takes the mean value based on the full $x$-range of the mass histogram (\textbf{ave}).
The other uses a more limited mass window for averaging (\textbf{itg}).

In this paper, the top mass shift analyses are performed separately with all the five estimators.
The same estimator is always used for estimating all three resonance positions.
The final result considers a combination of all the estimators, taking into account their relative errors.

\subsection{Analysis Workflow}

In Eq.~\eqref{eq:fullcircle} $K_{\text{JES}}^{\text{Res}}$ refers to the corresponding maximum likelihood value.
Nonetheless, while evaluating the $\hat{m}$ values, it is convenient to interpret $K_{\text{JES}}^{\text{Res}}$ as a free parameter.
This allows fixing the maximum likelihood value at a later point.
In a pseudo-experiment based on a simulation, also the value of $m_t^{gen}$ is a free parameter.
By evaluating the $\hat{m}$ values on a grid of $m_t^{gen}$ and $K_{\text{JES}}^{\text{Res}}$ values, dependencies between the $\hat{m}$'s and the underlying parameters are found.
We expect that $\hat{m}_W$ is sensitive to $K_{\text{JES}}^{\text{Res}}$, and that the $\hat{m}_t$'s are sensitive to both $K_{\text{JES}}^{\text{Res}}$ and $m_t^{gen}$.
This is a direct analogy to the D\O\ likelihood function of Eq.~\eqref{eq:mainlikeli}.

To make numerical analysis more convenient, $\hat{m}_W$ is translated into a $K_{\text{JES}}$ estimator:
\begin{equation} \label{eq:hatdef}
 \hat{K}_{\text{JES}} = \frac{\hat{m}_{W}}{80.4~\text{GeV}}.
\end{equation}
It is sensible to expect that the dependence between $\hat{K}_{\text{JES}}$ and $K_{\text{JES}}^{\text{Res}}$ is linear.
This can be expressed in the functional form
\begin{equation} \label{eq:linkdef}
 \hat{K}_{\text{JES}} \left( K_{\text{JES}}^{\text{Res}} \right) = \frac{\hat{m}_{W} \left( K_{\text{JES}}^{\text{Res}} \right)}{80.4~\text{GeV}} = q_0 + q_1 K_{\text{JES}}^{\text{Res}}.
\end{equation}
In analogy to Eq.~\eqref{eq:linkdef}, we expect that the $\hat{m}_t$'s depend linearly on $m_t^{gen}$.
Moreover, the $K_{\text{JES}}^{\text{Res}}$ dependence is estimated using a linear term.
The resulting 2D function reads as
\begin{equation} \label{eq:linearity}
 \hat{m}_t \left( m_t^{gen}, K_{\text{JES}}^{\text{Res}}\right) = p_0 + p_1 \times m_t^{gen} + p_2 \times K_{\text{JES}}^{\text{Res}}.
\end{equation}
The parameters $q_i$ and $p_i$ of Eqs.~(\ref{eq:linkdef},\ref{eq:linearity}) are calibrated on a grid of $m_t^{gen}$ and $K_{\text{JES}}^{\text{Res}}$ values.
For $\hat{K}_{\text{JES}}$ a degeneracy is present, as the position of the W boson resonance does not depend on $m_t^{gen}$.
This degeneracy improves the $K_{\text{JES}}^{\text{Res}}$ fit quality, as all the data points are used in the fit.

To proceed with the measurement, the maximum likelihood value of $K_{\text{JES}}^{\text{Res}}$ is determined.
In Eq.~\eqref{eq:meanmw} we concluded that the maximum likelihood value of $K_{\text{JES}}$ is determined by forcing the average invariant diquark mass to be equal with $m_W$.
Moreover, we known by Eq.~\eqref{eq:liteq} that for light quarks the interpreted $E_q$ of a measurement $E_j$ is constant, independent of the $F_{\text{Corr}}$ values.
That is, the relationship between the dijet and diquark resonance positions is constant.
The same argument can be made after swapping the $E_j$ energies to $E_j''$ energies: the maximum likelihood value of $\hat{m}_W$ should be constant, independent of the $F_{\text{Corr}}$ values.
The argument may be written as
\begin{equation} \label{eq:hatconst}
 \hat{m}_W^{ML} = \mathcal{A} \times 80.4~\text{GeV},
\end{equation}
where $\mathcal{A}$ is a scale-parameter close to unity.
As the $E_j''$ energies are abstract, there is no distinct correct choice for $\mathcal{A}$.
This contrasts to the $E_j$ energies, where a physically motivated value for $\mathcal{A}$ exists.

By employing Eqs.~(\ref{eq:linkdef},\ref{eq:hatconst}), the maximum likelihood value of $K_{\text{JES}}^{\text{Res}}$ is found:
\begin{equation} \label{eq:kjesres}
 K_{\text{JES}}^{\text{Res},ML} = \frac{\frac{\mathcal{A} \times 80.4~\text{GeV}}{80.4~\text{GeV}} - q_0}{q_1}
 = \frac{\mathcal{A} - q_0}{q_1}.
\end{equation}
Placing this value into Eq.~\eqref{eq:linearity} leaves a pure linear dependence between $\hat{m}_t$ and $m_t^{gen}$.
In this fashion, the relationship between the $m_t^{gen}$ and $\hat{m}_t^{had}$ or $\hat{m}_t^{lep}$  can be evaluated using both the D\O\ $F_{\text{Corr}}$ values and the recalibrated ones.

In the analysis chain we first produce a $\hat{m}_t$ value by employing the D\O\ $F_{\textrm{corr}}$ and $m_t^{\text{calib}}$ values.
The former are used for scaling jets, and $m_t^{gen}$ is set to be equal with the latter.
This $\hat{m}_t$ is re-interpreted using the new $F_{\textrm{corr}}$ values.
The resulting $m_t^{gen}$ is interpreted as the $m_t^{\text{calib}}$ value that D\O\ should have found.
The procedure must be performed for both the hadronic and leptonic $\hat{m}_t$'s, and the total $m_t^{\text{calib}}$ shift is found through a combination of the results.

Within the D\O\ analysis, it is possible that the hadronic and leptonic resonances prefer different $m_t^{\text{calib}}$ values.
The likelihood function of Eq.~\eqref{eq:mainlikeli} makes a compromise between these two information sources.
The relative importance of the two $m_t$ resonances in this process is unknown, and we express it using the parameter $\alpha \in [0,1]$:
\begin{equation} \label{eq:alpharel}
 m_t^{\text{calib}} = \alpha \times m_{t,had}^{\text{calib}} + \left( 1 - \alpha \right) \times m_{t,lep}^{\text{calib}}
\end{equation}

The $m_t^{\text{calib}}$ shifts need to be evaluated separately for the electron and muon measurement channels.
A~priori these should be almost identical, but in the D\O\ measurements, the phase spaces and $m_t^{\text{calib}}$ values differ.
Moreover, separate $m_t^{\text{calib}}$ and $F_{\textrm{corr}}$ values are employed for all the four Run~II eras: IIa, IIb1, IIb2 and IIb34.
The electron and muon channels need to be considered in all the four Run~II eras.
Hence the combined $m_t^{\text{calib}}$ consists of a total of 8 measurements, each of which involves the shifts of the hadronic and the leptonic $m_t$ resonances.

\subsection{Equivalence to the Phenomenological Approach}

To show that the present approach leads to correct results, it is necessary to understand Eq.~\eqref{eq:kjesres}, i.e. the constants $q_0$, $q_1$ and $\mathcal{A}$.
Intuitively, $q_0 \approx 0$ would seem as the most natural choice in Eq.~\eqref{eq:linkdef}.
However, a non-zero $q_0$ value can be motivated by the features of some resonance position estimators.
The value $\hat{m}$ may consists of a sum of the signal (fraction $\beta$) and the background (fraction $1-\beta$):
\begin{equation} \label{eq:signbak}
 \hat{m} = \beta \times \hat{m}^{\text{Sgn}} + \left(1-\beta\right) \times \hat{m}^{\text{Bkg}}.
\end{equation}
The presence of a background term is obvious especially for the integral-based estimators, ave and itg.
In a histogram, the invariant mass extracted from some event candidates can be close to the resonance position accidentally.
An integral over a mass window in the histogram considers also these background events.
For the itg estimator the integration window is typically balanced so that the value of $q_0$ is close to zero.
For the ave estimator the value of $q_0$ is out of control, as the histogram widths are not varied.

The background term is expected to be a constant that conveys no information about the resonance position.
We can apply Eq.~\eqref{eq:fullcircle} for the signal component of $\hat{m}_W$:
\begin{equation} \label{eq:signalk}
 \hat{m}_{W}^{\text{Sgn}} \left( K_{\text{JES}}^{\text{Res}} \right) = K_{\text{JES}}^{\text{Res}} \otimes F_{\text{Corr}}^{lq} \otimes \hat{m}_{W}^{\text{Sgn},gen}.
\end{equation}
The term $\hat{m}_{W}^{\text{Sgn},gen}$ is a constant, independent of the $F_{\text{Corr}}$ and $K_{\text{JES}}^{\text{Res}}$ values.
We can substitute the result of Eq.~\eqref{eq:signbak} into Eq.~\eqref{eq:linkdef}, and express the signal component using Eq.~\eqref{eq:signalk}:
\begin{equation} \label{eq:kterms}
 \hat{K}_{\text{JES}} \left( K_{\text{JES}}^{\text{Res}} \right)
 = \frac{\beta \times \hat{m}_{W}^{\text{Sgn}}}{80.4~\text{GeV}} + \frac{\left(1-\beta\right) \times \hat{m}_{W}^{\text{Bkg}}}{80.4~\text{GeV}}
 = \frac{\beta \times F_{\text{Corr}}^{lq} \otimes \hat{m}_{W}^{\text{Sgn},gen}}{80.4~\text{GeV}}  \otimes  K_{\text{JES}}^{\text{Res}} + \frac{\left(1-\beta\right) \times \hat{m}_{W}^{\text{Bkg}}}{80.4~\text{GeV}}.
\end{equation}
By comparing Eq.~\eqref{eq:kterms} with the parametrization of Eq.~\eqref{eq:linkdef}, we note that
\begin{eqnarray}
 q_0 = & \left(1-\beta\right) \times \frac{\hat{m}_{W}^{\text{Bkg}}}{80.4~\text{GeV}} & = \left(1-\beta\right) \times \hat{K}_{\text{JES}}^{\text{Bkg}}, \label{eq:kq0} \\
 q_1 = & \beta \times F_{\text{Corr}}^{lq} \otimes \frac{\hat{m}_{W}^{\text{Sgn},gen}}{80.4~\text{GeV}} & = \beta \times F_{\text{Corr}}^{lq} \otimes \hat{K}_{\text{JES}}^{\text{Sgn},gen}. \label{eq:kq1}
\end{eqnarray}
Thus, an explicit form was found for Eq.~\eqref{eq:linkdef}.
For a certain estimator $\hat{m}$, the fraction $\beta$ is expected to be constant.
Moreover, the background term $\hat{K}_{\text{JES}}^{\text{Bkg}}$ remains constant, making $q_0$ constant for the given estimator.
On the other hand, $q_1$ depends only on $F_{\text{Corr}}^{lq}$, as the generator-level mass term is constant.

By referring to Eq.~\eqref{eq:kq1}, Eq.~\eqref{eq:kjesres} turns into
\begin{equation} \label{eq:kjesres2}
 K_{\text{JES}}^{\text{Res},ML}
 = \frac{\mathcal{A} - q_0}{\beta \times F_{\text{Corr}}^{lq} \otimes \hat{K}_{\text{JES}}^{\text{Sgn},gen}}
 = \frac{\mathcal{B}}{F_{\text{Corr}}^{lq}},
\end{equation}
where
\begin{equation}
 \mathcal{B} = \frac{\mathcal{A} - q_0}{\beta \times \hat{K}_{\text{JES}}^{\text{Sgn},gen}}
\end{equation}
is a constant.
Now, applying the $K_{\text{JES}}^{\text{Res}}$ value from Eq.~\eqref{eq:kjesres2} on Eq.~\eqref{eq:fullcircle}, we find the light quark and b quark jet energies:
\begin{eqnarray}
 E_{j,lq}''  = & \frac{\mathcal{B}}{\cancel{F_{\text{Corr}}^{lq}}} \otimes \cancel{F_{\text{Corr}}^{lq}}  \otimes  E_j^{gen} = \mathcal{B}~\otimes &  E_j^{gen}, \\
 E_{j,b}'' = & \frac{\mathcal{B}}{F_{\text{Corr}}^{lq}} \otimes F_{\text{Corr}}^{b}   \otimes  E_j^{gen} = \mathcal{B}~\otimes & \frac{F_{\text{Corr}}^{b}}{F_{\text{Corr}}^{lq}} \otimes E_j^{gen}. \\
\end{eqnarray}
These expressions are valid for both the recalibrated and D\O\ $F_{\text{Corr}}$ values in all D\O\ Run~II eras.
To shift the D\O\ mass results, we interpret that the $E_j''$ energies in a single era are the same with both $F_{\text{Corr}}$ values.
However, the relationship to $E_j^{gen}$ can differ.
By equating the $E_j''$ energies, we find
\begin{eqnarray}
 \cancel{\mathcal{B}} \otimes E_{j,lq}^{gen,\text{D\O}} & = & \cancel{\mathcal{B}} \otimes  E_{j,lq}^{gen}, \\
 \cancel{\mathcal{B}} \otimes \left(\frac{F_{\text{Corr}}^{b,\text{D\O}}}{F_{\text{Corr}}^{lq,\text{D\O}}}\right) \otimes E_{j,b}^{gen,\text{D\O}} & = & \cancel{\mathcal{B}} \otimes \left(\frac{F_{\text{Corr}}^{b}}{F_{\text{Corr}}^{lq}}\right) \otimes E_{j,b}^{gen},
\end{eqnarray}
By referring to Eq.~\eqref{eq:ejgen} and noting that the hadronization and showering steps do not depend on $F_{\text{Corr}}$, we can make a direct inversion from $E_{j}^{gen}$ to quark energies.
Moving all the scaling constants to the right-hand-side and performing the inversion, we find
\begin{eqnarray}
 E_{lq}^{\text{D\O}} & = & E_{lq}, \label{eq:ljets} \\
 E_{b}^{\text{D\O}}  & = & \left(\frac{F_{\text{Corr}}^{lq,\text{D\O}}}{F_{\text{Corr}}^{lq}}\right) \otimes \left(\frac{F_{\text{Corr}}^{b}}{F_{\text{Corr}}^{b,\text{D\O}}}\right) \otimes E_{b}.
 \label{eq:bjets}
\end{eqnarray}
This is exactly the result of the phenomenological proof of Eqs.~(\ref{eq:liteq},\ref{eq:bq}).
That is, the presented numerical approach is equivalent to the phenomenological one.

\subsection{Evaluating the Maximum Likelihood Value of $K_{\text{JES}}^{\text{Res}}$}

The coefficient $\mathcal{A}$ remains yet unknown, and therefore it is not possible to evaluate Eq.~\eqref{eq:kjesres} numerically.
While deriving Eqs.~(\ref{eq:ljets},\ref{eq:bjets}), the constant $\mathcal{B}$ was completely canceled out.
All the $\mathcal{A}$-dependence was given in this term.
This implies that the value of $\mathcal{A}$ only sets a reference scale, and it can be chosen quite freely.
To make the reference scale behave flexibly with different $\hat{m}$ evaluators, we choose
\begin{equation} \label{eq:aconstant}
 \mathcal{A} = \frac{\hat{m}_{W}^{gen}}{80.4~\text{GeV}}.
\end{equation}
This sets the maximum likelihood value of the resonance position estimator to be equal with the corresponding generator-level value.
By this choice, Eq.~\eqref{eq:kjesres} for $K_{\text{JES}}^{\text{Res}}$ determination reads
\begin{equation} \label{eq:kjesres3}
 K_{\text{JES}}^{\text{Res},ML}
 = \frac{\frac{\hat{m}_{W}^{gen}}{80.4~\text{GeV}} - q_0}{q_1}
 = \frac{\hat{K}_{\text{JES}}^{gen} - q_0}{q_1}.
\end{equation}
To evaluate Eq.~\eqref{eq:kjesres3}, only the values of $q_0$, $q_1$ and $\hat{K}_{\text{JES}}^{gen}$ need to be determined.

\subsection{Evaluating the Combination Parameter $\alpha$}

Determining the value of $\alpha$ requires turning to the D\O\ measurements.
The three likelihood plots in Figs.~23 and 25 in Ref.~\cite{ref:d02015} show a falling linear maximum likelihood slope of $m_t$ as a function of $K_{\text{JES}}$.
We can understand this behavior by fixing the resonance position estimate $\hat{m}_t$ in Eq.~\eqref{eq:linearity} to a constant value.
In other words, by looking at such pairs of $(m_t^{gen},K_{\text{JES}}^{\text{Res}})$ values that could correspond to a given measurement,
before fixing the maximum likelihood value of $K_{\text{JES}}^{\text{Res}}$:
\begin{equation} \label{eq:linearity2}
 m_t^{gen} = - \frac{p_0 + \hat{m}_t}{p_1} - \frac{p_2}{p_1} \times K_{\text{JES}}^{\text{Res}}.
\end{equation}
This result displays directly the falling linear slope of the likelihood plots.
It must be noted that $K_{\text{JES}}^{\text{Res}}$ has the same characteristic behavior as plain $K_{\text{JES}}$ in the D\O\ measurement.
Hence, even if Eq.~\eqref{eq:linearity} is designed for the $E_j''$ level energies, the slope of Eq.~\eqref{eq:linearity2} is equally valid to the $E_j$ level energies.

The result of Eq.~\eqref{eq:linearity2} can be written separately for the hadronic and the leptonic mass resonances, and the results will differ.
In contrast, the plots in Ref.~\cite{ref:d02015} show the combined effect of both of the resonances.
We can make the difference between the resonances more concrete by looking at the slope of the hadronic resonance position:
\begin{equation} \label{eq:hadronslope}
 \frac{dm_{t,had}^{gen}}{dK_{\text{JES}}} = - \frac{p_2}{p_1} \approx - m_{t,had}^{gen}.
\end{equation}
This is due to the fact that the hadronic peak consists purely of jets, and the reconstructed energy of each of these is proportional to $K_{\text{JES}}$.
On the other hand, for the leptonic resonance position only the b jet and the neutrino reconstruction carry $K_{\text{JES}}$-dependence.
Hence, the slope of the leptonic resonance is not as steep as that of the hadronic one.

To extract the value of $\alpha$, we need to combine the hadronic and the leptonic slopes.
The combined slope should be equal to the one found in the D\O\ likelihood plots.
Therefore, we need to take the weighted vectorial sum of the normalized hadronic and leptonic slopes, and set this to be equal with the D\O\ likelihood slope after normalization.
Mathematically speaking, this vectorial version of Eq.~\eqref{eq:alpharel} reads as
\begin{equation} \label{eq:slopecombo}
 \text{Normalize}\left\{\alpha \times \hat{v}_{had} + (1-\alpha) \times \hat{v}_{lep}\right\} 
 = \alpha_1  \times \hat{v}_{had} + \alpha_2  \times \hat{v}_{lep}
 = \hat{v}_{\text{D\O}},
\end{equation}
where $\hat{v}$'s are unit vectors.
We can take the dot product of Eq.~\eqref{eq:slopecombo} with $\hat{v}_{\text{D\O}}$ and a unit vector perpendicular to it ($\hat{v}_{\text{D\O}}^\bot$):
\begin{eqnarray}
 \alpha_1 \times \hat{v}_{had} \cdot \hat{v}_{\text{D\O}} + \alpha_2 \times \hat{v}_{lep} \cdot \hat{v}_{\text{D\O}} & = & 1 \\
 \alpha_1 \times \hat{v}_{had} \cdot \hat{v}_{\text{D\O}}^\bot + \alpha_2 \times \hat{v}_{lep} \cdot \hat{v}_{\text{D\O}}^\bot & = & 0
\end{eqnarray}
Solving for $\alpha_2$ in the latter equation, the former equation yields
\begin{equation} \label{eq:alpha1sol}
 \alpha_1 = \left( \hat{v}_{had} \cdot \hat{v}_{\text{D\O}} - \hat{v}_{lep} \cdot \hat{v}_{\text{D\O}} \left[\frac{\hat{v}_{had} \cdot \hat{v}_{\text{D\O}}^\bot}{\hat{v}_{lep} \cdot \hat{v}_{\text{D\O}}^\bot}\right] \right)^{-1}
\end{equation}
It turns out that the coefficient $p_2$ is generally much larger than $p_1$ in Eq.~\eqref{eq:linearity}, as was discussed in the hadronic example case of Eq.~\eqref{eq:hadronslope}.
Hence $\hat{v}_{had} \cdot \hat{v}_{\text{D\O}} \approx 1$ and $\hat{v}_{lep} \cdot \hat{v}_{\text{D\O}} \approx 1$.
This leads to $\alpha_1 \approx \alpha$ and $\alpha_2 \approx 1-\alpha$, and Eq.~\eqref{eq:alpha1sol} reduces to
\begin{equation} \label{eq:alphasol}
 \alpha \approx \alpha_1 \approx \left( 1 - \frac{\hat{v}_{had} \cdot \hat{v}_{\text{D\O}}^\bot}{\hat{v}_{lep} \cdot \hat{v}_{\text{D\O}}^\bot} \right)^{-1}.
\end{equation}
This implies that the value of $\alpha$ is determined by the small vectorial deviations in the direction perpendicular to $\hat{v}_{\text{D\O}}$.
The $F_{\text{Corr}}$ values have generally little impact on the value of $\alpha$, and hence it is calculated as an average from the D\O\ and recalibrated $F_{\text{Corr}}$ scenarios.

It is likely that the optimal hadronic and leptonic $m_t^{\text{calib}}$ values were not equal in the D\O\ measurement, but it is not possible to determine their values.
The least biased choice is to set
\begin{equation} \label{eq:egalite}
 m_t^{\text{D\O}} \equiv m_t^{\text{calib},\text{D\O}} = m_{t,had}^{\text{calib},\text{D\O}} = m_{t,lep}^{\text{calib},\text{D\O}}.
\end{equation}
Thus, the parameter $\alpha$ is only present while producing the combined shifted top mass value.
With the choice of Eq.~\eqref{eq:egalite}, Eq.~\eqref{eq:alpharel} implies
\begin{equation} \label{eq:firsthift}
 m_t^{\text{calib}} = \alpha \times m_{t,had}^{\text{calib}} + \left( 1 - \alpha \right) \times m_{t,lep}^{\text{calib}}
 = m_t^{\text{D\O}} + \alpha \times \Delta m_{t,had}^{\text{calib}} \left(m_t^{\text{D\O}}\right) + \left( 1 - \alpha \right) \times \Delta m_{t,lep}^{\text{calib}} \left(m_t^{\text{D\O}}\right),
\end{equation}
where we have expressed the shifted $m_t^{\text{calib}}$ values as the difference to their origin, $m_t^{\text{D\O}}$.
If the original values of $m_{t,had}^{\text{calib},\text{D\O}}$ and $m_{t,lep}^{\text{calib},\text{D\O}}$ were known, the corresponding result would be
\begin{equation} \label{eq:secondshift}
 m_t^{\text{calib}} = \alpha \times m_{t,had}^{\text{D\O}} + \left( 1 - \alpha \right) \times m_{t,lep}^{\text{D\O}} +
  + \alpha \times \Delta m_{t,had}^{\text{calib}} \left(m_{t,had}^{\text{D\O}}\right) + \left( 1 - \alpha \right) \times \Delta m_{t,lep}^{\text{calib}} \left(m_{t,lep}^{\text{D\O}}\right).
\end{equation}
Referring to Eq.~\eqref{eq:alpharel}, the first two terms can be identified as $m_t^{\text{D\O}}$.
Furthermore, we may assume that slight variations in the point of origin $m_{t}^{\text{calib},\text{D\O}}$ have little effect on the shifts $\Delta m_{t}^{\text{calib}}$.
Following from this equality, the statements of Eqs.~(\ref{eq:firsthift},\ref{eq:secondshift}) become equal.
Therefore, the choice of Eq.~\eqref{eq:egalite} produces the same shifted $m_t^{\text{calib}}$ value as the original values of $m_{t,had}^{\text{calib},\text{D\O}}$ and $m_{t,lep}^{\text{calib}}$.

\subsection{Analysis Setup and Error analysis}

For the main measurement, we use the event generator Pythia~6 (P6) \cite{ref:pythia6} with the same tuning parameters as D\O\ used for the l+jets measurement of Ref.~\cite{ref:d02015}.
Hence, all the generator-related error sources and magnitudes are estimated to be the same as D\O\ presented.
Moreover, we will assume that the $F_{\textrm{corr}}$-related errors are the same as D\O\ claimed.
This is sensible, as the $F_{\textrm{corr}}$ recalibration of Ref.~\cite{ref:tonimaster} demonstrates a systematic unaccounted shift in the top mass value.
In total, this means that the original D\O\ error estimates should be applicable for the shifted top mass.
On top of this error, we must add the \emph{method error}, related to the presented $m_t$ shifting method.
The total method error is found by considering the individual steps of the mass shifting procedure.

According to Eqs.~(\ref{eq:linkdef},\ref{eq:linearity}), the error of the re-interpreted $m_t^{\text{calib}}$ reads as
\begin{equation} \label{eq:caliberr}
 \delta m_t^{\text{calib}}
 = \delta m_t^{gen}
 = \frac{\delta \hat{m}_t + p_2 \times \delta K_{\text{JES}}^{\text{Res}}}{p_1}
 = \frac{\delta \hat{m}_t}{p_1} + \frac{p_2}{p_1} \times \frac{\delta \hat{K}_{\text{JES}}}{q_1}.
\end{equation}
In these error estimates we generally take the fit coefficients as they are, and interpret all fit errors using the variables $\hat{m}_t$ and $\hat{K}_{\text{JES}}$.
For the mean values produced by linear fits, the error estimates are found as shown in Appendix~\ref{app:regress}.
We note that the term $\delta \hat{m}_t$ must carry the fit error of $\hat{m}_t$, and the total error of the value $\hat{m}_t^{\text{D\O}}$:
\begin{equation} \label{eq:caliberr2}
 \delta \hat{m}_t = \delta \hat{m}_t^{\text{Fit}} + \delta \hat{m}_t^{\text{Tot,D\O}}
  = \delta \hat{m}_t^{\text{Fit}} + \delta \hat{m}_t^{\text{Fit,D\O}} + \delta \hat{m}_t^{\text{D\O}}
\end{equation}
The term $\delta \hat{m}_t^{\text{D\O}}$ can be calculated inverting Eq.~\eqref{eq:caliberr}:
\begin{equation} \label{eq:caliberr3}
 \delta \hat{m}_t^{\text{D\O}} = - p_2^{\text{D\O}} \times \frac{\delta \hat{K}_{\text{JES}}^{\text{D\O}}}{q_1^{\text{D\O}}} + p_1^{\text{D\O}} \times \delta m_t^{\text{D\O}}.
\end{equation}
Putting Eqs.~(\ref{eq:caliberr},\ref{eq:caliberr2},\ref{eq:caliberr3}) together, we find
\begin{equation} \label{eq:calibs}
 \delta m_t^{\text{calib}} = \frac{1}{p_1} \left(\delta \hat{m}_t^{\text{Fit}} + \delta \hat{m}_t^{\text{Fit,D\O}} 
 + \frac{p_2 \times \delta \hat{K}_{\text{JES}}}{q_1}
 - \frac{p_2^{\text{D\O}} \times \delta \hat{K}_{\text{JES}}^{\text{D\O}}}{q_1^{\text{D\O}}}\right) + \cancel{\frac{p_1^{\text{D\O}}}{p_1}} \times \delta m_t^{\text{D\O}}
 = \delta m_t^{\text{Stat}} + \delta m_t^{\text{D\O}}.
\end{equation}
The values of $p_1$ vary very little between the D\O\ parametrization and the new one, and any differences between them are mostly statistical fluctuations.
Thus, the term $\frac{p_1^{\text{D\O}}}{p_1}$ is set to unity.
Hence, the last term in Eq.~\eqref{eq:calibs} performs the transition of the original D\O\ error to the shifted top mass value.
The remaining term is the total statistical method-related error, which consists of four separate terms.
For the numerical evaluation of an error such as that in Eq.~\eqref{eq:calibs}, the separate error terms are summed in quadrature.

Care must be taken while combining the errors of the four D\O\ Run~II eras and the electron and muon channels.
Generally, the mass shift measurement is performed for all the eight measurements from the four eras and two channels, considering both the hadronic and leptonic resonances.
The combination of the hadronic and the leptonic channels is postponed as much as possible.
This is the most volatile step of the mass shifting procedure, and notable uncertainty on the parameter $\alpha$ is present.
If the combination of the hadronic and leptonic channels is performed at an early phase of the measurement, handling the combination errors becomes more complicated.

For the pre-combination $m_t$ resonances, the electron and muon Run~II combinations are calculated using the total statistical errors of Eq.~\eqref{eq:calibs} as weights.
Similarly, the pairs of electron and muon measurements can be combined to full l+jets measurements.
The method error is found by substracting in quadrature the original D\O\ statistical error from the combined statistical error.

As the combination of Eq.~\eqref{eq:alpharel} is considered, the combined error term must be determined.
By variational calculus, it is found:
\begin{equation} \label{eq:dalpharel1}
 \delta m_{t,comb}^{\text{calib}} = \alpha \times \delta m_{t,had}^{\text{calib}} + \left( 1 - \alpha \right) \times \delta m_{t,lep}^{\text{calib}} + \delta \alpha \times \left( m_{t,had}^{\text{calib}} - m_{t,lep}^{\text{calib}} \right).
\end{equation}
Considering Eq.~\eqref{eq:calibs}, this result can be written as
\begin{equation} \label{eq:dalpharel2}
 \delta m_{t,comb}^{\text{calib}} = \delta m_t^{\text{D\O}} + \alpha \times \delta m_{t,had}^{\text{Stat}}
 + \left( 1 - \alpha \right) \times \delta m_{t,lep}^{\text{Stat}} + \delta \alpha \times \left( m_{t,had}^{\text{calib}} - m_{t,lep}^{\text{calib}} \right).
\end{equation}
The first term is the transferred (statistical) D\O\ error, and the next two terms are the statistical method errors of the two resonances.
The last term is based on the uncertainty $\delta \alpha$, which does not behave similarly as the statistical error terms.
Furthermore, it turns out, the $\delta \alpha$ term is the dominating method-related error.

The combination of the hadronic and leptonic results is performed for the the 8+2 electron and muon measurements.
For these, Eq.~\eqref{eq:dalpharel2} gives the total error.
In the final step, these are turned into five combined l+jets results.
This electron-muon combination is performed using using the errors of Eq.~\eqref{eq:dalpharel2} as weights, excluding the $\delta \alpha$ term.
This is motivated by the fact that the error based on $\delta \alpha$ is systematical in its nature, and does not reduce in a combination similarly as the statistical errors.
The $\delta \alpha$ term for the combination is found as the weighted average of the corresponding electron and muon terms.

The values of $\alpha$ and $\delta \alpha$ can be estimated only in the cases of the combined electron channel, combined muon channel and the total combination.
The difference of a statistical combination of the electron and muon $\alpha$ values to the total combination $\alpha$ value is used to estimate systematical errors in $\alpha$ extraction.
In general, the errors related to $\alpha$ determination are complicated, starting from the accuracy of the linear approximation of Eq.~\eqref{eq:alpharel}.
It should also be noted that if the calibration error presented in Appendix~\ref{app:likeli} is present, an additional error in the $\alpha$ calibration will result.
The volatile nature of the $\alpha$ parameter requires conservative error analysis.

\newpage

\section{Measurements}

For the measurements, we use three P6 l+jets samples with $10^7$ events in each, at the three parameter values $m_t^{gen} = \{172.5, 173.7, 175.0\}$.
The mass values are selected so that these are approximately evenly spaced and so that most of the $m_t^{\text{calib}}$ values can be interpolated between these.
Furthermore, the five values $K_{\text{JES}}^{\text{Res}} = \{1.000,1.005,1.010,1.015,1.020\}$ are used for the calibrations.
In the analyses, we use the D\O\ Tune~A and the kinematic selections presented in Ref.~\cite{ref:d02015}.
Moreover, we use the D\O\ jet cone with FastJet~\cite{ref:fastjet} to cluster the particles into jets.

In the analyses, we apply $F_{\text{Corr}}$ values derived with two separate simulation software packages: P6 and Herwig~7~(H7)~\cite{ref:h71,ref:h72}.
Furthermore, two separate sets of $F_{\text{Corr}}$ parametrizations are used to probe systematical errors.
The separating factor between the two parametrizations is a subtle non-trivial choice within the $F_{\text{Corr}}$ calibration process.
The choice seems to have no practical effect on the calibration quality, and hence it provides a method for assessing systematical $F_{\text{Corr}}$ errors.
The $F_{\text{Corr}}$ results with a less D\O\ like choice is denoted with the index 1, and the ones with a more D\O\ like choice with the index 2.
It turns out that there is generally little difference between these.
Considering the two generators (P6, H7) and the two parametrizations, there are in total 4 separate $F_{\text{Corr}}$ parameter sets.
All these were provided by the author of Ref.~\cite{ref:tonimaster}.

The logistical chain of the analysis is the following.
The event information including jets and leptons is saved into a ROOT \cite{ref:root} file using the software handle from Ref.~\cite{ref:genhandle}.
These are further analyzed with the software from Ref.~\cite{ref:genanalysis}.
The latter software package applies the $F_{\text{Corr}}$ values for the different D\O\ eras and separates the electron and muon channels.
This software tool uses the simulation truth to distinguish the various resonances.
It also emulates the reconstruction of a neutrino from the missing transverse momentum and the leptonic W boson resonance.
The results are saved mainly into mass histograms, which are turned into resonance position estimators $\hat{m}$ in the next step.
Evaluating the values of these estimators involves e.g. fitting and taking mean values of the histograms.
Finally, the linear dependence between the $\hat{m}$'s and $K_{\text{JES}}^{\text{Res}}$ and $m_t^{gen}$ is fitted using the method of least squares.
Based on these fits and the D\O\ measurements, the shifted top mass values and their errors are found.

In this section, the individual steps of the measurement are performed.
First, the values of the maximum likelihood $K_{\text{JES}}^{\text{Res}}$ is determined.
Then, the hadronic and leptonic top mass shifts are derived separately for all estimators, and for the 4 $F_{\text{Corr}}$ sets.
The results of the five resonance position estimators are combined according to their method errors for both the hadronic and leptonic resonances.
In the following step, the combination parameter $\alpha$ is determined.
Using the values of $\alpha$, the results for the hadronic and leptonic channels are combined.
Finally, the results based on the four $F_{\text{Corr}}$ parameter sets are compared and combined.

\subsection{Maximum Likelihood $K_{\text{JES}}^{\text{Res}}$}

The maximum likelihood $K_{\text{JES}}^{\text{Res}}$ extraction procedure is based on hadronic W boson resonance histograms.
From these all the resonance position estimators are extracted.
In Appendix~\ref{app:wexamples} some examples of these are shown.
It is not meaningful to provide all of these:
there are 4 runs,
times 3 $m_t^{gen}$ values,
times 5 $K_{\text{JES}}^{\text{Res}}$ values,
times two channels (electron/muon),
times two $F_{\text{Corr}}$ versions (D\O\ and recalibrated),
times two sets of $F_{\text{Corr}}$ parametrization,
times two generators (P6, H7) in $F_{\text{Corr}}$ determination.
Also counting the generator level resonance histograms at the three $m_t^{gen}$ values, this makes a total of 963 histograms.

The exact values of $K_{\text{JES}}^{\text{Res}}$ depend on the resonance position estimator type, and are not informative.
This is due to the freedom of choice of the $K_{\text{JES}}^{\text{Res}}$ reference scale, $\mathcal{A}$.
However, the relative changes in $K_{\text{JES}}^{\text{Res}}$ provide useful information.

\begin{figure}[H]
\centering
 \includegraphics[width=0.82\textwidth]{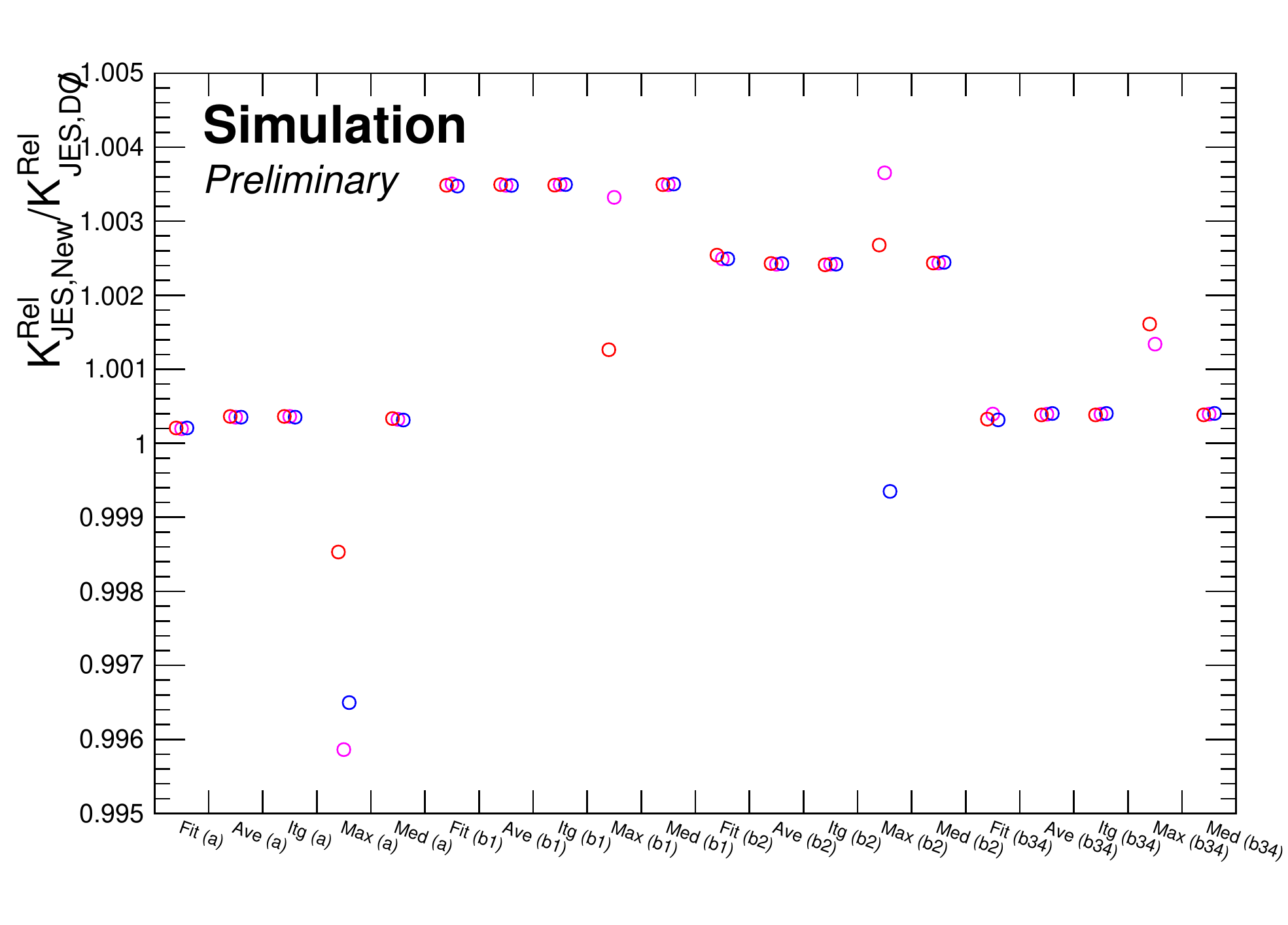}
 \includegraphics[width=0.82\textwidth]{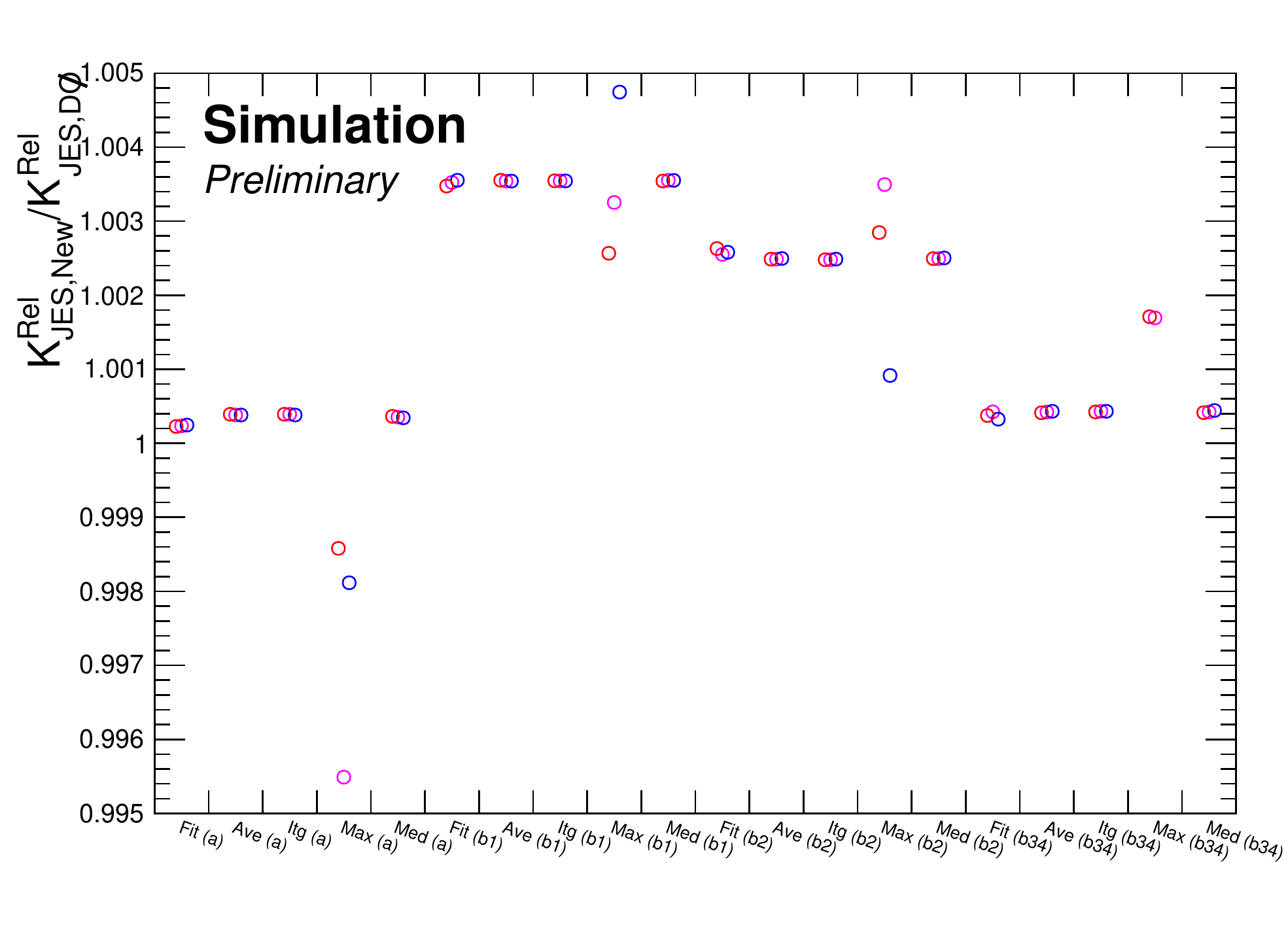}
 \caption{Relative maximum likelihood values of $K_{\text{JES}}^{\text{Res}}$ with P6, $F_{\text{Corr}}$ set 1 (top), set 2 (bottom). Electron (blue), muon (red) and combination (magenta) shown separately.}
 \label{fig:kjesrel1}
\end{figure}

\begin{figure}[H]
\centering
 \includegraphics[width=0.82\textwidth]{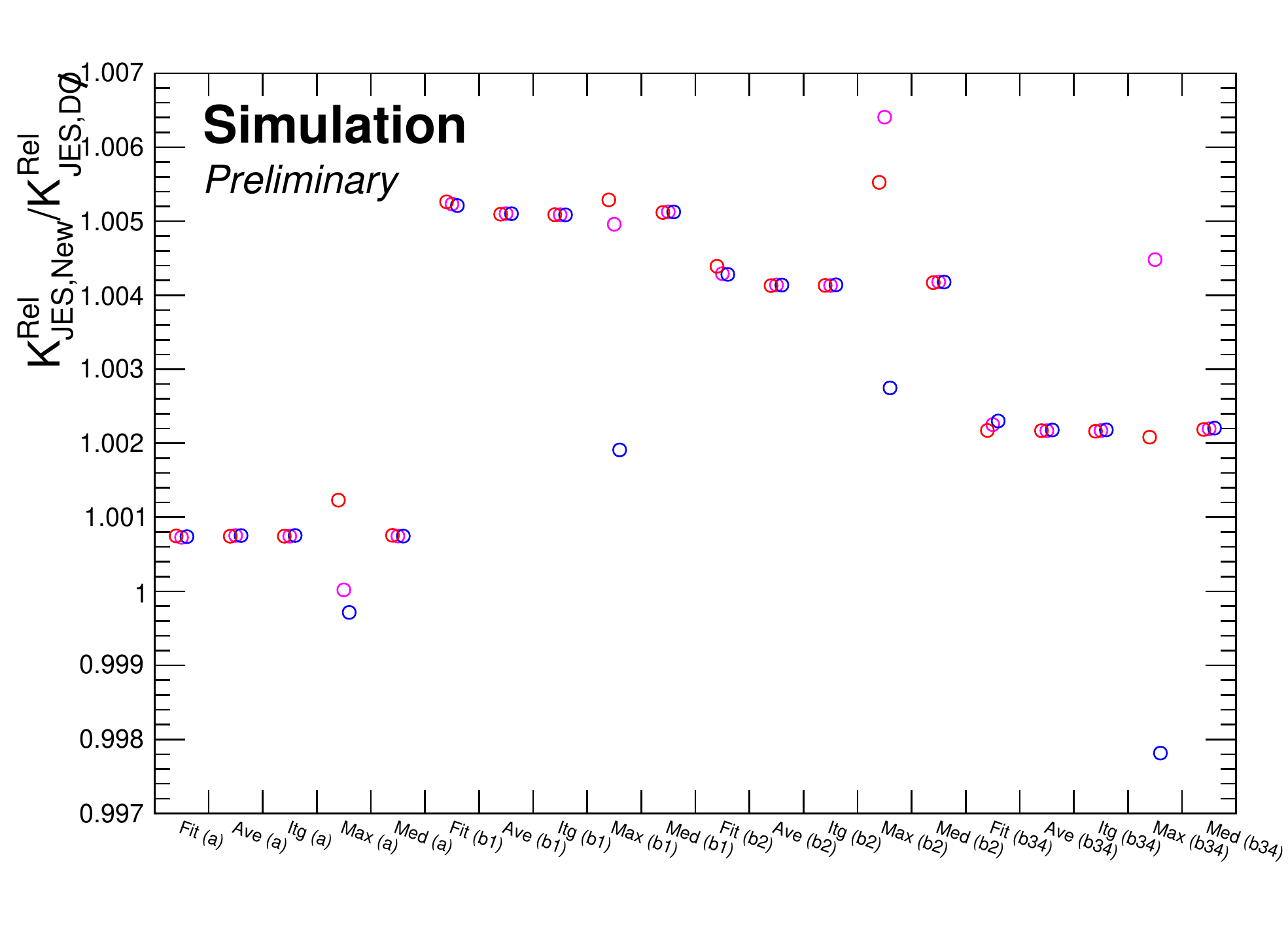}
 \includegraphics[width=0.82\textwidth]{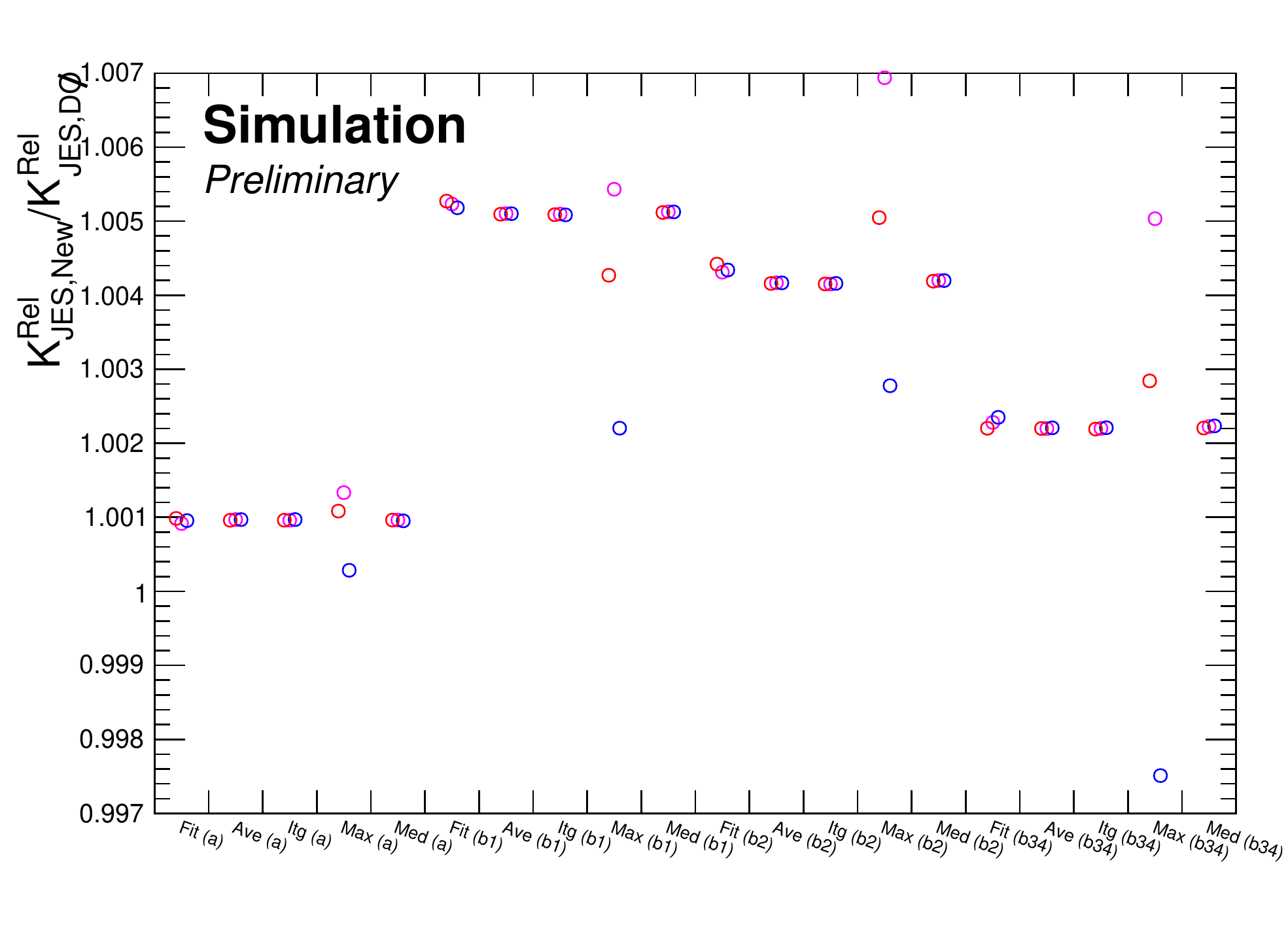}
 \caption{Relative maximum likelihood values of $K_{\text{JES}}^{\text{Res}}$ with H7, $F_{\text{Corr}}$ set 1 (top), set 2 (bottom). Electron (blue), muon (red) and combination (magenta) shown separately.}
 \label{fig:kjesrel2}
\end{figure}

In Fig.~\ref{fig:kjesrel1} the fraction of the maximum likelihood $K_{\text{JES}}^{\text{Res}}$ obtained with new $F_{\text{Corr}}$ values and that obtained with the D\O\ ones is shown on P6.
In Fig.~\ref{fig:kjesrel2} the same results are given for H7.

In the Figs.~\ref{fig:kjesrel1} and \ref{fig:kjesrel2}, a fraction close to unity implies that there has been very little change in $F_{\text{Corr}}^{lq}$.
A large deviation from unity implies the opposite.
In these figures and a range of test cases the maximum estimator displays unstable behavior.
The combined value is not always between the electron and muon channels, as the combined histogram can have a maximum elsewhere.

The remaining four methods show generally a good agreement.
The fit-based estimator does not work as well as wished, even after extensive tuning of the fit functions.
This is due to the fact that especially in the asymmetric top mass distributions, small variations in the peak shape can have a large impact on the fit parameters.
Undesired exchange can occur between the resonance position parameters and the resonance width parameters.

Notably, the three remaining estimators yield similar, but more stable values as the fit-based estimator.
A range of example cases has shown that especially the ave and med methods exhibit a great precision.

The era IIa shows little difference between the original D\O\ $F_{\text{Corr}}$'s and the re-calibrated ones with both P6 and H7.
The same is true for IIb34 with P6, as H7 shows slightly more scaling.
In the era IIb1 the $K_{\text{JES}}^{\text{Res}}$ scaling effect is the greatest, and in IIb2 almost as great.
In general, H7 shows larger scaling effects that P6.
The two $F_{\text{Corr}}$ parameter sets produce generally very similar results.

\subsection{Hadronic Top Resonance Measurement}

The measurement of the top mass shift in the hadronic channel is based on the hadronic top resonance histograms.
All the 963 histograms do not fit in this document, but Appendix.~\ref{app:htexamples} shows some example cases of these.

\begin{figure}[H]

 \includegraphics[width=0.245\textwidth]{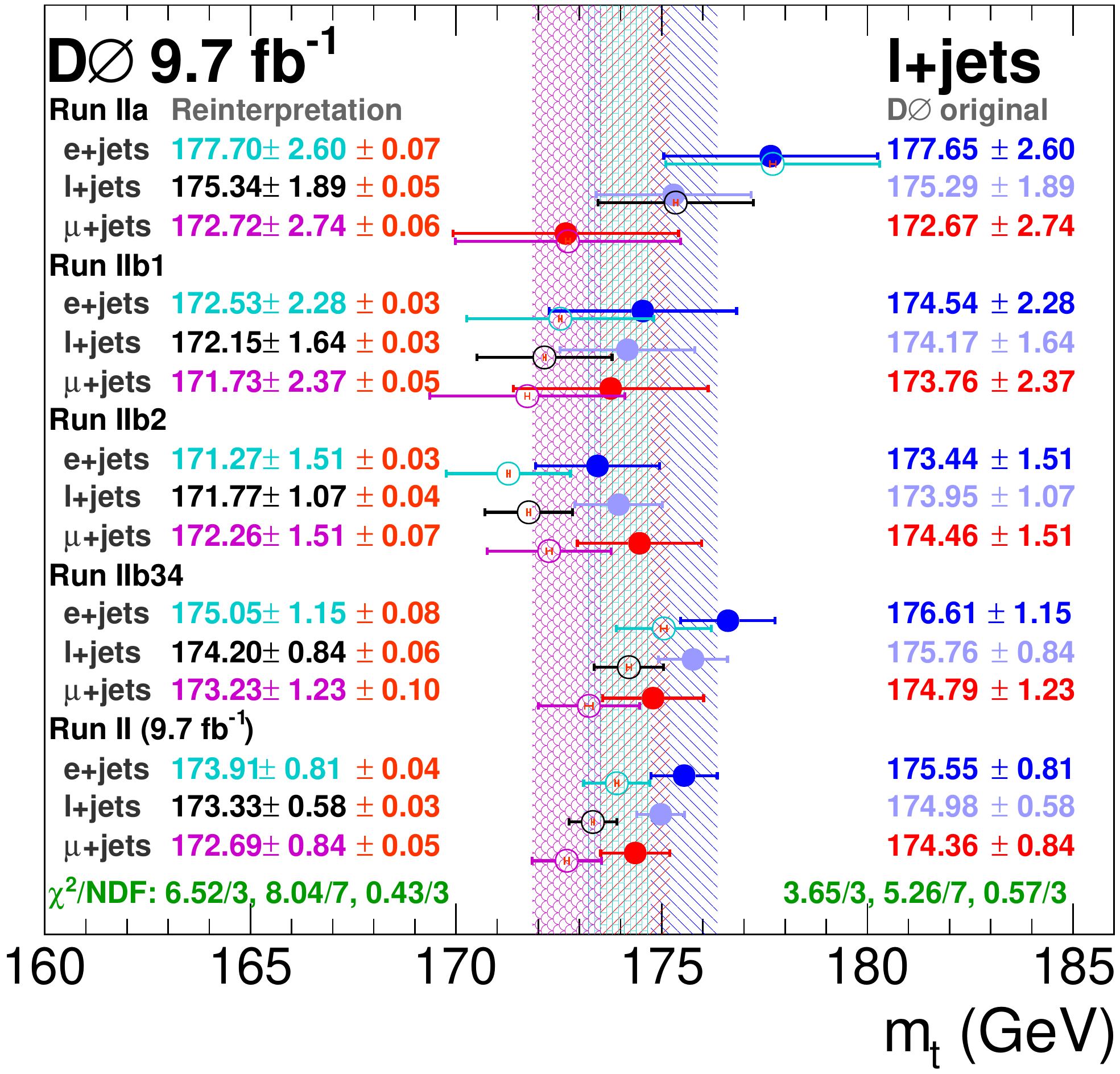}
 \includegraphics[width=0.245\textwidth]{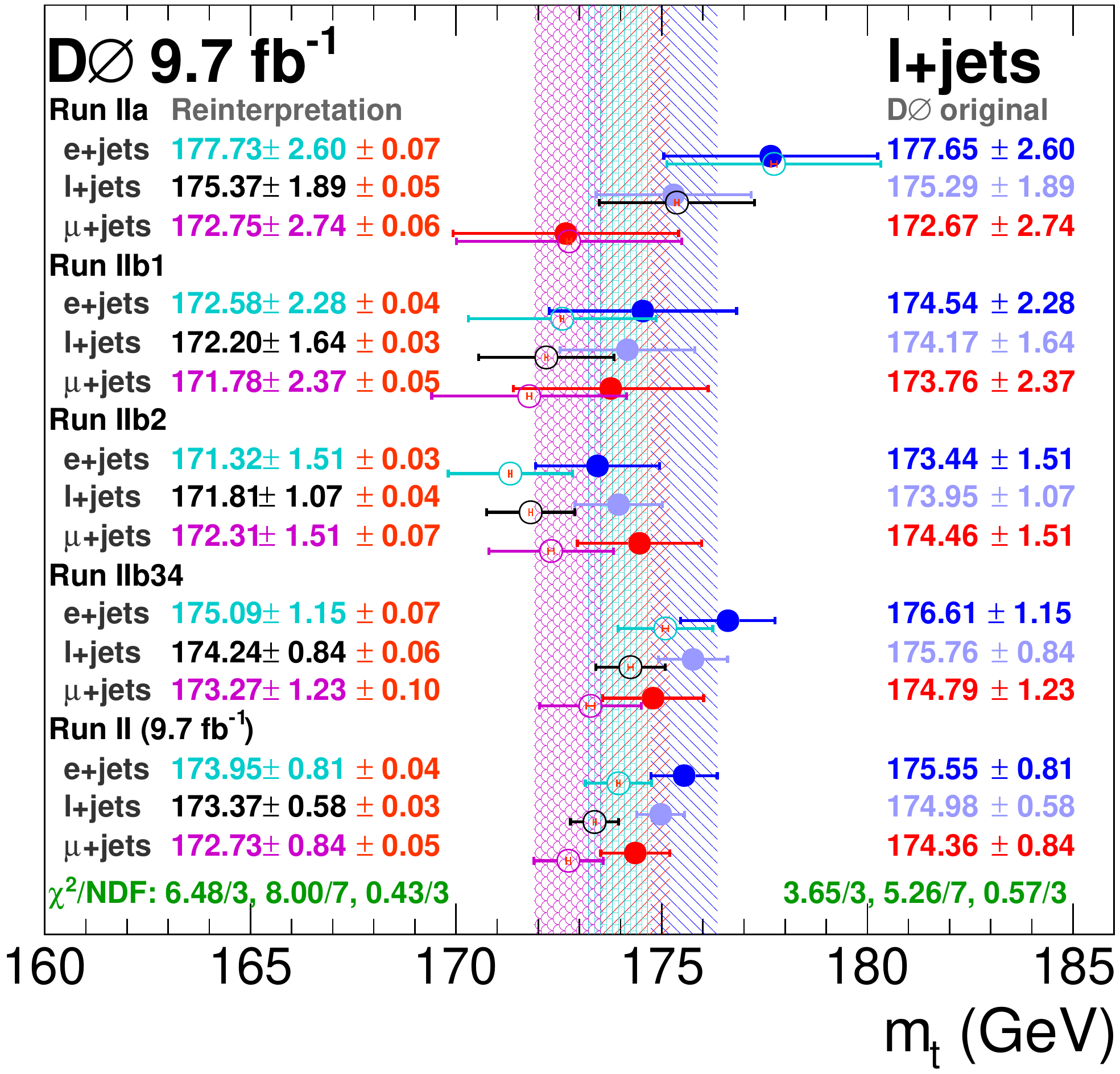}
 \includegraphics[width=0.245\textwidth]{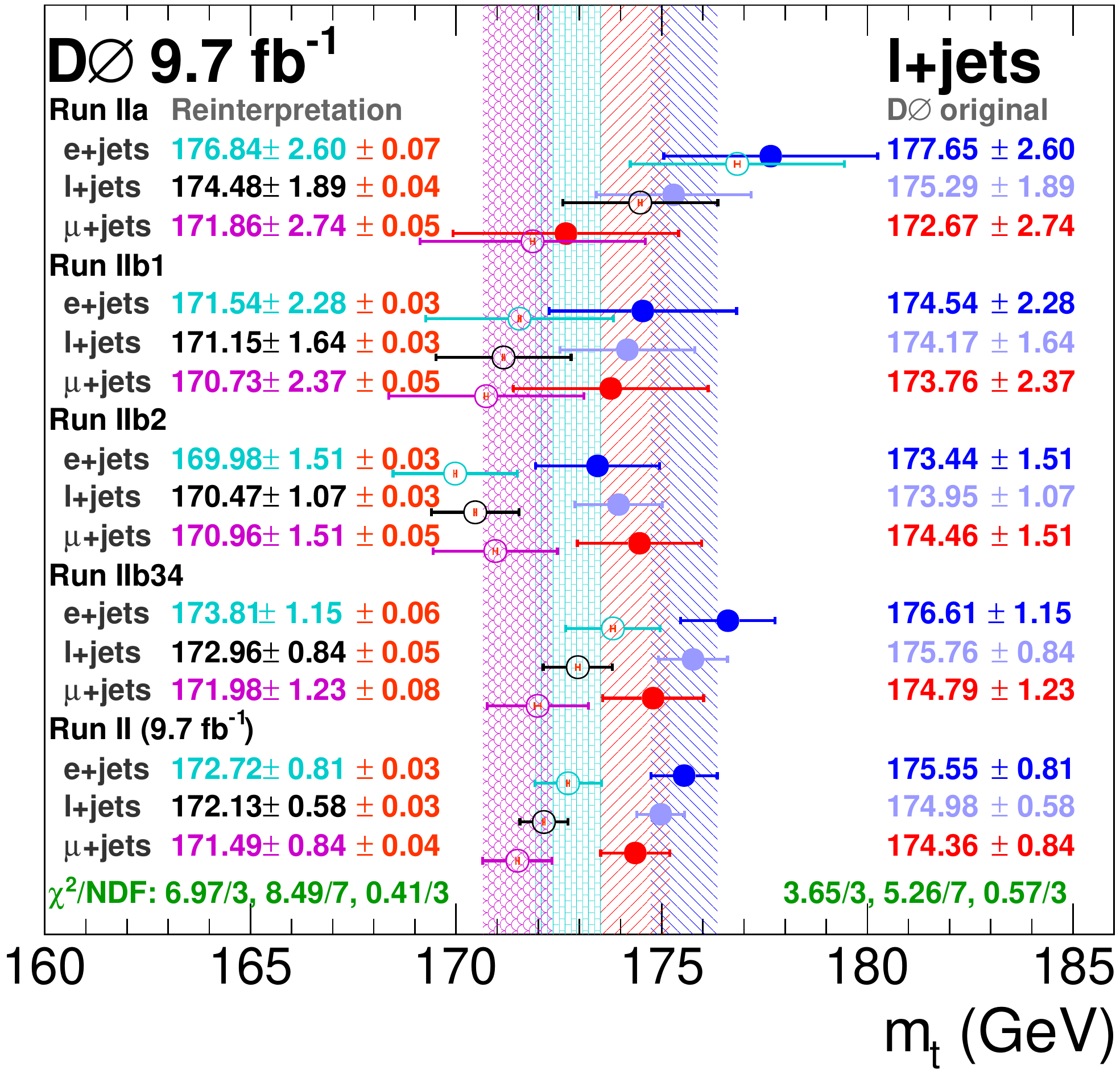}
 \includegraphics[width=0.245\textwidth]{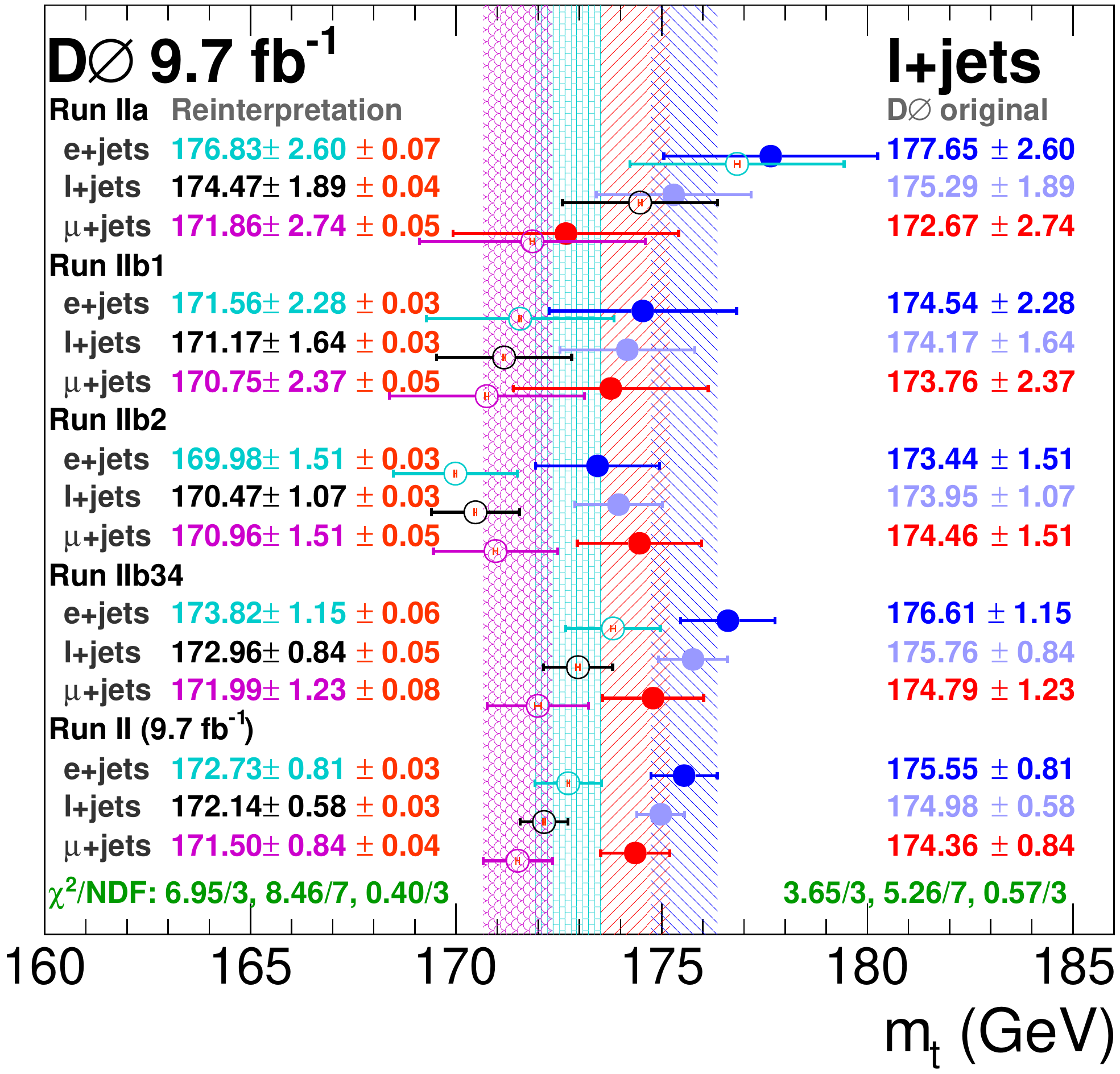}
  \caption{Shifted hadronic top mass values for the median estimator in all the measurements and combinations.
  From left to right: P6 $F_{\text{Corr}}$ set 1 and 2, H7 $F_{\text{Corr}}$ set 1 and 2.}
 \label{fig:hadronic1}
\end{figure}

\begin{figure}[H]
\centering
 \includegraphics[width=0.245\textwidth]{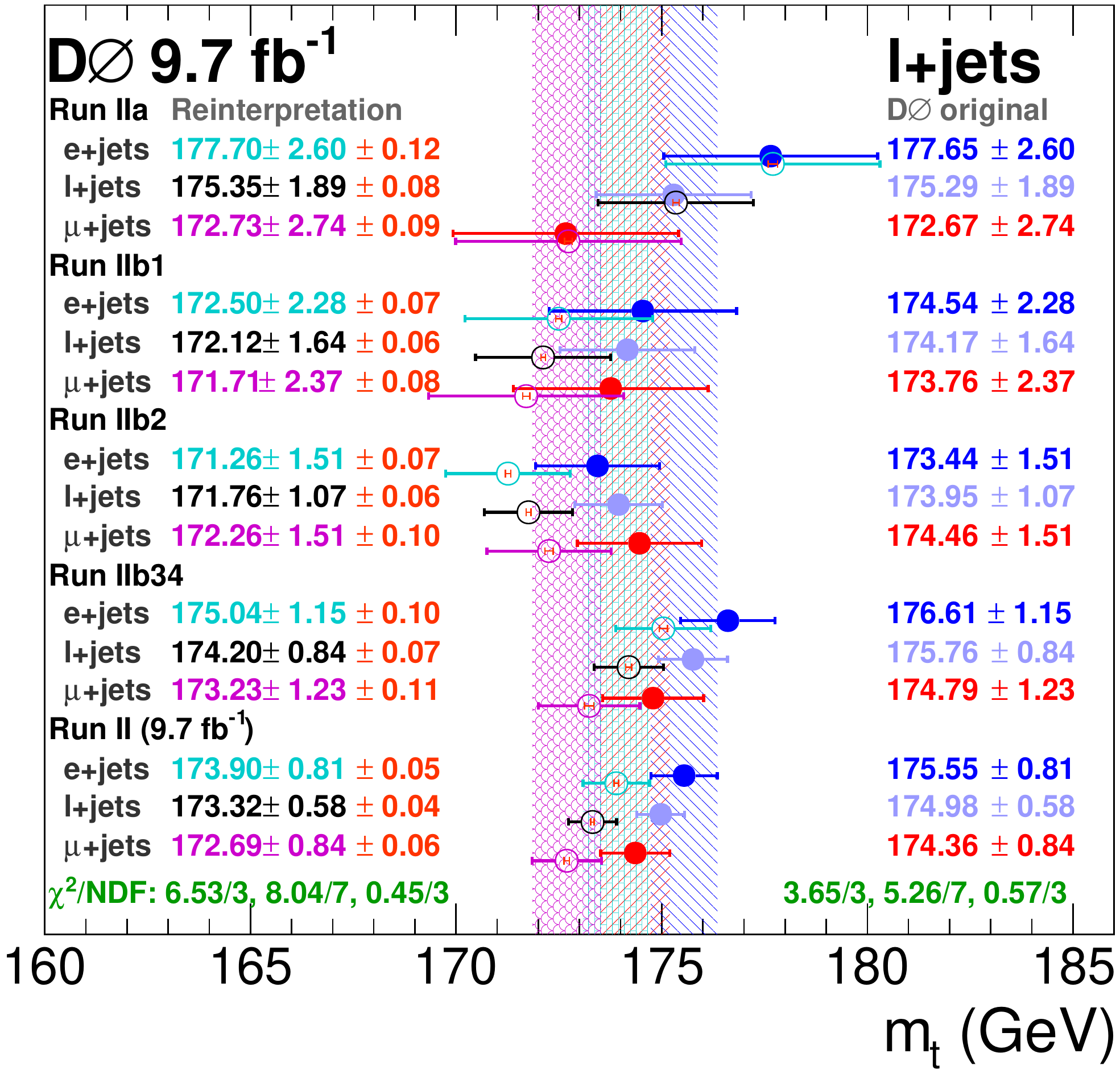}
 \includegraphics[width=0.245\textwidth]{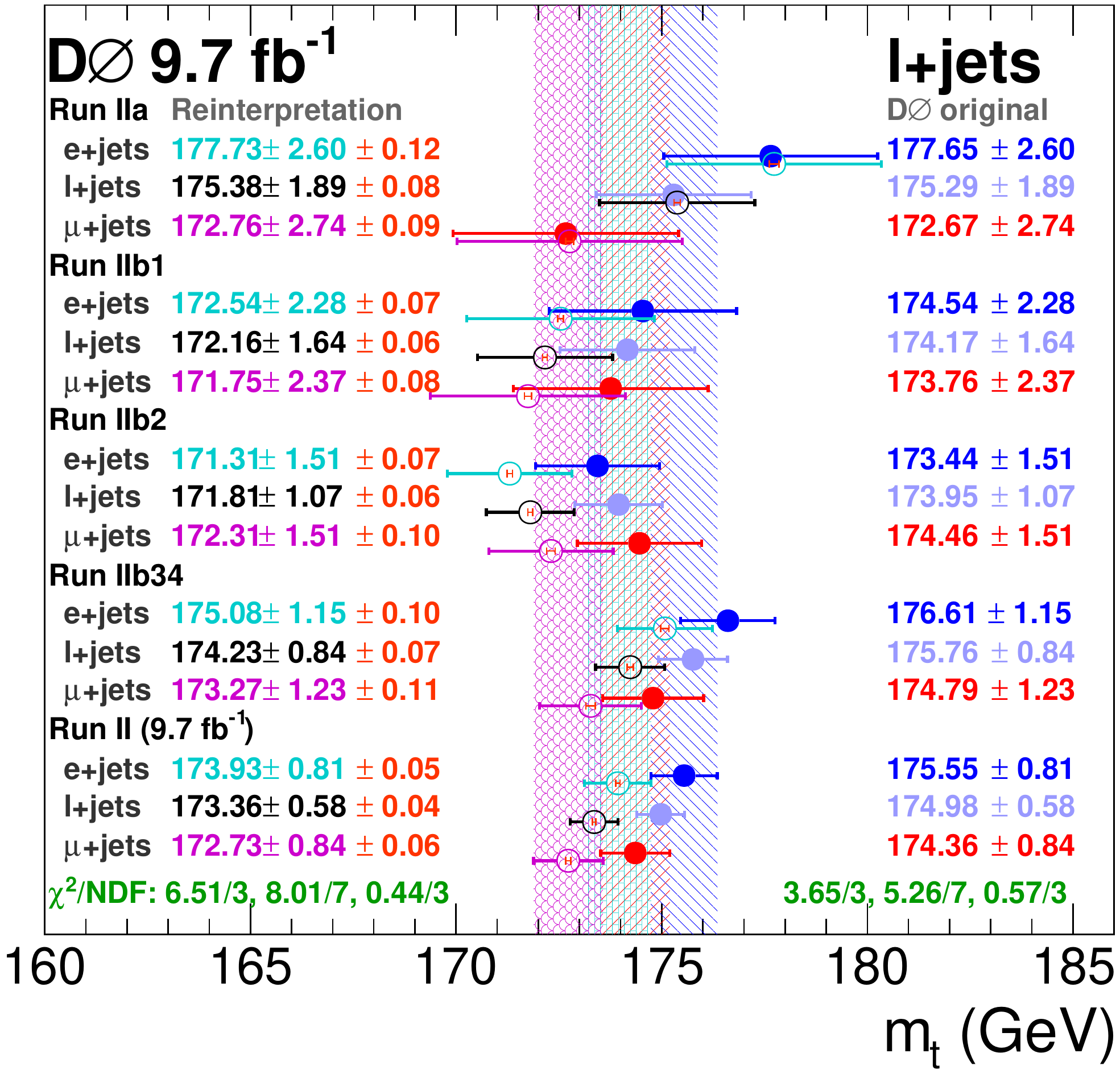}
 \includegraphics[width=0.245\textwidth]{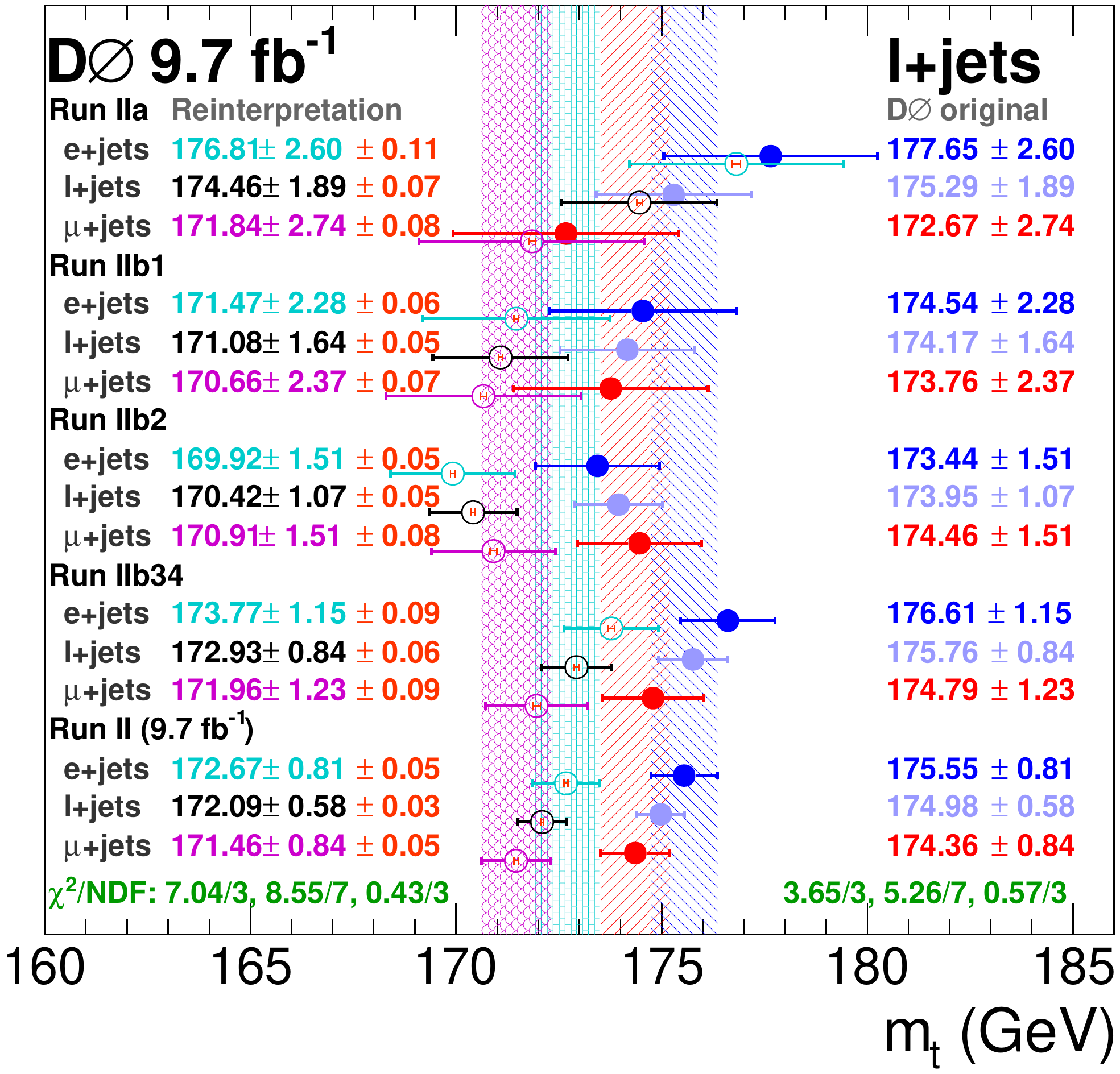}
 \includegraphics[width=0.245\textwidth]{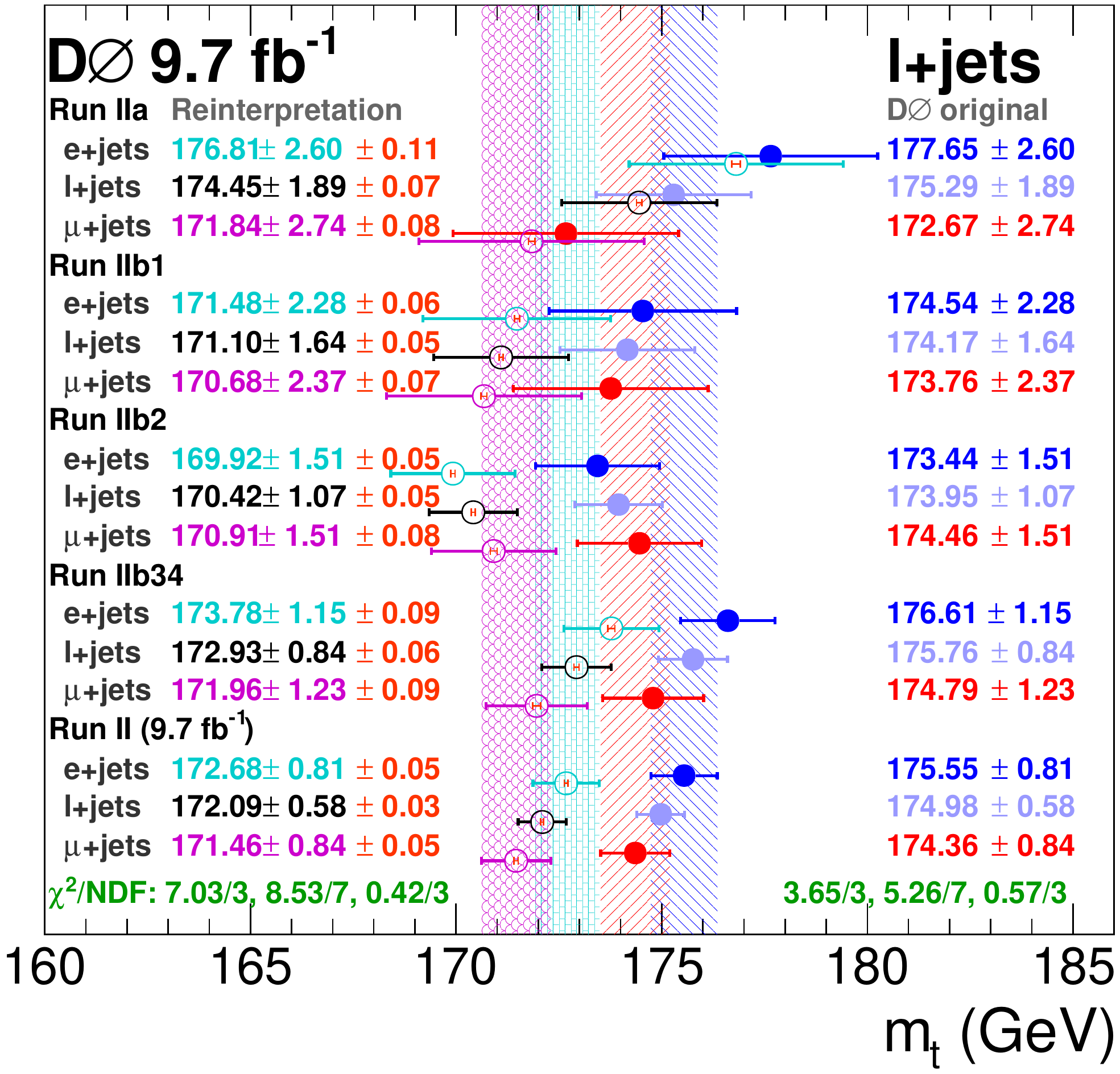}
 \caption{Shifted hadronic top mass values for the average estimator in all the measurements and combinations.
 From left to right: P6 $F_{\text{Corr}}$ set 1 and 2, H7 $F_{\text{Corr}}$ set 1 and 2.}
 \label{fig:hadronic2}
  \vspace{0.2cm}
  
 \includegraphics[width=0.245\textwidth]{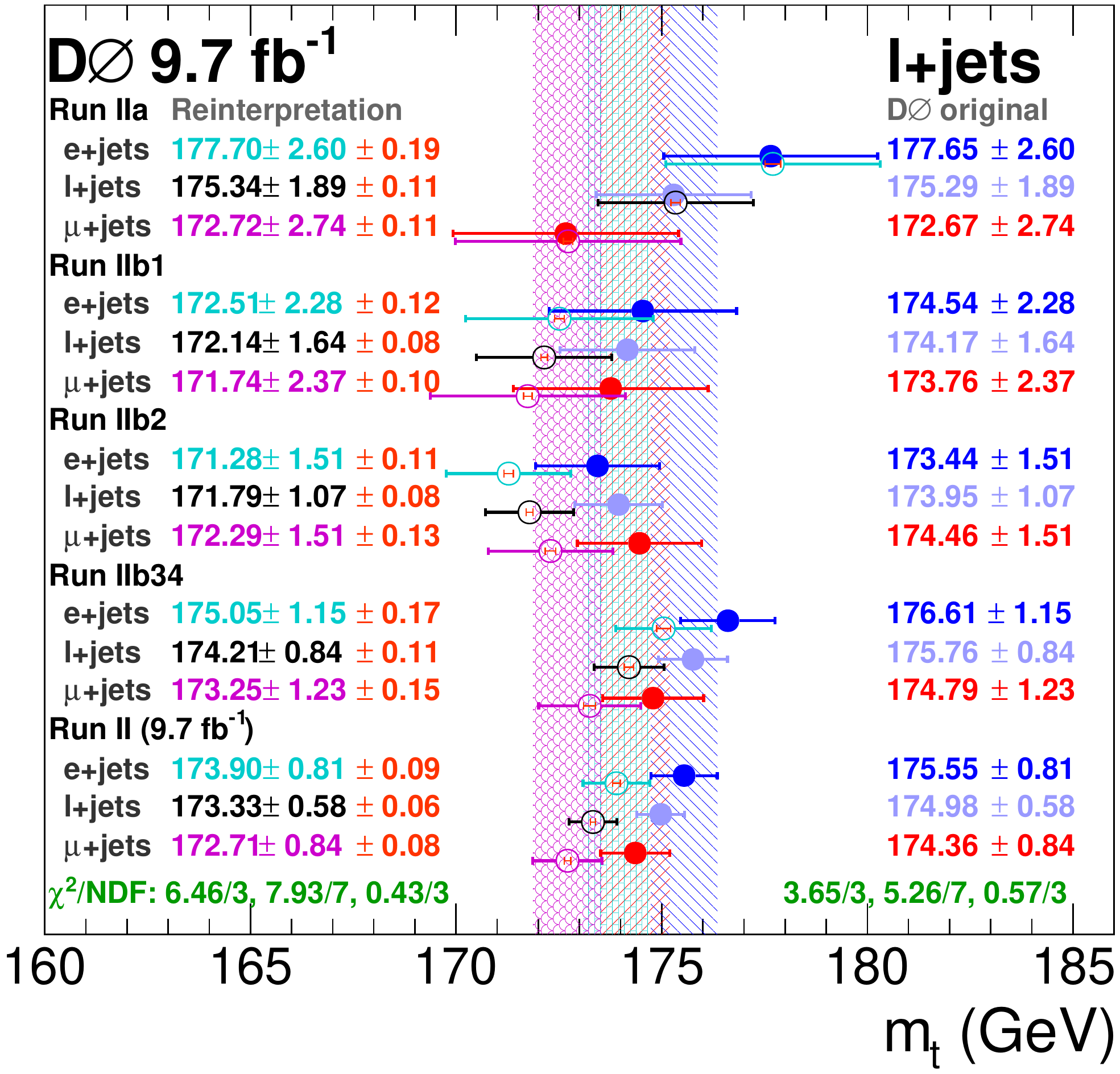}
 \includegraphics[width=0.245\textwidth]{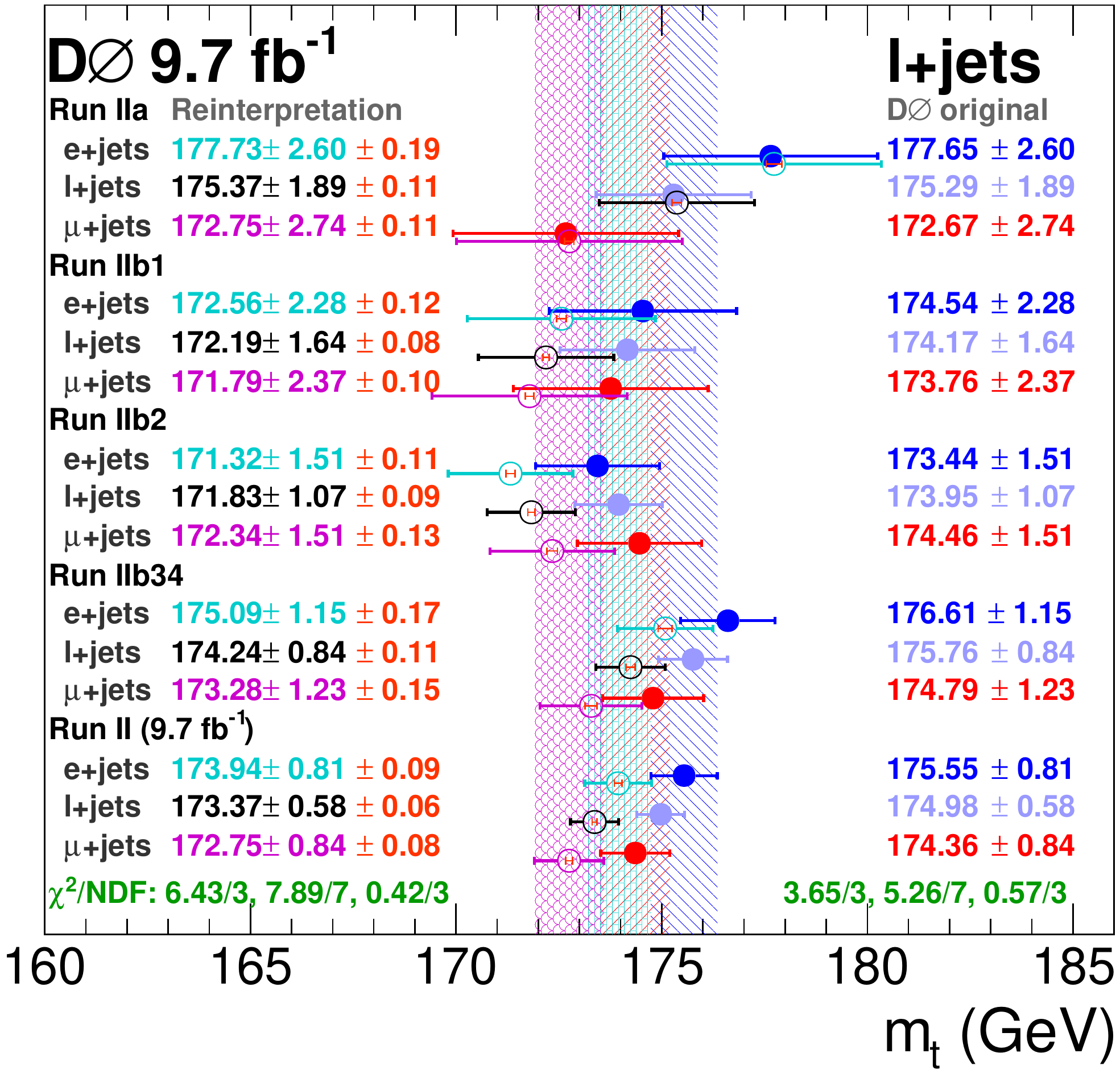}
 \includegraphics[width=0.245\textwidth]{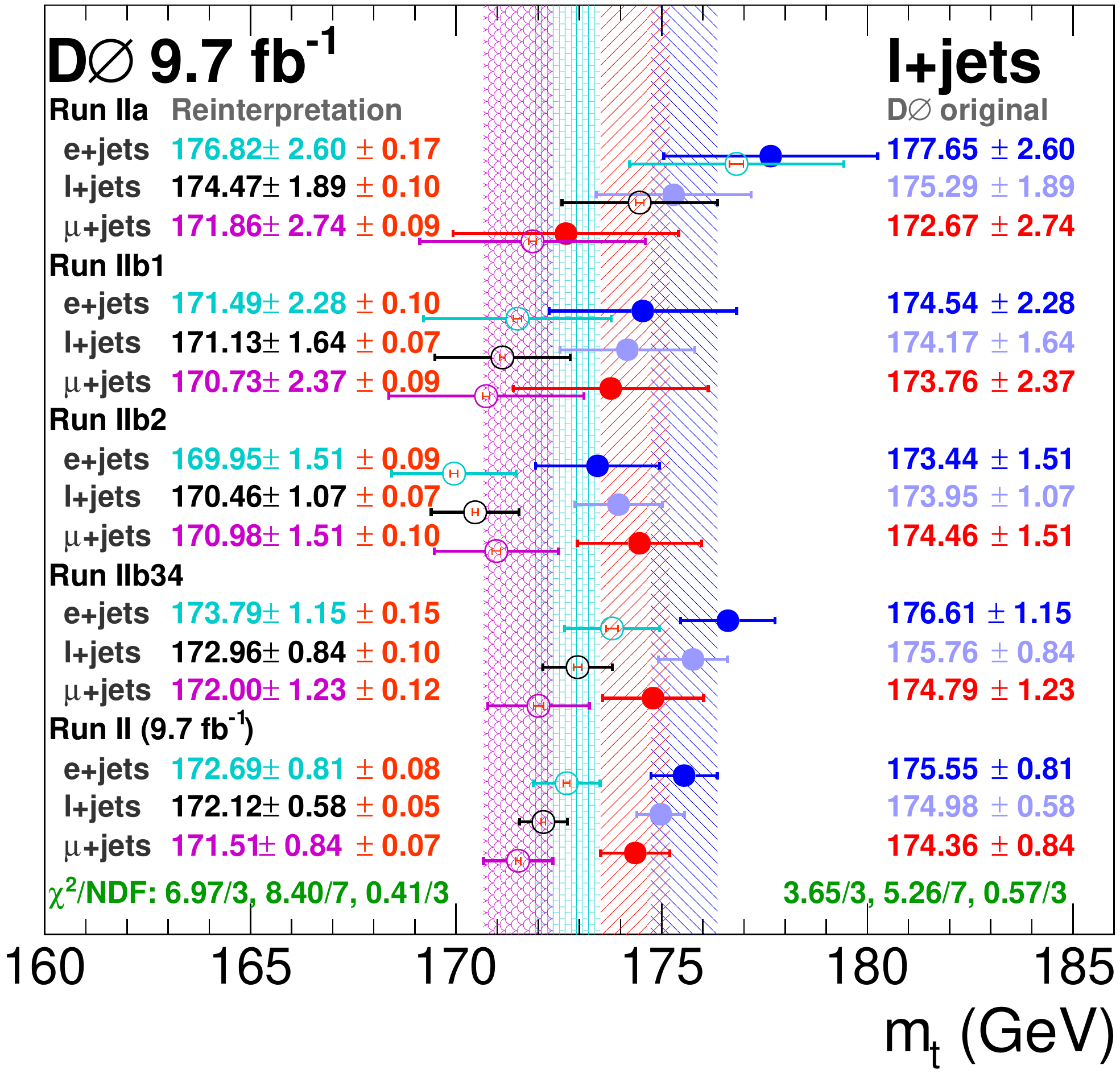}
 \includegraphics[width=0.245\textwidth]{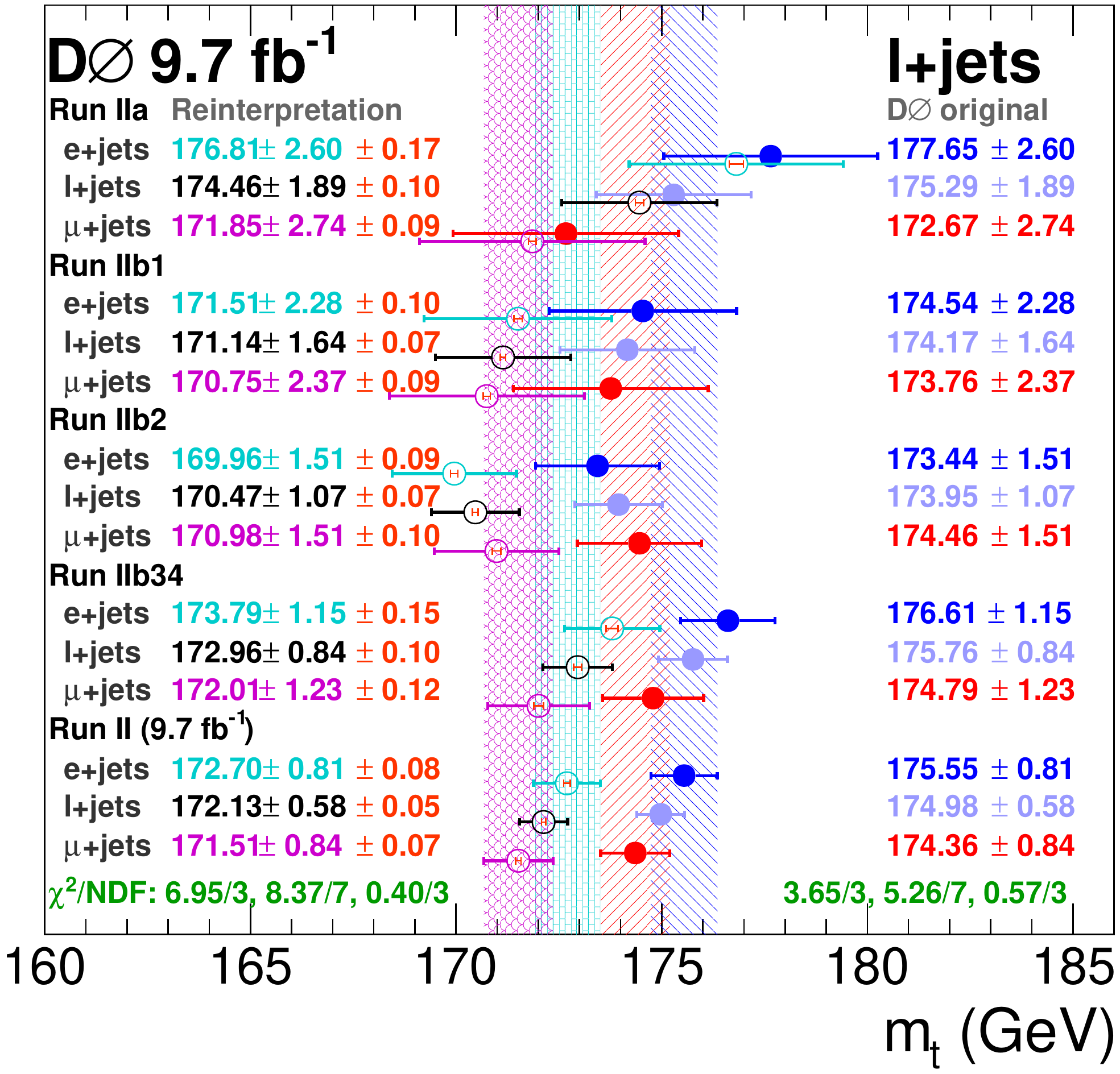}
\caption{Shifted hadronic top mass values for the itg estimator in all the measurements and combinations.
From left to right: P6 $F_{\text{Corr}}$ set 1 and 2, H7 $F_{\text{Corr}}$ set 1 and 2.}
\label{fig:hadronic3}
\vspace{0.2cm}

 \includegraphics[width=0.245\textwidth]{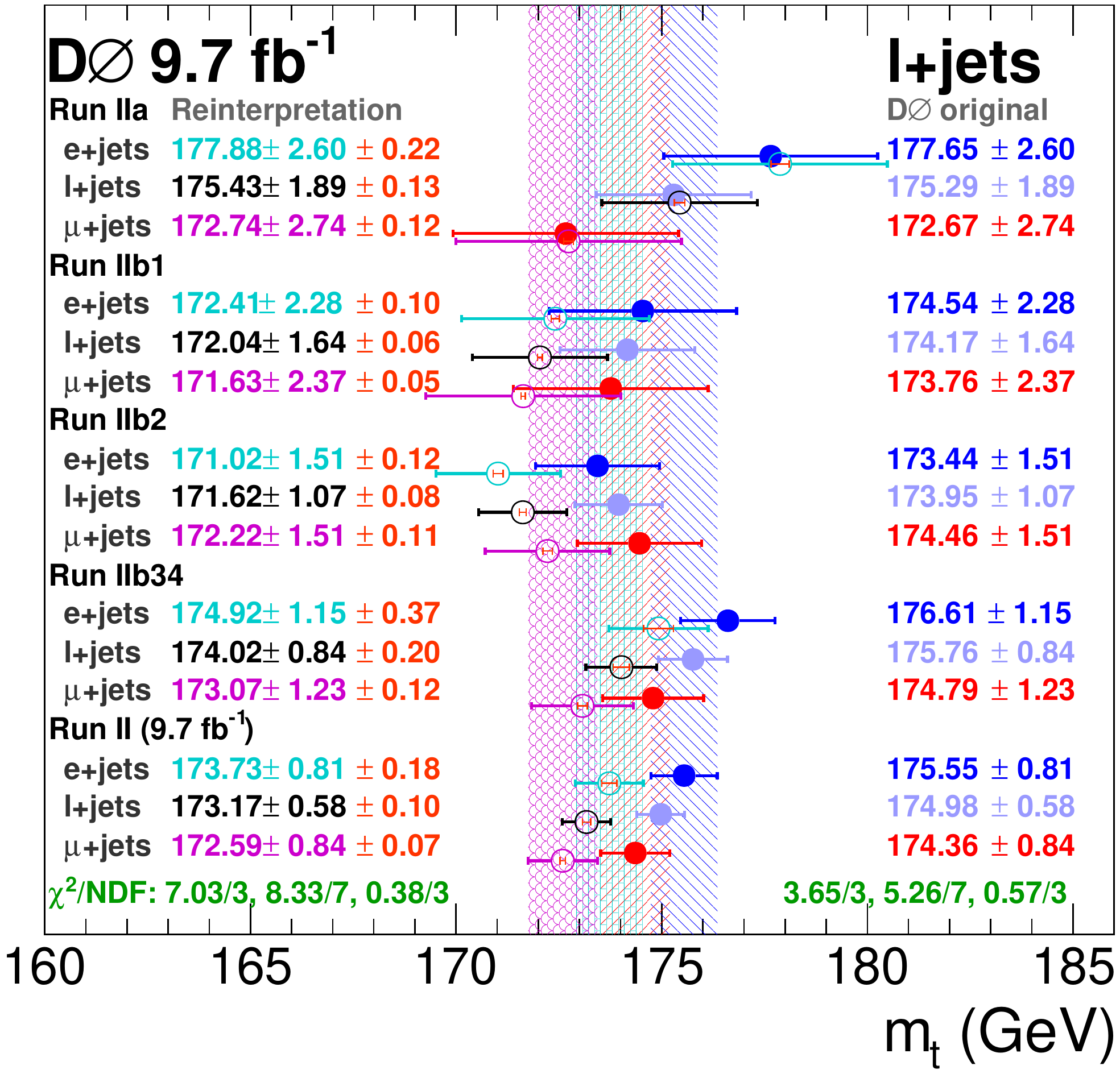}
 \includegraphics[width=0.245\textwidth]{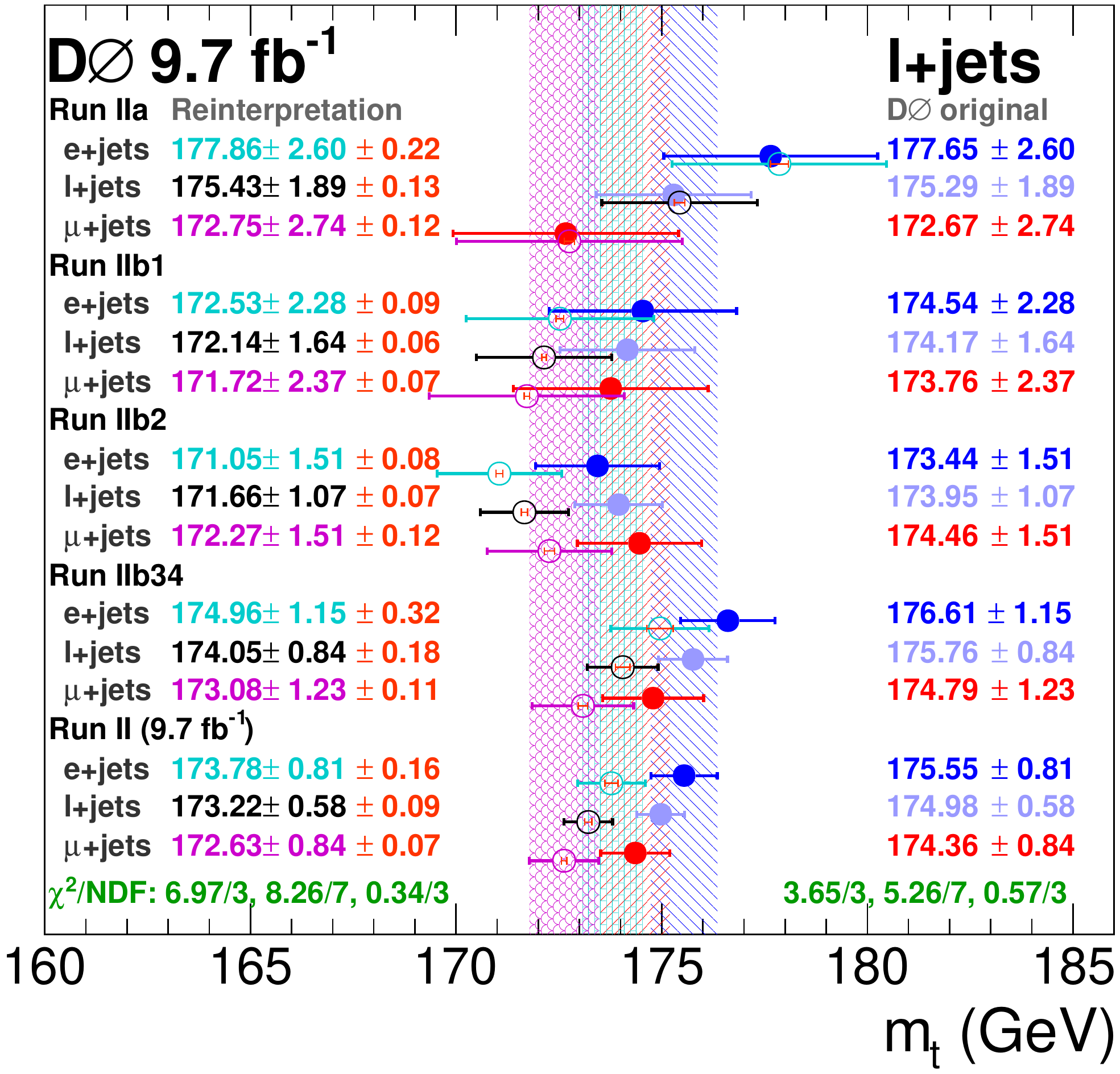}
 \includegraphics[width=0.245\textwidth]{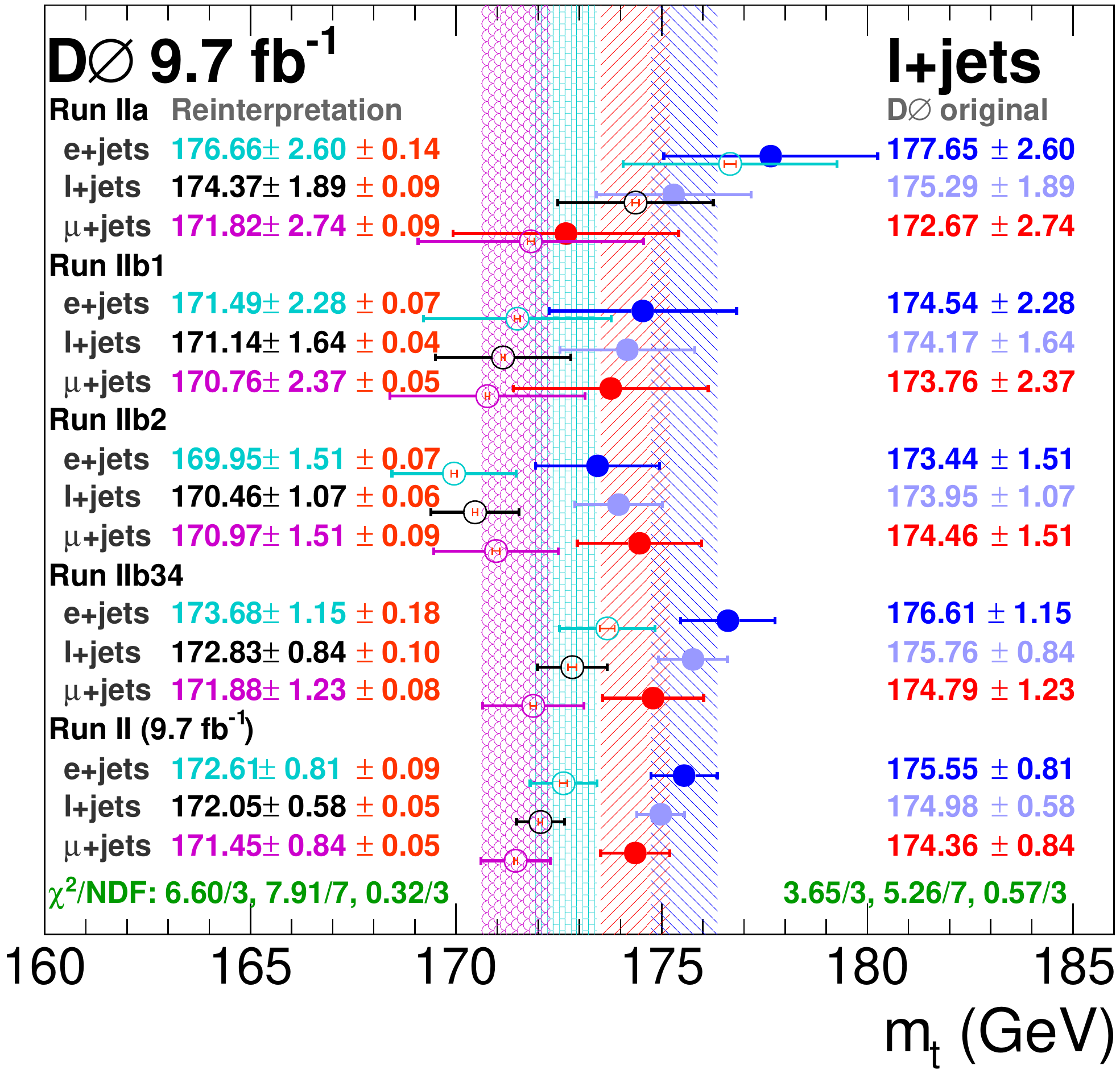}
 \includegraphics[width=0.245\textwidth]{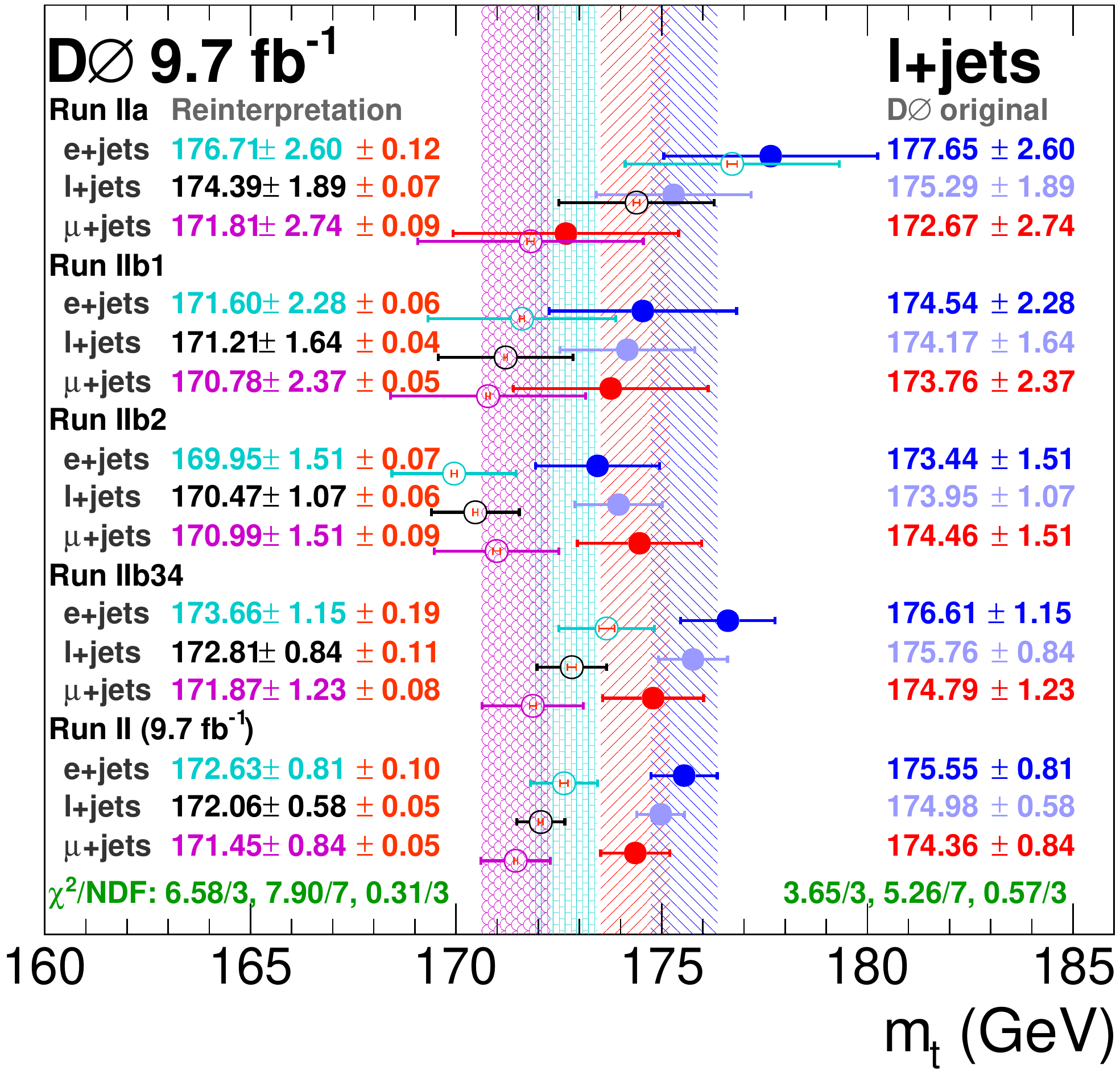}
 \caption{Shifted hadronic top mass values for the fit estimator in all the measurements and combinations.
 From left to right: P6 $F_{\text{Corr}}$ set 1 and 2, H7 $F_{\text{Corr}}$ set 1 and 2.}
 \label{fig:hadronic4}
 \vspace{0.2cm}

 \includegraphics[width=0.245\textwidth]{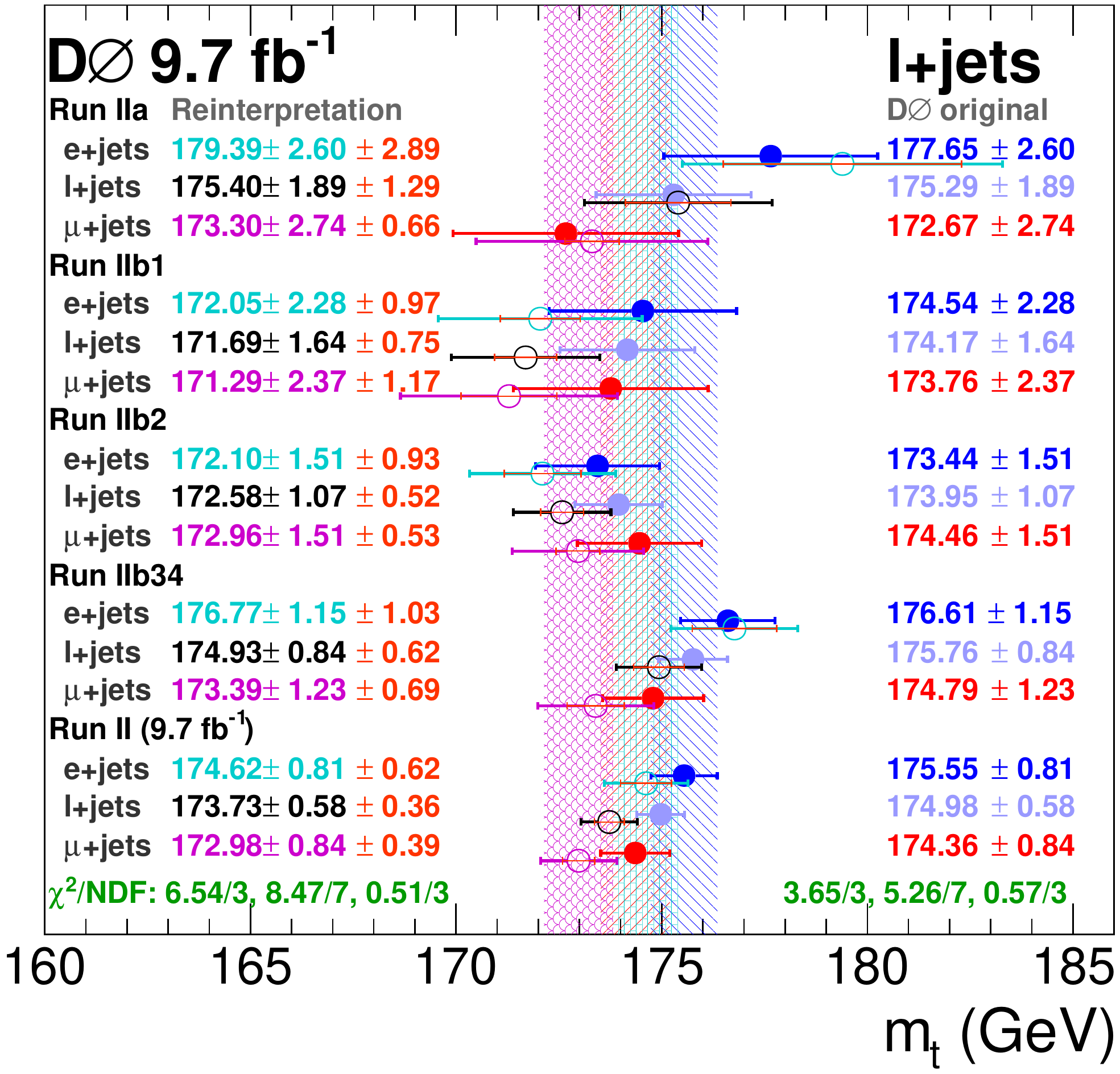}
 \includegraphics[width=0.245\textwidth]{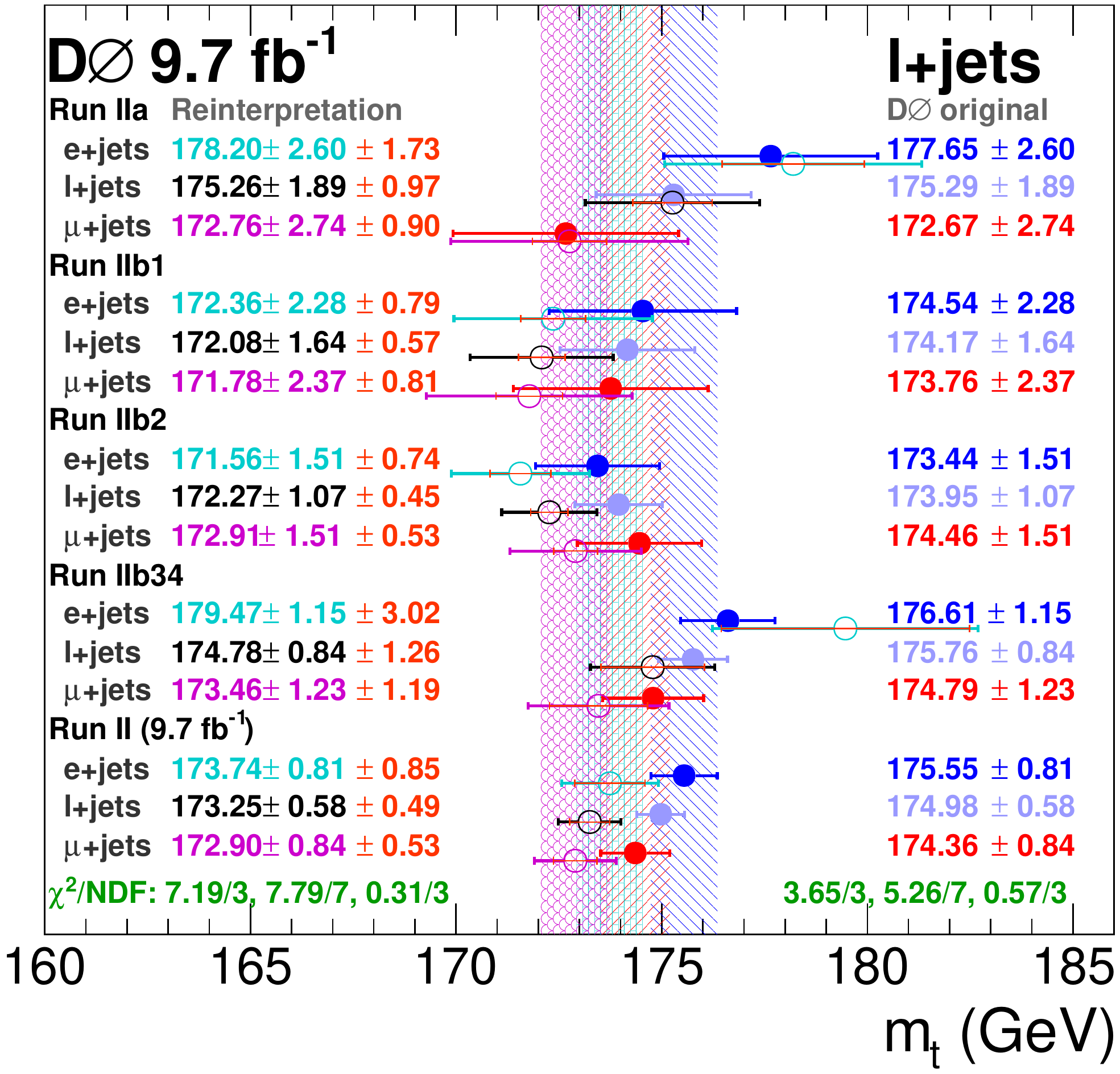}
 \includegraphics[width=0.245\textwidth]{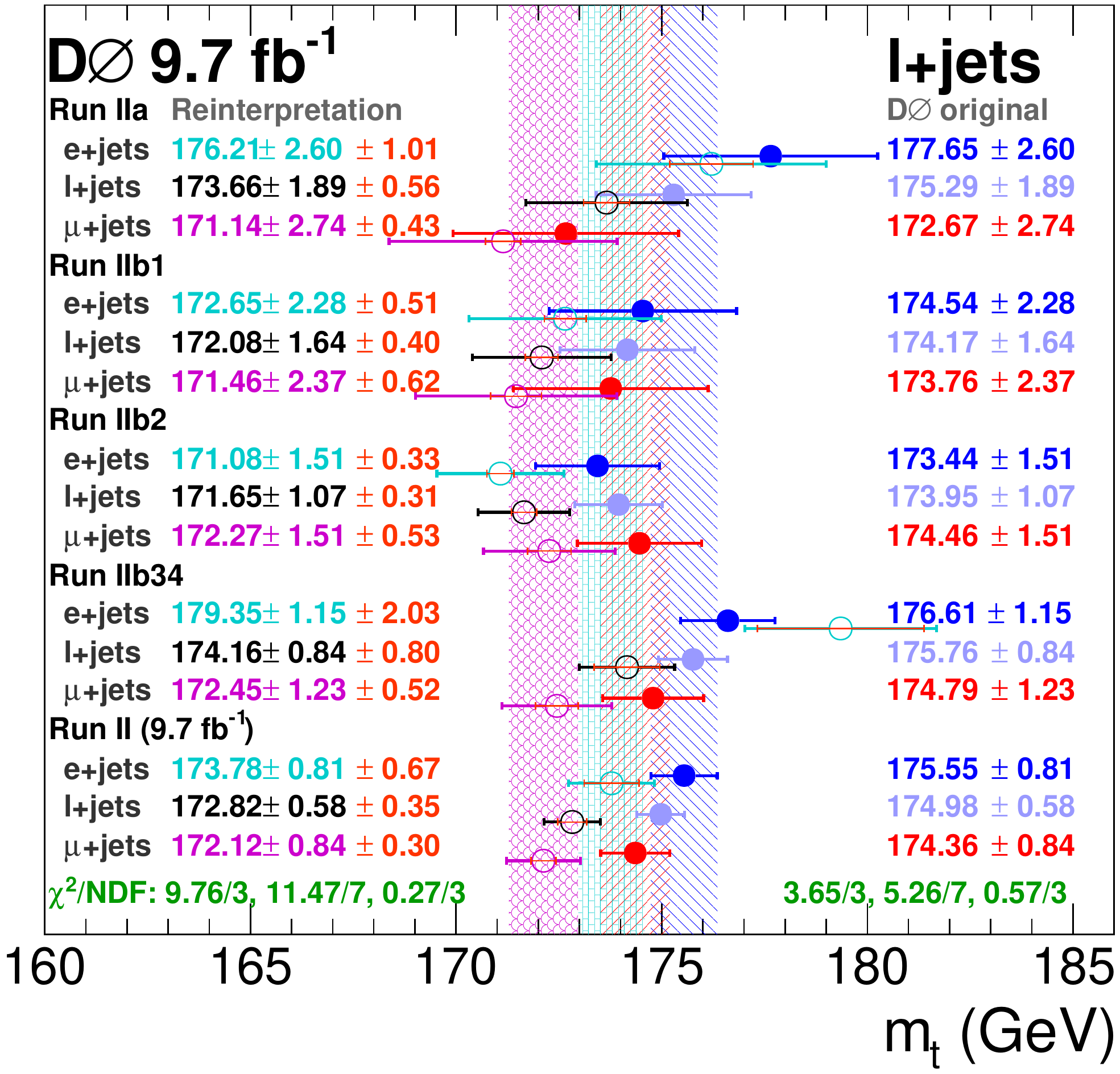}
 \includegraphics[width=0.245\textwidth]{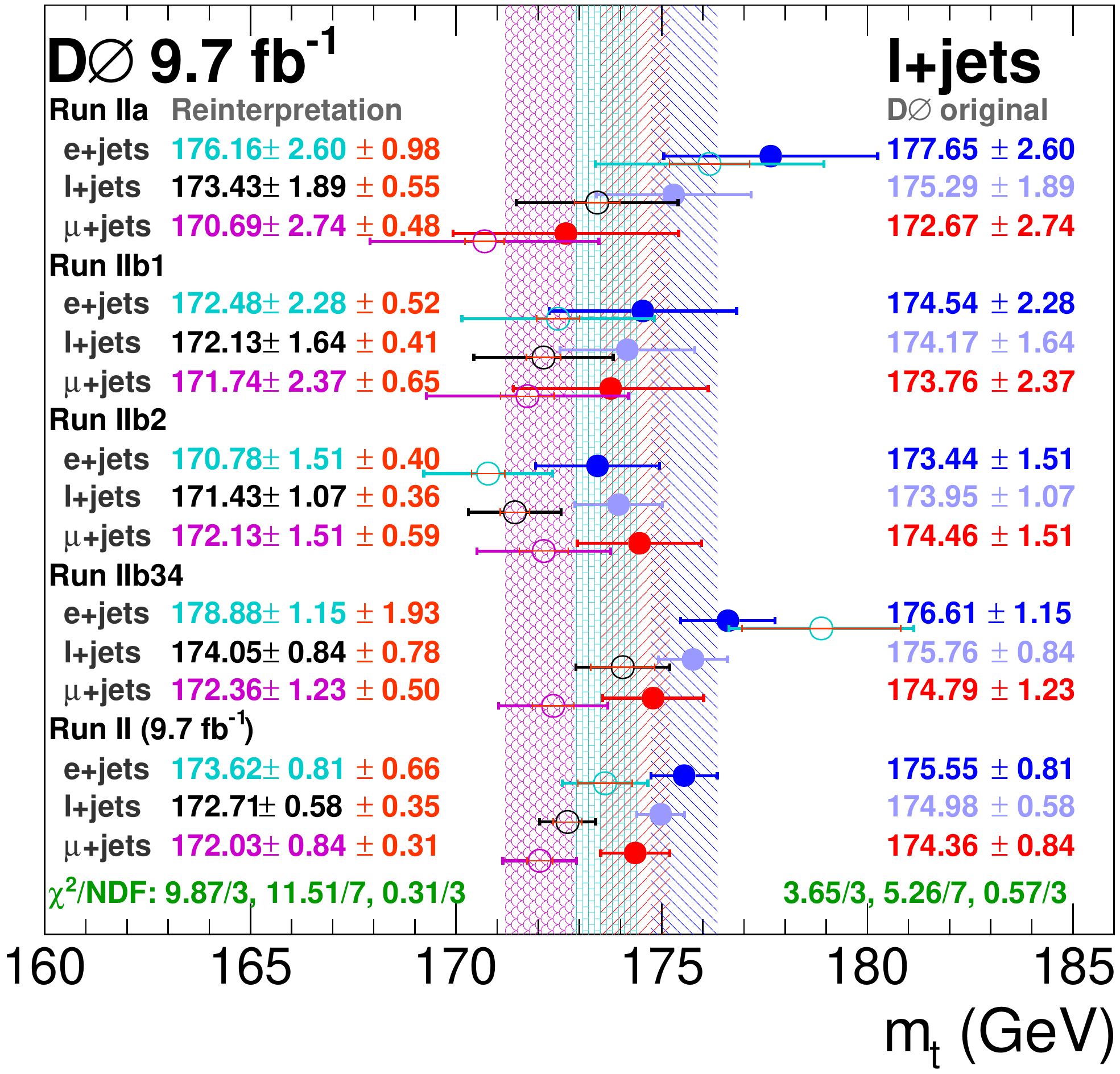}
 \caption{Shifted hadronic top mass values for the maximum estimator in all the measurements and combinations.
 From left to right: P6 $F_{\text{Corr}}$ set 1 and 2, H7 $F_{\text{Corr}}$ set 1 and 2.}
 \label{fig:hadronic5}
\end{figure}

In Figs.~\ref{fig:hadronic1},\ref{fig:hadronic2},\ref{fig:hadronic3},\ref{fig:hadronic4},\ref{fig:hadronic5}, the shift results are given for all the resonance position estimators.
These are ordered from the smallest method error to the largest one.
The method error of the shifted results is displayed in an orange font, as an addition to the original D\O\ error.
The method errors have mostly a small impact on the total error.
The results are given for both P6 and H7, and the two $F_{\text{Corr}}$ parameter sets.
The latter have little impact on the results, but H7 shows generally a greater shift that P6.

\subsection{Leptonic Top Resonance Measurement}

The measurement of the top mass shift in the leptonic channel is based on the leptonic top resonance histograms.
All the 963 histograms do not fit in this document, but Appendix.~\ref{app:hlexamples} shows some example cases of these.

\begin{figure}[H]
 \includegraphics[width=0.245\textwidth]{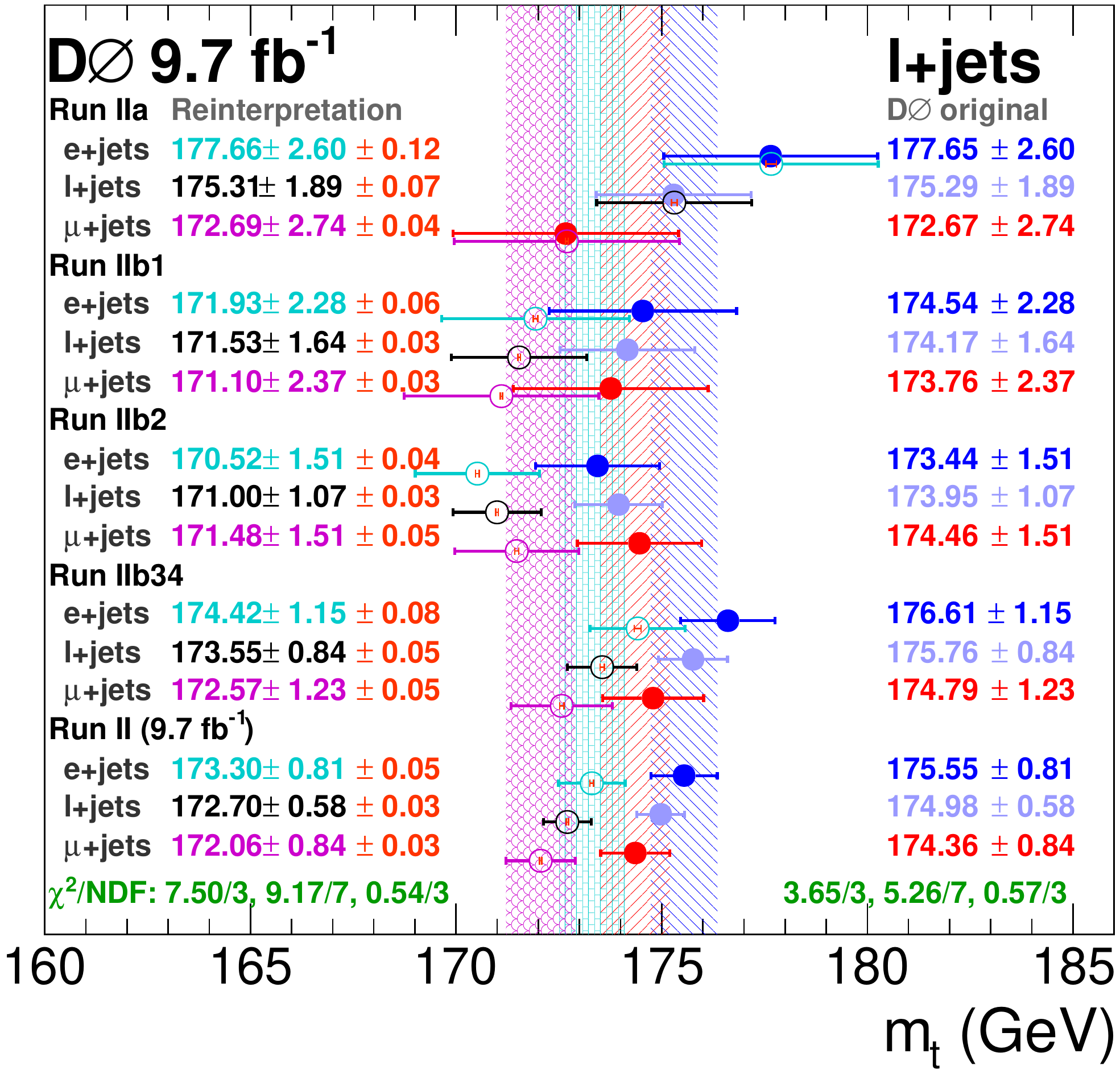}
 \includegraphics[width=0.245\textwidth]{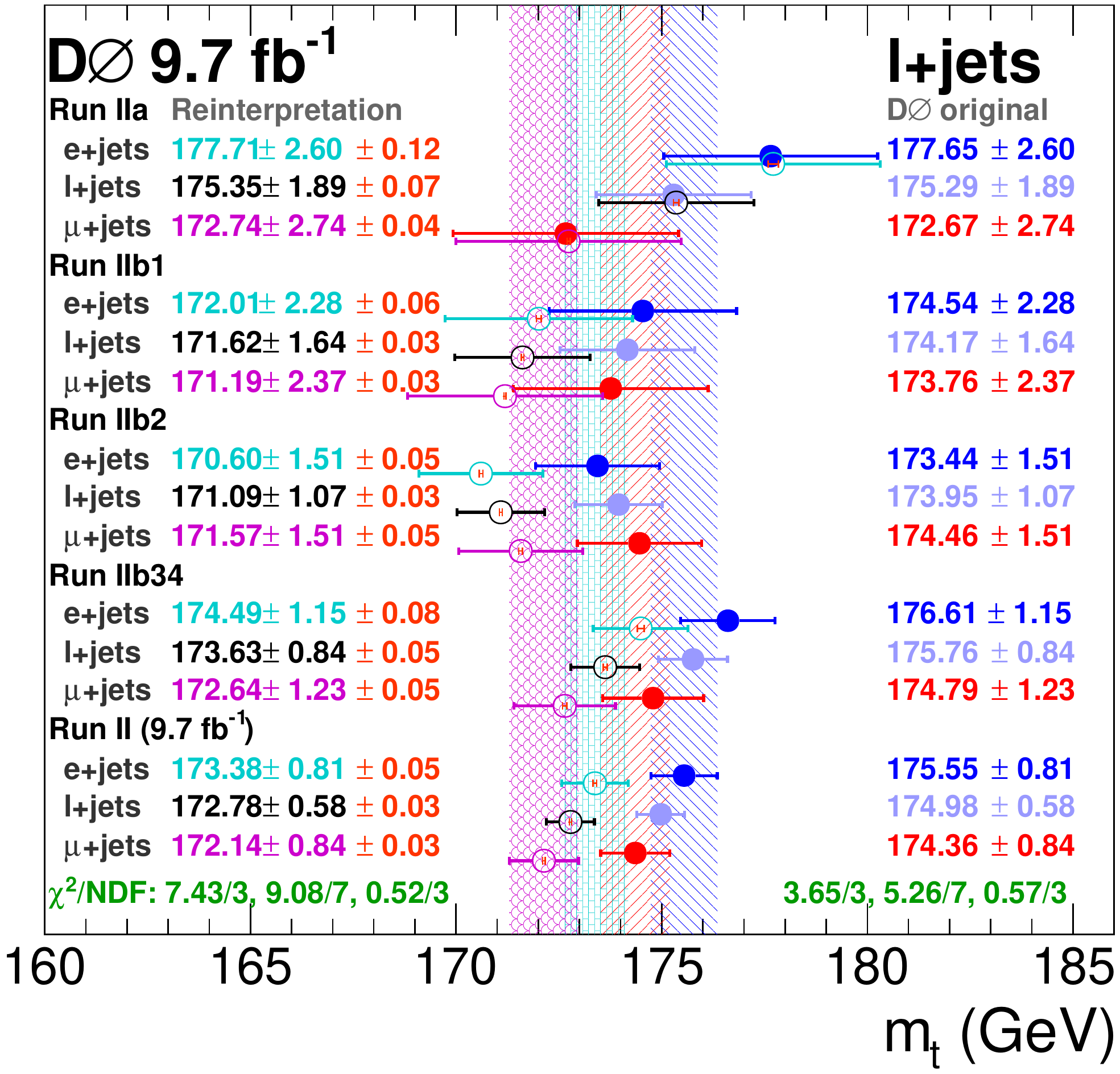}
 \includegraphics[width=0.245\textwidth]{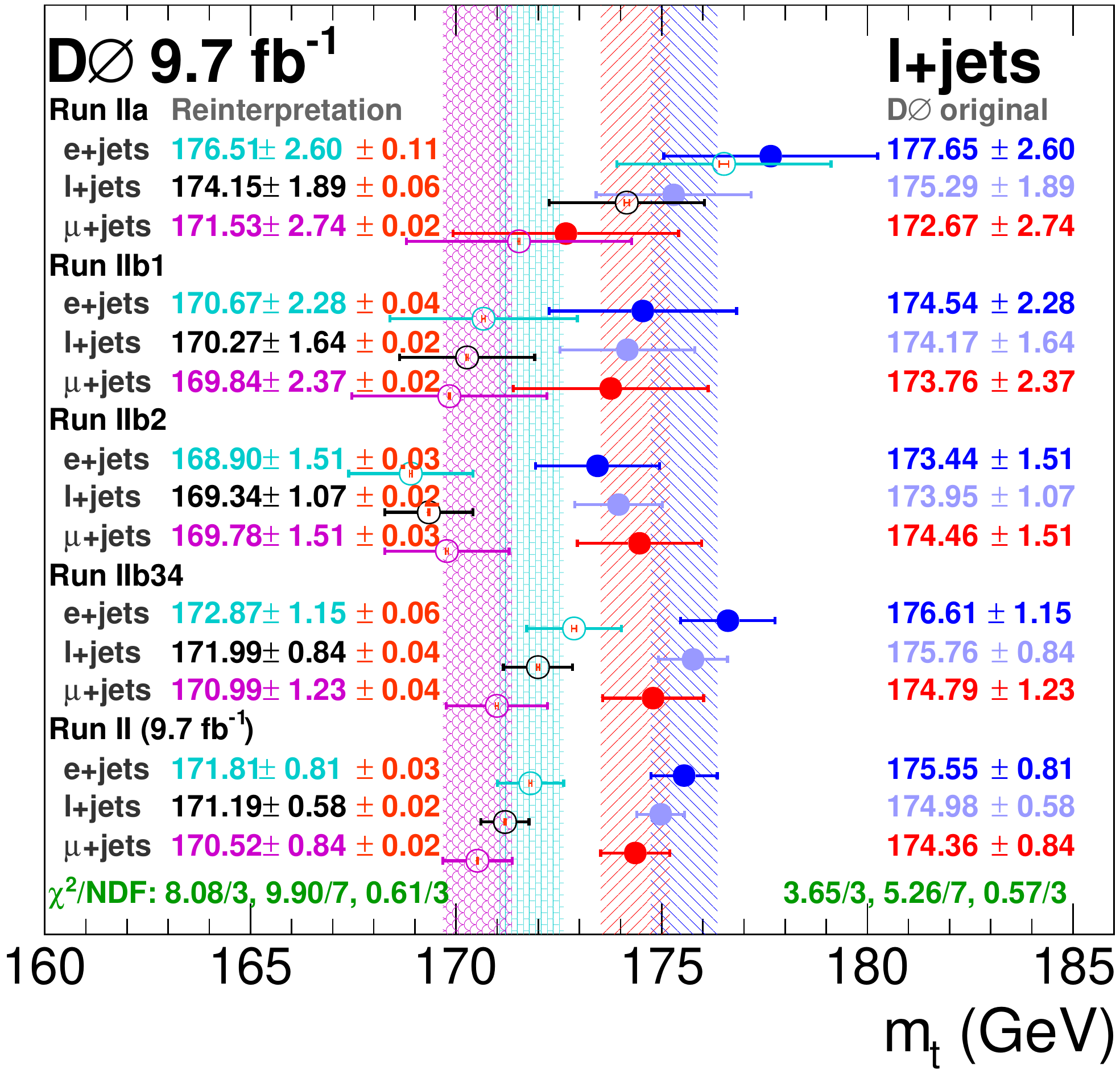}
 \includegraphics[width=0.245\textwidth]{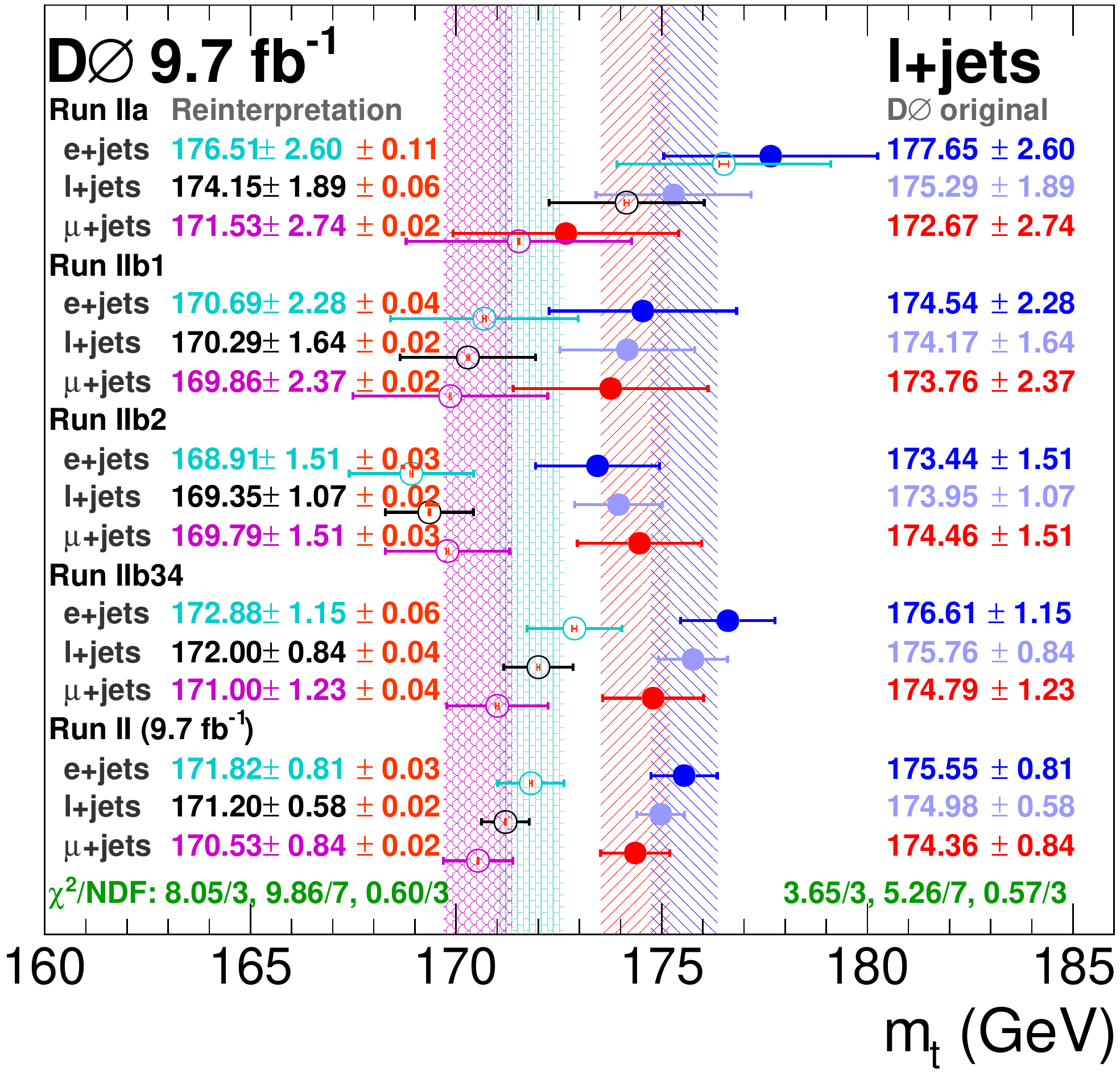}
 \caption{Shifted leptonic top mass values for the median estimator in all the measurements and combinations. P6-based $F_{\text{Corr}}$ shift on the left and H7-based on the right.}
  \label{fig:leptonic1}
   \vspace{0.2cm}
 \includegraphics[width=0.245\textwidth]{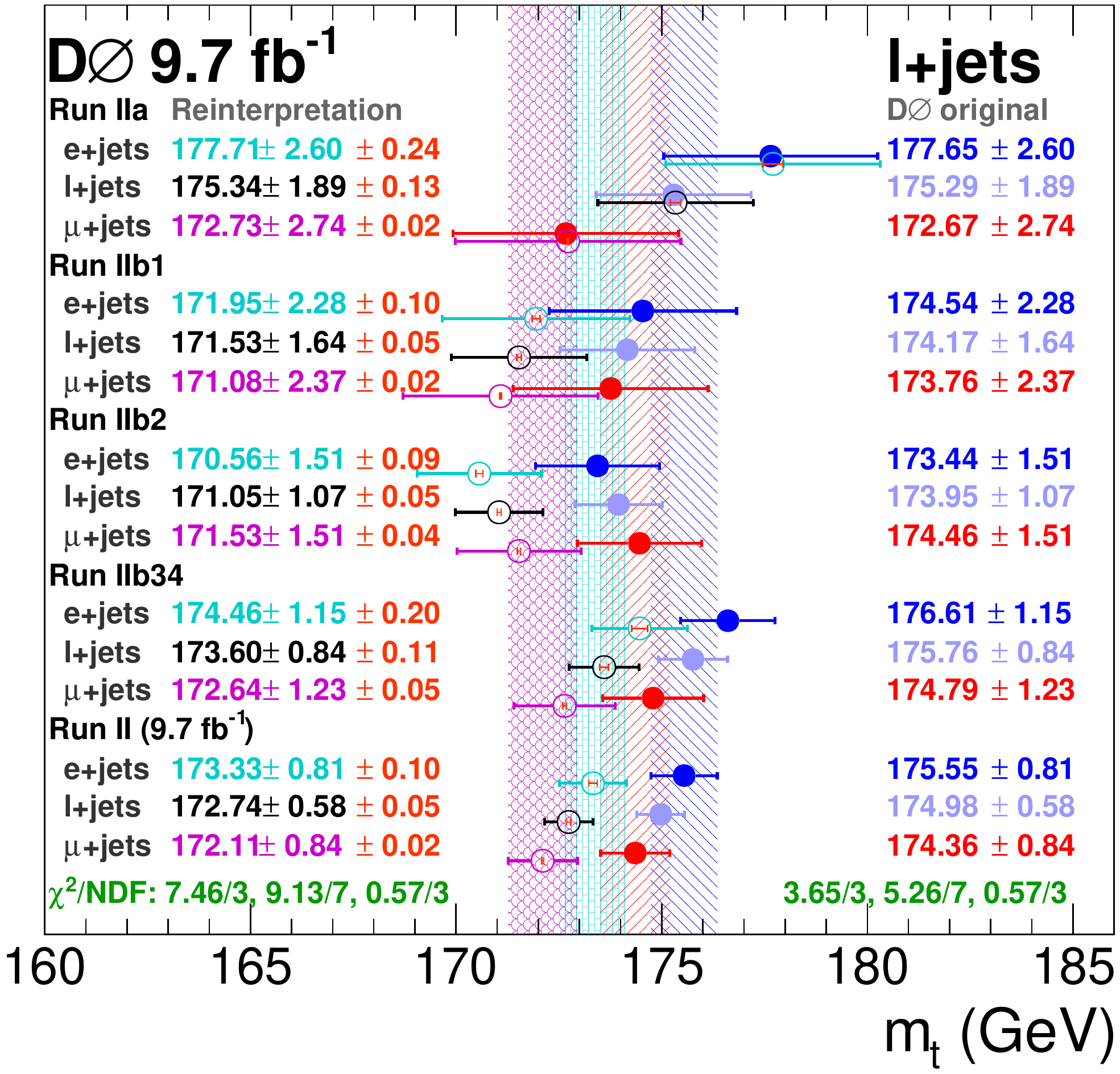}
 \includegraphics[width=0.245\textwidth]{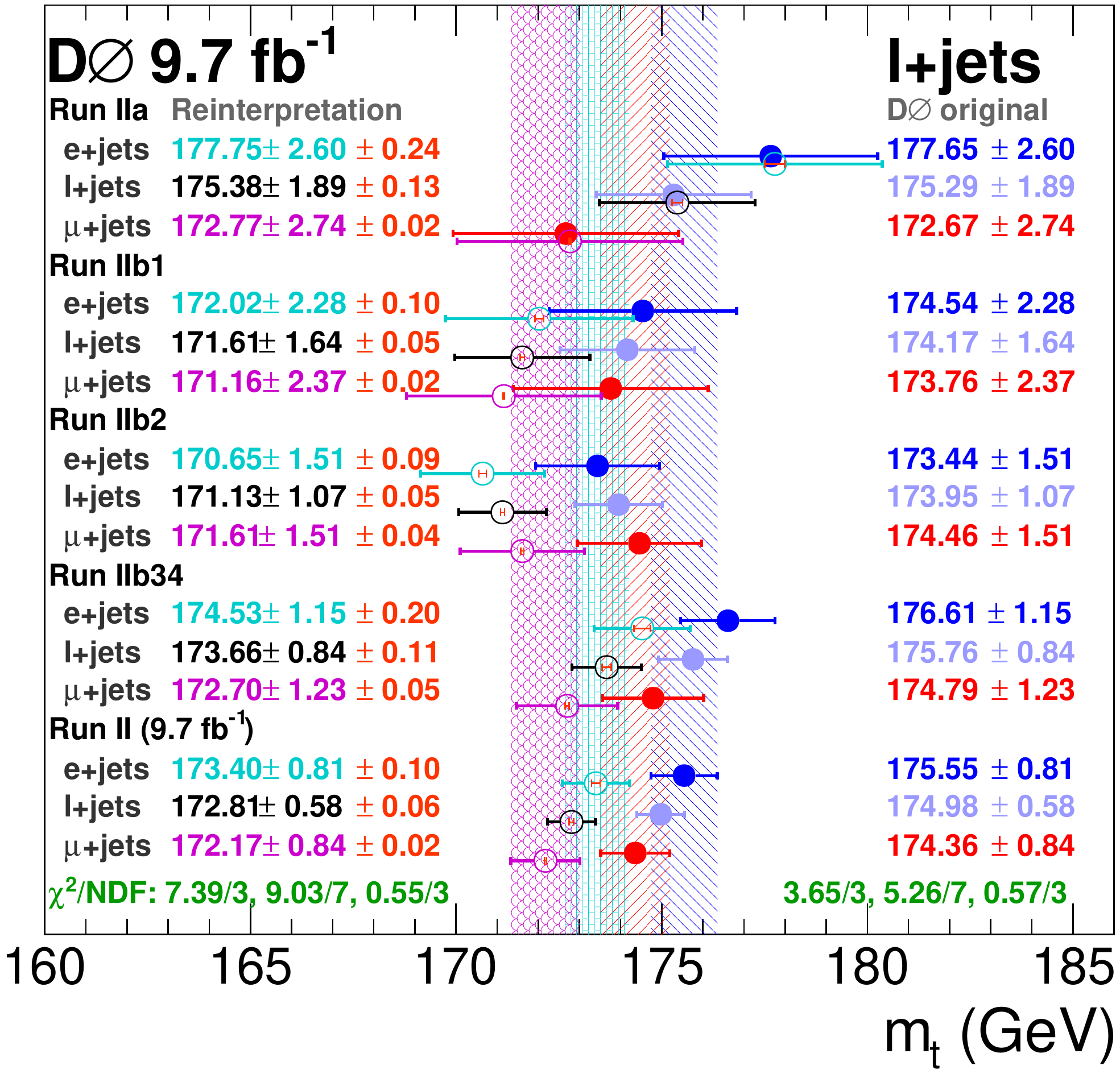}
 \includegraphics[width=0.245\textwidth]{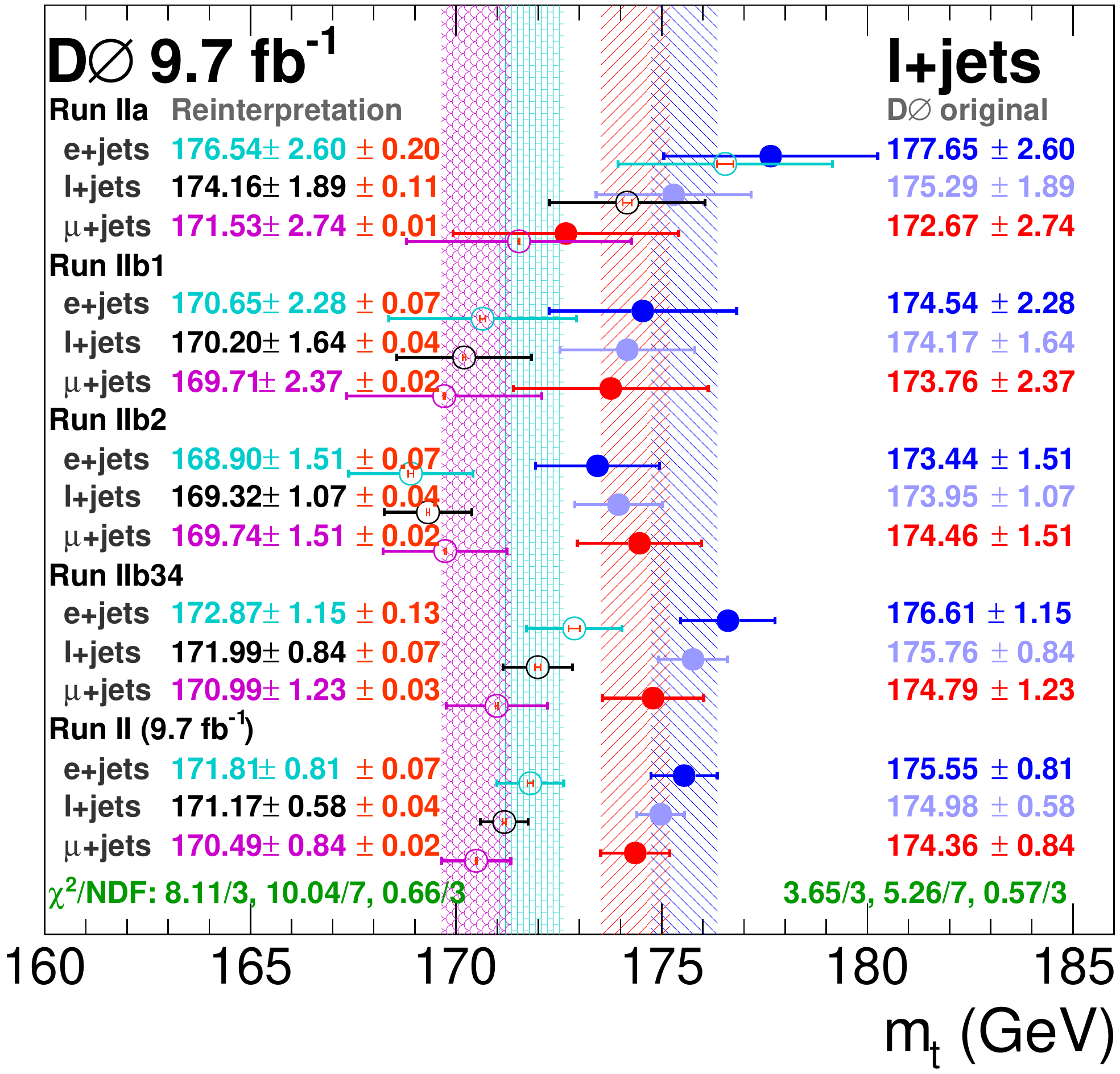}
 \includegraphics[width=0.245\textwidth]{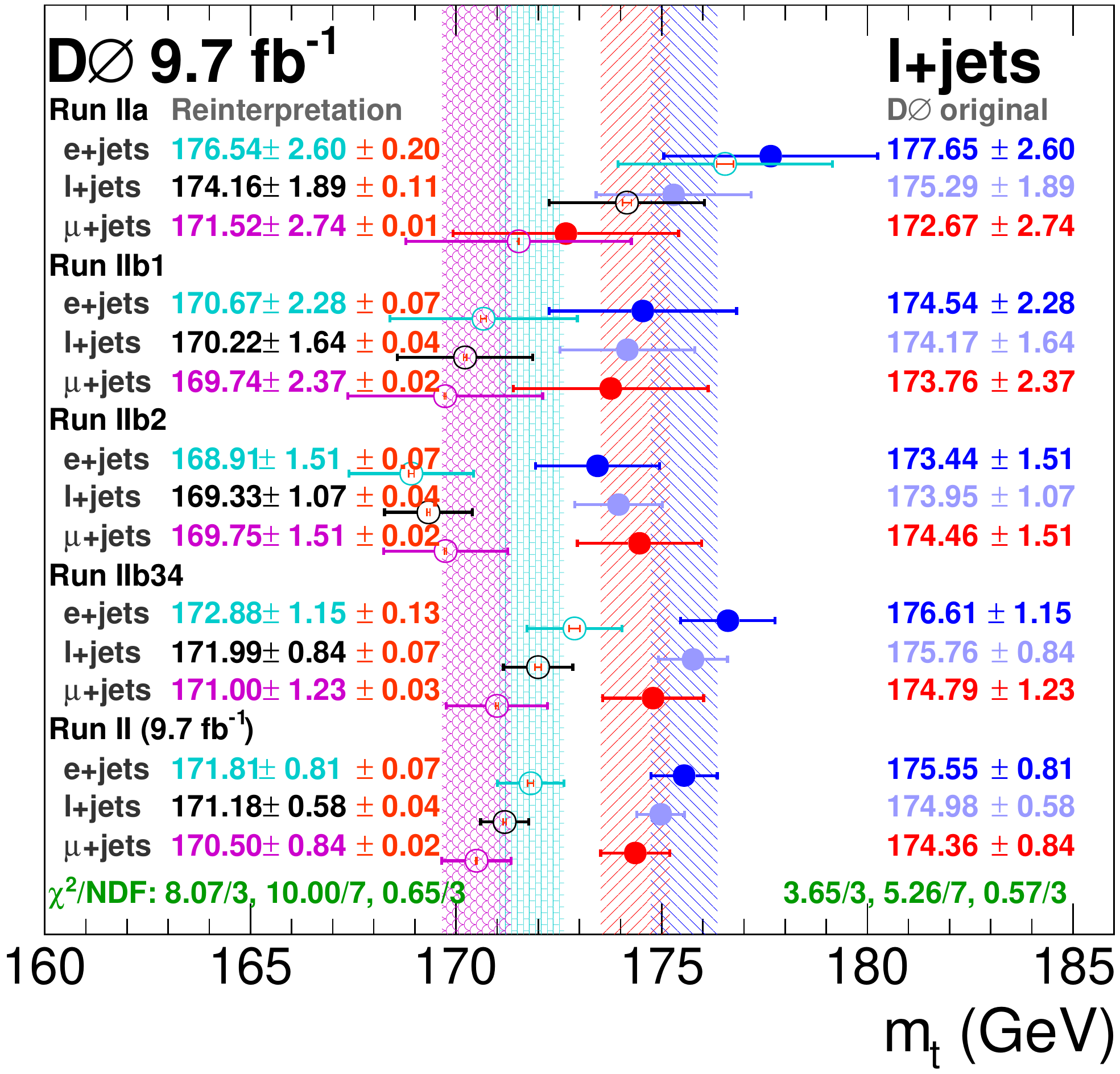}
 \caption{Shifted leptonic top mass values for the average estimator in all the measurements and combinations. P6-based $F_{\text{Corr}}$ shift on the left and H7-based on the right.}
 \label{fig:leptonic2}
    \vspace{0.2cm}
 \includegraphics[width=0.245\textwidth]{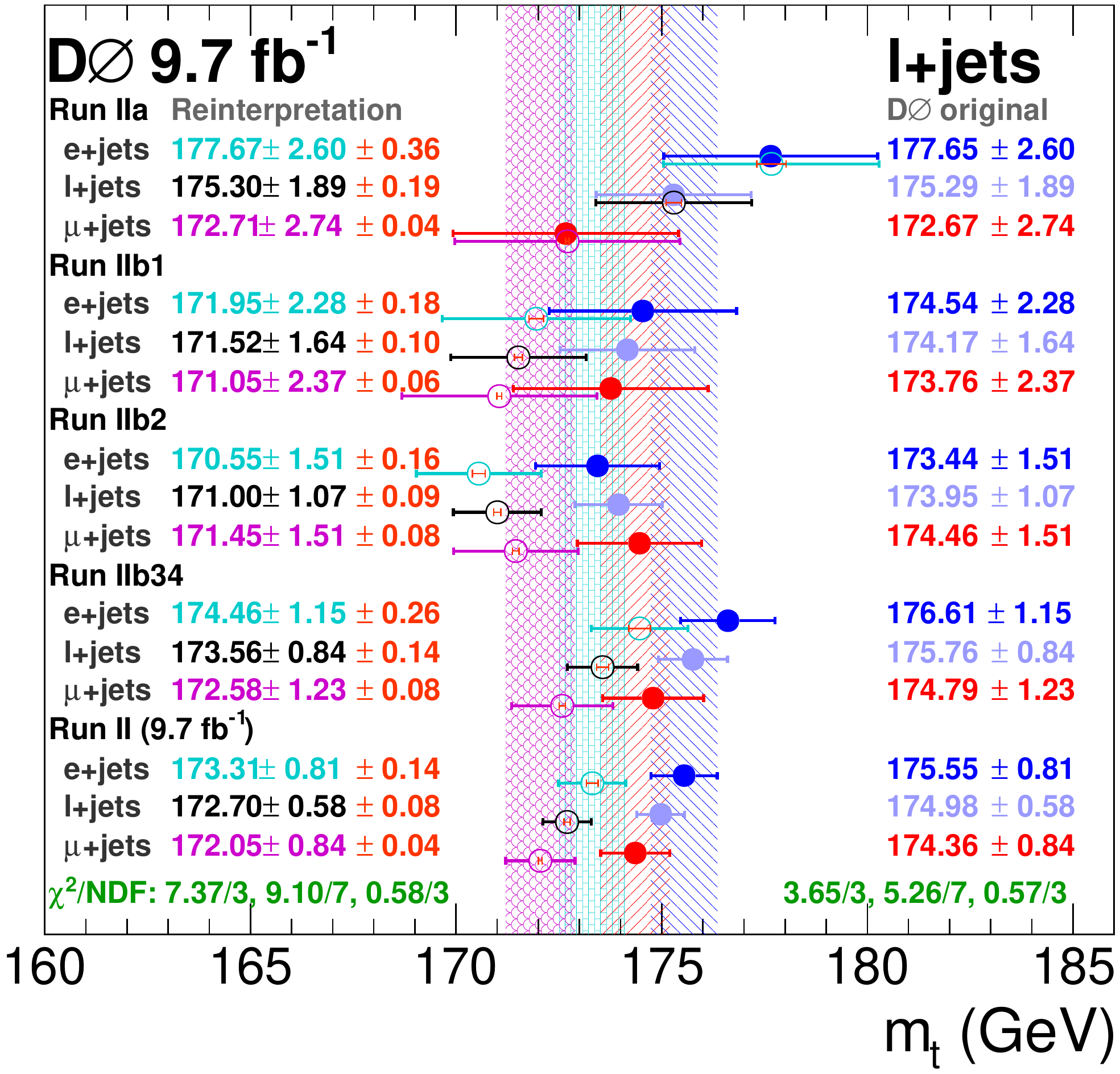}
 \includegraphics[width=0.245\textwidth]{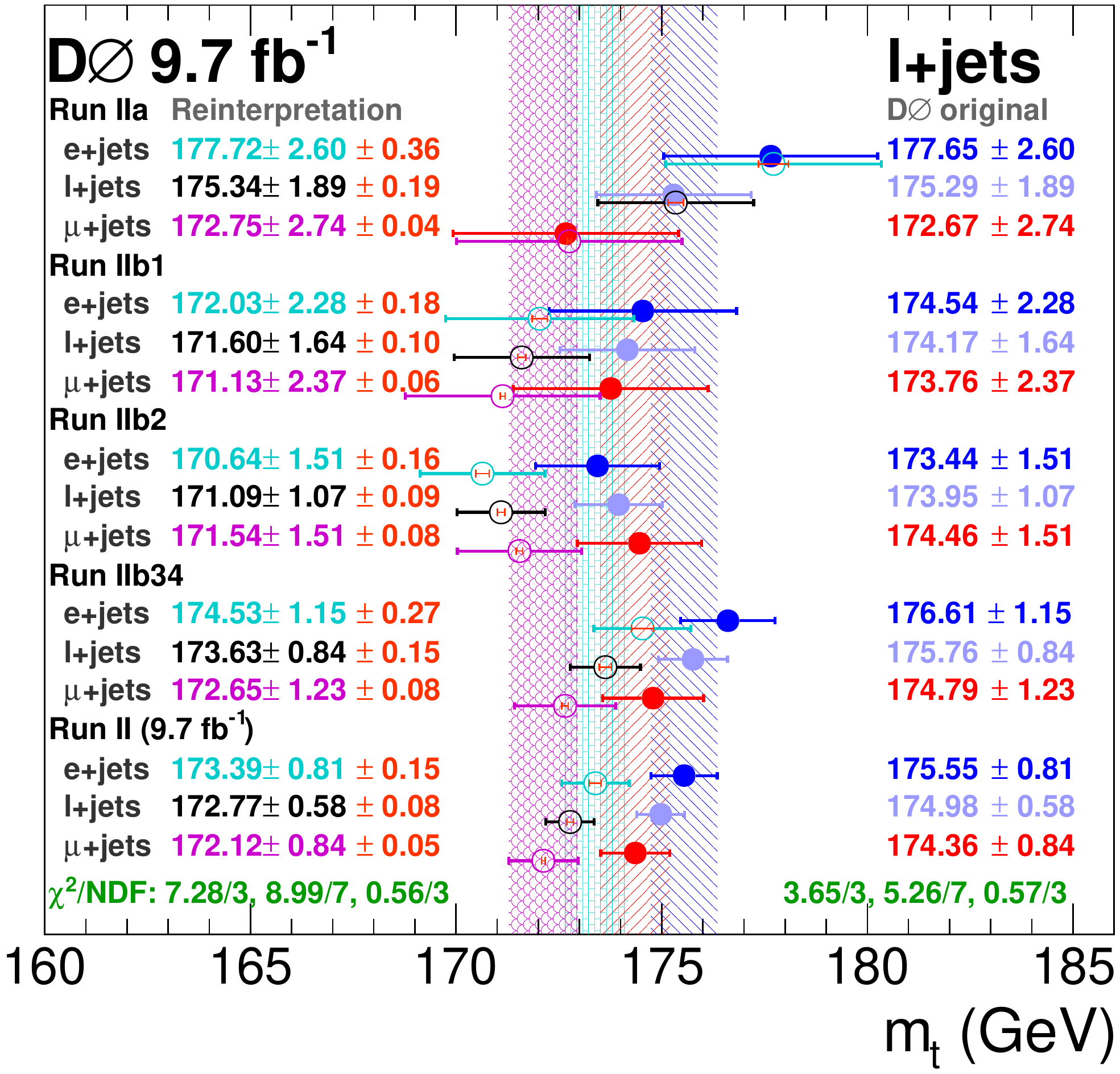}
 \includegraphics[width=0.245\textwidth]{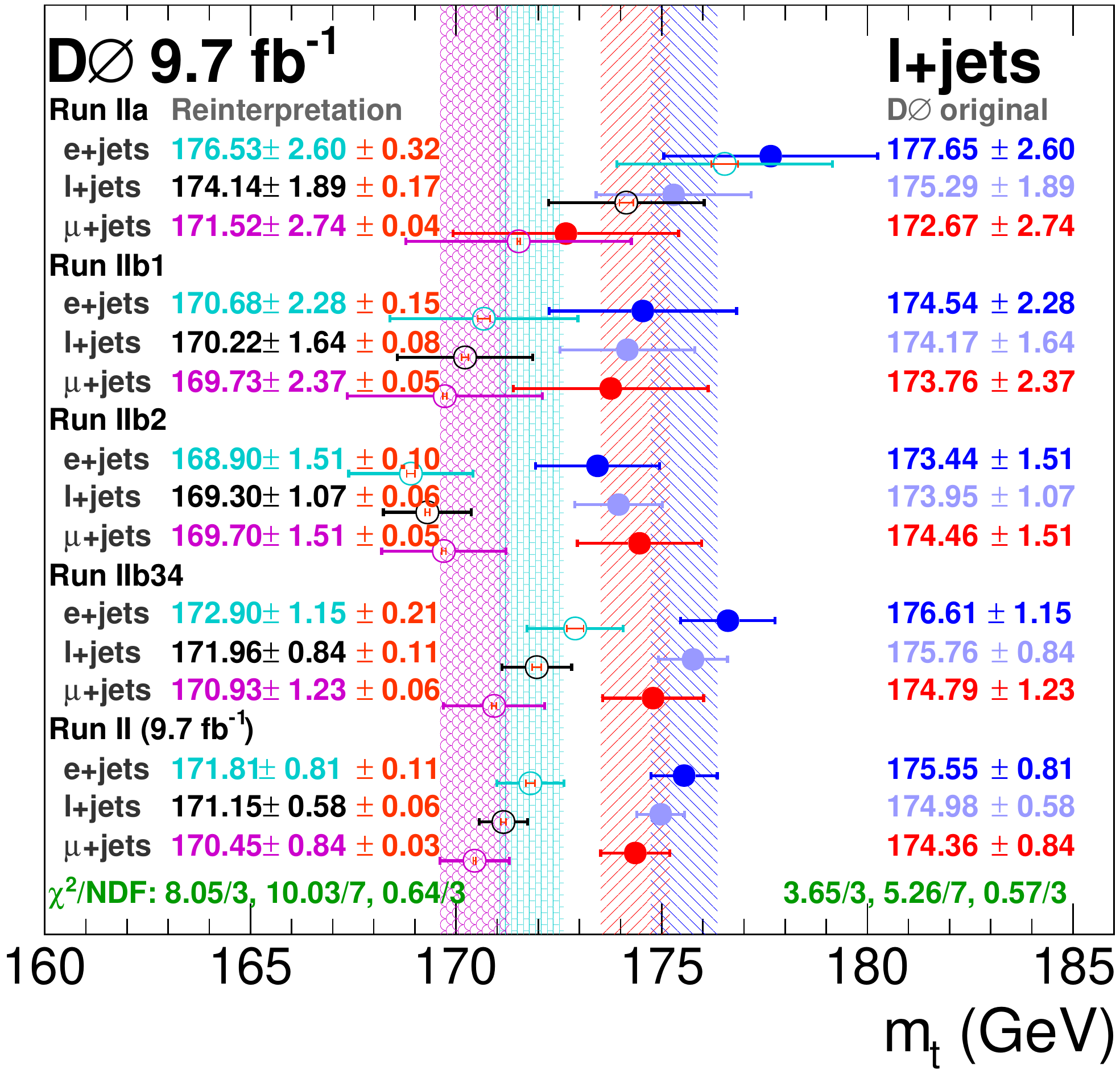} 
 \includegraphics[width=0.245\textwidth]{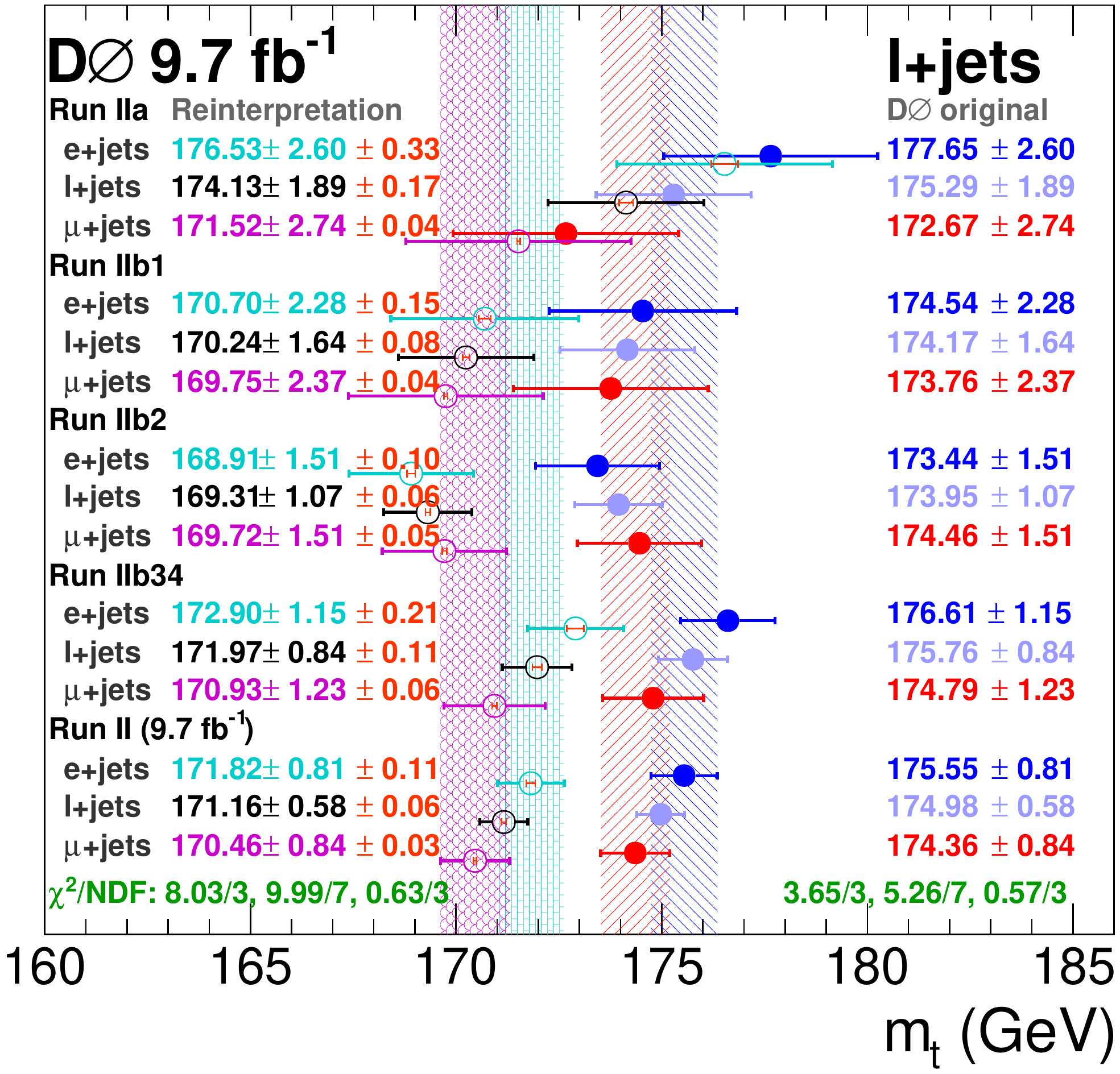}
 \caption{Shifted leptonic top mass values for the itg estimator in all the measurements and combinations. P6-based $F_{\text{Corr}}$ shift on the left and H7-based on the right.}
 \label{fig:leptonic3}
\end{figure}

\begin{figure}[H]
 \includegraphics[width=0.245\textwidth]{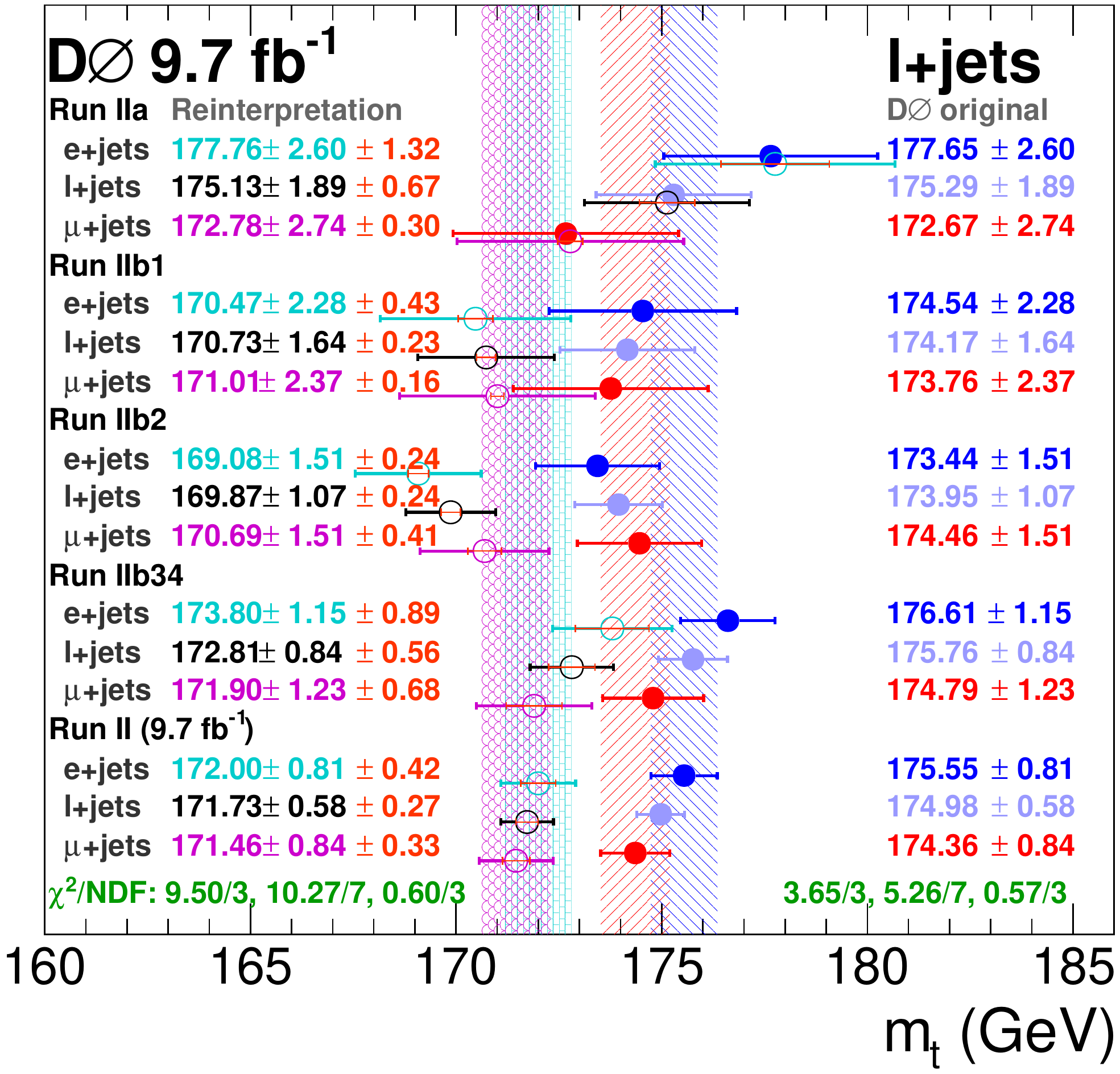}
 \includegraphics[width=0.245\textwidth]{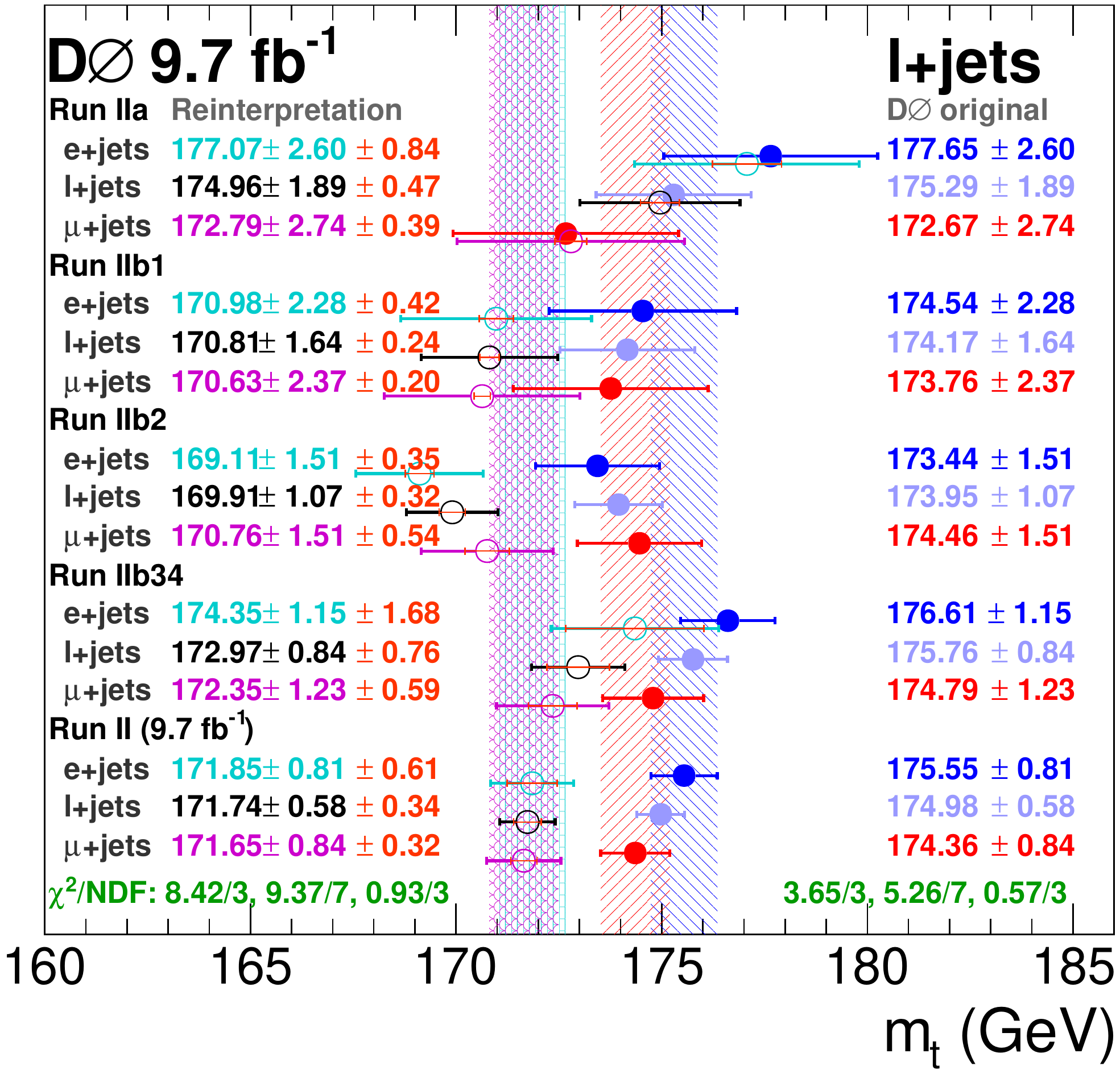}
 \includegraphics[width=0.245\textwidth]{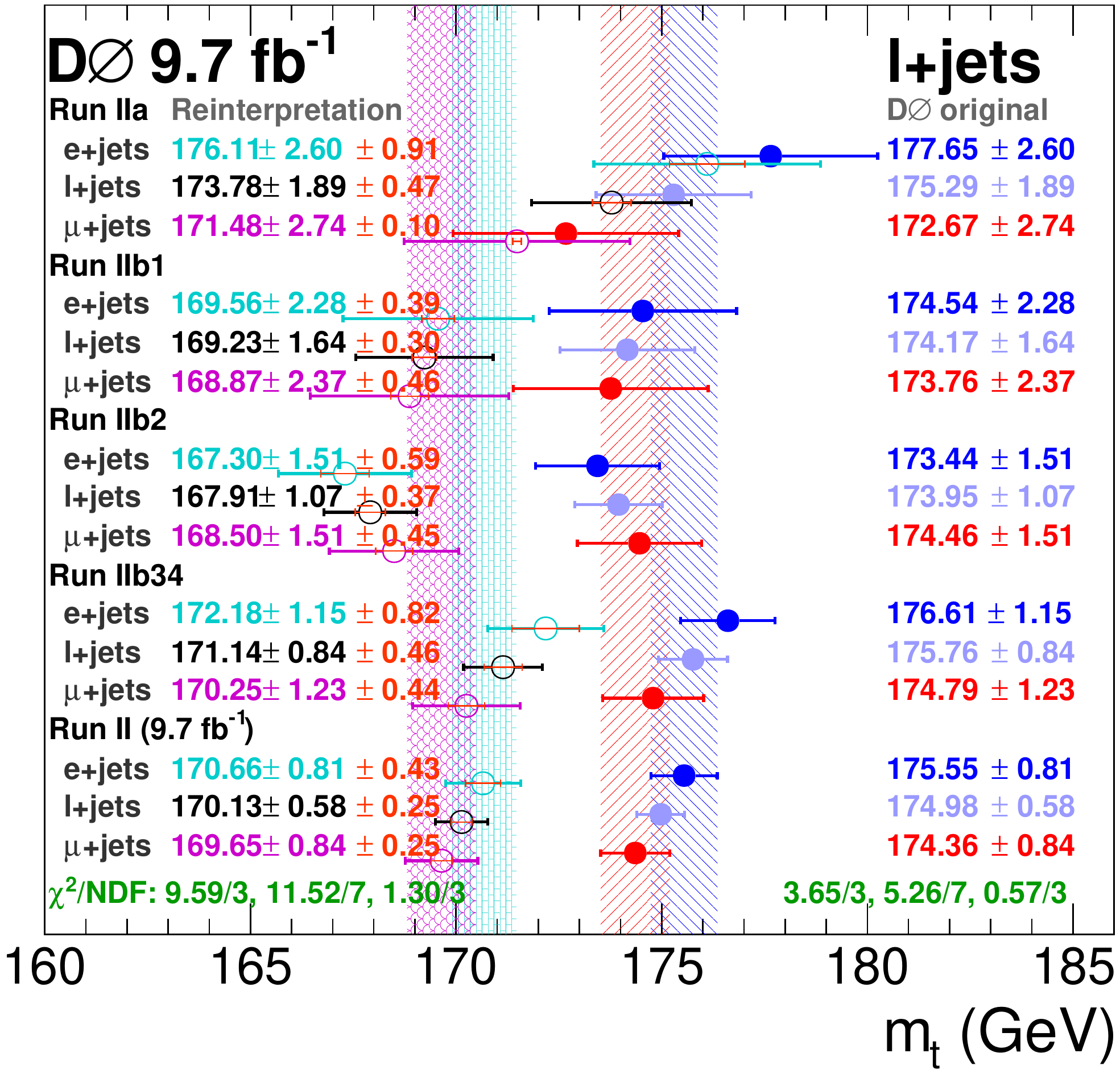}
 \includegraphics[width=0.245\textwidth]{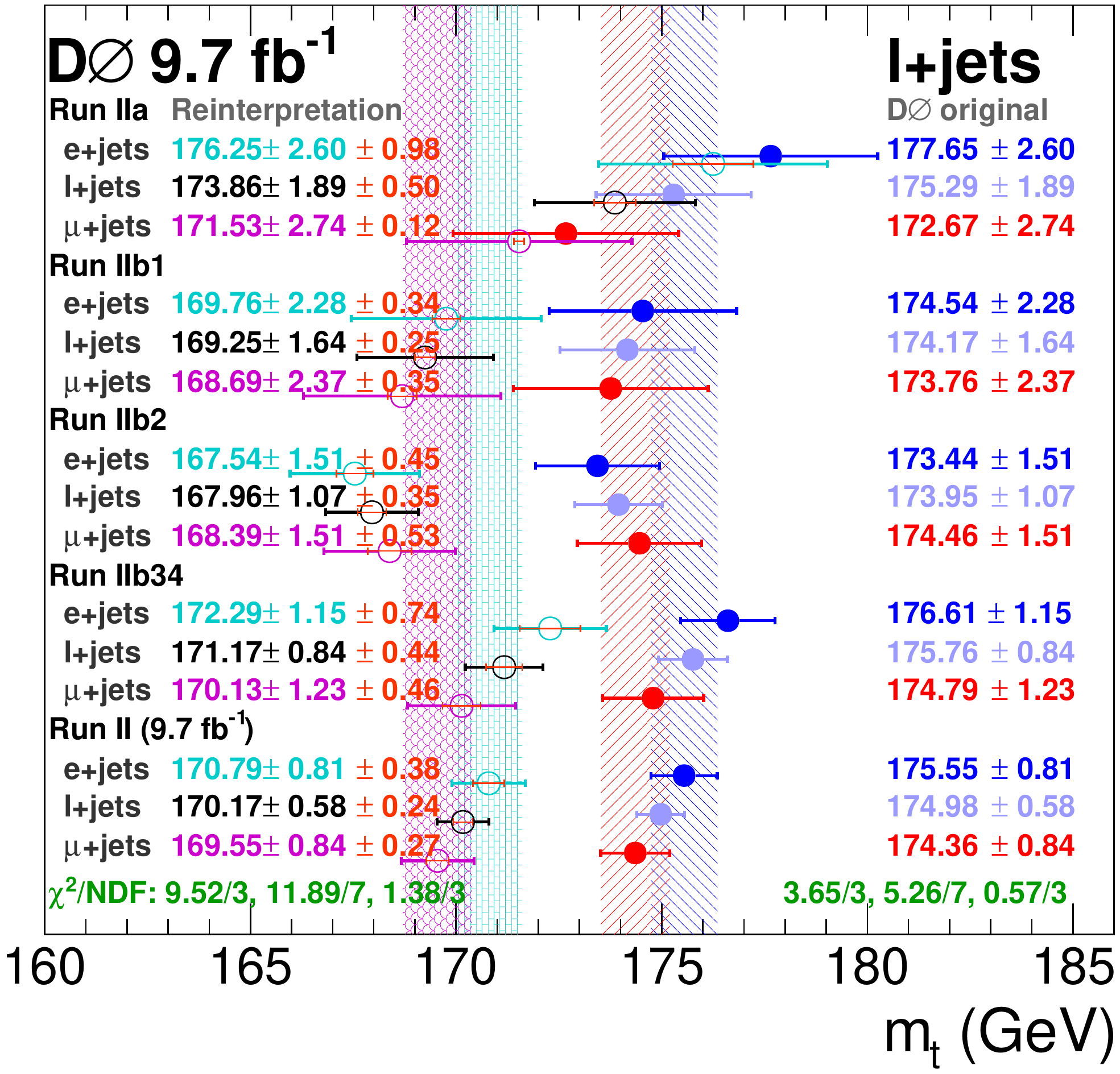}
 \caption{Shifted leptonic top mass values for the fit estimator in all the measurements and combinations. P6-based $F_{\text{Corr}}$ shift on the left and H7-based on the right.}
 \label{fig:leptonic4}
    \vspace{0.2cm}
 \includegraphics[width=0.245\textwidth]{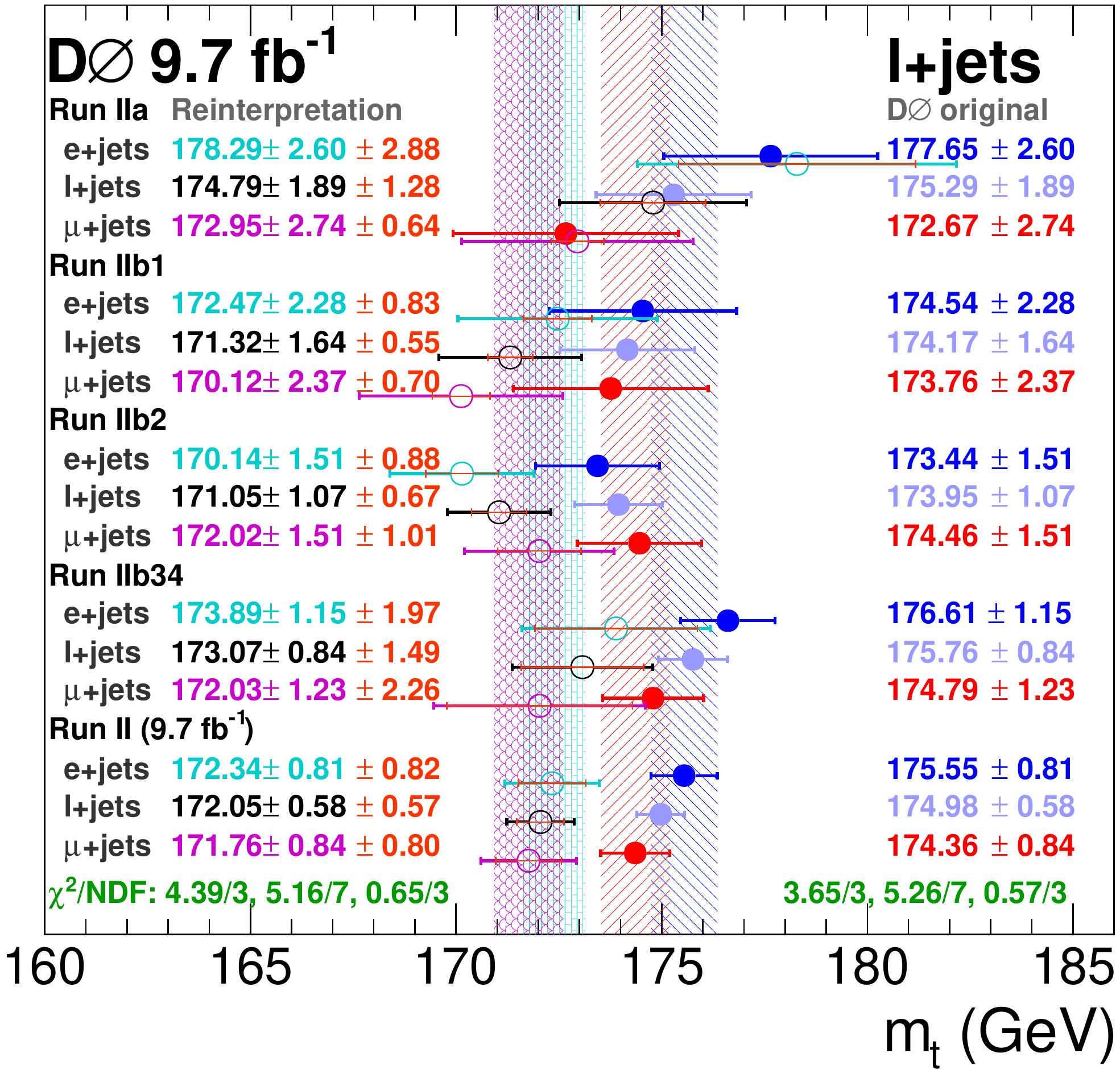}
 \includegraphics[width=0.245\textwidth]{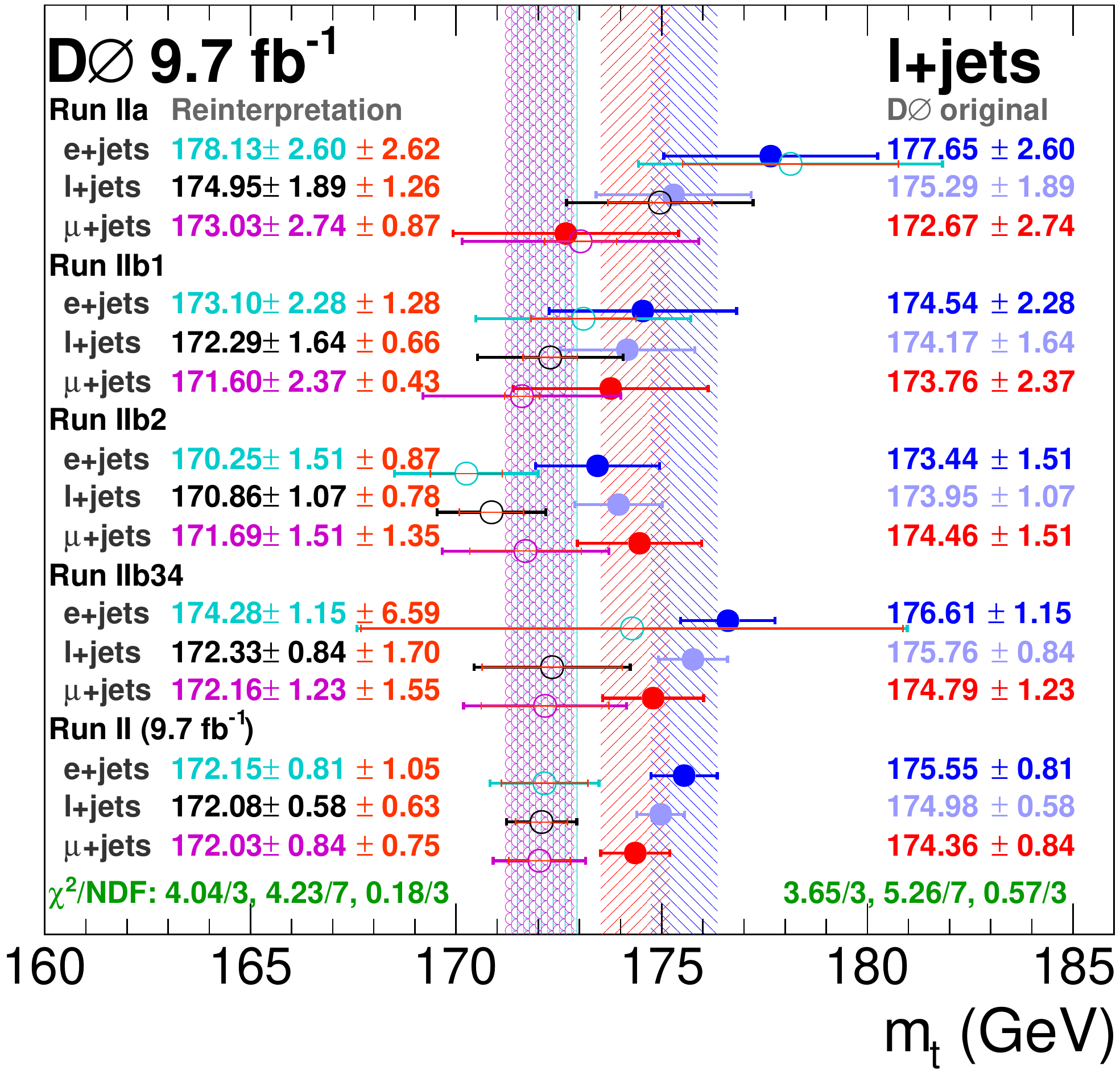}
 \includegraphics[width=0.245\textwidth]{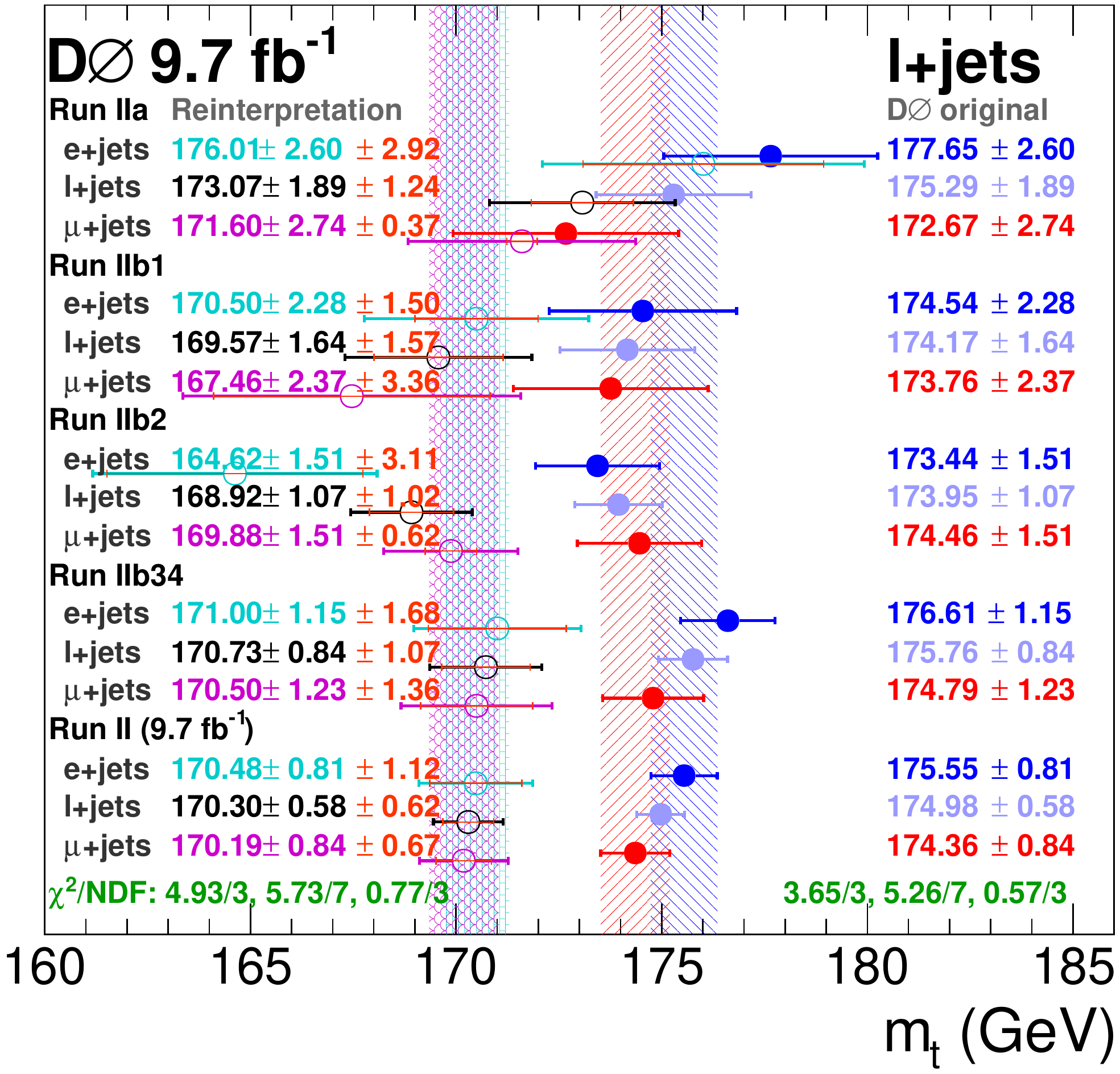}
 \includegraphics[width=0.245\textwidth]{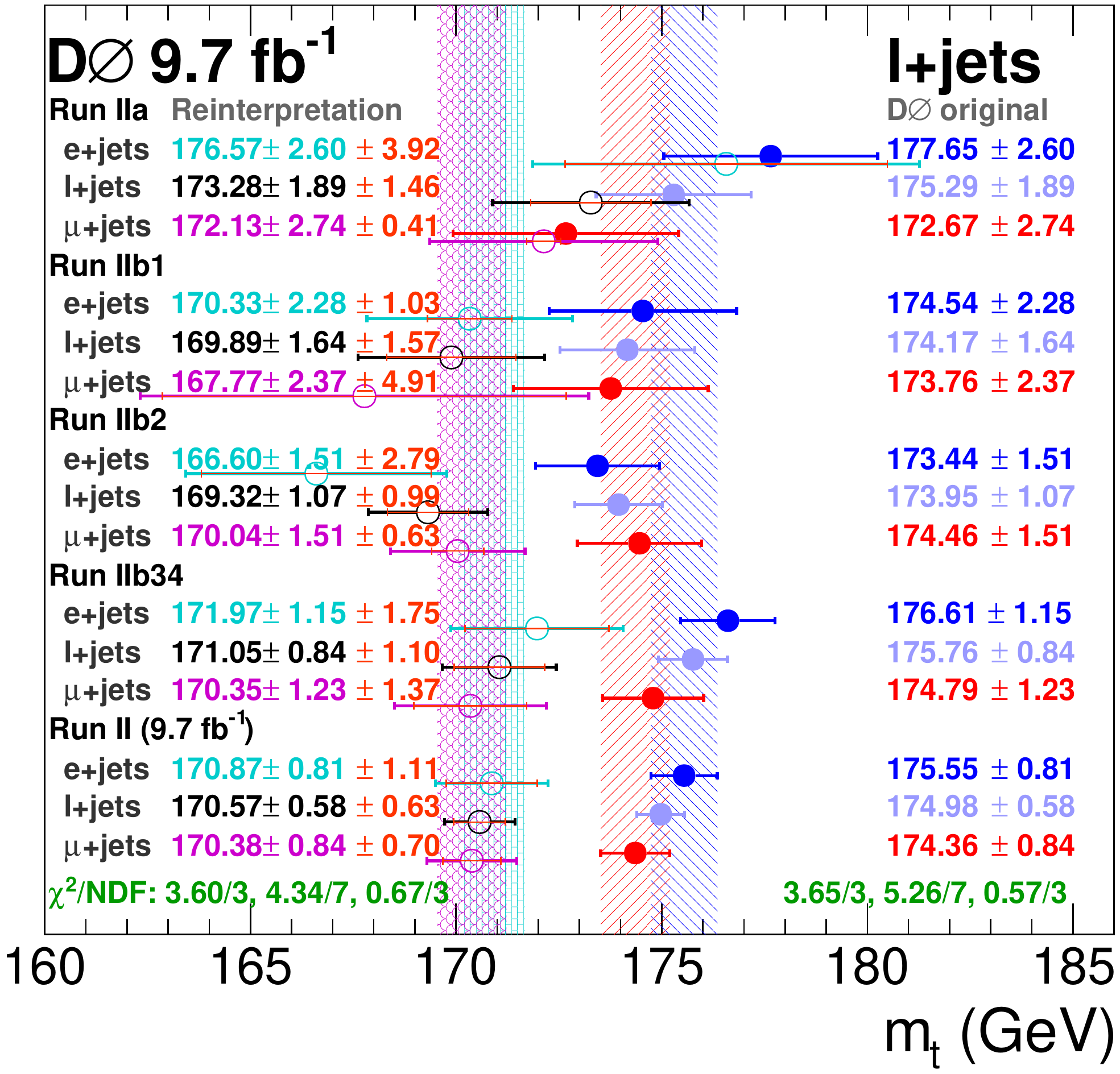}
 \caption{Shifted leptonic top mass values for the maximum estimator in all the measurements and combinations. P6-based $F_{\text{Corr}}$ shift on the left and H7-based on the right.}
 \label{fig:leptonic5}
\end{figure}

In Figs.~\ref{fig:leptonic1},\ref{fig:leptonic2},\ref{fig:leptonic3},\ref{fig:leptonic4},\ref{fig:leptonic5} the shift results in the leptonic measurement channel are shown.
These are ordered from the smallest method error to the largest one.
The method error of the shifted results is displayed in an orange font, as an addition to the original D\O\ error.
The results are given with both P6 and H7, and the two $F_{\text{Corr}}$ parameter sets.
Analogously to the hadronic channel, H7 shows a larger shift.
Moreover, the shift is systematically larger in comparison to the hadronic measurement channel.

\subsection{Combination of the Methods}

\begin{figure}[H]
 \includegraphics[width=0.5\textwidth]{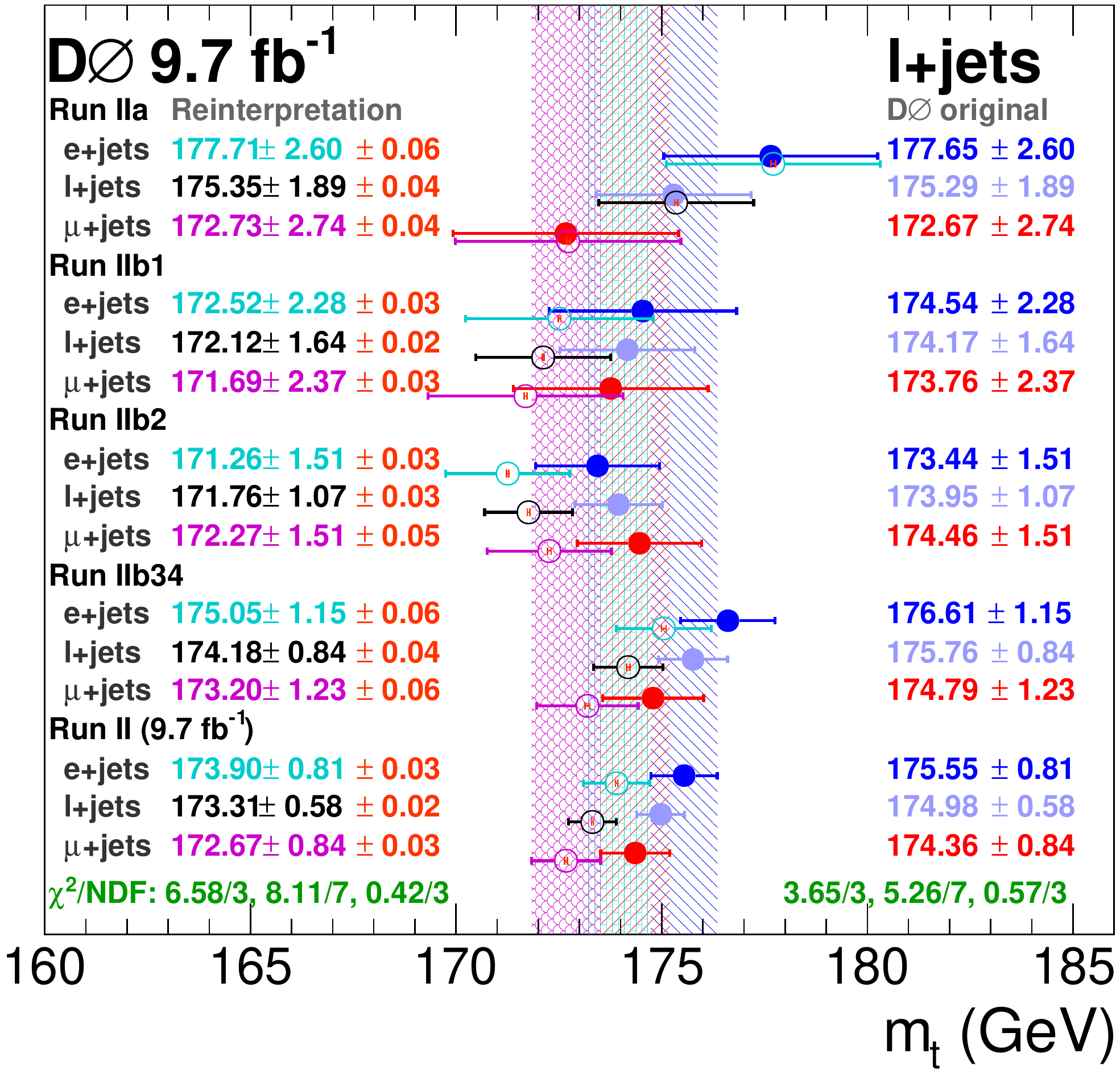}
 \includegraphics[width=0.5\textwidth]{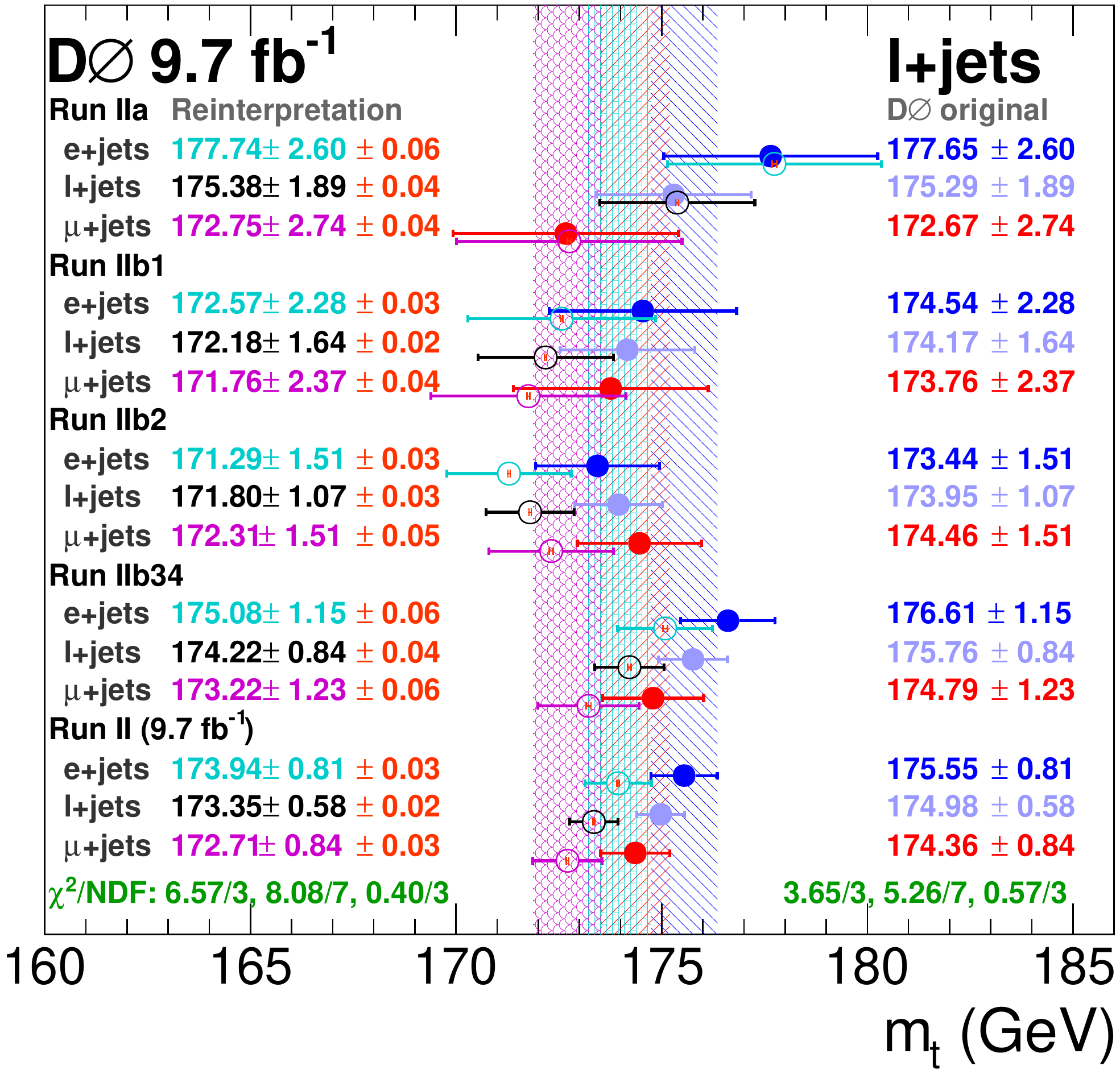}
 \caption{Shifted hadronic top mass values, combined for all estimators. P6-based $F_{\text{Corr}}$, set 1 (left) and set 2 (right).}
 \label{fig:comb1}
   \vspace{0.2cm}
 \includegraphics[width=0.5\textwidth]{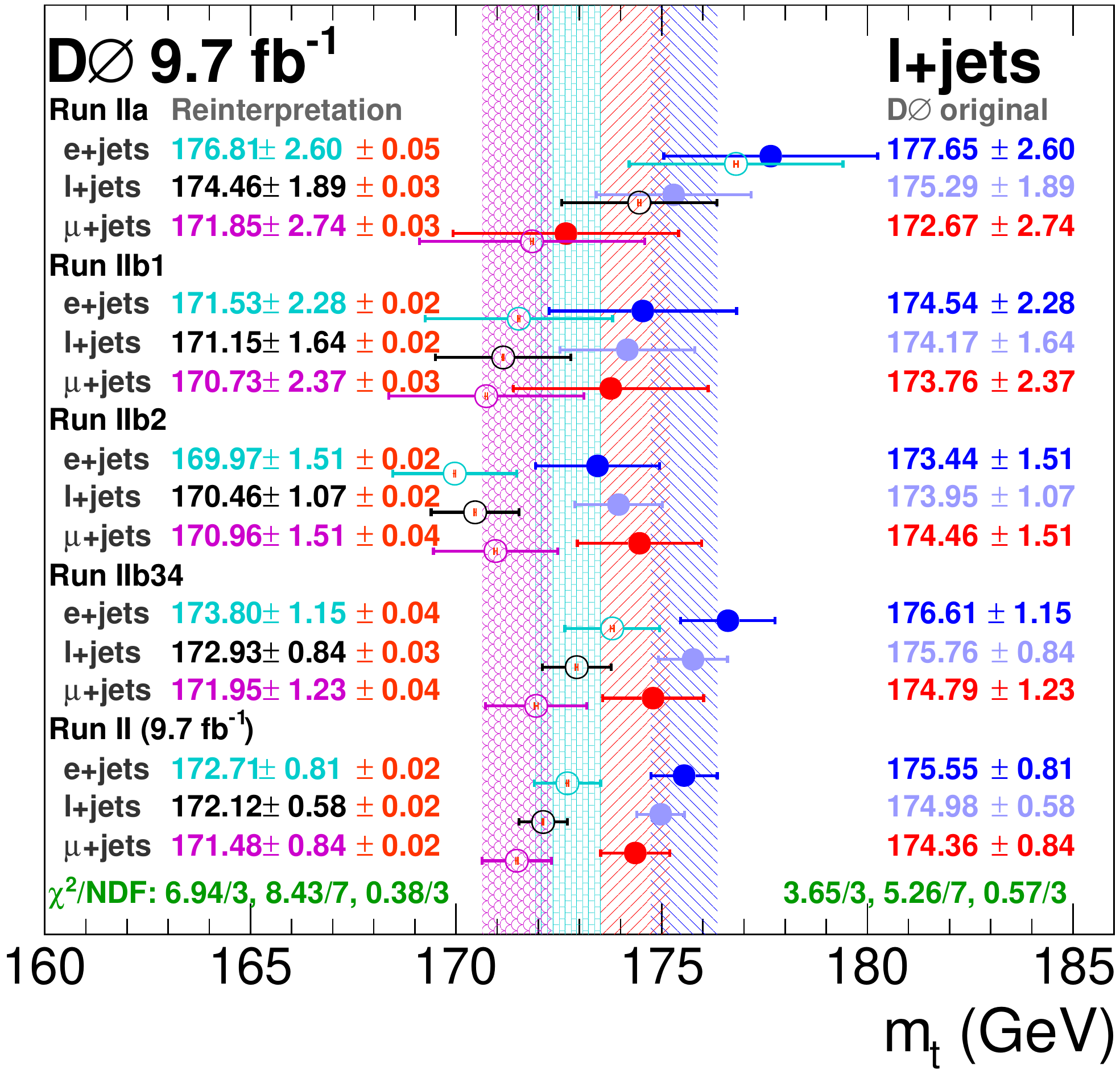}
 \includegraphics[width=0.5\textwidth]{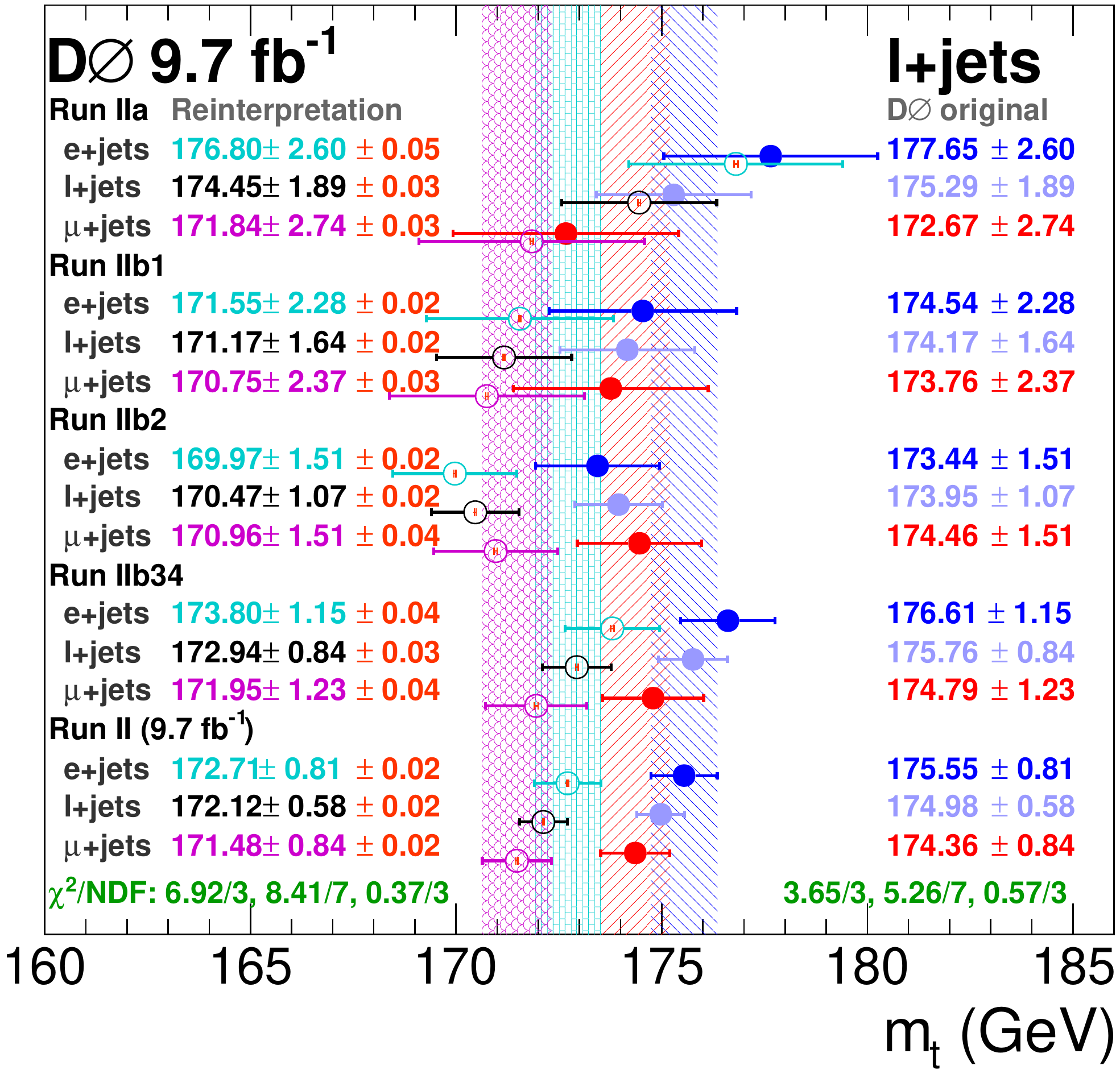}
 \caption{Shifted hadronic top mass values, combined for all estimators. H7-based $F_{\text{Corr}}$, set 1 (left) and set 2 (right).}
 \label{fig:comb2}
\end{figure}

Table~\ref{tab:shifts} shows the summary of the shift results for the different measurement channels, resonance position estimators and their combinations.
The results with different estimators are combined according to the method errors.
Figs.~\ref{fig:comb1},\ref{fig:comb2} show the intermediate results of this process.
The method error of the shifted results is displayed in an orange font, as an addition to the original D\O\ error.
The resulting values are close to those of the best estimators, med and ave.

\begin{table}[H]
\centering
 \caption{Summary of the Run~II l+jets averages with their statistical method errors in the hadronic and leptonic channels.
 Results for the two $F_{\text{Corr}}$ sets, and based on P6 and H7.}
 \label{tab:shifts}
 \vspace{0.5cm}
\resizebox{\textwidth}{!}{
\begin{tabular}{ c | c | c | c | c | c | c  }
  Channel & med & ave & itg & fit & max & combination \\
  \hline
  Had P6 1 & $173.33 \pm 0.03$ & $173.32 \pm 0.04$ & $173.33 \pm 0.06$ & $173.17 \pm 0.10$ & $173.73 \pm 0.36$ & $173.31 \pm 0.02$ \\
  Had P6 2 & $173.37 \pm 0.03$ & $173.36 \pm 0.04$ & $173.37 \pm 0.06$ & $173.22 \pm 0.09$ & $173.25 \pm 0.49$ & $173.35 \pm 0.02$ \\
  Had H7 1 & $172.13 \pm 0.03$ & $172.09 \pm 0.03$ & $172.12 \pm 0.05$ & $172.05 \pm 0.05$ & $172.82 \pm 0.35$ & $172.12 \pm 0.02$ \\
  Had H7 2 & $172.14 \pm 0.03$ & $172.09 \pm 0.03$ & $172.13 \pm 0.05$ & $172.06 \pm 0.05$ & $172.71 \pm 0.35$ & $172.12 \pm 0.02$ \\
  Lep P6 1 & $172.70 \pm 0.03$ & $172.74 \pm 0.05$ & $172.70 \pm 0.08$ & $171.73 \pm 0.27$ & $172.05 \pm 0.57$ & $172.71 \pm 0.02$ \\
  Lep P6 2 & $172.78 \pm 0.03$ & $172.81 \pm 0.06$ & $172.77 \pm 0.08$ & $171.74 \pm 0.34$ & $172.08 \pm 0.63$ & $172.79 \pm 0.02$ \\
  Lep H7 1 & $171.19 \pm 0.02$ & $171.17 \pm 0.04$ & $171.15 \pm 0.06$ & $170.13 \pm 0.25$ & $170.30 \pm 0.62$ & $171.18 \pm 0.02$ \\
  Lep H7 2 & $171.20 \pm 0.02$ & $171.18 \pm 0.04$ & $171.16 \pm 0.06$ & $170.17 \pm 0.24$ & $170.57 \pm 0.63$ & $171.19 \pm 0.02$
\end{tabular}}
\end{table}

\begin{figure}[H]
 \includegraphics[width=0.5\textwidth]{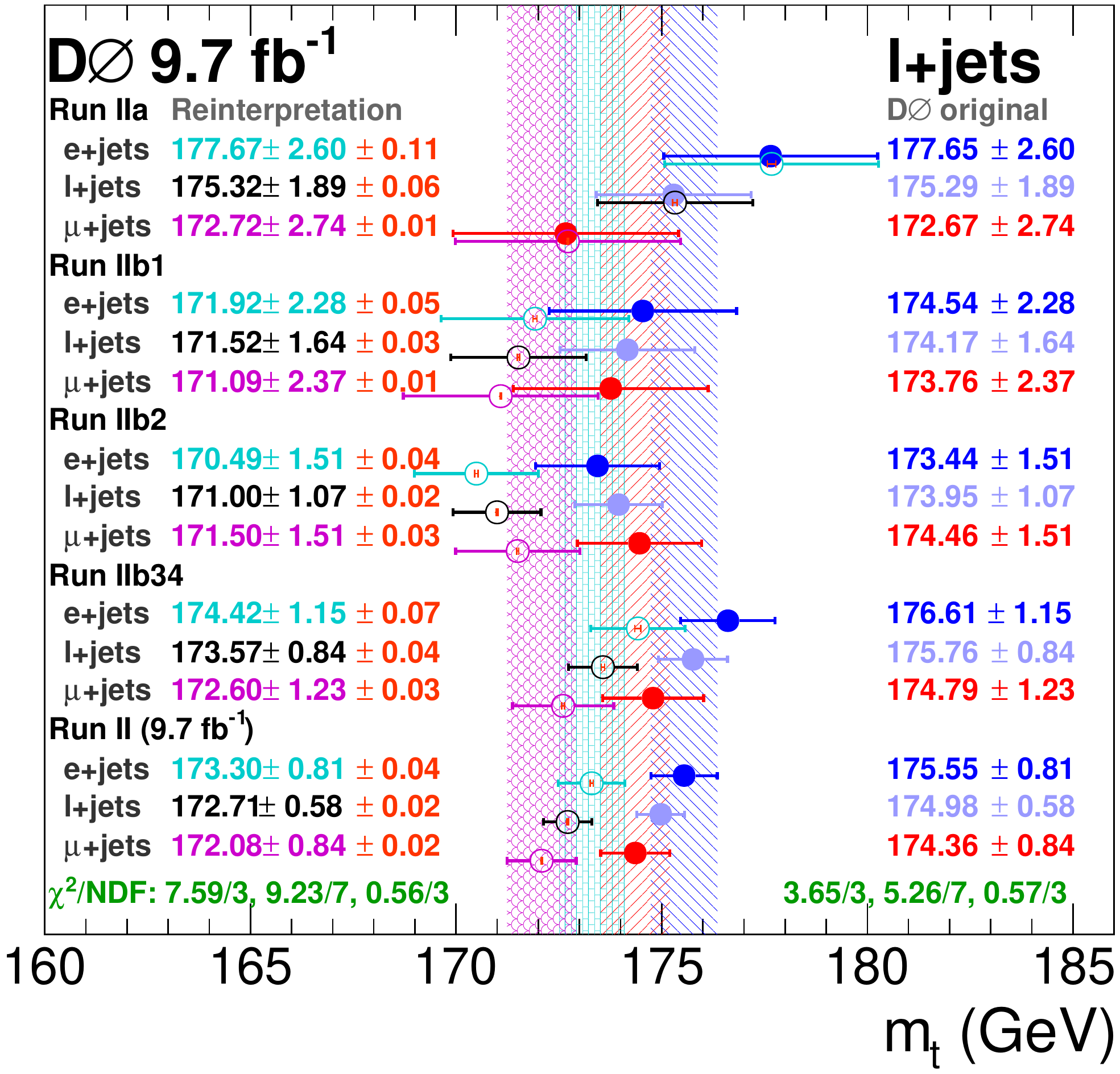}
 \includegraphics[width=0.5\textwidth]{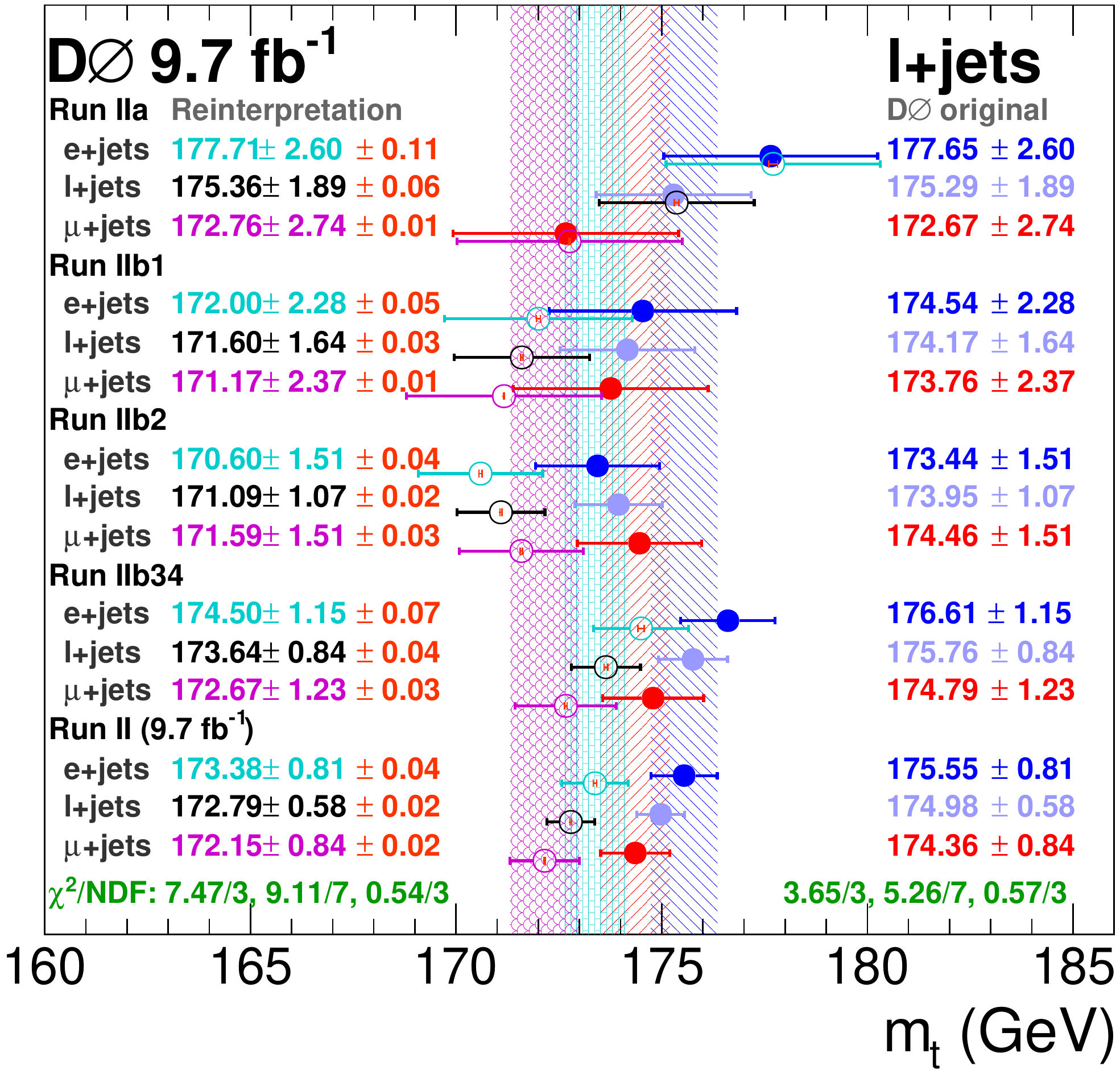}
 \caption{Shifted leptonic top mass values, combined for all estimators. P6-based $F_{\text{Corr}}$, set 1 (left) and set 2 (right).}
 \label{fig:comb3}
   \vspace{0.2cm}
 \includegraphics[width=0.5\textwidth]{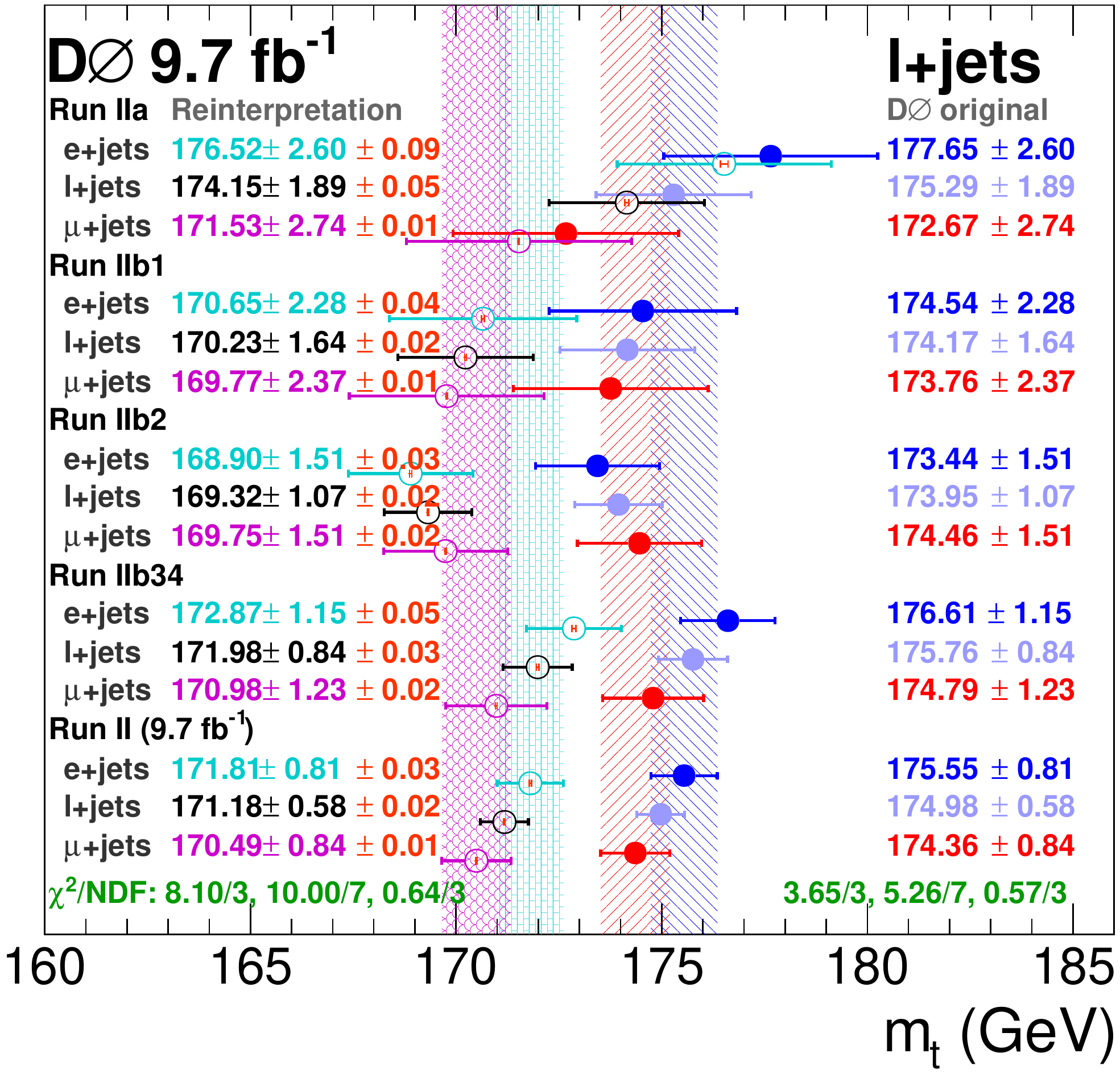}
 \includegraphics[width=0.5\textwidth]{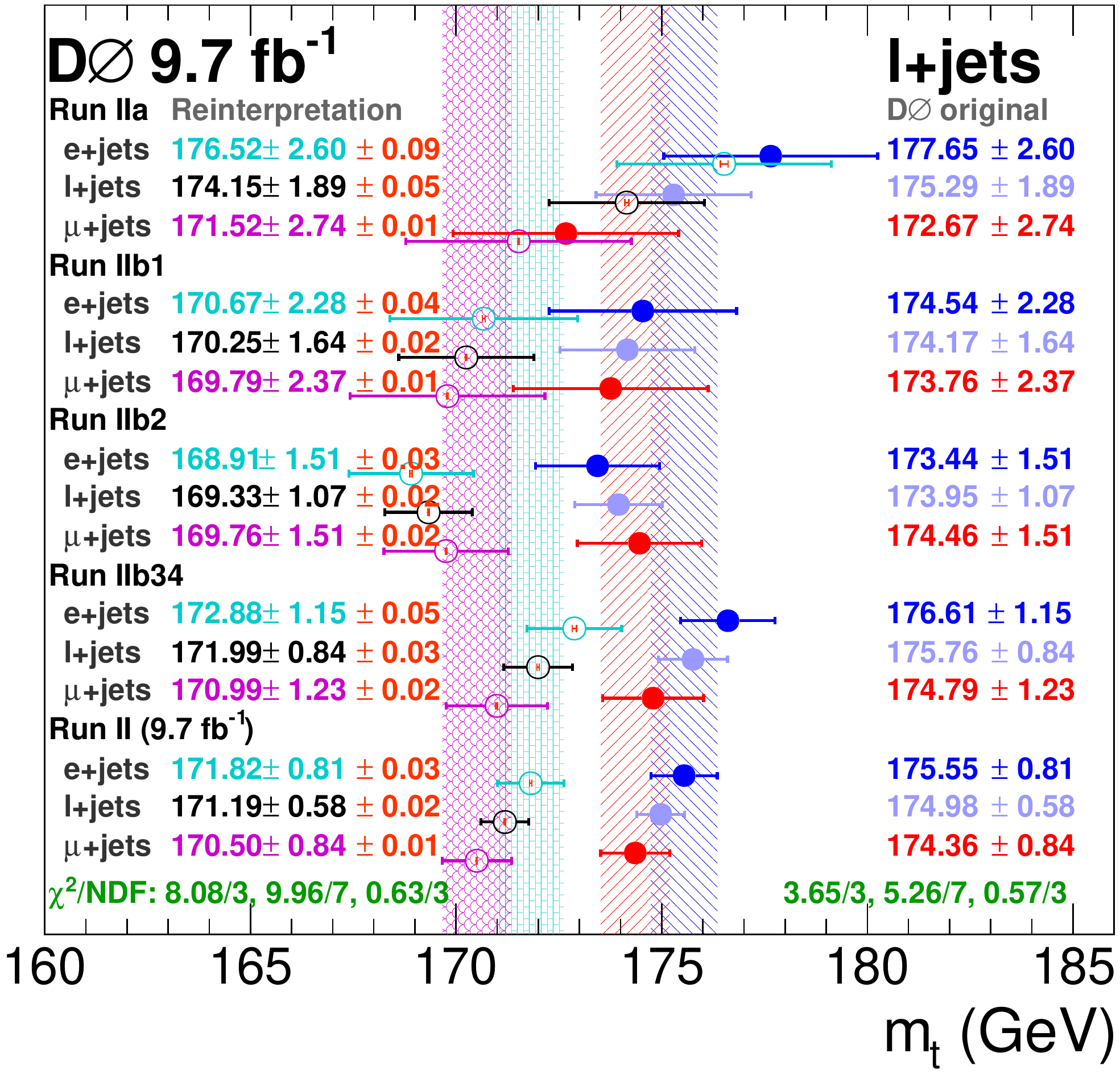}
 \caption{Shifted leptonic top mass values, combined for all estimators. H7-based $F_{\text{Corr}}$, set 1 (left) and set 2 (right).}
 \label{fig:comb4}
\end{figure}

\subsection{The Combination Parameter $\alpha$}

For $\alpha$ extraction, we compare the likelihood slopes extracted from the D\O\ paper~\cite{ref:d02015}, and those produced using the resonance position estimators.
The calibration process of $\alpha$ is very sensitive, and hence the effect of any errors is magnified.
Hence, the results vary by resonance position estimators, and with the lowest quality estimators they are very poor.
For a reliable estimation of the $K_{\text{JES}}-m_t^{\text{calib}}$ slope, we use the two estimators with the best quality: ave and med.
The final $\alpha$ value is considered from the average of these two.
Moreover, we average over all the four $F_{\text{Corr}}$ sets to find generally applicable values for $\alpha$.
The preference is to use common central values of $\alpha$ in all the four measurements, and explain the variations observed here as a larger systematical $\alpha$ error.

The results for P6 and H7 with the average estimator are given in Table~\ref{tab:tbl1} and the ones with the median estimator in Table~\ref{tab:tbl2}.
Here, the D\O\ results act as a stabilizing agent, as the values of these are almost constant.
The small numerical fluctuations are explained by the fact that for each $F_{\text{Corr}}$ set, also the interpretation of jet $p_T$ within the $F_{\text{Corr}}$'s is re-parametrized.
The statistical combination of the electron and muon $\alpha$'s is performed using $f_e = 0.538339$, obtained from the statistical errors of D\O.
The two combined $\alpha$'s differ by a factor $\delta \alpha_{comb}$, which is interpreted as a systematical error source.

\begin{table}[H]
\centering
 \caption{Measured values of $\alpha$ using the average resonance position estimator.}
 \label{tab:tbl1}
 \vspace{0.5cm}
\begin{tabular}{ c | c | c | c | c | c  }
  Measurement & $\alpha_e$ & $\alpha_\mu$ & $\alpha_{comb}$ & $f_e\times \alpha_e + (1-f_e)\times \alpha_\mu$ & $\delta \alpha_{comb}$ \\
  \hline
  P6 1 New  & 0.7532 & 0.5882 & 0.7017 & 0.6770 & 0.0247 \\
  P6 2 New  & 0.7521 & 0.5861 & 0.7003 & 0.6755 & 0.0248 \\
  H7 1 New  & 0.7460 & 0.5562 & 0.6851 & 0.6584 & 0.0267 \\
  H7 2 New  & 0.7460 & 0.5560 & 0.6850 & 0.6583 & 0.0268 \\
  P6 1 D\O\ & 0.7561 & 0.5951 & 0.7063 & 0.6818 & 0.0246 \\
  P6 2 D\O\ & 0.7561 & 0.5952 & 0.7064 & 0.6818 & 0.0246 \\
  H7 1 D\O\ & 0.7558 & 0.5946 & 0.7060 & 0.6814 & 0.0246 \\
  H7 2 D\O\ & 0.7558 & 0.5946 & 0.7060 & 0.6814 & 0.0246
\end{tabular}
 \caption{Measured values of $\alpha$ using the median resonance position estimator.}
 \label{tab:tbl2}
 \vspace{0.5cm}
\begin{tabular}{ c | c | c | c | c | c  }
  Measurement & $\alpha_e$ & $\alpha_\mu$ & $\alpha_{comb}$ & $f_e\times \alpha_e + (1-f_e)\times \alpha_\mu$ & $\delta \alpha_{comb}$ \\
  \hline
  P6 1 New  & 0.8131 & 0.6504 & 0.7612 & 0.7380 & 0.02317 \\
  P6 2 New  & 0.8120 & 0.6486 & 0.7598 & 0.7366 & 0.02318 \\
  H7 1 New  & 0.8058 & 0.6292 & 0.7490 & 0.7243 & 0.02469 \\
  H7 2 New  & 0.8054 & 0.6293 & 0.7488 & 0.7241 & 0.02471 \\
  P6 1 D\O\ & 0.8209 & 0.6492 & 0.7641 & 0.7417 & 0.02245 \\
  P6 2 D\O\ & 0.8208 & 0.6493 & 0.7641 & 0.7417 & 0.02242 \\
  H7 1 D\O\ & 0.8211 & 0.6483 & 0.7638 & 0.7413 & 0.02246 \\
  H7 2 D\O\ & 0.8211 & 0.6483 & 0.7638 & 0.7413 & 0.02246
\end{tabular}
  \caption{Combined $\alpha$ values and error terms.}
 \label{tab:tbl3}
 \vspace{0.5cm}
\begin{tabular}{ c | c | c | c | c | c }
  Measurement & $\alpha_e \pm \delta \alpha_e$ & $\alpha_\mu \pm \delta \alpha_\mu$ & $\delta \alpha_{comb}$ & $\delta \alpha^{tot}_e$ & $\delta \alpha^{tot}_\mu$ \\
  \hline
  ave         & $0.7526 \pm 0.0044$ & $0.5833 \pm 0.0171$ & $0.0252$ & 0.0255 & 0.0304 \\
  med         & $0.8150 \pm 0.0069$ & $0.6441 \pm 0.0092$ & $0.0232$ & 0.0242 & 0.0250 \\
  combination & $0.7838 \pm 0.0327$ & $0.6137 \pm 0.0341$ & $0.0242$ & 0.0408 & 0.0418
\end{tabular}
\end{table}

The merged results with their sample variances are given in Table~\ref{tab:tbl3}.
For further error estimation, these plain sample variances are used.
They do not suppress systematical differences between the $F_{\text{Corr}}$ sets, in contrast to the error of the mean.
The most notable systematical difference can be noted between the ave and med estimators, but there also seems to exist some systematical difference e.g. between P6 and H7.

The total errors are found by combining the sample variances with the average $\delta \alpha_{comb}$ terms in quadrature.
To make a conservative estimation of the total error, these ($\delta \alpha^{tot}_e$ and $\delta \alpha^{tot}_\mu$) are finally multiplied by the factor of 2.
As a result, the electron and muon channels yield the approximate values $0.78 \pm 0.08$ and $0.61 \pm 0.08$.
The difference between these is explained by the differences in phase spaces.

\subsection{Combination of the Hadronic and Leptonic Results}

\begin{figure}[H]
 \includegraphics[width=0.5\textwidth]{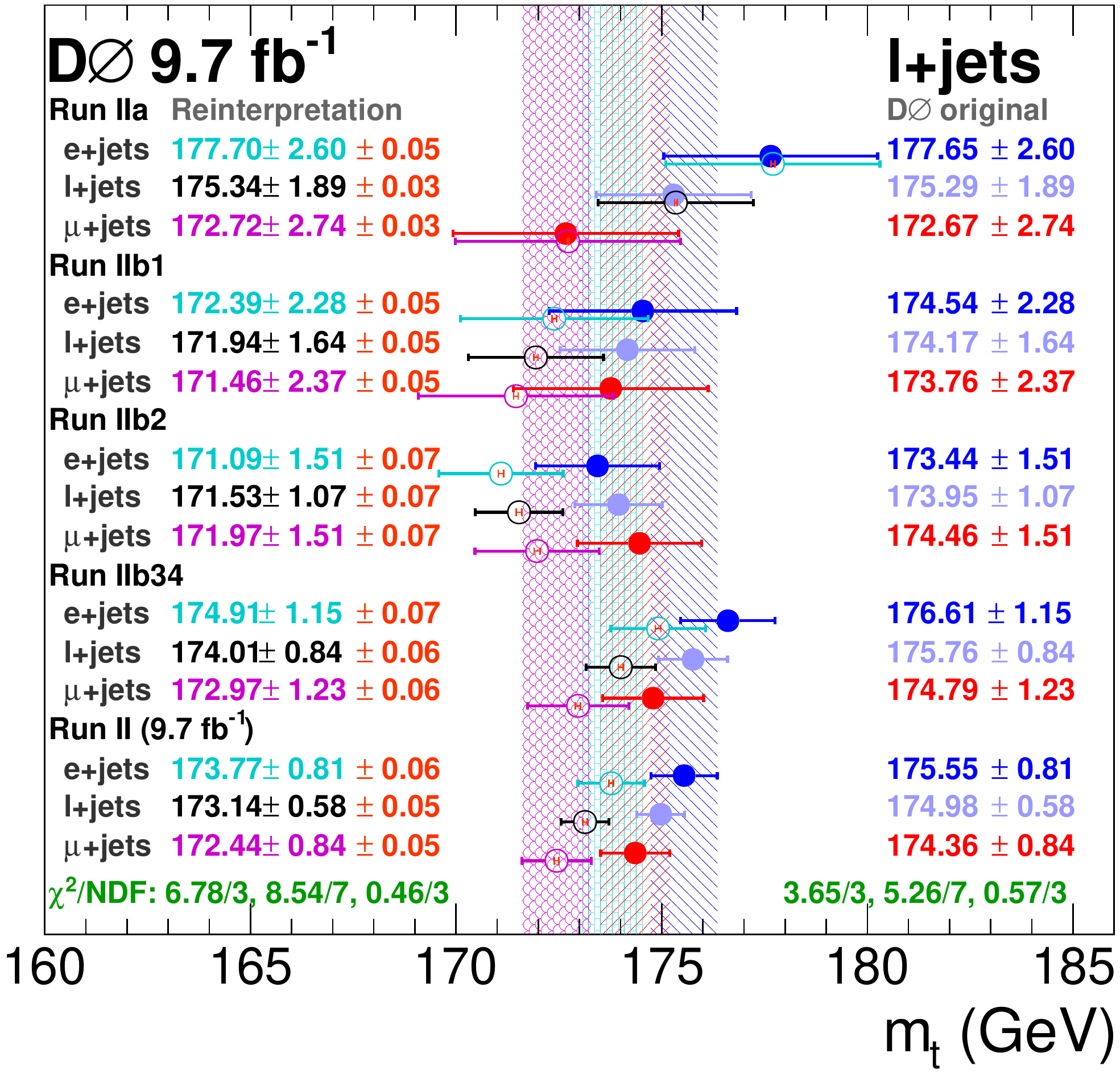}
 \includegraphics[width=0.5\textwidth]{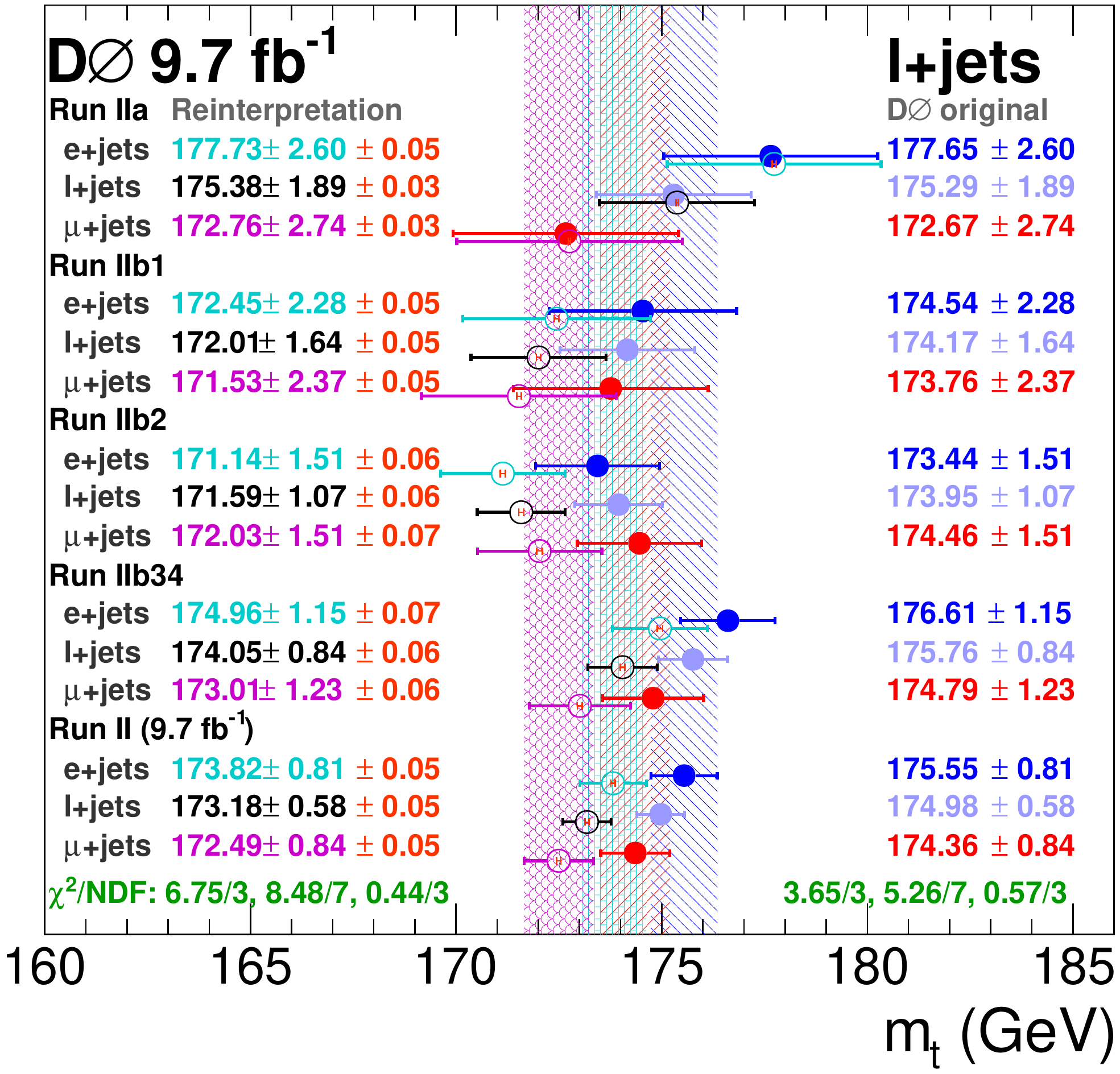}
 \caption{Shifted combined top mass values for the P6 combinations: $F_{\text{Corr}}$ set 1 (left) and set 2 (right).}
 \label{fig:tcomb1}
   \vspace{0.2cm}
 \includegraphics[width=0.5\textwidth]{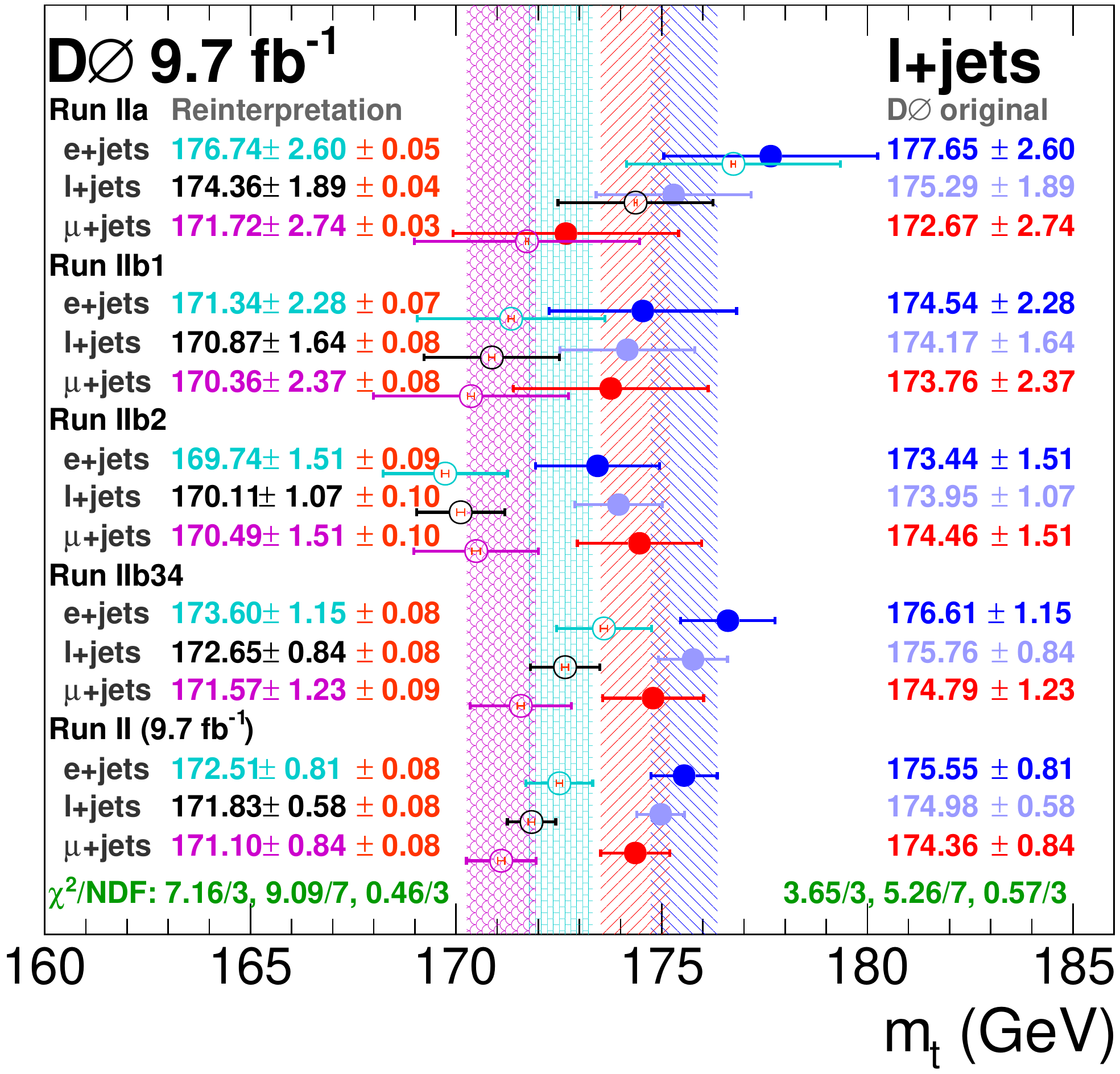}
 \includegraphics[width=0.5\textwidth]{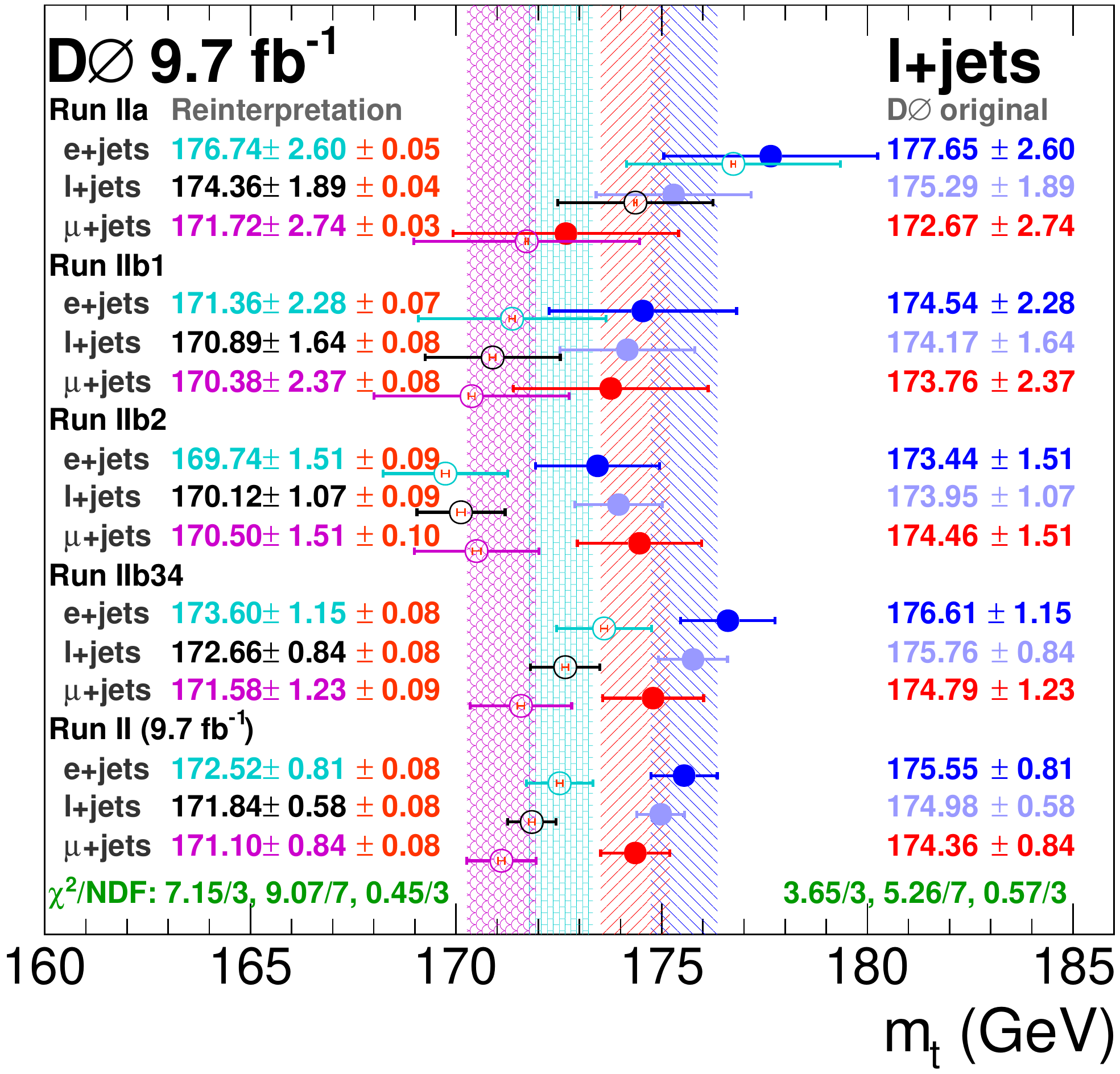}
 \caption{Shifted combined top mass values for the H7 combinations: $F_{\text{Corr}}$ set 1 (left) and set 2 (right).}
 \label{fig:tcomb2}
\end{figure}

The combination of the hadronic and leptonic channels is given in Figs.~\ref{fig:tcomb1} and \ref{fig:tcomb2}.
The results from the two $F_{\text{Corr}}$ parametrizations are very much in line, but the P6 and H7 results are more in tension.
In Table~\ref{tab:tulokset}, the summary of the shift results is shown.
For both P6 and H7, the combinations are given, and also a final combination of these two values is performed.
The combinations between the four $F_{\text{Corr}}$ sets are performed using simple linear averages.

\begin{table}[H]
\centering
\caption{A summary of the shifted $m_t$ values and the corresponding method errors.}
\label{tab:tulokset}
\begin{tabular}{ c | c | c | c | c | c | c | c }
  & P6 1 & P6 2 & H7 1 & H7 2 & P6 & H7 & Total \\
  \hline
 $m_t$ & $173.14$ & $173.18$ & $171.83$ & $171.84$ & $173.16$ & $171.84$ & $172.50$ \\
 $\delta m_t$ & $0.05$ & $0.05$ & $0.08$ & $0.08$ & $0.05$ & $0.08$ & $0.06$
\end{tabular}
\end{table}

We note that the difference between the two $F_{\text{Corr}}$ sets is $0.04$~GeV for P6 and $0.01$~GeV for H7 in the average value.
By a closer inspection we note that single estimators exhibit similar fluctuations in both P6 and H7, but in H7 these are mostly cancelled out.
To make a conservative estimate of the systematic error of $F_{\text{Corr}}$-determination, we pick the maximal fluctuation of $0.04$~GeV.
This is added to the combined method error.
This is a valuable way of assessing the systematic errors related to $F_{\text{Corr}}$, and the general stability of the mass shifting method.

\begin{figure}[ht!]
\centering
 \includegraphics[width=\textwidth]{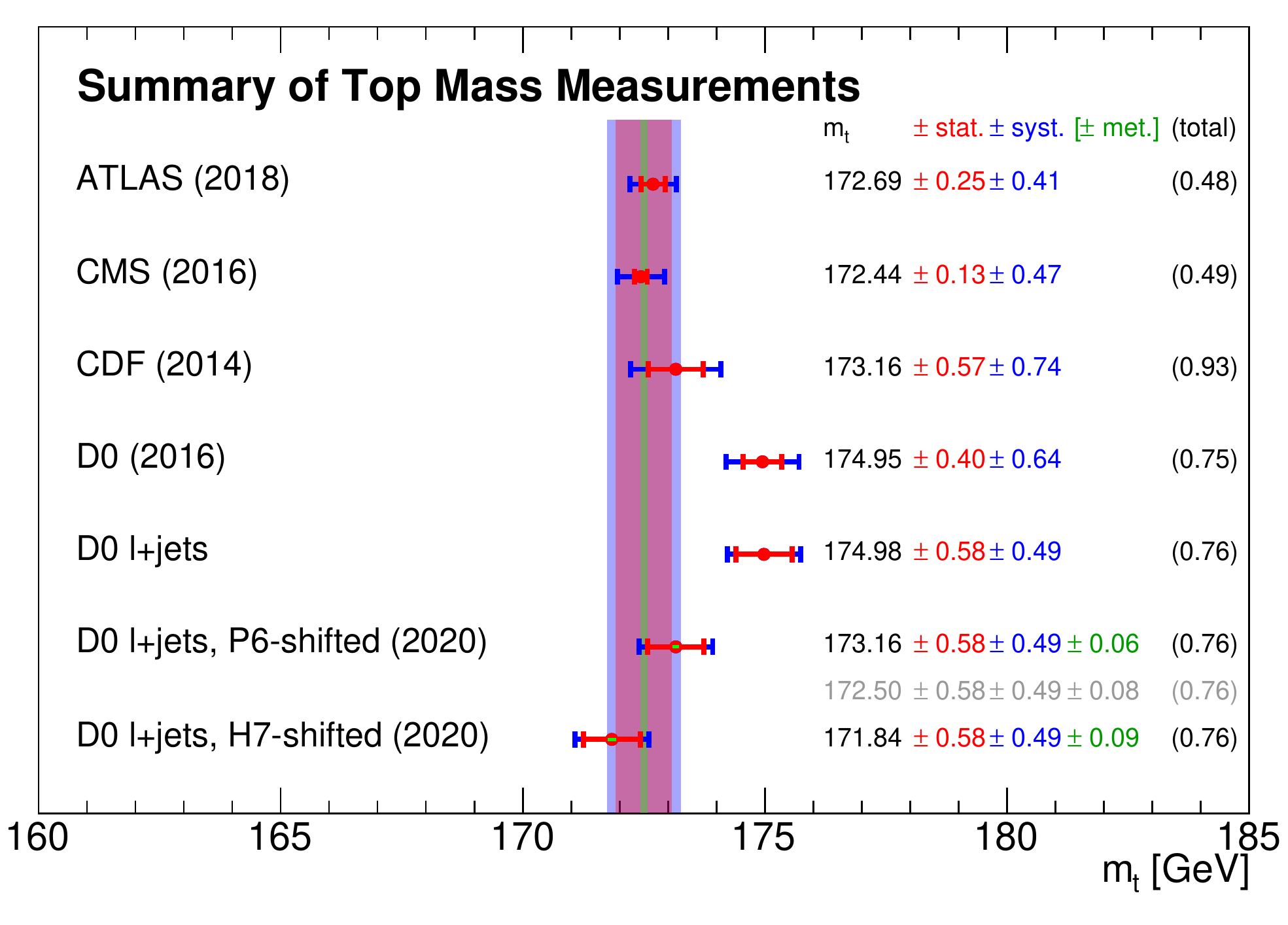}
 \caption{A comparison of the shifted top mass values with recent measurements.}
 \label{fig:combglobus}
\end{figure}

\newpage

In Fig.~\ref{fig:combglobus} the final results are combined to recent notable top mass measurements.
For D\O, the total average and l+jets measurement are displayed separately to underline the fact that the l+jets measurement dominates the average.
The combination of the method error and systematic $F_{\text{Corr}}$ error is given in green font.
The linear average of P6 and H7 is shown in the background color bars and gray text.

\section{Discussion}

We have derived values for the top mass shift necessitated by the re-calibration of the D\O\ $F_{\text{Corr}}$ values.
This was done with a variety of resonance position estimators, $F_{\text{Corr}}$ values derived using both P6 and H7, and two separate derivations for the $F_{\text{Corr}}$ values.
As a remarkable finding, the results found with the two separate $F_{\text{Corr}}$ sets, given in Table~\ref{tab:tulokset}, were very close to each other.
This hints of the stability of the method, and of small systematic errors in $F_{\text{Corr}}$ determination.

However, a notable difference between the P6 and H7 $F_{\text{Corr}}$ based results was observed.
This needs to be considered as a significant systematical error, as the two generators employ different heuristics for the collision dynamics.
Nonetheless, these two results are not completely on equal footing.
For P6, a D\O\ tune was used, but for H7 such was not available.
Thus, the H7-based values might be less reliable than the P6 ones.
For the sake of simplicity, the combination of P6 and H7 was performed as a simple average, yielding the value $172.499$~GeV.
A more sophisticated combination method might give more weight for P6, shifting the combination result well above $172.5$~GeV.

The results obtained with the different resonance position estimators agree remarkably well, as shown in Table~\ref{tab:shifts}.
Even the quite unstable maximum method yielded good results.
The fit method tended to produce results that are shifted by approximately one (method-related) sigma downwards from the central value.
This was, however, explained by the larger method error.
Moreover, the fit method experiences a significant increase in the method error for the leptonic resonance in comparison to the hadronic resonance.
Such a behavior is not expressed by the other resonance position estimators.
This implies difficulties in fitting on the convoluted leptonic resonance.
This can be noted by comparing the example cases of Appendix~\ref{app:hlexamples} to those of Appendix~\ref{app:htexamples}

In Fig.~\ref{fig:combglobus} the final results were displayed.
The, the D\O\ average differs significantly of that given by the CDF, ATLAS and CMS collaborations.
In contrast, the shifted l+jets result is in a good agreement with the other collaborations.
Shifting the D\O\ l+jets result shifts the whole D\O\ average.

It is necessary to underline that a downward shift of $m_t$ is a direct follow-up of the $F_{\text{Corr}}$ recalibration.
The phenomenological proof of the second section shows that the D\O\ measurements should have produced a lower $m_t$ value.
From the magnitude of the change in $F_{\text{Corr}}$ in the recalibration, we knew that this error is in the order of GeV's, which functioned as a motivation for further research.
Thus, the motivation of this paper was not shifting the D\O\ $m_t$ value, but only propagating the inevitable consequences of the $F_{\text{Corr}}$ recalibration to the $m_t$ measurement.

Using the methods presented in this paper, the resulting top mass shift should be reproducible both phenomenologically, and with numerical evaluation.
For the sake of openness, the complete source code of the measurement is provided in Refs.~\cite{ref:genhandle,ref:genanalysis}.
The source code includes all the numerical parameter values.
The found numerical values are completely based on the series of linear fits made on top of the resonance histograms, produced using simulations.

\newpage

\section{Conclusions}

This paper was motivated by the observation that the D\O\ calibration of $F_{\text{Corr}}$ jet energy corrections was not optimal.
A suggested re-calibration of the corrections has a great impact on the top quark mass measurements.
To demonstrate a shift caused by the change in $F_{\text{Corr}}$ values, an intricate method was designed.

In the first section of this paper we dissected analytically the steps between the quark and the jet level in the D\O\ measurements.
Using the obtained results, the second section demonstrated phenomenologically, how the changes in $F_{\text{Corr}}$ are conveyed to the top mass values.
Finally, in the third section a numerical simulation-driven method for estimating the shift in the $m_t$ values was designed.
In the fourth section, this method was applied, producing estimates for the $m_t$ value that D\O\ should have obtained considering the $F_{\text{Corr}}$ recalibration.

For the complete top mass shift, an average was taken of two separate $F_{\text{Corr}}$ parameter sets.
Moreover, the shift results were derived with $F_{\text{Corr}}$ values based on P6 and on H7.
The combination of P6 and H7 results was performed by a simple average, yielding a shifted top mass value around $172.5$~GeV.
A more profound combination method might favor the P6 results more, but such considerations are out of scope here.
Nevertheless, there is a significant difference between P6 and H7.
This difference appears to fit within the error bars, so there was no reason to revise these.

The result of Fig.~\ref{fig:combglobus} shows the impact of the D\O\ l+jets $m_t$ shift in comparison to other global results.
Here it was clearly demonstrated that the $F_{\text{Corr}}$ recalibration leads to a better agreement with other measurements.
The D\O\ average is almost completely driven by the l+jets measurement.

The fact that the $m_t$ value shifted by the $F_{\text{Corr}}$ recalibration agrees with other collaborations is encouraging.
The two peculiarities of the D\O\ measurement: the Run~IIb $F_{\text{Corr}}$ values and the high $m_t$ value appear to have the same root cause.
To avoid the necessity of a shift in the top mass value, also the need for a $F_{\text{Corr}}$ recalibration should be refuted.
Future top mass world combinations should consider these results profoundly.

In total, there is still much to be understood in the measurement of the top quark mass.
The differences between simulation softwares are under an increased scrutiny.
An important point of interest is the difference between the MC-based mass and the physical top quark pole mass.
In this paper we have shed some light on the $m_t$ disagreement between D\O\ and other collaborations.
As this difference is better understood, an agreement on the value of $m_t$ is one step closer.

\newpage

\appendix

\section{D\O\ Likelihood Calibration Ambiguity} \label{app:likeli}
 
In the 2011 \cite{ref:d02011} and 2015 \cite{ref:d02015} D\O\ measurements, the analysis chain of the l+jets top mass measurement is slightly altered from its previous incarnations.
As the 2011 paper \cite{ref:d02011} states above Eq. (21),
''The likelihoods \ldots~are calibrated by replacing $m_t$ and $K_{\text{JES}}$ by parameters fitted to the response plots``.
Correspondingly, the 2015 paper \cite{ref:d02015} explains above Eq. (35),
''The linear calibrations are applied through the likelihoods \ldots~by transforming $m_t$ and $K_{\text{JES}}$ parameters according to \ldots``.
The replacement suggested by both papers is of the form $m_t \rightarrow g^{-1} (m_t)$ and $K_{\text{JES}} \rightarrow f^{-1}(K_{\text{JES}})$.

We will show using $m_t$, where this transformation leads (analogous derivation is valid for $K_{\text{JES}}$).
Starting by making the substitution $m_t \rightarrow g^{-1} (m_t)$ for the likelihoods in Eq. \eqref{eq:mtfit}:
\begin{equation} \label{eq:d0mt}
 m_t^{\text{D\O}} = \left< m_t \right>_{\text{Transformed}} = \frac{\int_\mathbb{R} dm_t m_t \mathcal{L} (g^{-1}(m_t))}{\int_\mathbb{R} dm_t \mathcal{L} (g^{-1}(m_t))}.
\end{equation}
Remembering the parametrization Eq.~\eqref{eq:mtfitappro} of Eq.~\eqref{eq:gunction}, the inverse function attains the parametrized form
\begin{equation} \label{eq:mtgenfit}
 m_t^{gen} = g^{-1}(m_t^{fit}) = \frac{\left( m_t^{fit} - 172.5 \right) - \mathcal{O}}{\mathcal{S}} + 172.5.
\end{equation}
Let us perform a change of variables: $m_t' = g^{-1}(m_t)$, i.e. $m_t = g(m_t')$.
By Eq. \eqref{eq:mtgenfit}, we find the transformation
\begin{equation} \label{eq:dmtd0}
dm_t = (dg(m_t')/dm_t')~dm_t' = \mathcal{S}~dm_t'.
\end{equation}
After the transformation, the integration limits are still over the whole $\mathbb{R}$.
Eq. \eqref{eq:d0mt} reads now
\begin{eqnarray*}
 \left< m_t \right>_{\text{Transformed}} & = & \frac{\int_\mathbb{R} \cancel{\mathcal{S}}~dm_t' g(m_t') \mathcal{L} (m_t')}{\int_\mathbb{R} \cancel{\mathcal{S}}~dm_t' \mathcal{L} (m_t'))} \\
  & = & \frac{\int_\mathbb{R} dm_t' \left( \mathcal{S} \left( m_t' - 172.5 \right) + \mathcal{O} + 172.5 \right) \mathcal{L} (m_t')}{\int_\mathbb{R} dm_t' \mathcal{L} (m_t'))} \\
  & = & \mathcal{S} \frac{\int_\mathbb{R} dm_t' m_t' \mathcal{L} (m_t')}{\int_\mathbb{R} dm_t' \mathcal{L} (m_t'))}
   + \left( -\mathcal{S}~172.5 + \mathcal{O} + 172.5 \right) \frac{\cancel{\int_\mathbb{R} dm_t' \mathcal{L} (m_t')}}{\cancel{\int_\mathbb{R} dm_t' \mathcal{L} (m_t'))}} \\
  & = & \mathcal{S} \left( \left< m_t \right>_0 - 172.5 \right) + \mathcal{O} + 172.5
   =  f(\left< m_t \right>_0),
\end{eqnarray*}
where we have again utilized Eqs.~(\ref{eq:mtfit},\ref{eq:mtfitappro}).
The found result is
\begin{equation} \label{eq:d0mttransform}
 \left< m_t \right>_{transformed} = g(\left< m_t \right>_0) = g(m_t^{fit}).
\end{equation}
This is not the desired result from Eq. \eqref{eq:mtgenfitfit}: we should have obtained $g^{-1}(m_t^{fit})$ and not $g(m_t^{fit})$.
By interchanging $g$ and $g^{-1}$ in the equation chain above, it turns out that the substitution $m_t \rightarrow g(m_t)$ into $\mathcal{L}(m_t)$ produces the correct $m_t^{\text{calib}}$.
Now, utilizing Eqs. \eqref{eq:mtgenfitfit} and \eqref{eq:d0mttransform}, we notice that
\begin{equation} \label{eq:d0mtcalib}
 m_t^{\text{calib}} = g^{-1}(m_t^{fit}) = g^{-1}(g^{-1}(m_t^{D0})).
\end{equation}
Continuing with Eqs. \eqref{eq:mtgenfit} and \eqref{eq:d0mtcalib}, we find
\begin{equation} \label{eq:mtlikeli0}
 m_t^{\text{calib}} = \frac{\left( \frac{\left(m_t^{D0}- 172.5 \right) - \mathcal{O}}{\mathcal{S}} + \cancel{172.5} - \cancel{172.5} \right) - \mathcal{O}}{\mathcal{S}} + 172.5.
\end{equation}
That is,
\begin{equation} \label{eq:mtlikeli1}
 m_t^{\text{calib}} = \frac{\left(m_t^{D0}- 172.5 \right) - \mathcal{O} \left(1 + \mathcal{S}\right)}{\mathcal{S}^2} + 172.5.
\end{equation}
For $\sigma$'s we can derive similar statements, except that the translational terms are dropped out:
\begin{equation} \label{eq:mtlikeli2}
 \sigma_t^{\text{calib}} = \sigma_t^{D0}/\mathcal{S}^2
\end{equation}
All in all, if the 2011 and 2015 D\O\ l+jets measurements have performed the calibration step as they state in the papers, we find explicit corrections that should be performed to the measured values. 
The scale $\mathcal{S}$ is typically around 0.89 - 0.97, so the variances (statistical errors) found by D\O\ should be scaled slightly upwards.
In addition, the measured $m_t$ values attain non-trivial translational terms.

Applying the transformations presented in Eqs. \eqref{eq:mtlikeli1} and \eqref{eq:mtlikeli2}, we find for the 2011 paper \cite{ref:d02011}
\begin{equation*}
 m_t = 174.94 \pm 1.49 \rightarrow 174.126 \pm 1.55
\end{equation*}
and or the 2015 paper \cite{ref:d02015}, we find similarly 
\begin{eqnarray*}
 m_t^e & = & 175.55 \pm 0.81 \rightarrow 175.88 \pm 0.97 \\
 m_t^\mu & = & 174.36 \pm 0.84 \rightarrow 174.02 \pm 0.87 \\
 m_t^{tot} & = & 174.98 \pm 0.58 \rightarrow 174.87 \pm 0.65 \\
\end{eqnarray*}
The changes are subtle, yet notable.
As a side note: in the transformed values it has also been taken into account that in the electronic channel for IIb34 (according to Fig. (15) in \cite{ref:d02015}) the pull around $m_t = 175$~GeV is actually approximately 1.4, instead of the stated 1.16.

As a final interesting detail, the $\chi^2$ values of combining the four RunII eras (NDF = 3) are viewed.
For the electron and muon channels separately, these are found to transform as follows:
\begin{eqnarray*}
 {\chi^2/NDF}_e & = & 1.22 \rightarrow 3.71 \\
 {\chi^2/NDF}_\mu & = & 0.19 \rightarrow 1.02
\end{eqnarray*}
So if the presented kind of a transformation was required, the $\chi^2/NDF$ for electrons moves from reasonable to slightly unreasonable (too high).
Correspondingly, the muonic value would transform from unreasonable (too low) to reasonable.
This could suggest that the electron channel errors are underestimated.

\newpage

\section{Error Estimates of the Mean in Linear Regression} \label{app:regress}

For $\hat{K}_{\text{JES}}$, the one-parameter linear regression model of Eq.~\eqref{eq:linkdef} is used.
Correspondingly, for $\hat{m}_t$ we use the two-parameter linear regression of Eq.~\eqref{eq:linearity}.
The predictions produced by the fits are interpreted as the best estimates for the mean values of fits.
For the one-parameter regression of $\hat{K}_{\text{JES}}$ based on $N$ data points, this estimate reads~\cite{ref:statbook}
\begin{equation} \label{eq:kjeserr}
 \delta \hat{K}_{\text{JES}} = t_{\alpha/2}^{N-2}
 \times \sqrt{\frac{1}{N-2} \sum_i^N \left(q_0+q_1 \times K_{\text{JES}}^{\text{Res},i} - \hat{K}_{\text{JES}}^i\right)^2}
 \times \sqrt{\frac{1}{N} + \frac{\left(K_{\text{JES}}^{\text{Res}}-\overline{K}_{\text{JES}}^{\text{Res}}\right)^2}{\sum_{i=1}^N \left(K_{\text{JES}}^{\text{Res},i}-\overline{K}_{\text{JES}}^{\text{Res}}\right)^2}},
\end{equation}
where the summations are taken over the data points used for the regression.
By $\overline{K}_{\text{JES}}^{\text{Res}}$ we refer to the average $K_{\text{JES}}^{\text{Res}}$ value of the regression dataset.
The first square root term is the unbiased estimator of the regression error variance, and the second a $K_{\text{JES}}^{\text{Res}}$-dependent factor.
Moreover, $t_{\alpha/2}^{N-2}$ is the Student t score value with $N-2$ degrees of freedom.
Here, we select $\alpha \approx 68\%$ to retain a correspondence to a one sigma deviation.
With this choice, at the the infinite $N$ limit where the Student t distribution corresponds to a normal distribution,
\begin{equation}
 \lim_{N \rightarrow \infty} t_{\alpha/2}^{N-2} = 1. 
\end{equation}
For the two parameter $\hat{m}_t$ fit the corresponding expression becomes more complicated.
It stands as~\cite{ref:statbook}
\begin{equation} \label{eq:mterr}
 \delta \hat{m}_t = t_{\alpha/2}^{N-3}
 \times \sqrt{\frac{1}{N-3} \sum_i^N \left(p_0+p_1 \times m_t^{gen,i} + p_2 \times K_{\text{JES}}^{\text{Res},i} - \hat{m}_t^i\right)^2}
 \times \sqrt{\vec{z}^T \left(\mathbf{X}^T \mathbf{X}\right)^{-1} \vec{z}}.
\end{equation}
This is a generalization of Eq.~\eqref{eq:kjeserr}, where the degrees of freedom have been reduced by one, due to the one additional fit parameter.
The latter square root term conveys dependence on $m_t^{gen}$ and $K_{\text{JES}}^{\text{Res}}$.
The column vector term is defined as
\begin{equation}
 \vec{z} = [1,m_t^{gen},K_{\text{JES}}^{\text{Res}}]^T.
\end{equation}
The measurement matrix $\mathbf{X}$ holds the data points used for regression in its $N$ rows.
The $i$th row $x_i$ of $\mathbf{X}$ consists of the row vector
\begin{equation}
 x_i = [1,m_t^{gen,i},K_{\text{JES}}^{\text{Res},i}].
\end{equation}

\newpage

\section{Example Cases for the Hadronic W Boson Resonance}
\label{app:wexamples}

\begin{figure}[H]
\includegraphics[width=0.33\textwidth]{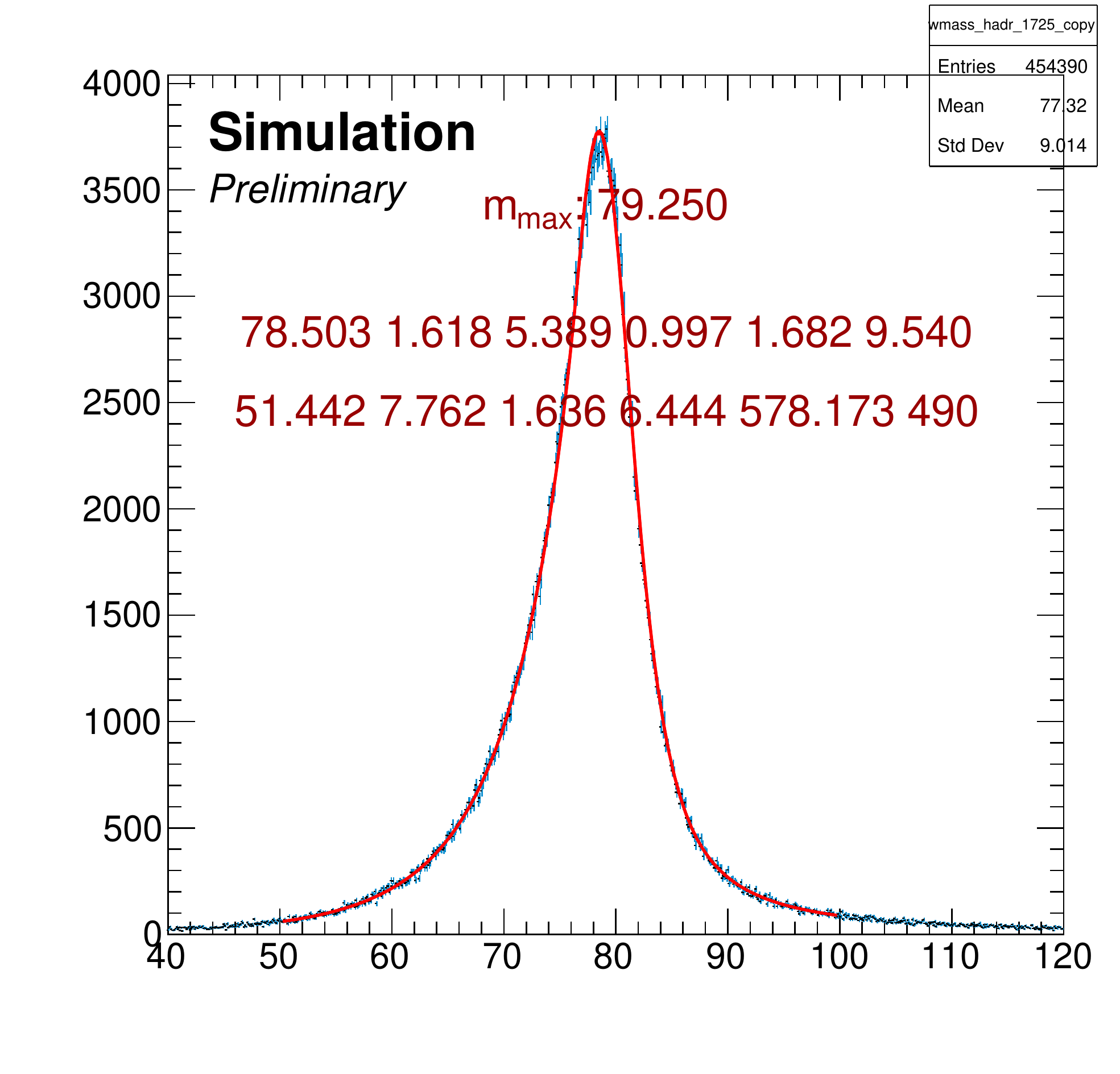}
\includegraphics[width=0.33\textwidth]{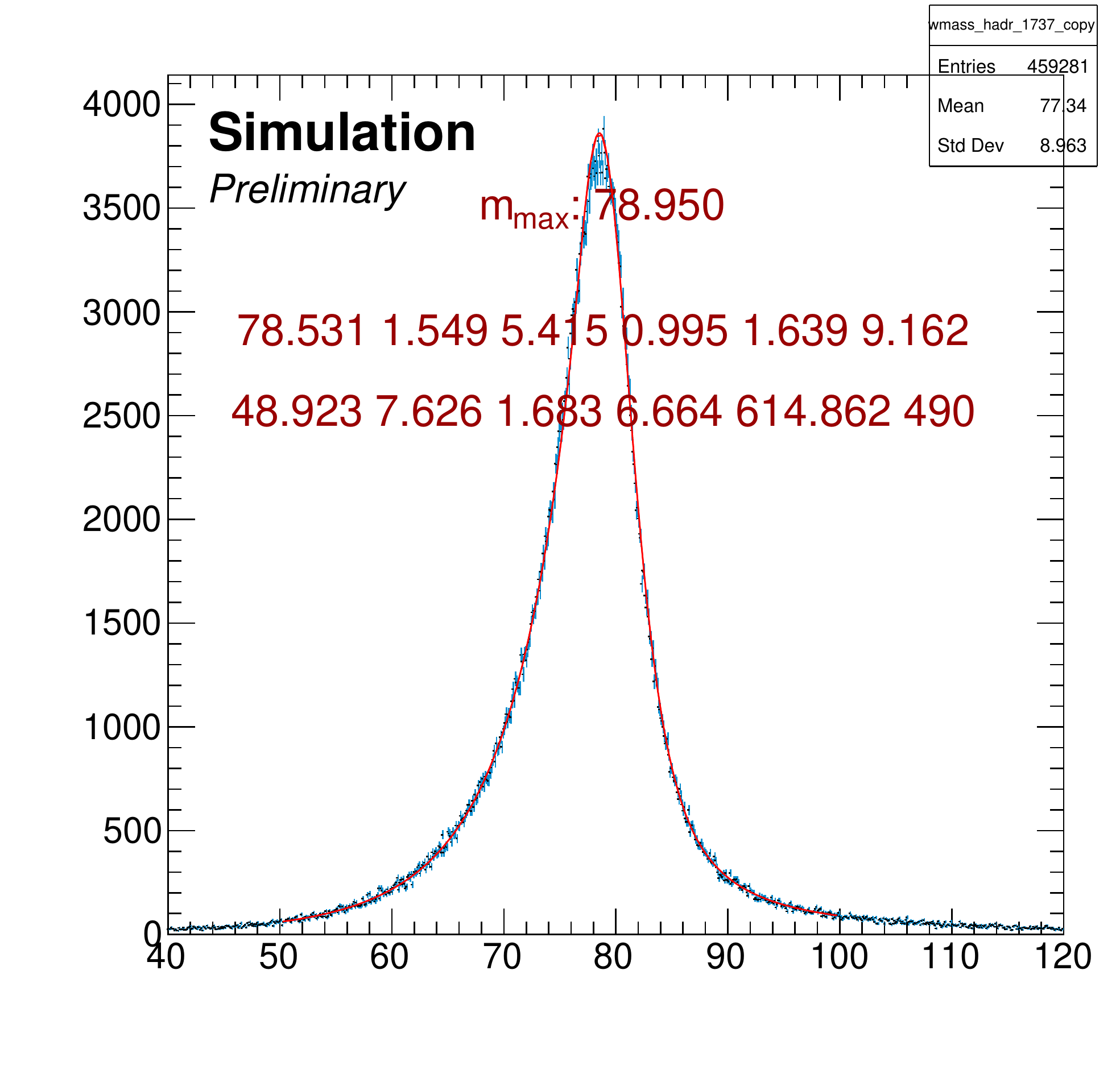}
\includegraphics[width=0.33\textwidth]{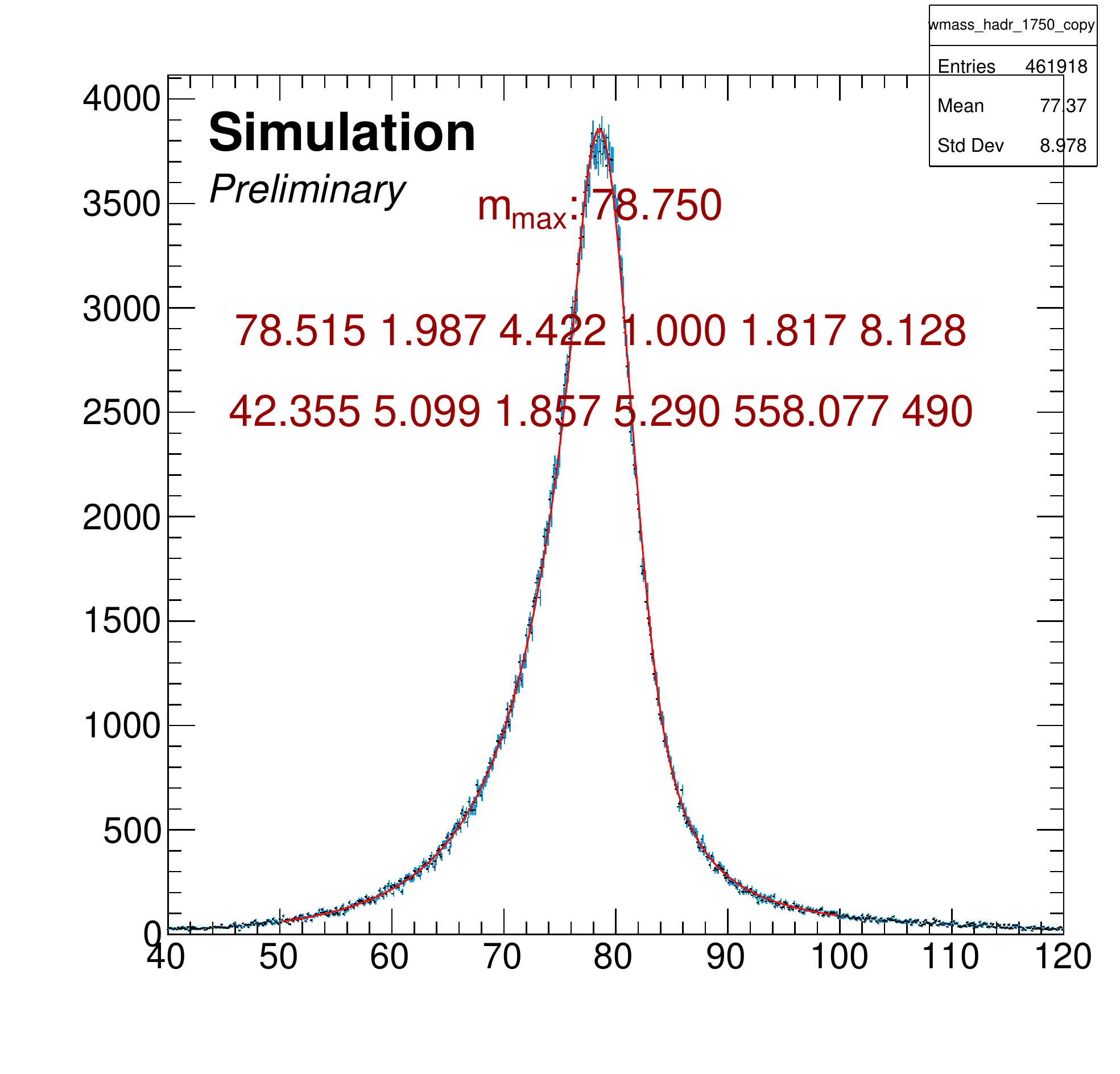}
\caption{Generator level (no $F_{\text{Corr}}$ nor $K_{\text{JES}}^{\text{Res}}$) hadronic W resonances in the electron channel with $m_t^{gen} = 172.5,173.7,175.0$~GeV from left to right.}
\label{fig:genws}
\end{figure}

\begin{figure}[H]
\includegraphics[width=0.195\textwidth]{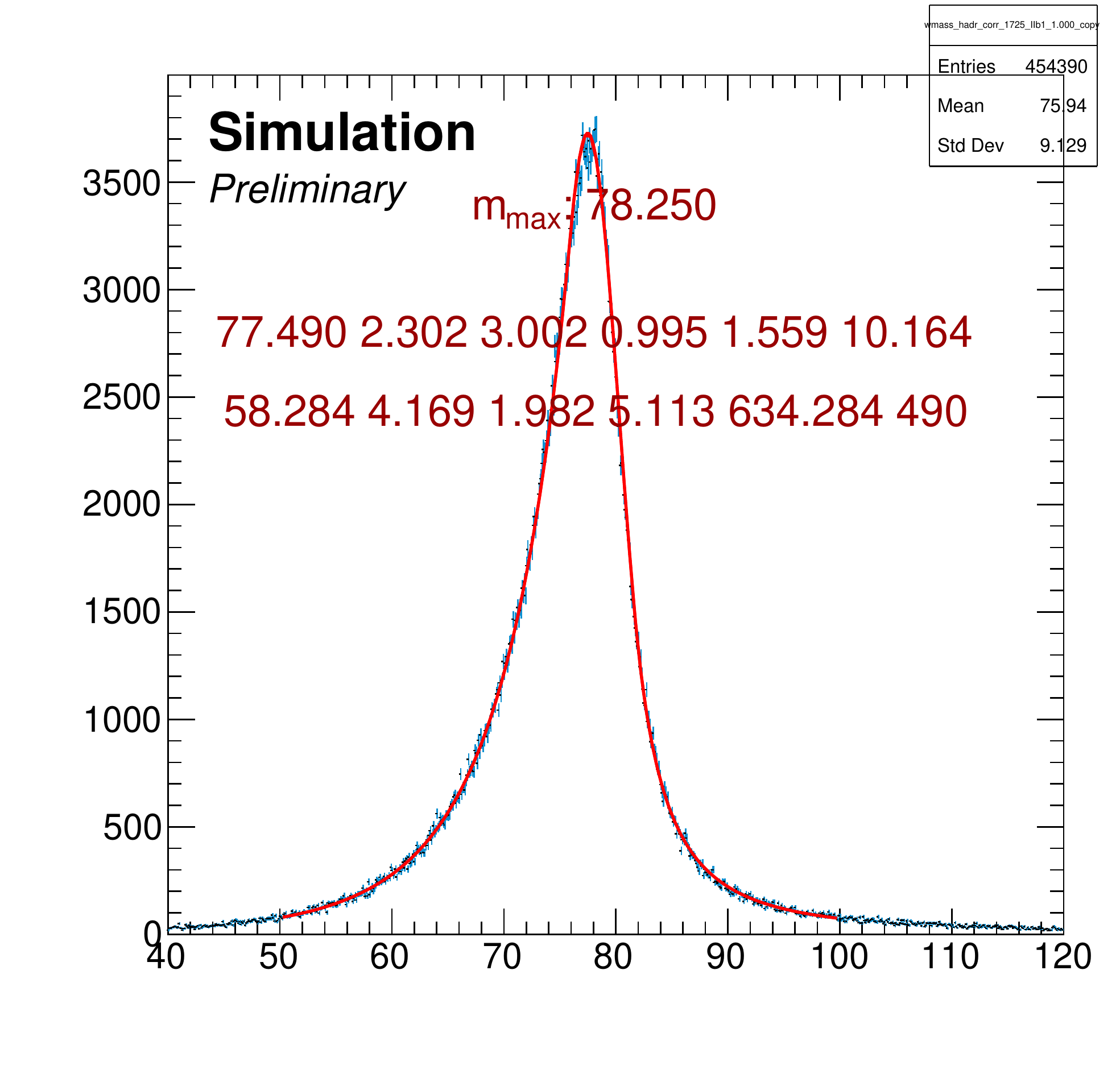}
\includegraphics[width=0.195\textwidth]{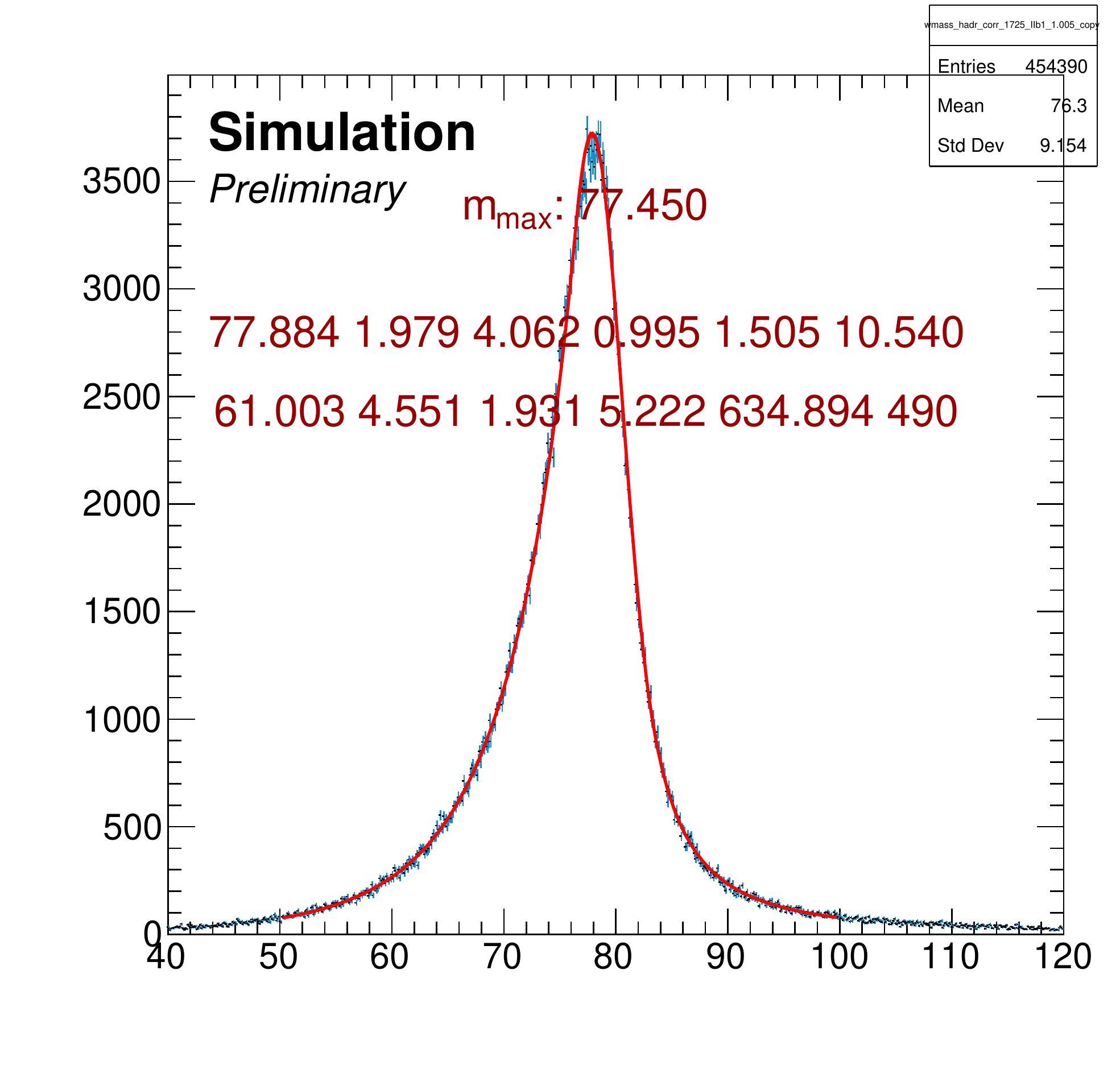}
\includegraphics[width=0.195\textwidth]{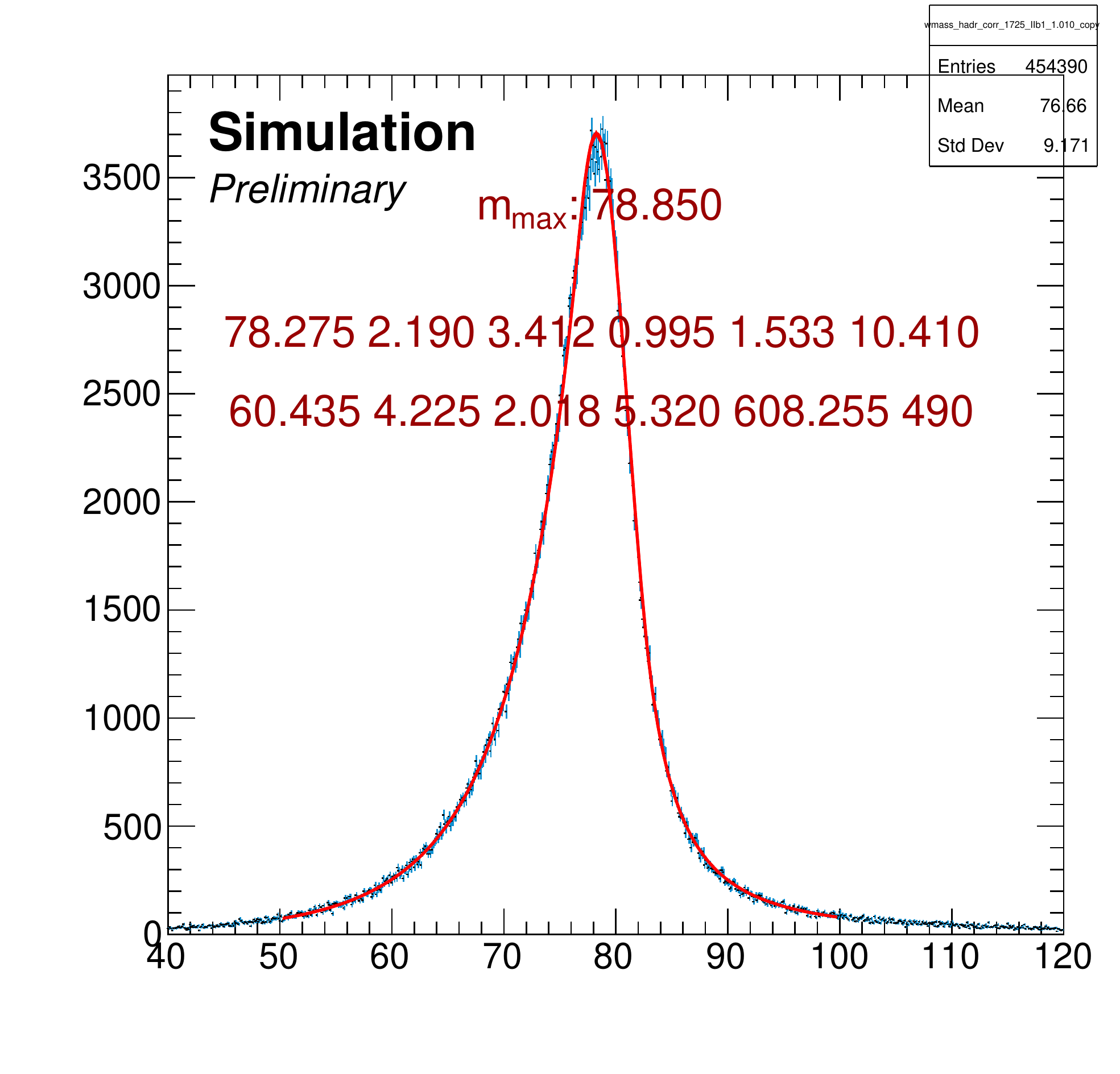}
\includegraphics[width=0.195\textwidth]{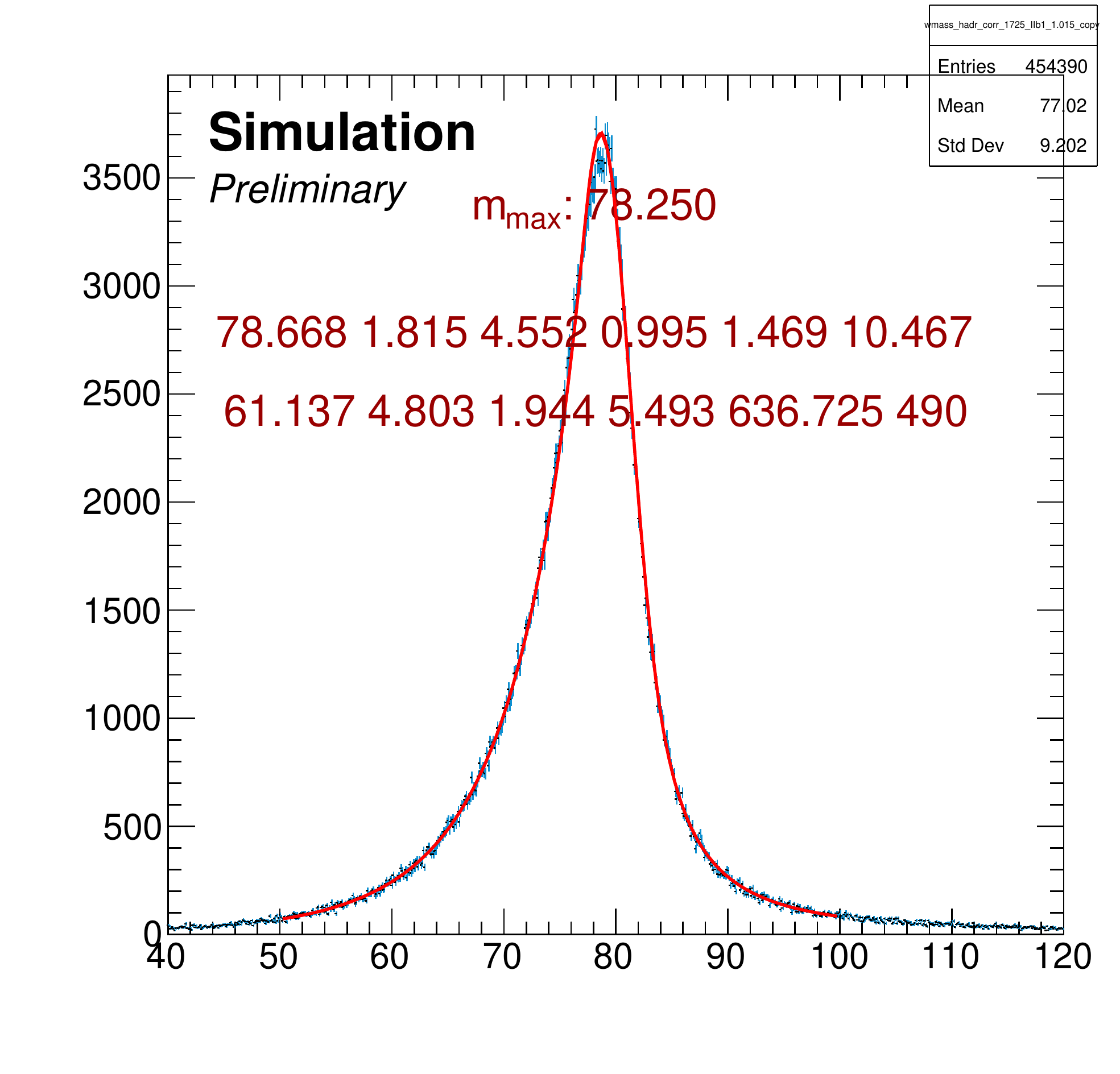}
\includegraphics[width=0.195\textwidth]{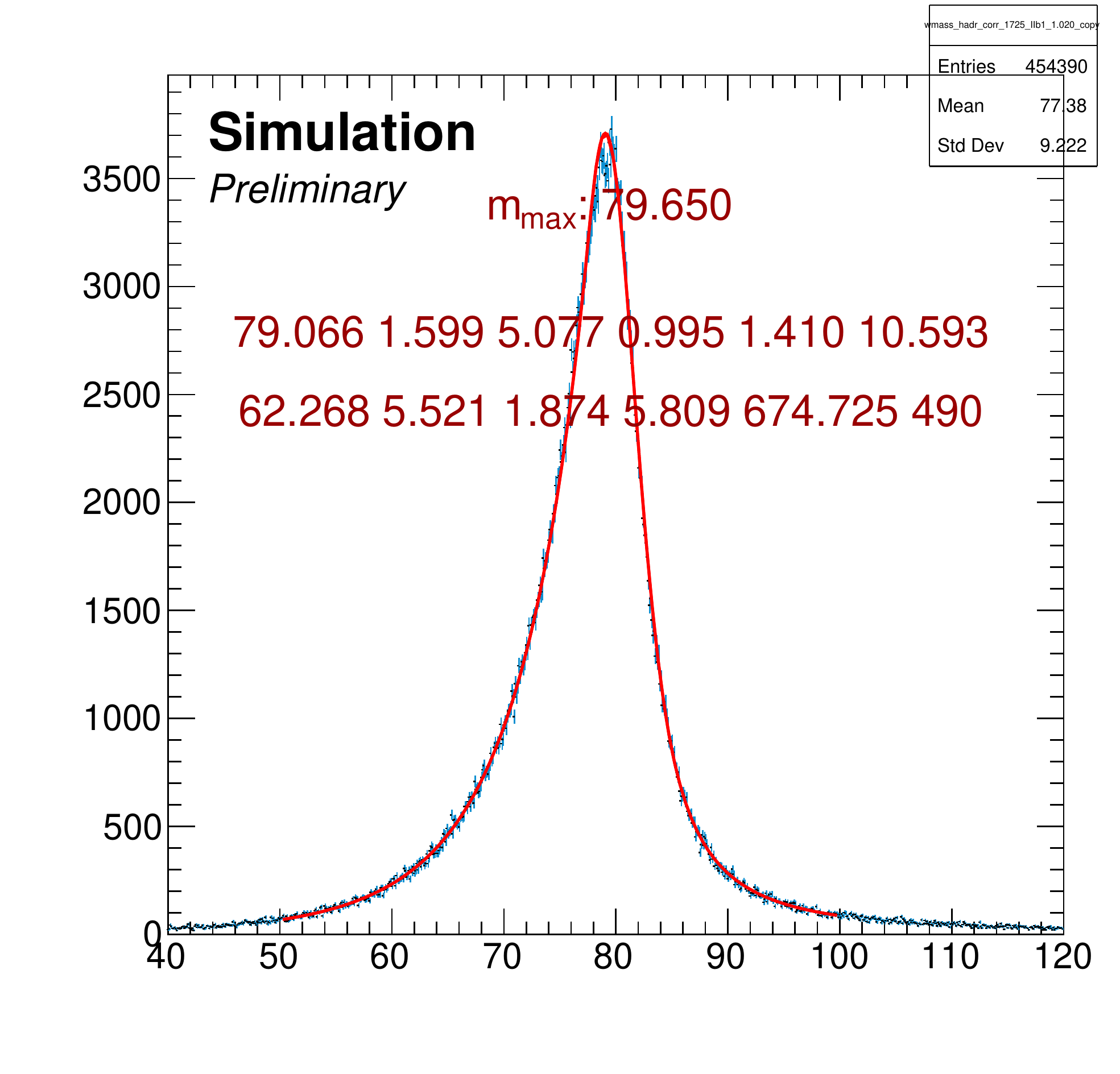}
\includegraphics[width=0.195\textwidth]{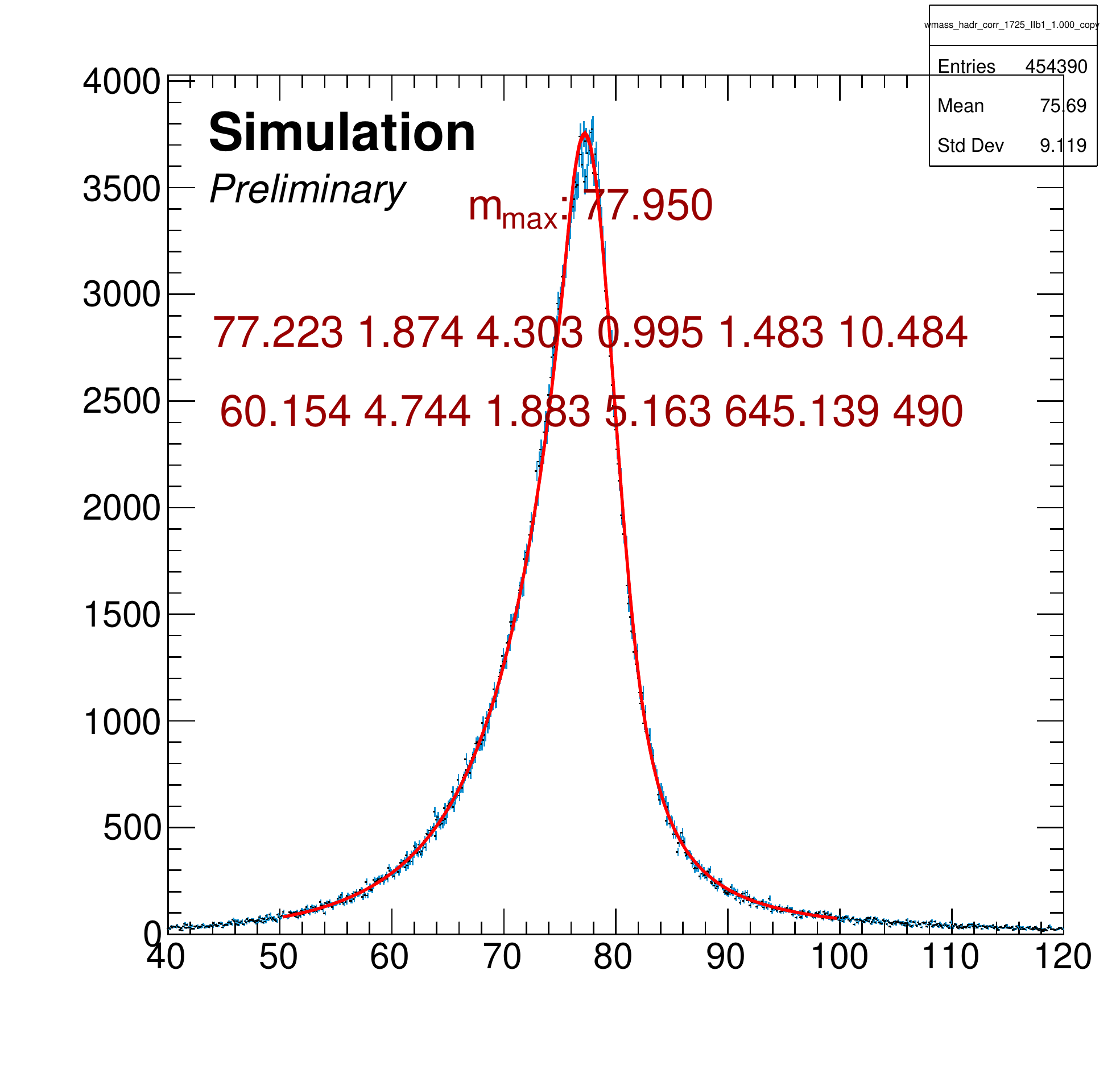}
\includegraphics[width=0.195\textwidth]{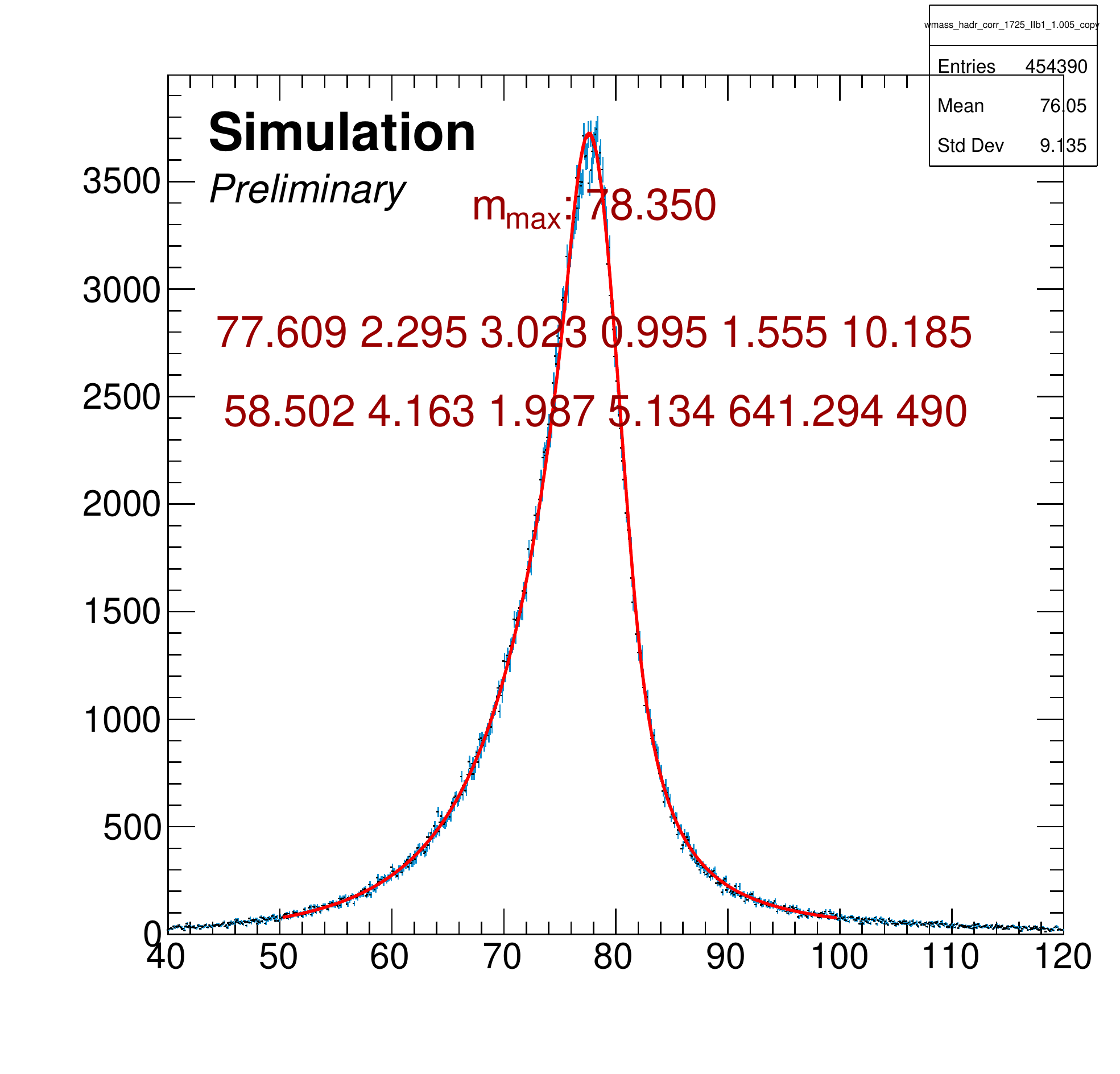}
\includegraphics[width=0.195\textwidth]{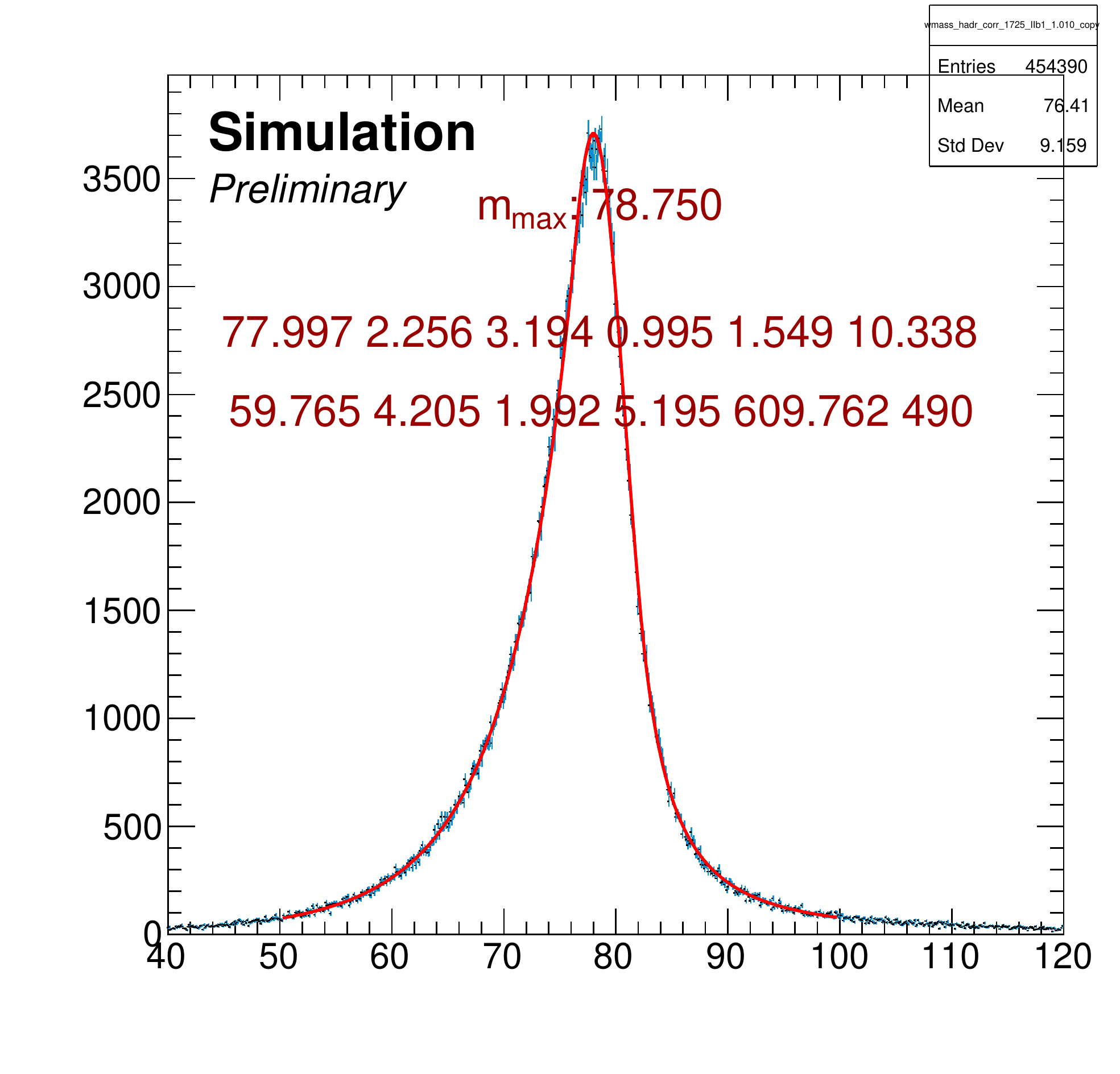}
\includegraphics[width=0.195\textwidth]{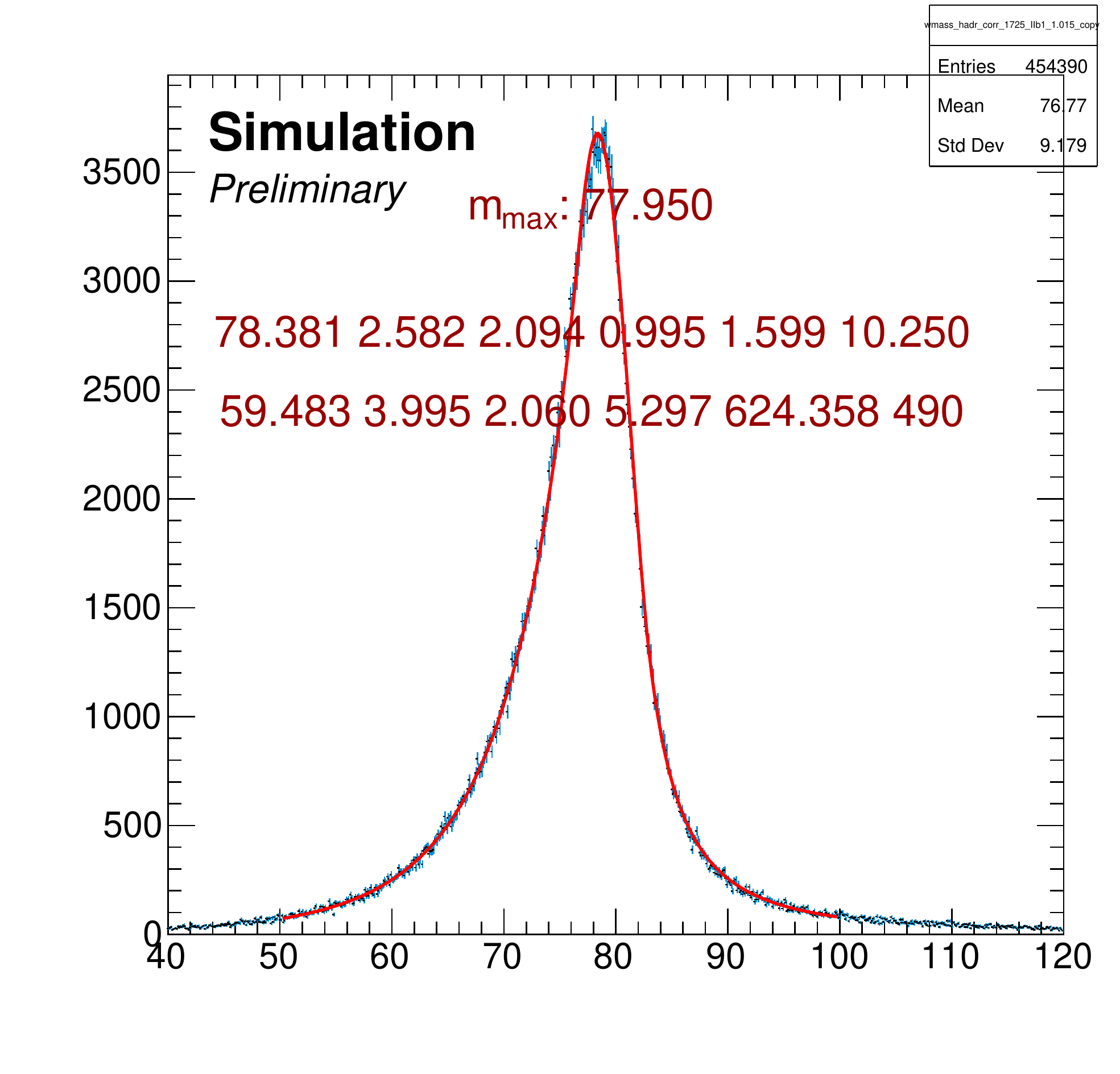}
\includegraphics[width=0.195\textwidth]{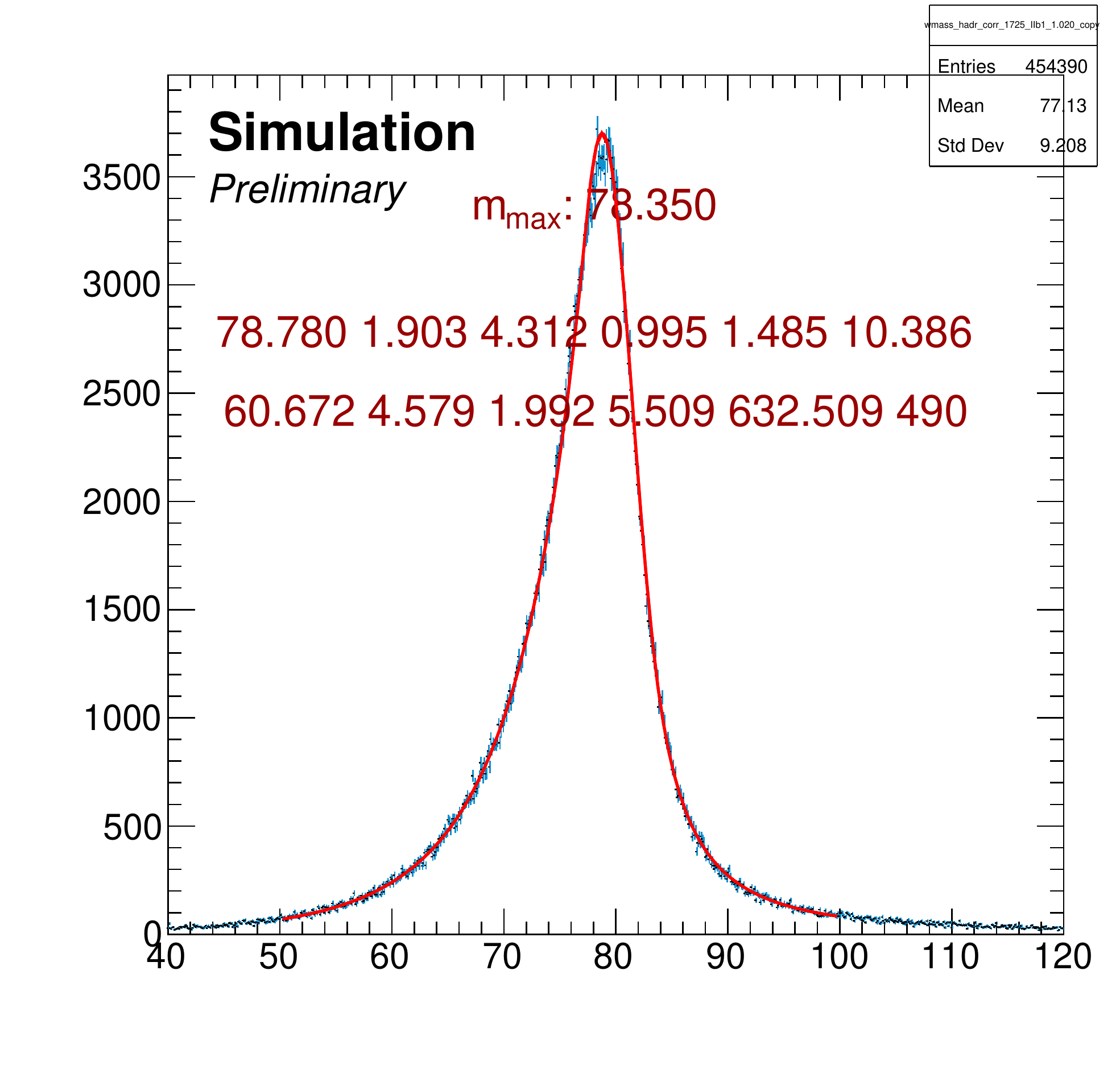}
\caption{Hadronic W resonances in the electron channel at $m_t^{gen} = 172.5$~GeV with Run~IIb1 $F_{\text{Corr}}$ values and $K_{\text{JES}}^{\text{Res}} = 1.0,1.005,1.01,1.015,1.02$ from left to right.
Upper row displays original D\O\ $F_{\text{Corr}}$ parameters, lower row the re-calibrated ones.}
\label{fig:kjesws}
\end{figure}

In Fig.~\ref{fig:genws} generator-level W resonances are shown at different $m_t^{gen}$ values.
In Fig.~\ref{fig:kjesws} the $K_{\text{JES}}^{\text{Res}}$ dependence of the W resonance is shown in specific example cases.
The red numbers show the maximum position, some fit parameters and fit statistics.

\newpage

\section{Example Cases for the Hadronic Top Mass Resonance}
\label{app:htexamples}

\begin{figure}[H]
\includegraphics[width=0.33\textwidth]{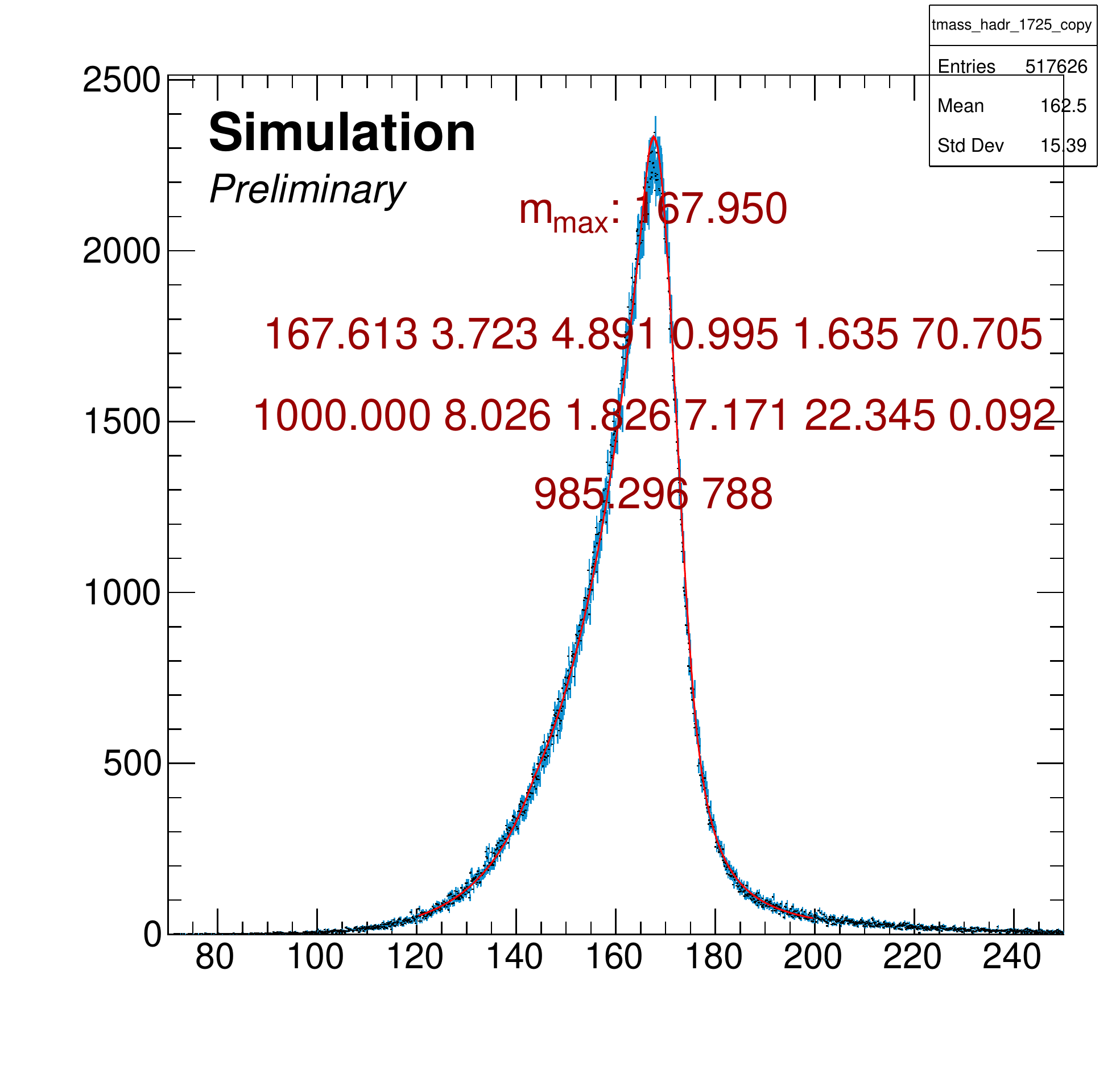}
\includegraphics[width=0.33\textwidth]{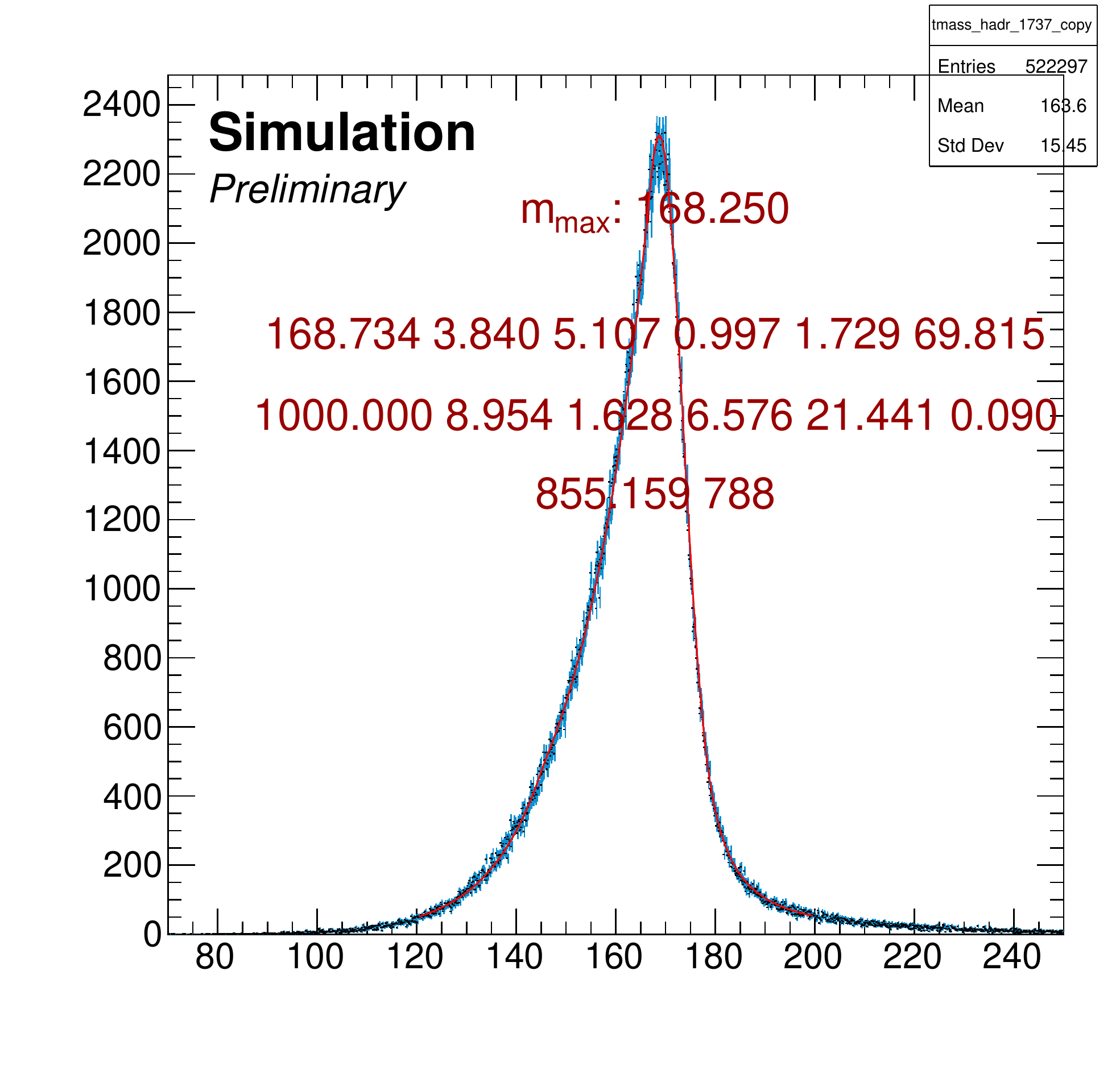}
\includegraphics[width=0.33\textwidth]{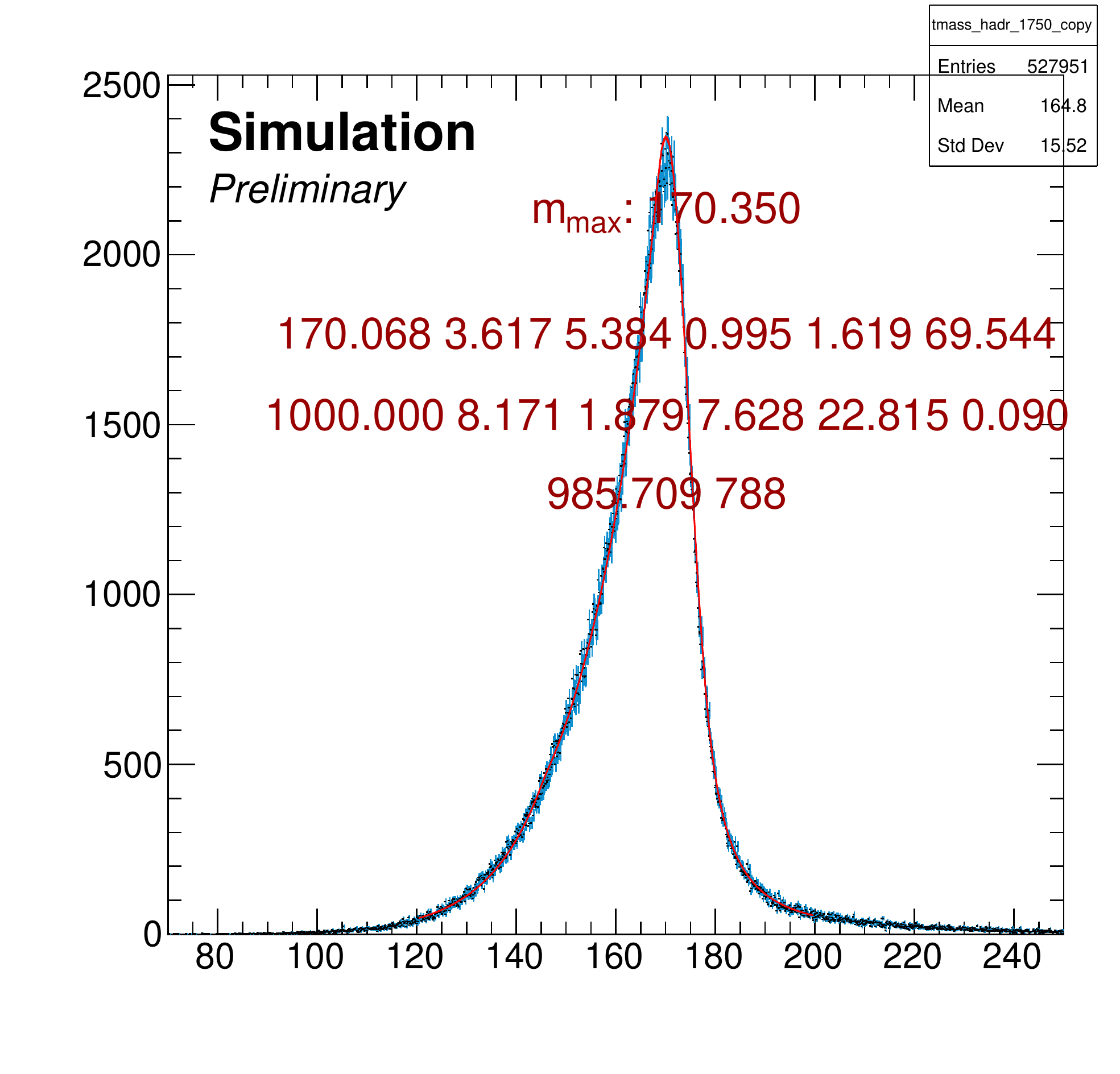}
\caption{Generator level (no $F_{\text{Corr}}$ nor $K_{\text{JES}}^{\text{Res}}$) hadronic top resonances in the muon channel with $m_t^{gen} = 172.5,173.7,175.0$~GeV from left to right.}
\label{fig:genhts}
\end{figure}

\begin{figure}[H]
\includegraphics[width=0.195\textwidth]{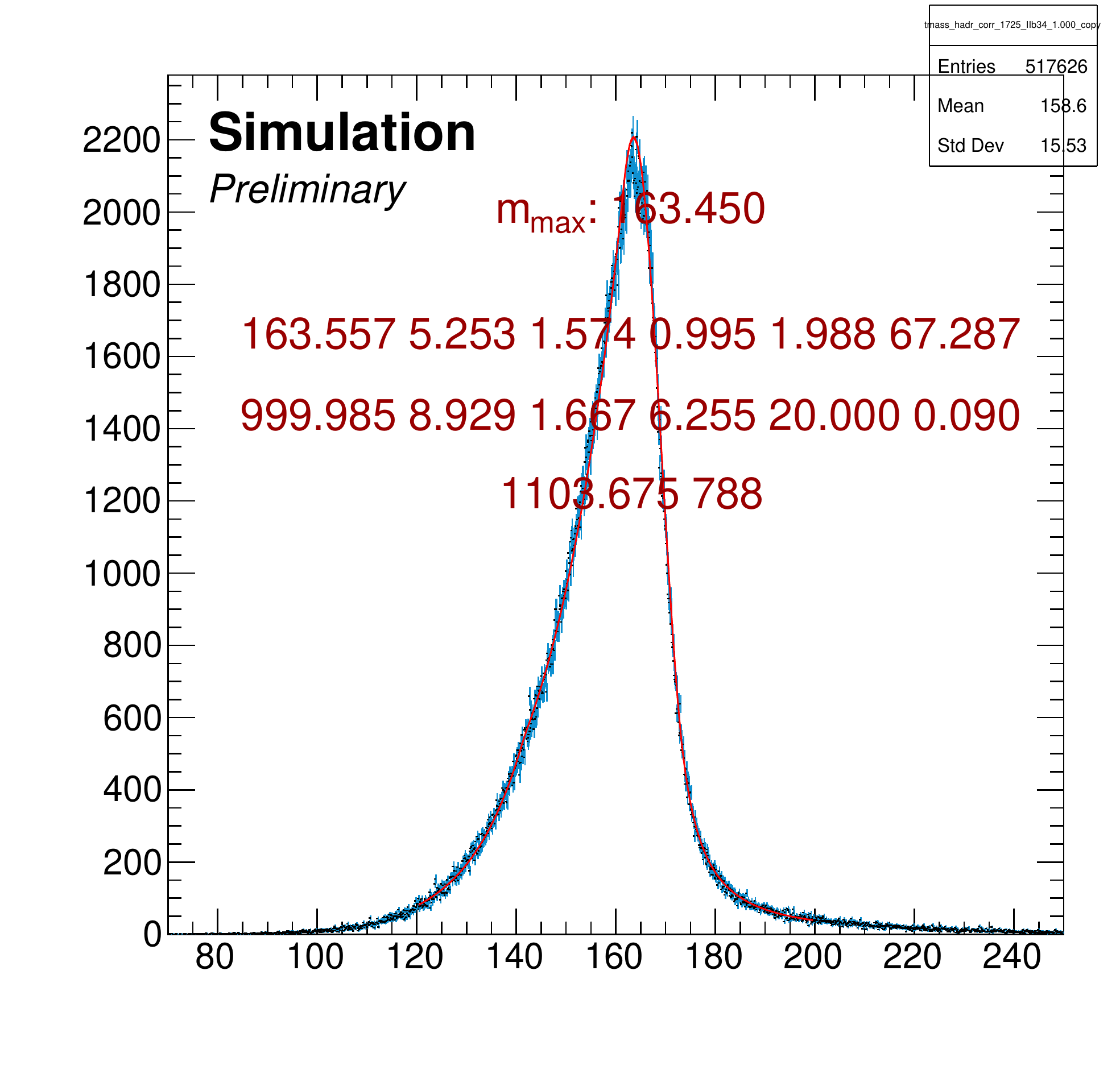}
\includegraphics[width=0.195\textwidth]{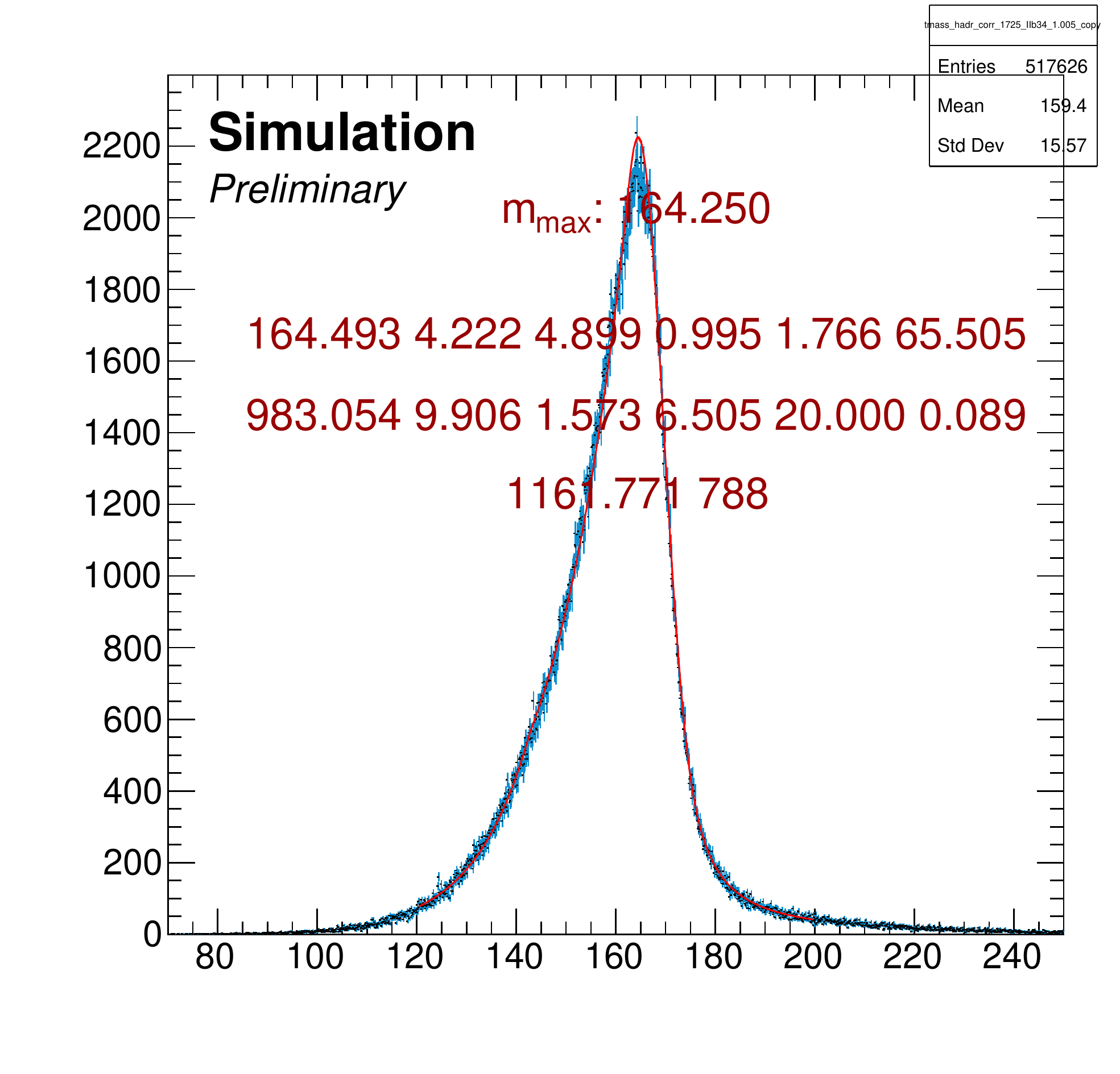}
\includegraphics[width=0.195\textwidth]{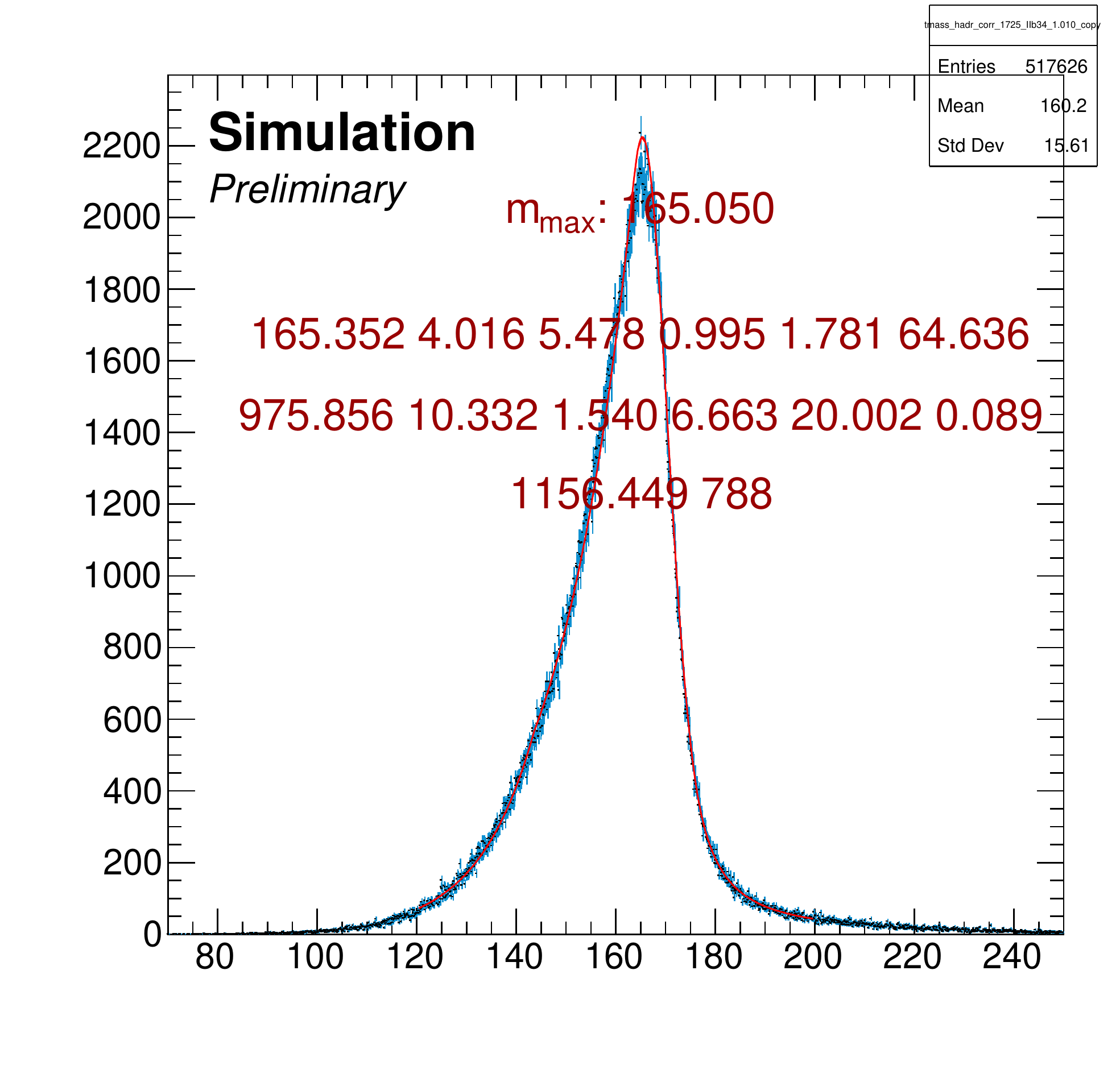}
\includegraphics[width=0.195\textwidth]{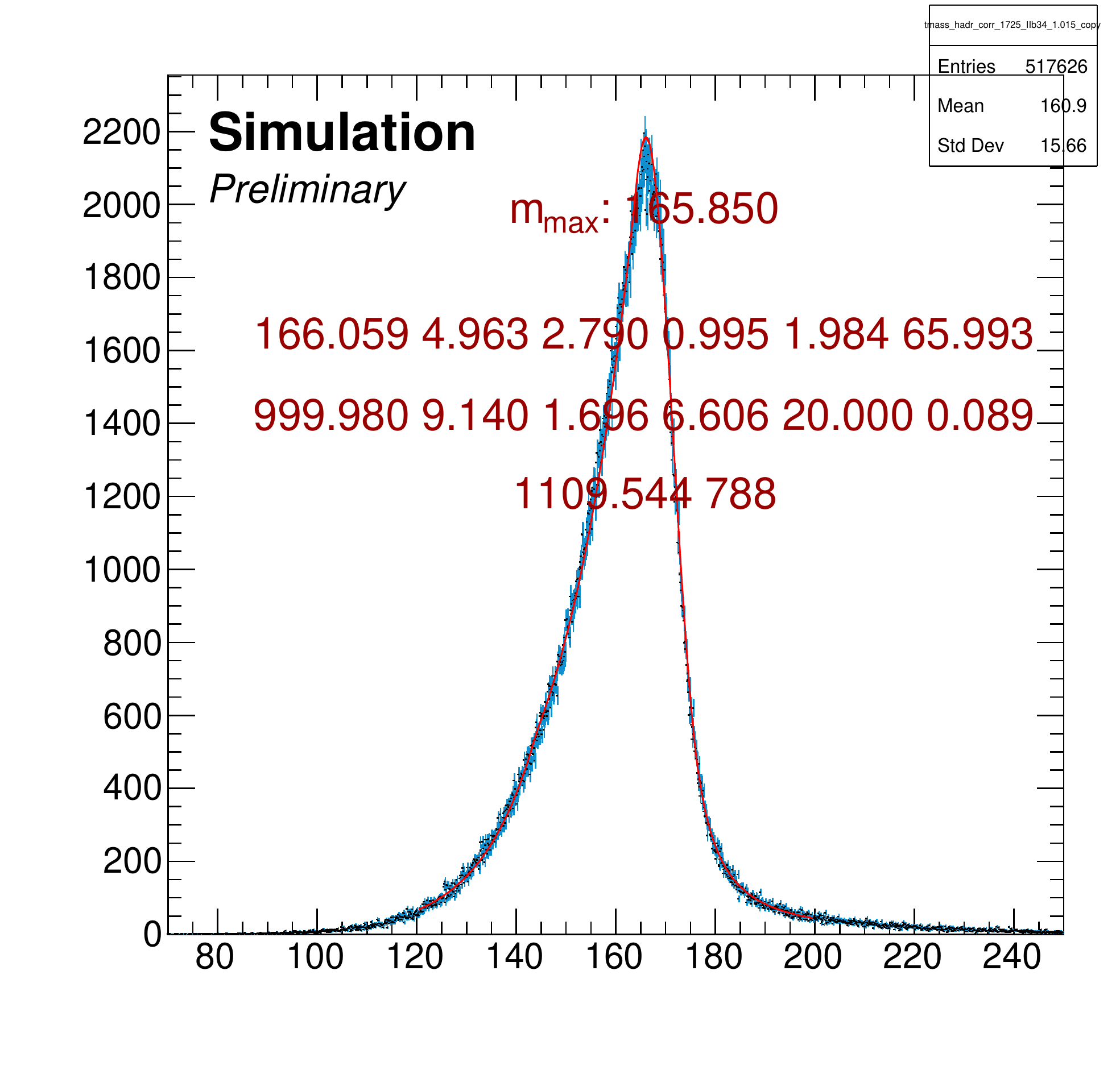}
\includegraphics[width=0.195\textwidth]{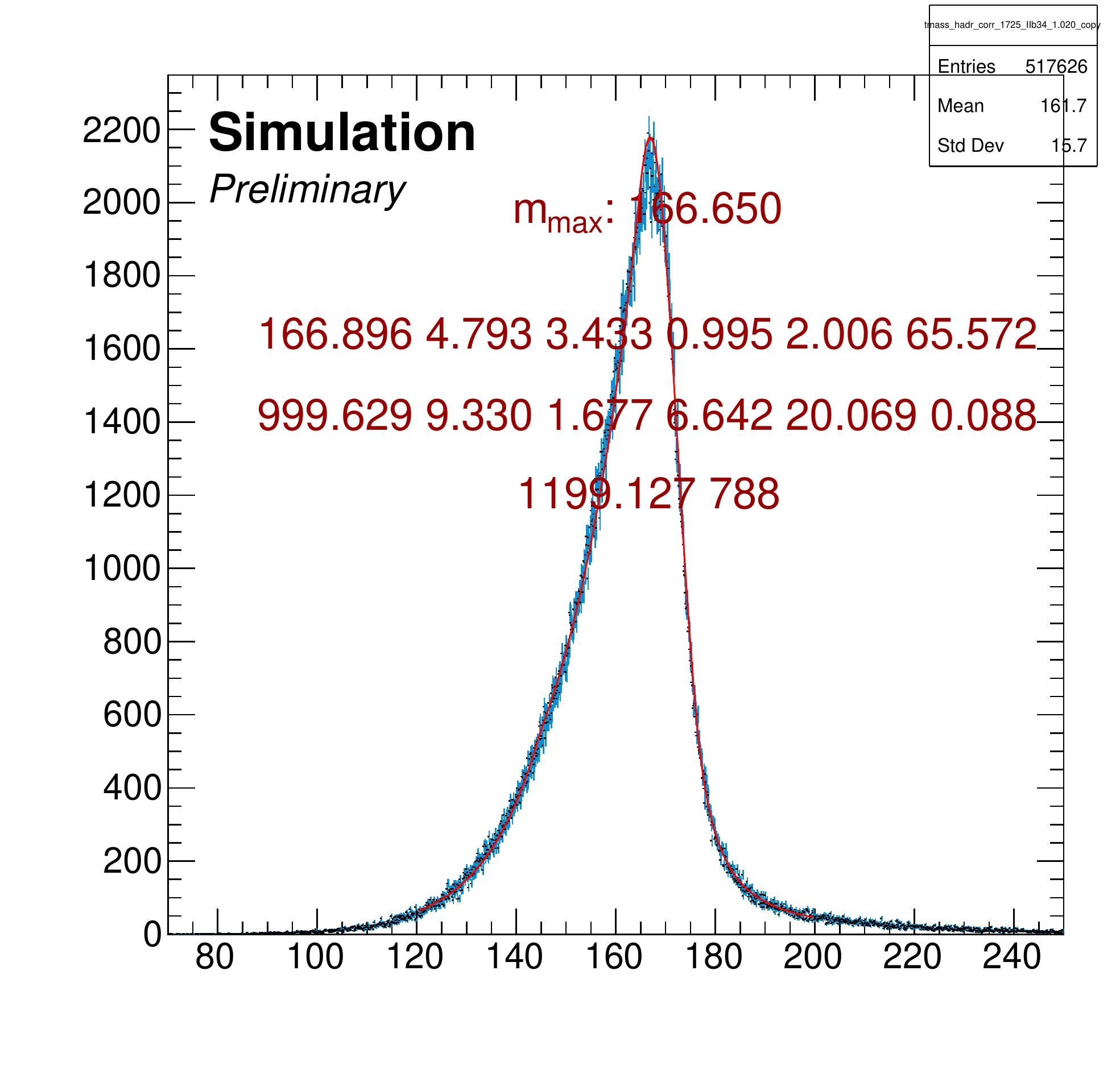}
\includegraphics[width=0.195\textwidth]{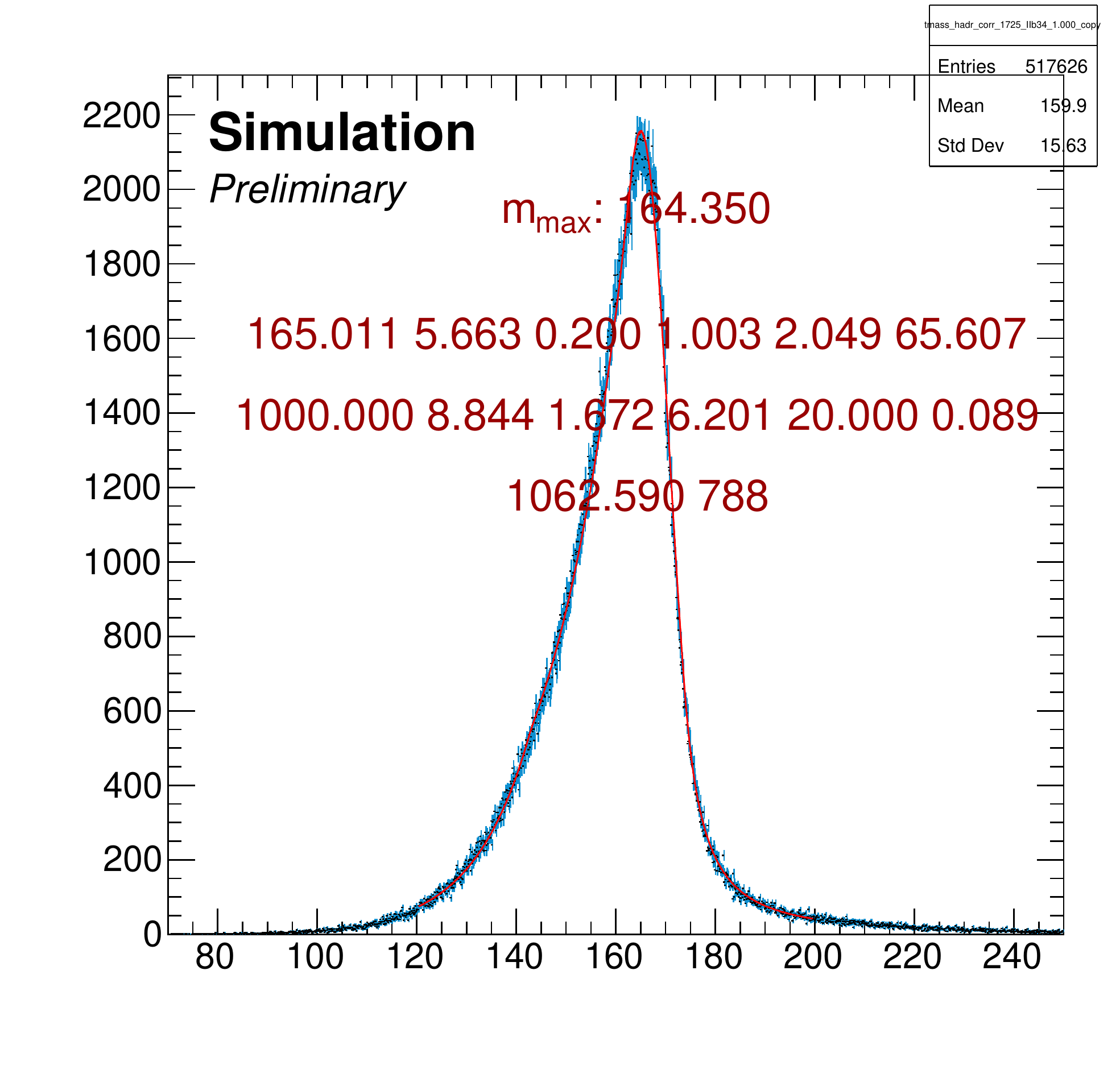}
\includegraphics[width=0.195\textwidth]{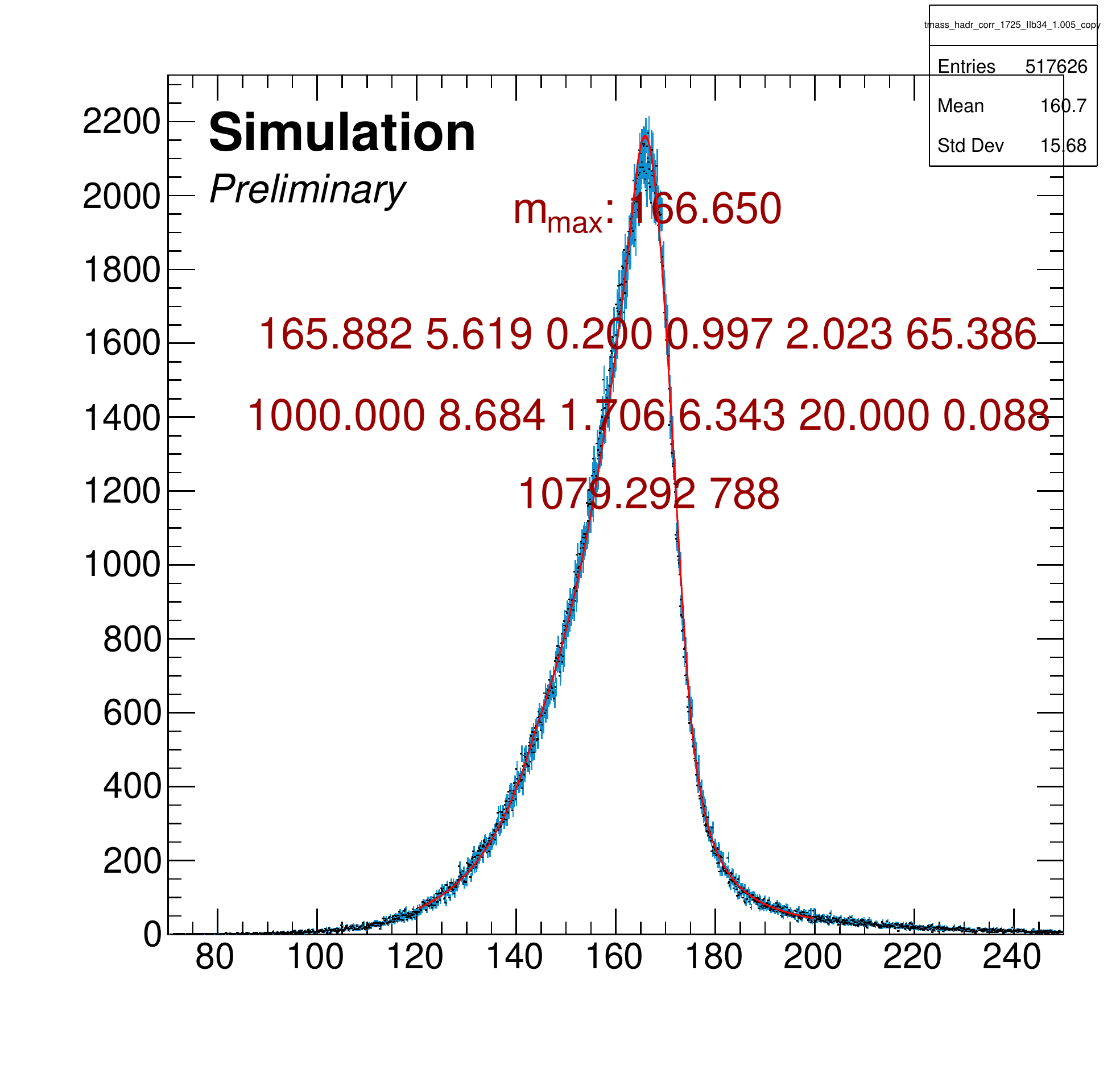}
\includegraphics[width=0.195\textwidth]{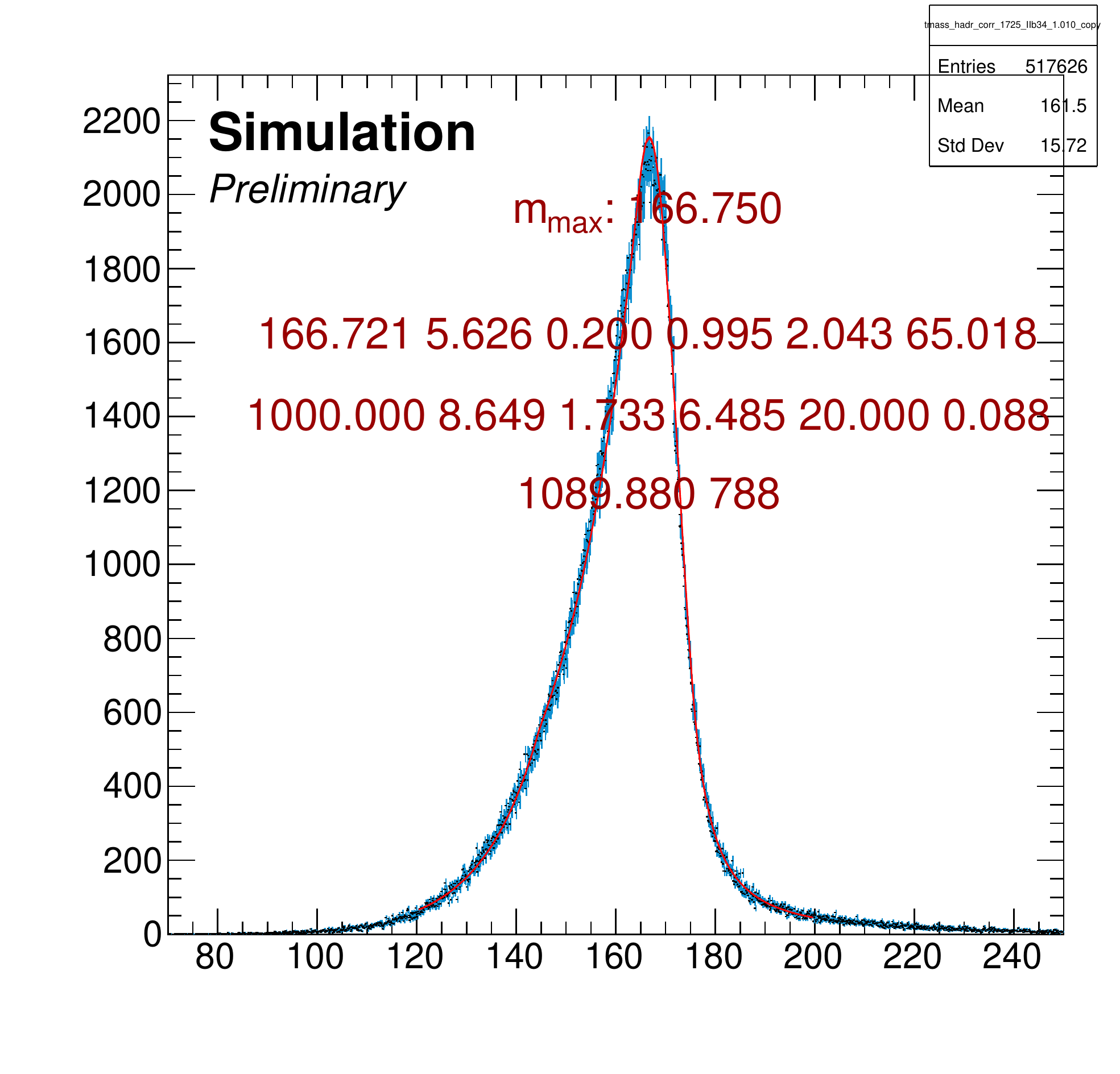}
\includegraphics[width=0.195\textwidth]{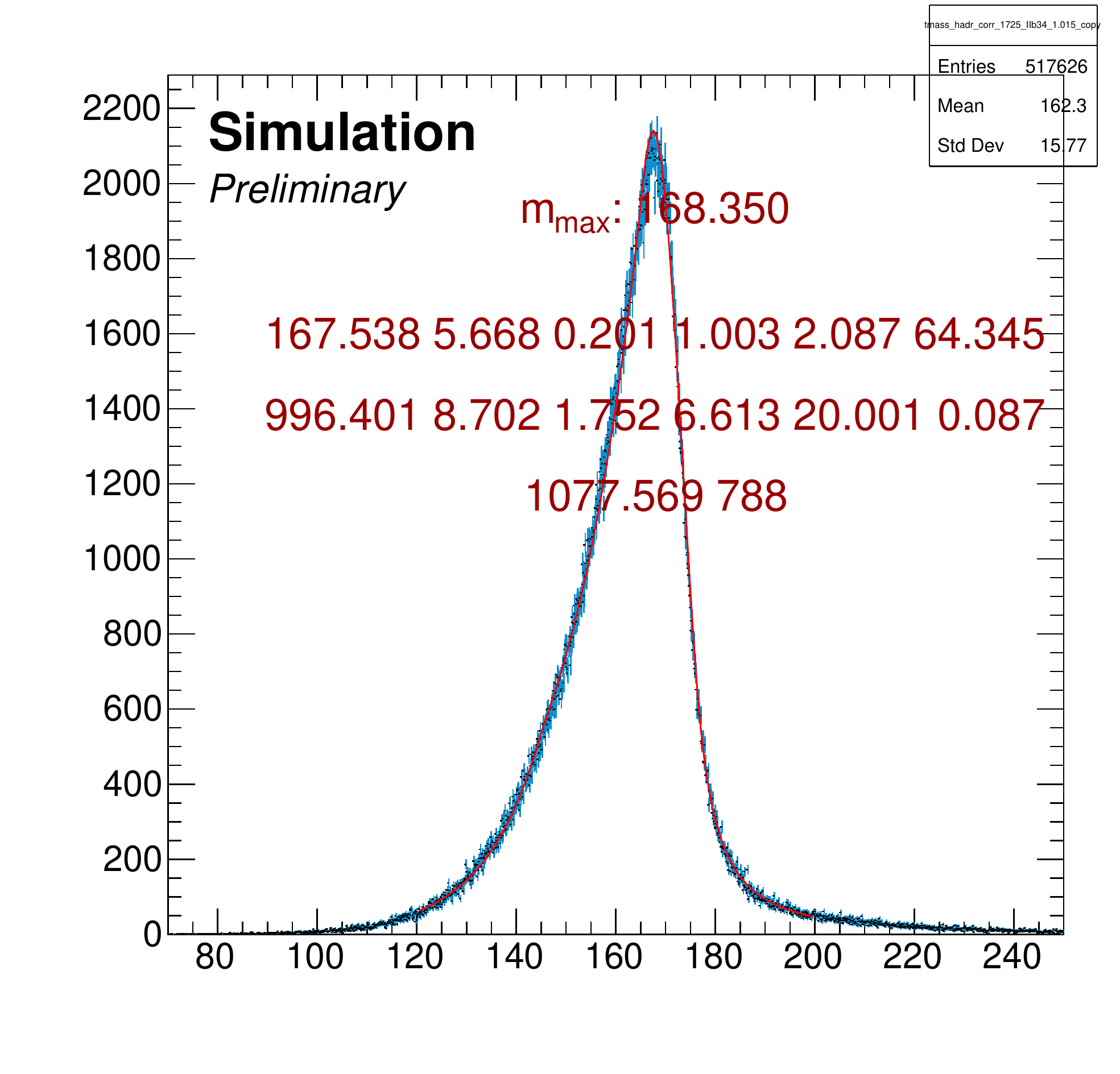}
\includegraphics[width=0.195\textwidth]{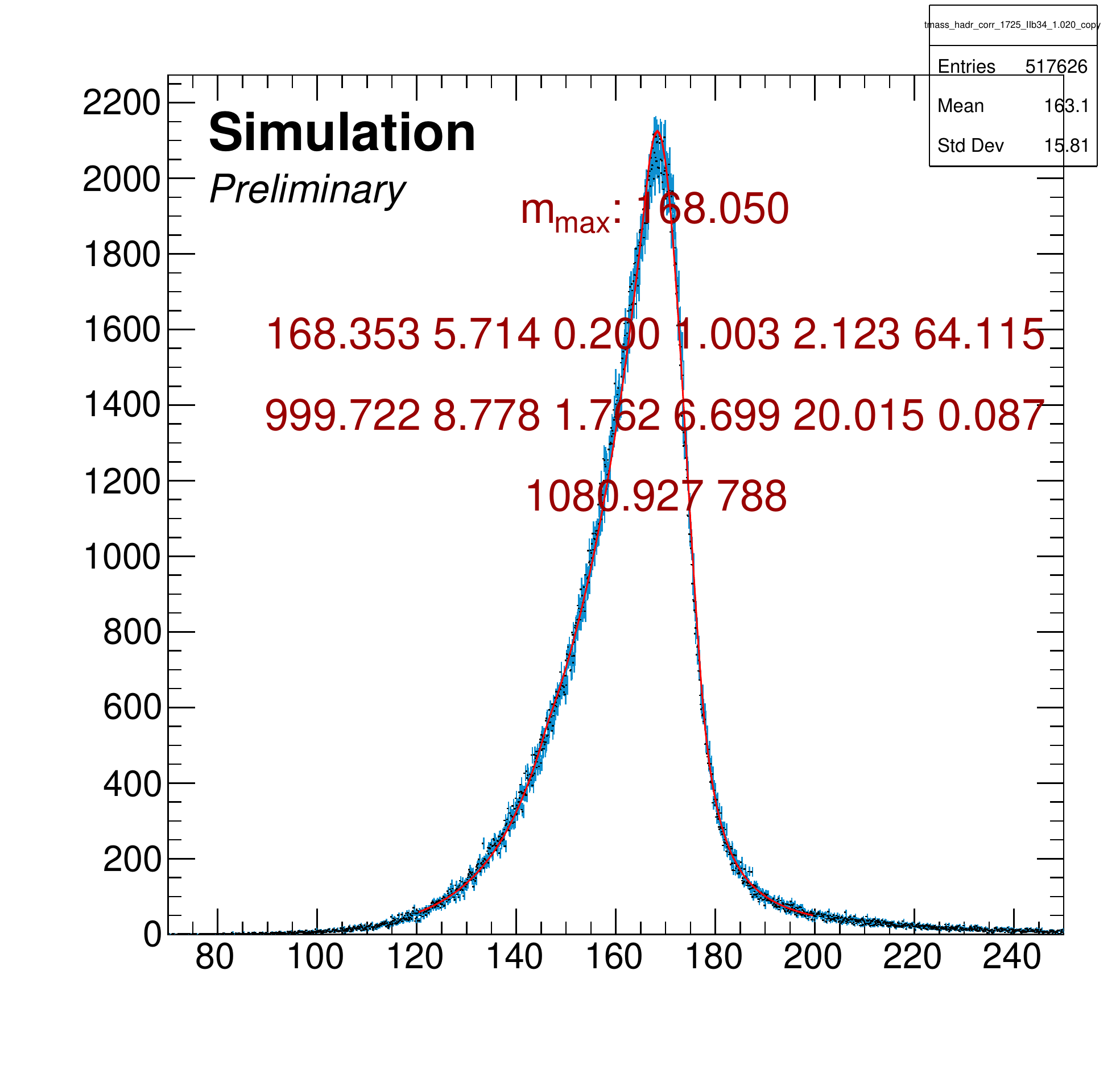}
\caption{Hadronic top resonances in the muon channel at $m_t^{gen} = 172.5$~GeV with Run~IIb34 $F_{\text{Corr}}$ values and $K_{\text{JES}}^{\text{Res}} = 1.0,1.005,1.01,1.015,1.02$ from left to right.
Upper row displays original D\O\ $F_{\text{Corr}}$ parameters, lower row the re-calibrated ones.}
\label{fig:kjeshts}
\end{figure}

In Fig.~\ref{fig:genhts} generator-level hadronic top resonances are shown at different $m_t^{gen}$ values.
In Fig.~\ref{fig:kjeshts} the $K_{\text{JES}}^{\text{Res}}$ dependence of the hadronic top resonance is shown in specific example cases.
The red numbers show the maximum position, some fit parameters and fit statistics.

\newpage

\section{Example Cases for the Leptonic Top Mass Resonance}
\label{app:hlexamples}

\begin{figure}[H]
\includegraphics[width=0.33\textwidth]{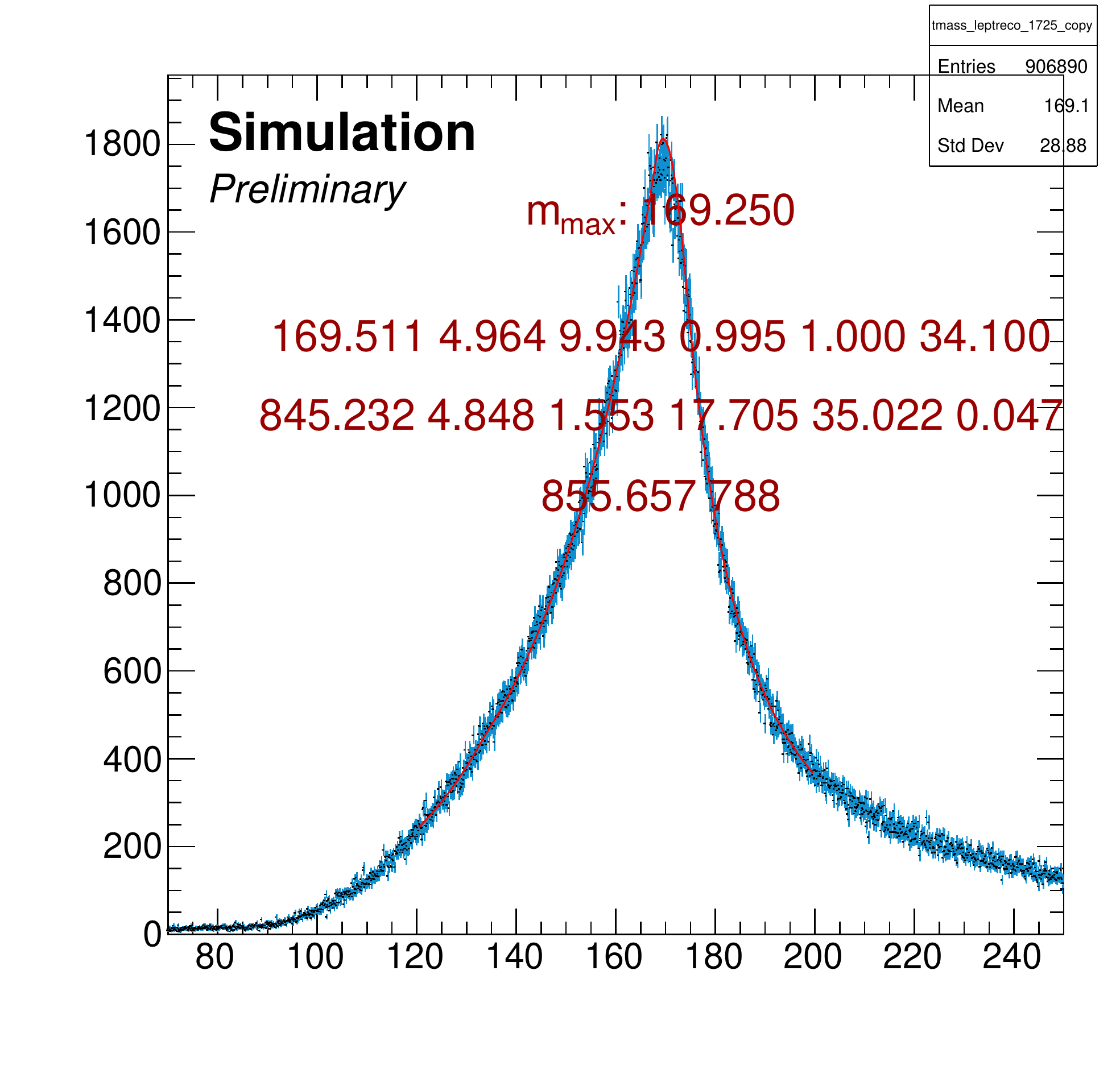}
\includegraphics[width=0.33\textwidth]{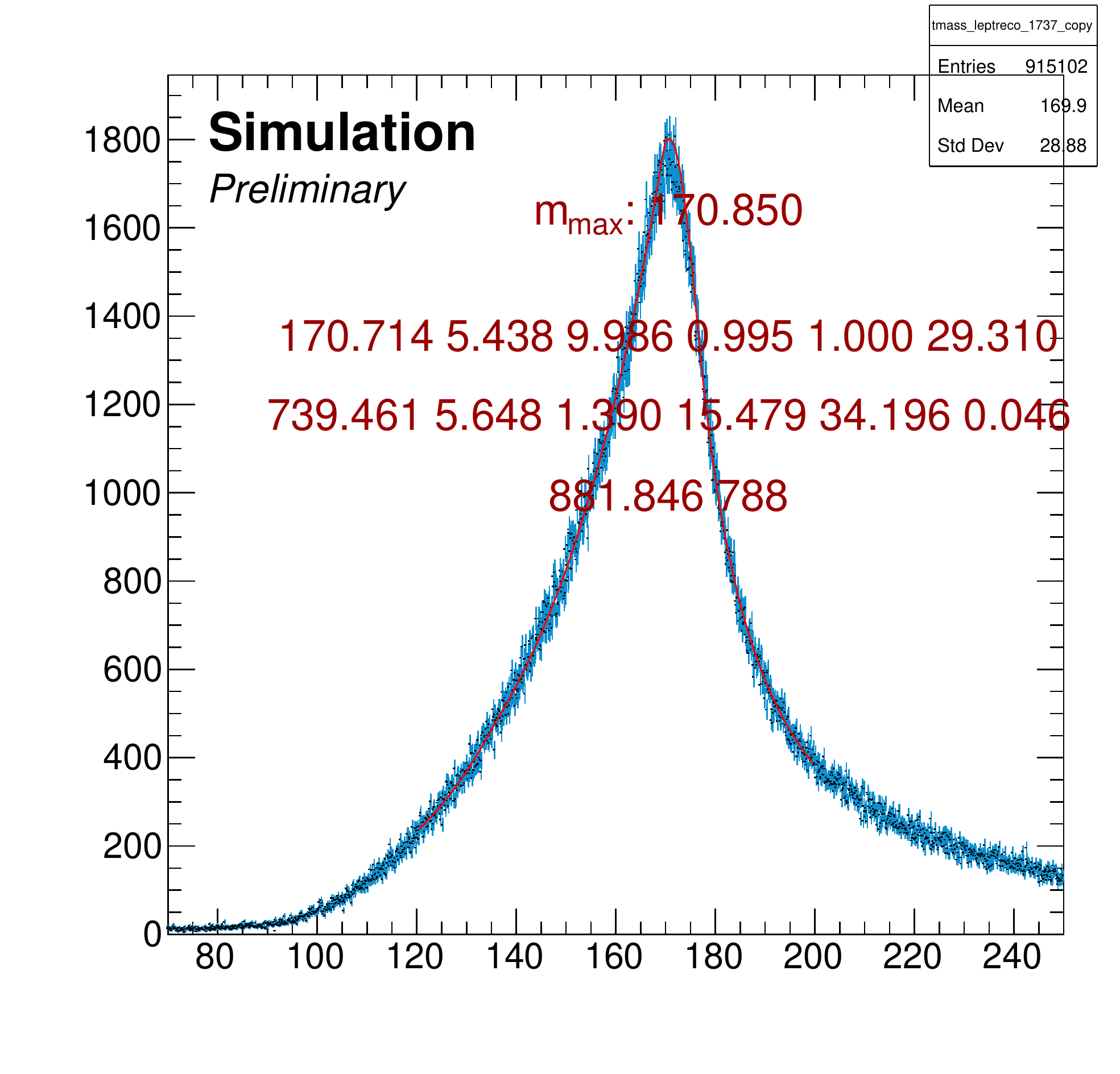}
\includegraphics[width=0.33\textwidth]{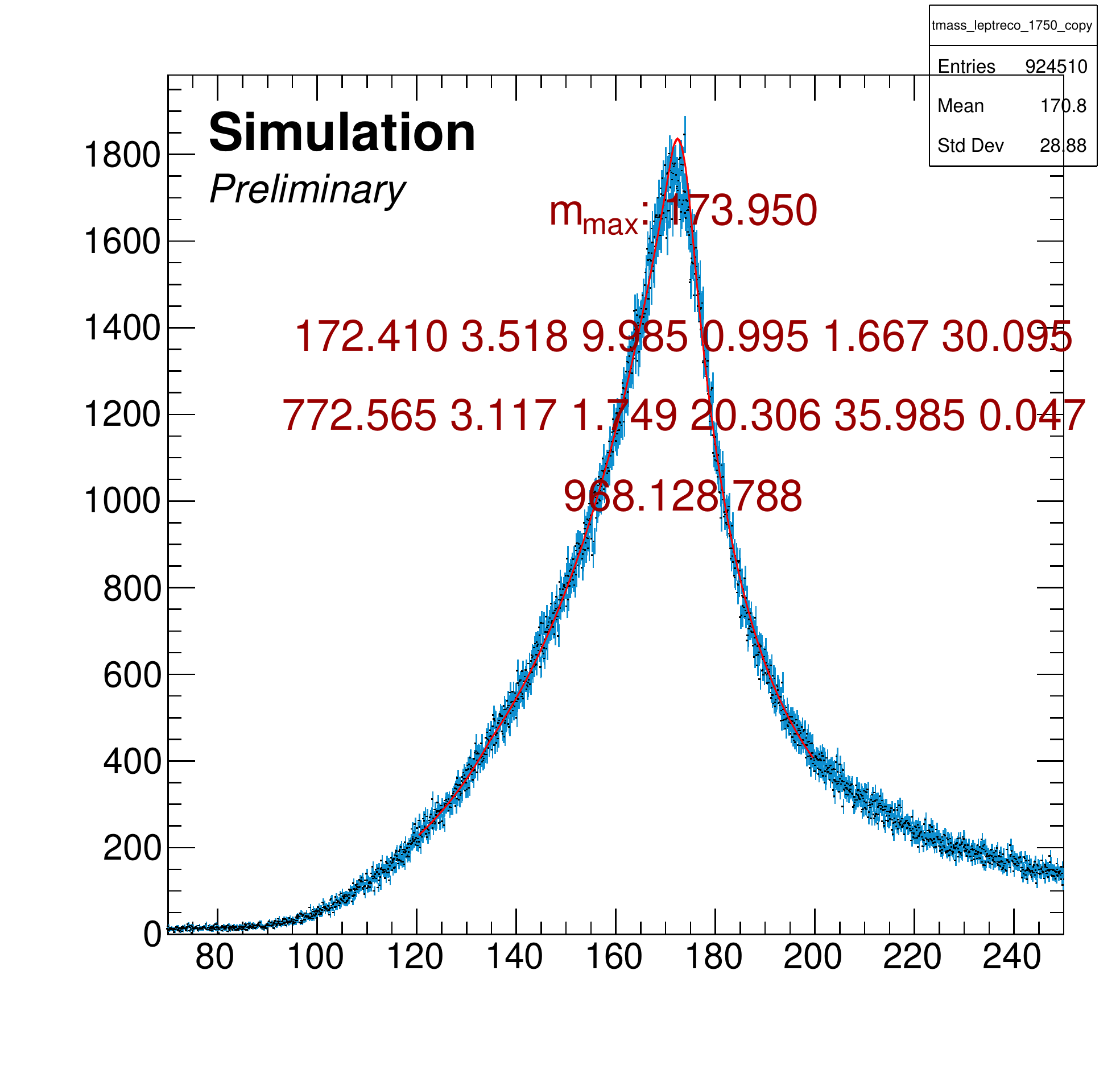}
\caption{Generator level (no $F_{\text{Corr}}$ nor $K_{\text{JES}}^{\text{Res}}$) leptonic top resonances in the muon channel with $m_t^{gen} = 172.5,173.7,175.0$~GeV from left to right.}
\label{fig:genlts}
\end{figure}

\begin{figure}[H]
\includegraphics[width=0.195\textwidth]{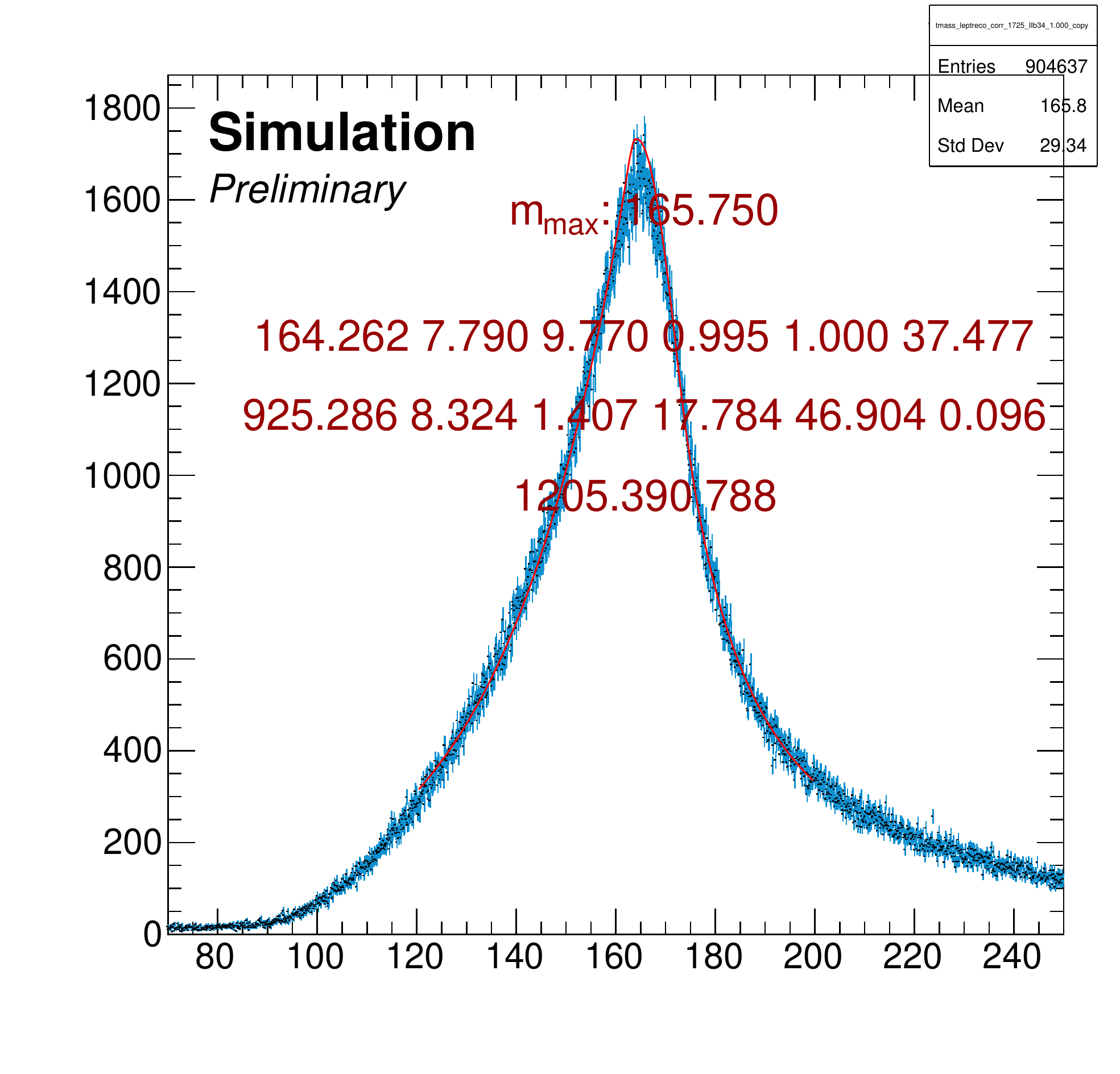}
\includegraphics[width=0.195\textwidth]{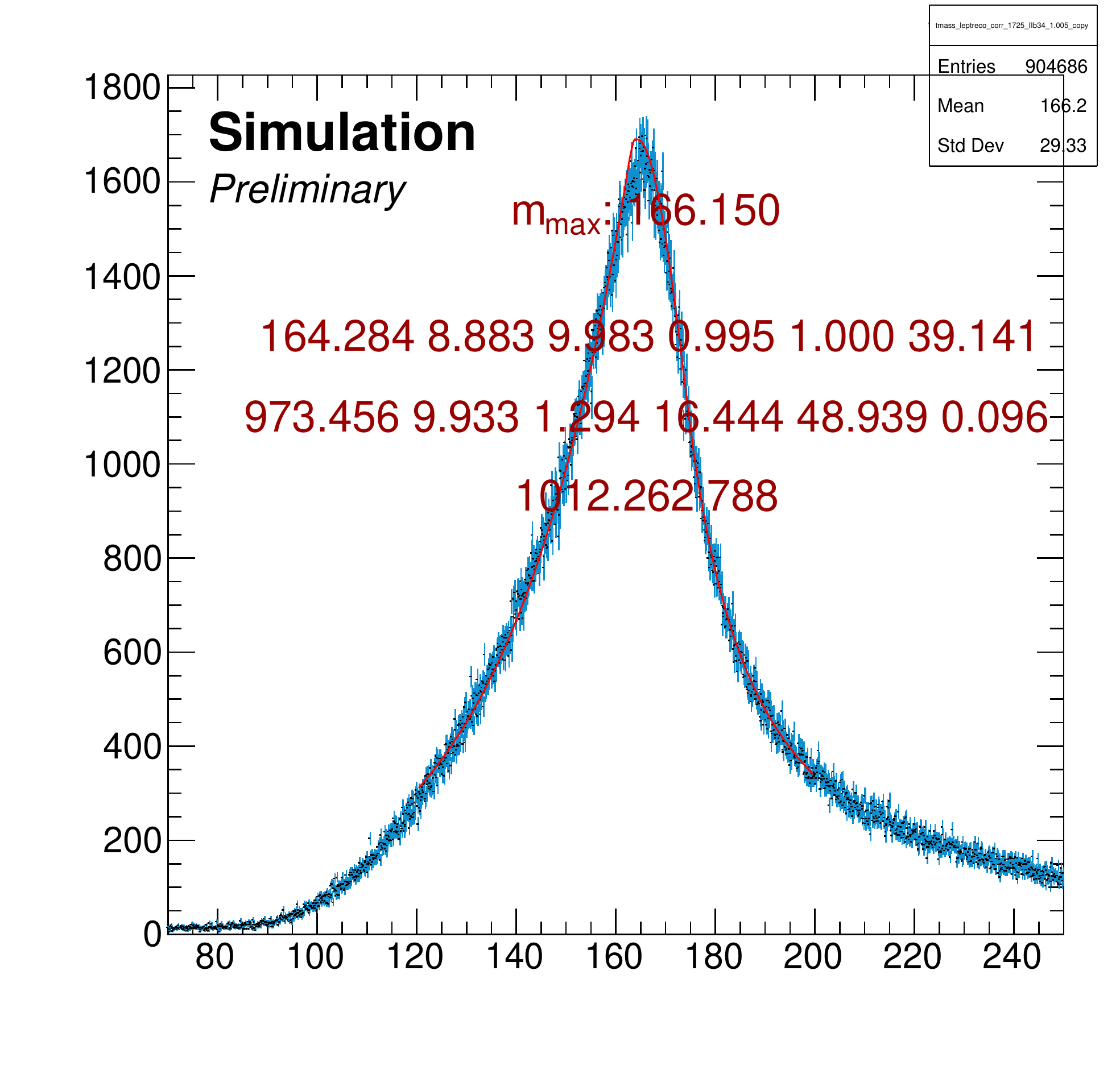}
\includegraphics[width=0.195\textwidth]{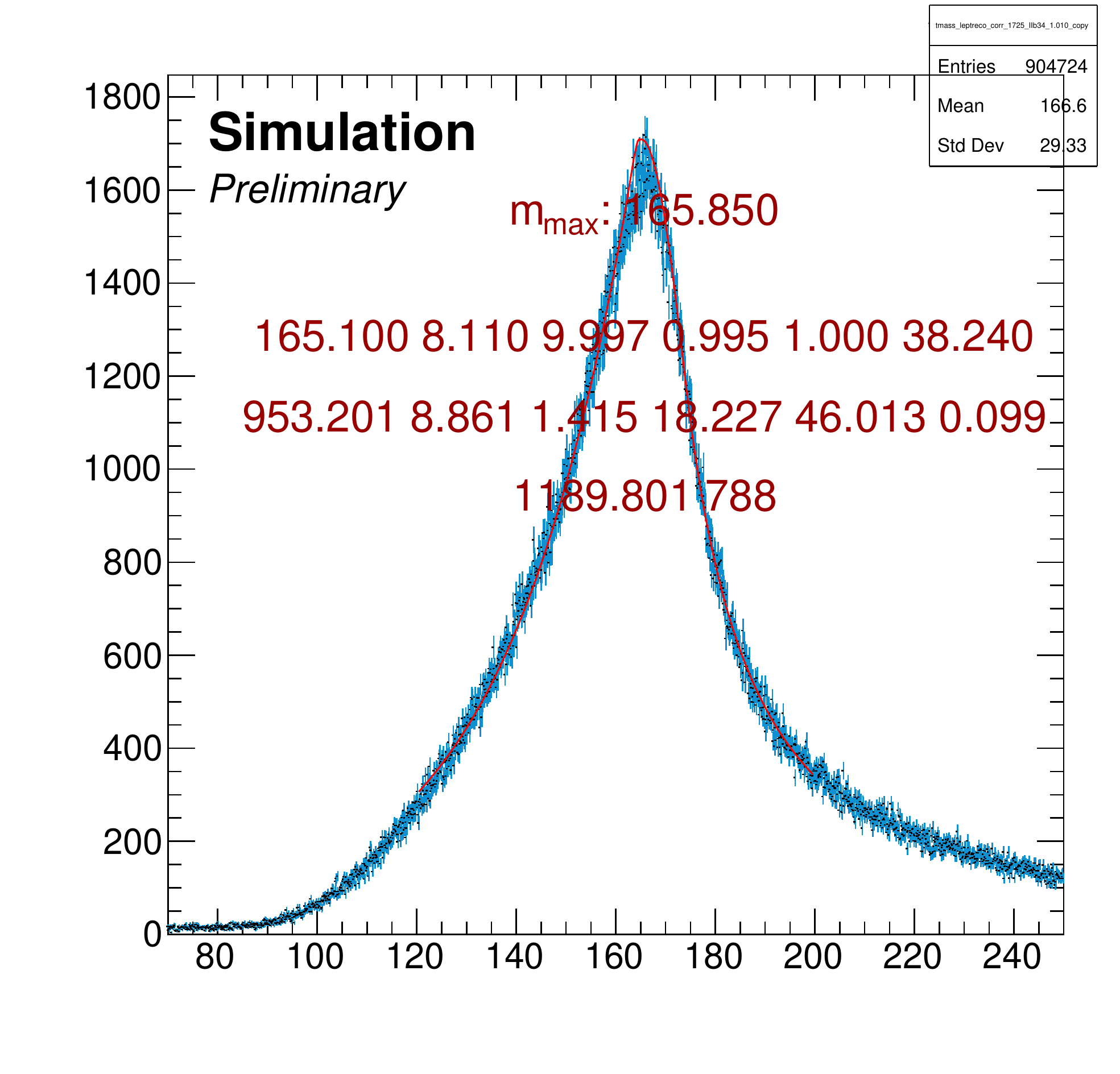}
\includegraphics[width=0.195\textwidth]{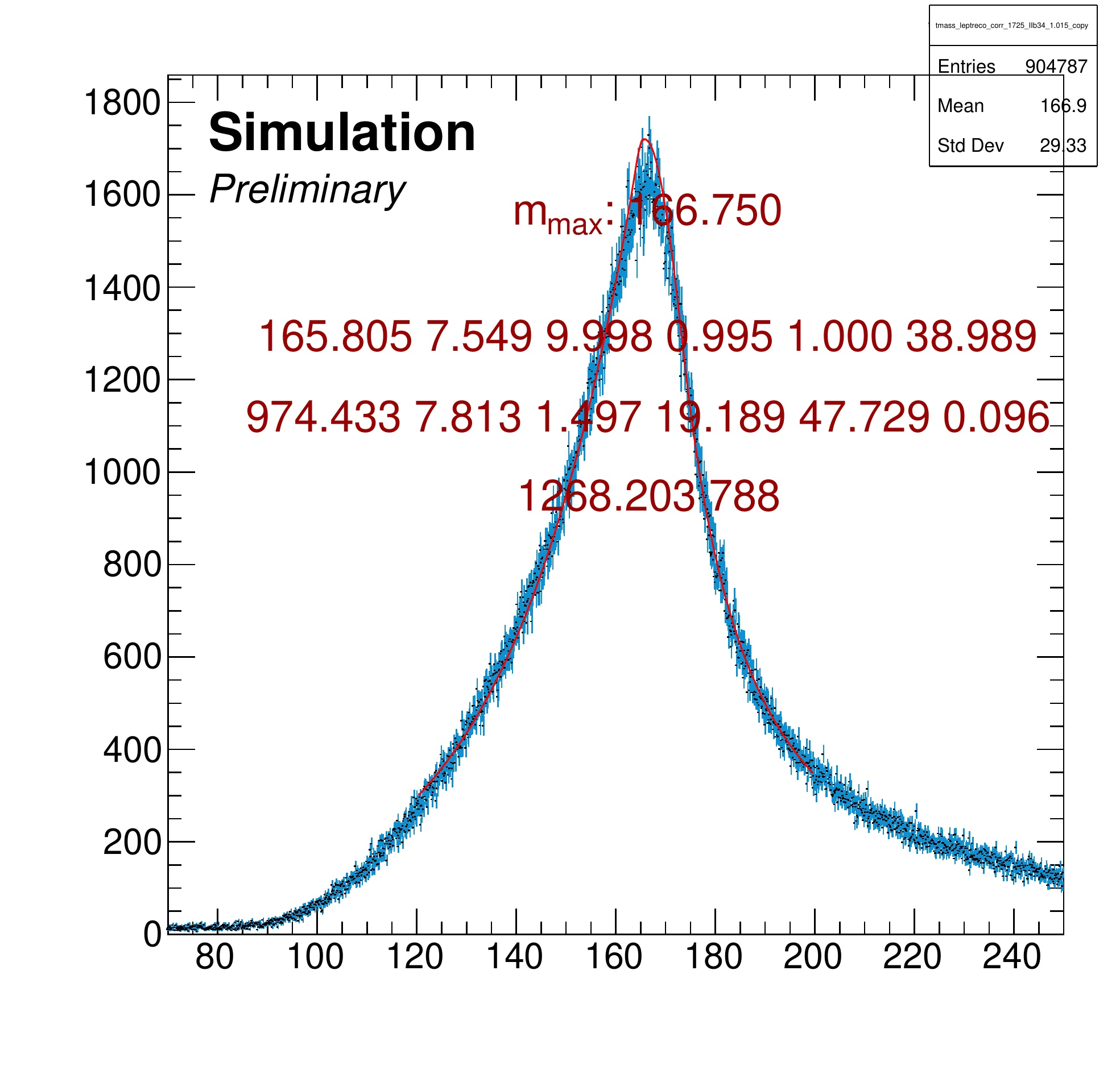}
\includegraphics[width=0.195\textwidth]{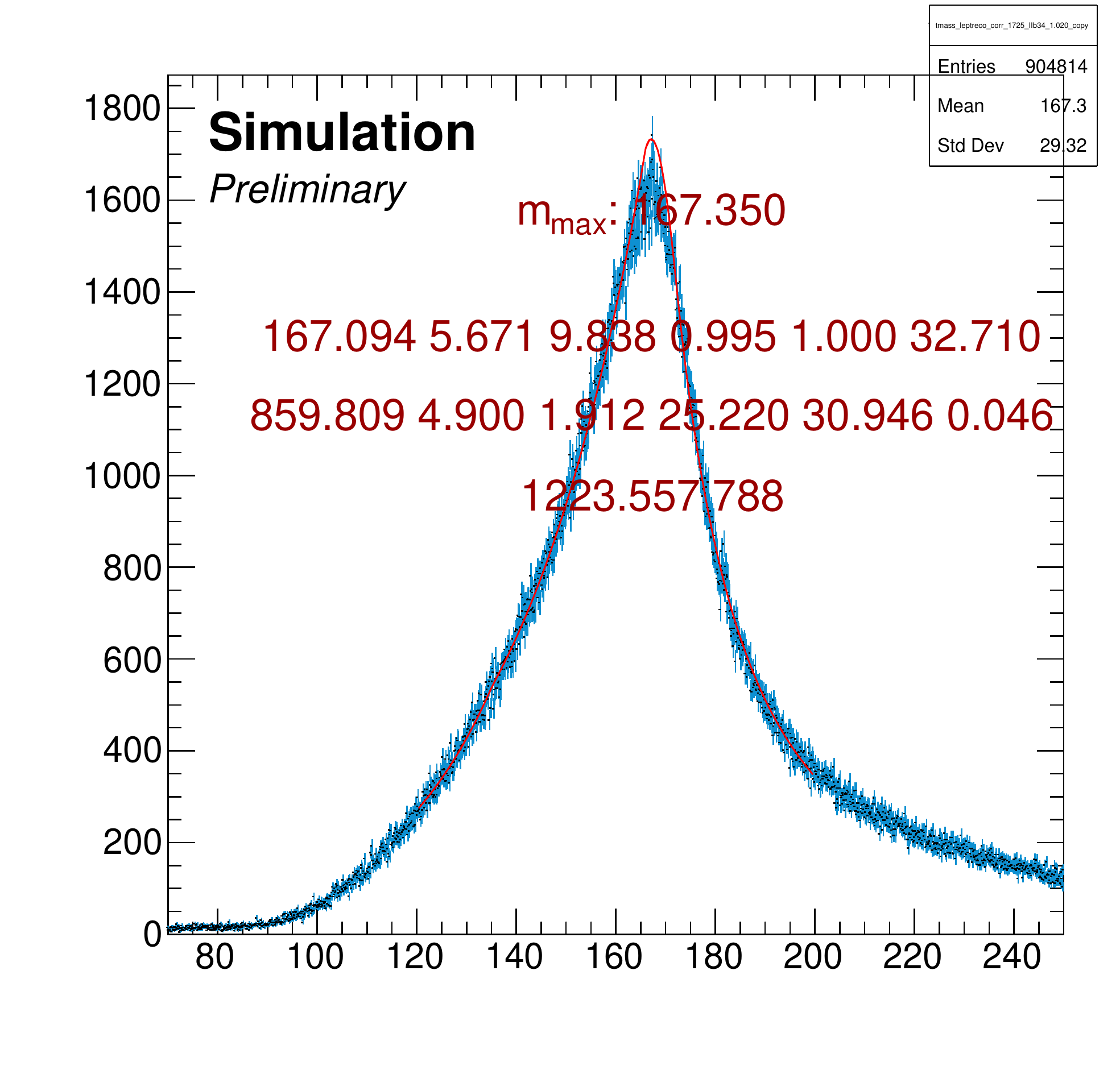}
\includegraphics[width=0.195\textwidth]{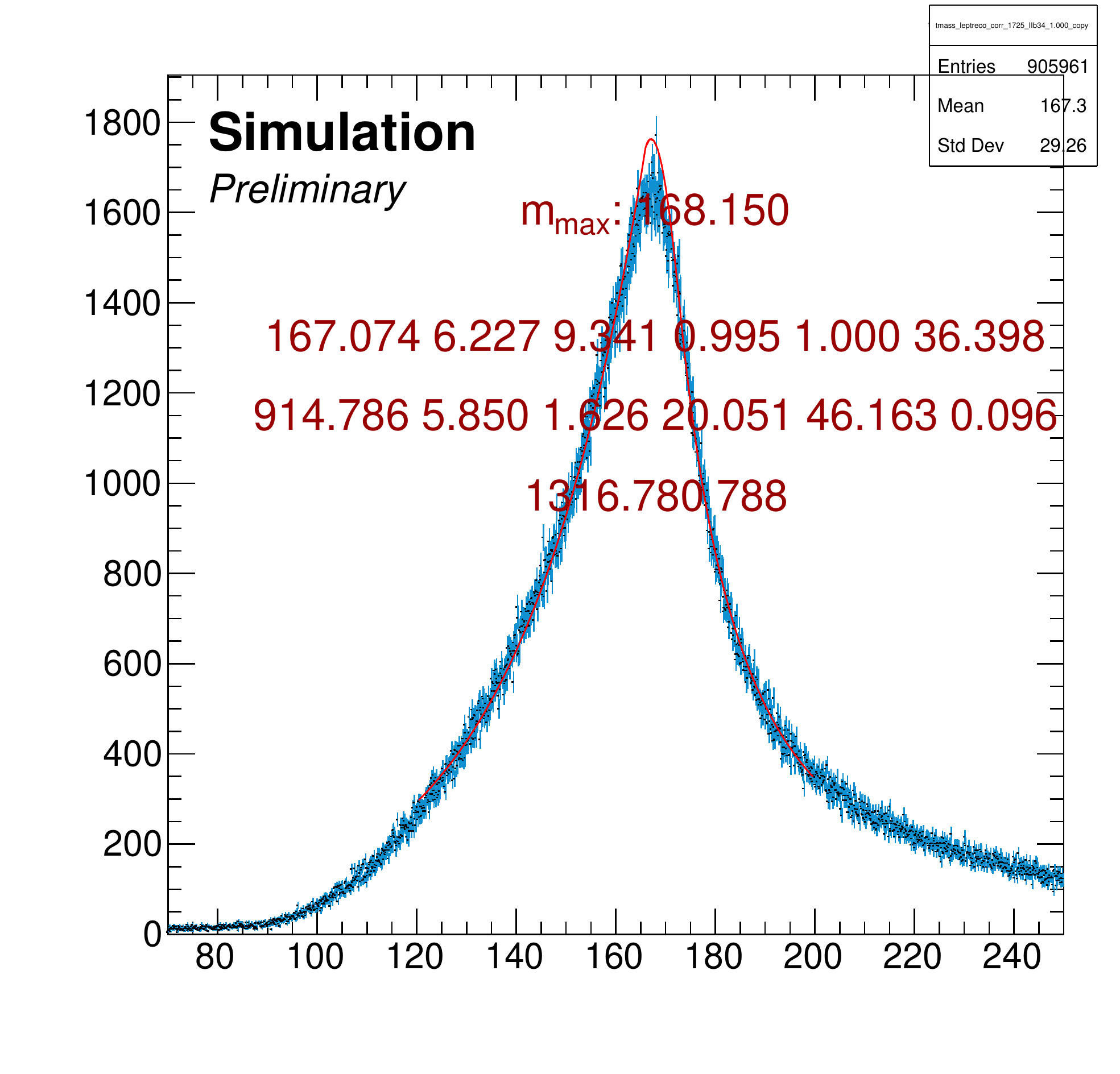}
\includegraphics[width=0.195\textwidth]{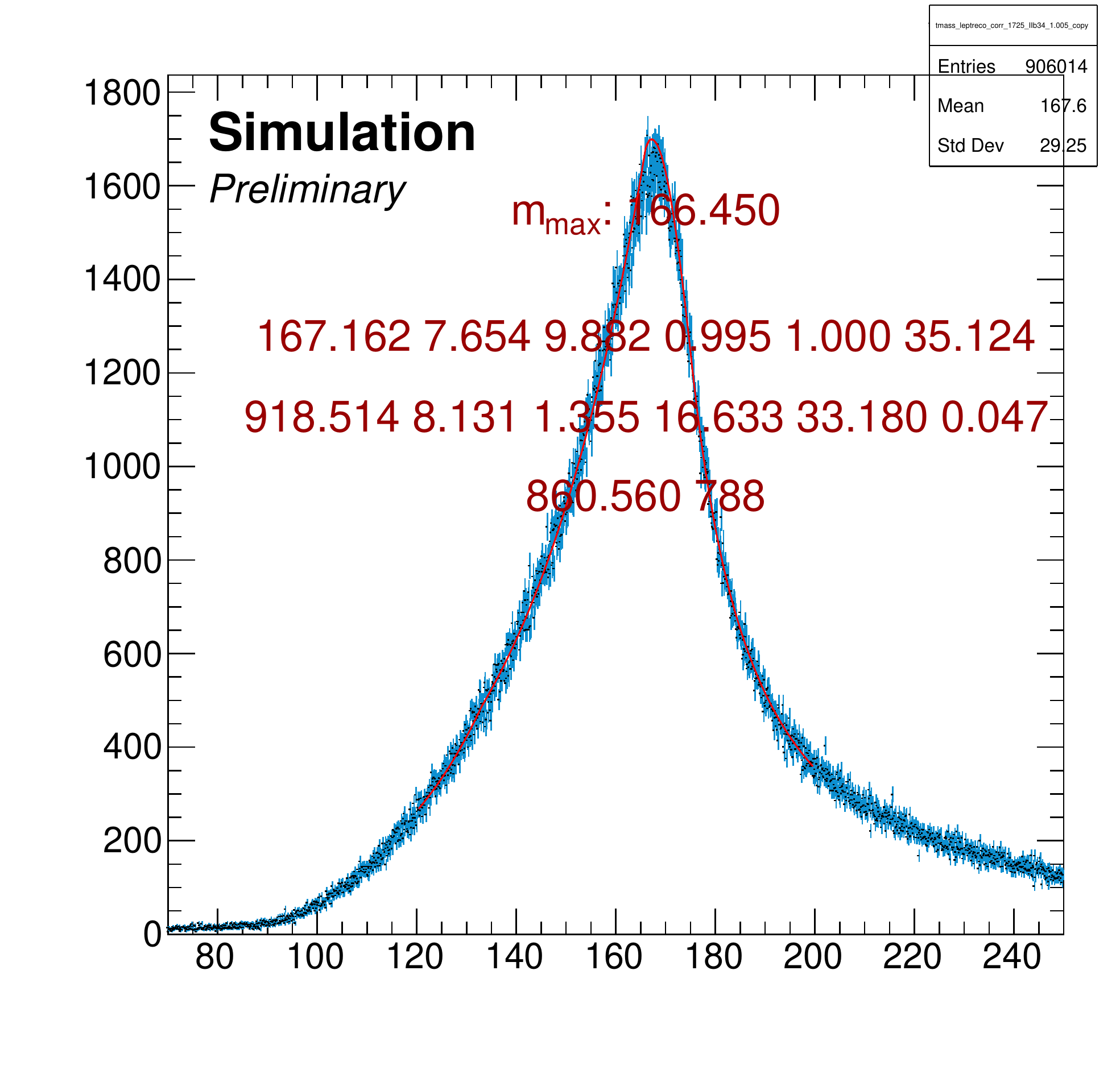}
\includegraphics[width=0.195\textwidth]{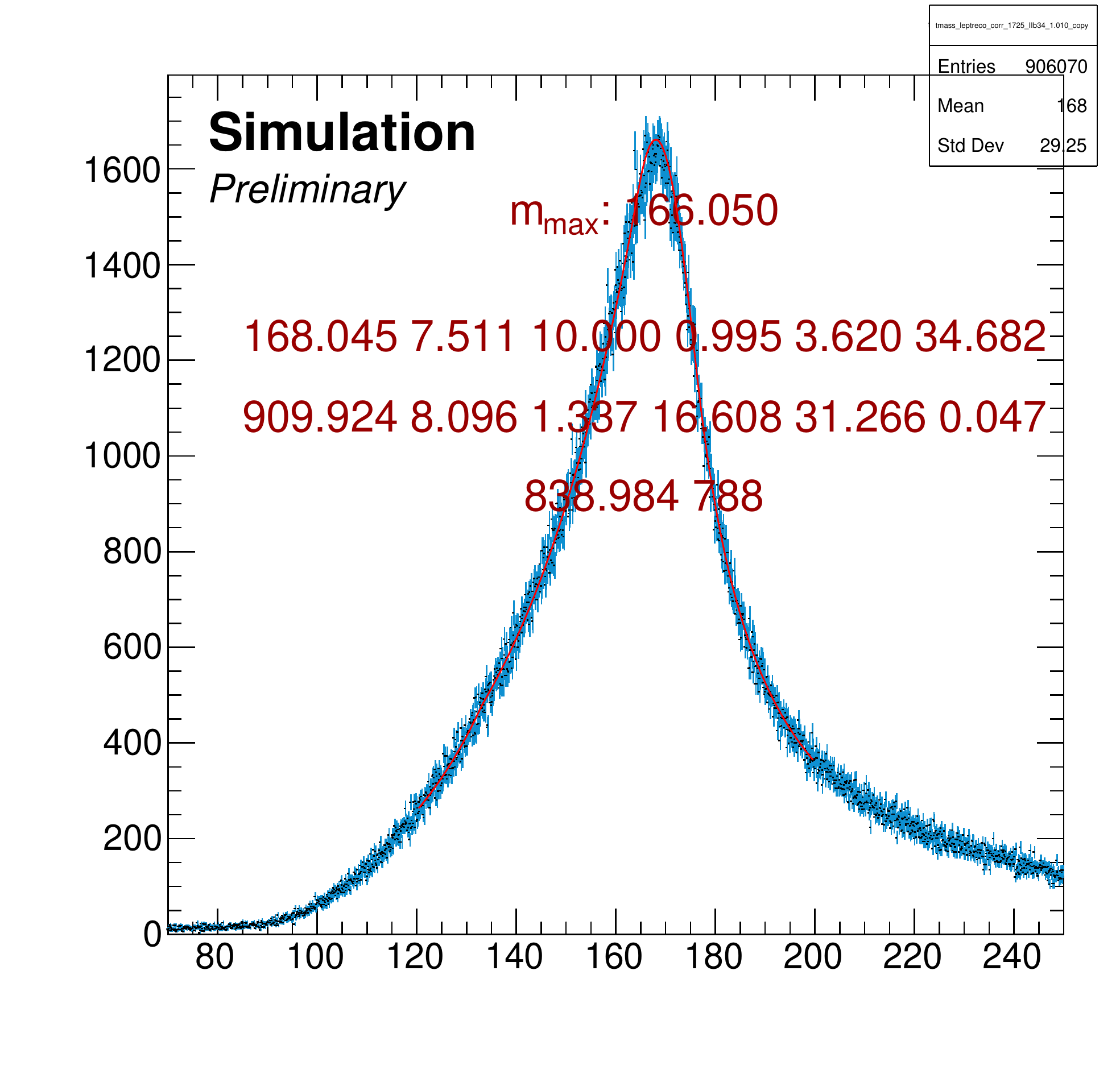}
\includegraphics[width=0.195\textwidth]{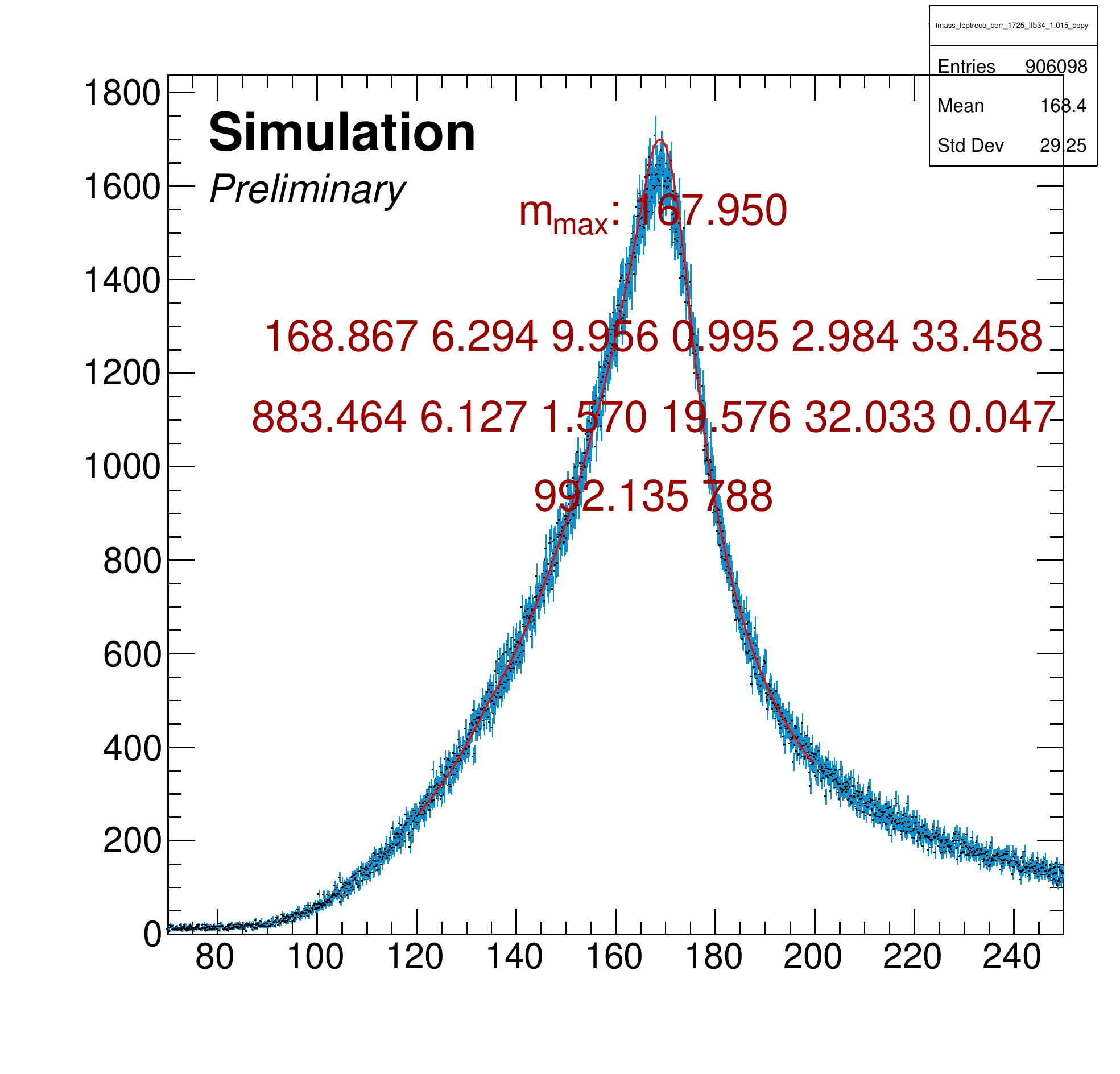}
\includegraphics[width=0.195\textwidth]{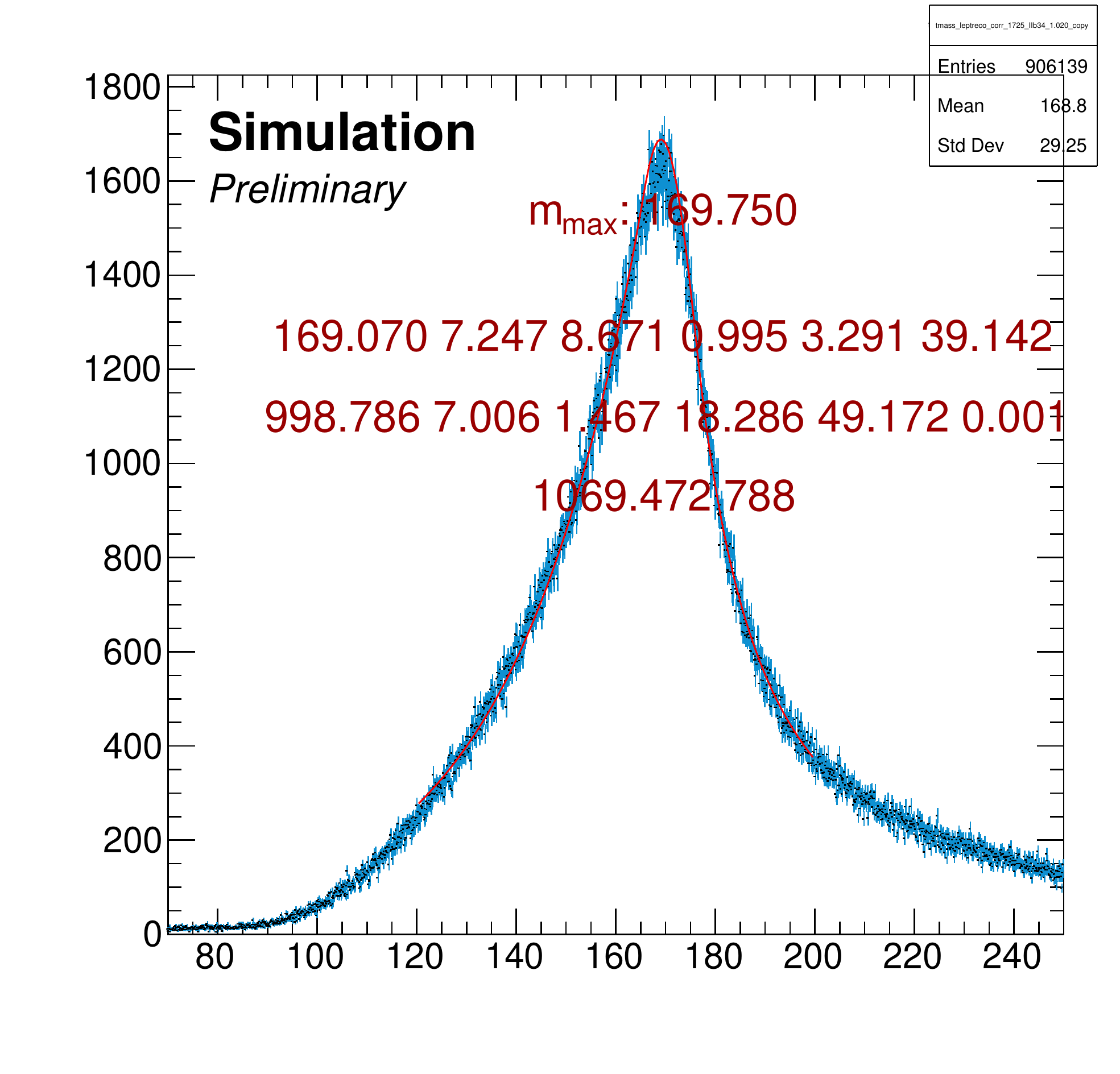}
\caption{Reconstructed leptonic top resonances in the muon channel at $m_t^{gen} = 172.5$~GeV with Run~IIb34 $F_{\text{Corr}}$ values and $K_{\text{JES}}^{\text{Res}} = 1.0,1.005,1.01,1.015,1.02$ from left to right.
Upper row displays original D\O\ $F_{\text{Corr}}$ parameters, lower row the re-calibrated ones.}
\label{fig:kjeslts}
\end{figure}

In Fig.~\ref{fig:genlts} generator-level reconstructed leptonic top resonances are shown at different $m_t^{gen}$ values.
In Fig.~\ref{fig:kjeslts} the $K_{\text{JES}}^{\text{Res}}$ dependence of the reconstructed leptonic top resonance is shown in specific example cases.
The red numbers show the maximum position, some fit parameters and fit statistics.

\end{document}